%% file: PhD_thesis.tex
\documentclass[openright,
11pt, 
oneside, 
english, 
singlespacing, 
headsepline, 
]{MastersDoctoralThesis} 
\usepackage[T1]{fontenc} 
\usepackage{amsmath}
\usepackage{tikz}
\usepackage{verbatim}
\usepackage{revsymb}

\usepackage{color, colortbl}
\definecolor{LightCyan}{rgb}{0.88,1,1}
\usepackage{enumitem}
\usepackage{textcomp}
\usepackage[autostyle=true]{csquotes} 
\usepackage[outdir=,suffix=]{epstopdf}
\usepackage{floatrow}
\usepackage{blindtext}
\usepackage{float}
\usepackage{multirow}
\usepackage{longtable}
\usepackage{supertabular}
\usepackage{rotating}
\usepackage{bbold}
\usepackage{bm}
\usepackage{cite}
\allowdisplaybreaks[1]

\usepackage{afterpage}
\newcommand\blankpage{%
  \null
  \thispagestyle{empty}%
      \newpage}

\def\beq{\begin{equation}}
\def\eeq{\end{equation}}

\newcommand{\xx}{\boldsymbol{x}}
\newcommand{\br}{\boldsymbol{r}}
\newcommand{\brr}{\boldsymbol{R}}

\newcommand{\hxx}{\hat{x}}

\newcommand{\ph}{\varphi}

\newcommand{\bmq}{\boldsymbol{q}}

\newcommand{\al}{\alpha}
\newcommand{\be}{\beta}
\newcommand{\gm}{\gamma}
\newcommand{\cali}{{\cal I}}
\newcommand{\sg}{\sigma}
\newcommand{\tp}{\tau}
\newcommand{\bra}{\langle}
\newcommand{\ket}{\rangle}
\newcommand{\Het}{{}^3{\rm He}}
\newcommand{\Ht}{{}^3{\rm H}}
\newcommand{\Heq}{{}^4{\rm He}}
\newcommand{\Li}{{}^6{\rm Li}}
\newcommand{\Be}{{}^7{\rm Be}}
\newcommand{\Lis}{{}^7{\rm Li}}
\newcommand{\kb}{{\overline{K}}}
\newcommand{\lb}{{\overline{L}}}
\newcommand{\sbb}{{\overline{S}}}
\newcommand{\tb}{{\overline{T}}}
\newcommand{\bb}{{\overline{\beta}}}
\newcommand{\lbb}{{\overline{l}}}
\newcommand{\phz}{\phantom{0}}
\def\m{\phantom{-}}

\thesistitle{Theoretical calculation of nuclear reactions of interest for
Big Bang Nucleosynthesis} 
\supervisor{Prof. Laura Elisa Marcucci\\
Prof. Michele Viviani} 
\tutor{}
\examiner{} 
\degree{Doctor of Philosophy} 
\author{Alex Gnech} 
\addresses{} 

\subject{Astroparticle Physics} 
\university{\href{http://www.gssi.it}{Gran Sasso Science Institute}} 
\department{\href{http://department.university.com}{Department or School Name}} 
\group{\href{http://researchgroup.university.com}{Research Group Name}} 
\faculty{\href{http://faculty.university.com}{Faculty Name}} 

\AtBeginDocument{
\hypersetup{pdftitle=\ttitle} 
\hypersetup{pdfauthor=\authorname} 
\hypersetup{pdfkeywords=\keywordnames} 
}

\begin{document}

\frontmatter 

\pagestyle{plain} 

\include{TitlePage}
\blankpage

\cleardoublepage







\include{Acknowledgments}

\blankpage
\cleardoublepage
%


\begin{abbreviations}{ll} 

  $\chi$EFT & Chiral Effective Field Theory \\
  ANC    & Asymptotic Normalization Coefficient\\
  AV18   & Argonne $v_{18}$ (potential)\\
  BBN    & Big Bang Nucleosynthesis \\
  BR     & Branching Ratio\\
  BSM    & Beyond Standard Model\\
  CD-B2k & Charge Dependent-Bonn2000 (potential)\\
  CFF    & Cluster Form Factor\\
  c.m.   & Center of Mass\\
  FES    & First Excited State\\
  GFMC   & Green's Function Monte Carlo\\
  GS     & Ground State\\
  HH     & Hyperspherical Harmonic\\
  LEC    & Low-Energy Constant\\
  LO     & Leading-Order \\
  N2LO/NNLO & Next-to-Next-to-Leading-Order\\
  N3LO   & Next-to-Next-to-Next-to-Leading-Order\\
  N4LO   & Next-to-Next-to-Next-to-Next-to-Leading-Order\\
  NCSM   & No-Core-Shell-Model\\
  NLO    & Next-to-Leading-Order\\
  NN     & Nucleon-Nucleon\\
  NSHH   & Non-Symmetrized Hyperspherical Harmonic\\
  QCD    & Quantum Chromodynamics \\
  RGM    & Resonanting Group Model \\
  SBBN   & Standard Big Bang Nucleosynthesis \\
  SM     & Standard Model \\
  SRG    & Similarity Renormalization Group\\
  TC     & Transformation Coefficient \\
  UIX    & Urbana IX (three-body potential)\\

\end{abbreviations}
\clearpage
\null
\thispagestyle{empty}
\newpage

\tableofcontents







%
%
%
%



\mainmatter 

\pagestyle{thesis} 
\cleardoublepage
\addtocontents{toc}{\setcounter{tocdepth}{2}}
\include{Introduction_v3}
\addtocontents{toc}{\setcounter{tocdepth}{2}}
\include{Chapter1_v3}

\addtocontents{toc}{\setcounter{tocdepth}{2}}
\include{Chapter2_v3} 
\addtocontents{toc}{\setcounter{tocdepth}{2}}
\include{Chapter3_v3}

\addtocontents{toc}{\setcounter{tocdepth}{2}}
\blankpage
\include{Chapter4}

\addtocontents{toc}{\setcounter{tocdepth}{2}}
\include{Conclusions_v2}
\cleardoublepage
 \phantomsection



\cleardoublepage

\appendix 
\input{trans_coeff}

\input{trans_HH}
\input{matrix_element}
\input{Lanczos}
\clearpage
\blankpage
\input{validation}

\input{em_calculation}
\clearpage
\blankpage
\input{app_cff}
\input{app_cc}
\input{cs_formula}
\clearpage
\blankpage


\bibliography{bibliography.bib}
\addcontentsline{toc}{chapter}{Bibliography}
\bibliographystyle{unsrt}

\end{document}

%% file: TitlePage.tex


\newcommand{\thetitle}{Theoretical calculation of nuclear reactions of interest for Big Bang Nucleosynthesis}

\newcommand{\candidatename}{Alex Gnech}

\newcommand{\dateofsubmission}{April 13, 2020}


	\thispagestyle{empty}
	\enlargethispage{6em}

	\vspace*{-2em}

	\begin{center} %
  		%
	    \begin{minipage}{0.75\textwidth}
  			\vspace{-36pt}
  			
  			\begin{center}
  				\textsc{\textbf{Gran Sasso Science Institute}} \par
	  			\vspace{-6pt}
  			
  				\rule{.7\textwidth}{0.1ex}
  				\par
  				\vspace{0pt}
  				
				\textsc{\small\textbf{School of advanced studies}}\par 
			\end{center}
		\end{minipage}
		%
	\end{center}

\vspace{\stretch{1.7}}

\begin{center} \huge \bf%
	\thetitle \par
\end{center}

\vspace{\stretch{1.0}}

\begin{center} %
  \Large Ph.D. thesis \\[2pt]
\end{center}

\vspace{-1.5em}

\begin{center}
  \large submitted \dateofsubmission \par
\end{center}

\vspace{-1.5em}
\vspace{\stretch{1.0}}

\begin{center}
  {\large \textbf{PhD Candidate}} \\[.5em]
  {\Large \candidatename}
\end{center}

\vspace{\stretch{1.2}}

\hspace{2em}
\begin{minipage}{0.5\textwidth}
{\large \textbf{Advisors}}
\vspace{1em}

{\large Prof. Laura Elisa Marcucci} \\
{\small Universit\`a di Pisa}\\
{\small INFN-Pisa}

\vspace{.7em}
{\large Prof. Michele Viviani} \\
{\small INFN-Pisa}
\end{minipage}
\hspace{1em}
\hspace{1.2cm}\begin{minipage}{0.4\textwidth}
\vspace{1em}

\vspace{3.4em}
\vspace{4.5em}
\end{minipage}

\vspace{\stretch{0.4}}

\begin{center}
  { \large 
  	\textbf{Gran Sasso Science Institute} \\[.5em]
  	XXXII Cycle -- A.Y. 2016-2019
  }
\end{center}


%% file: Acknowledgments.tex
\chapter*{Acknowledgements}

  This work would have not be possible without the support of the INFN-Pisa
  that hosted me during my visit in Pisa to my Advisors. In particular I want to
  thank Prof. A.\ Kievsky for the support.

  The {\it ab-initio} calculation of the 
  $\Li$ would have not be possible without the
  computing resources provided by CINECA. Moreover, I acknowledge useful discussion with
  P. Navr\'atil who provided me detailed informations on NCSM results on $\Li$.
  
  As regarding the work on the $p+\Li$ radiative capture,
  I am grateful to all LUNA Collaboration for their hospitality during
  the meetings and the data taking of the $p+\Li$ radiative capture.
  In particular, I would like to thank R.\ Depalo, L.\ Csedreki, D.\ Piatti and
  T.\ Chillery for comments and
  the useful discussion about the role of the photon angular distribution
  in the data analysis.
  I also acknowledge useful discussions with R.J. deBoer,
  who suggested to perform the theoretical investigation
  of the angular distribution.

  A personal thanks goes to my Advisors Laura and Michele that
  supported me both form the scientific and the human side
  during all the PhD period, keeping the
  doors always open.

%% file: Introduction_v3.tex
\chapter{Introduction}\label{ch:intro}
  Big Bang Nucleosynthesis (BBN) predicts the abundances
  of the light elements formed during the earliest time of the Universe
  from a fraction of a second to hundreds of seconds.
  By considering the physics of the Standard Model (SM),
  BBN is able to predict the abundances of $d$, $\Het$, $\Heq$ and
  $\Lis$ without free parameters.
  This theory, which is named Standard BBN (SBBN), 
  obtains a good overall agreement with the primordial abundances
  inferred from astrophysical observations.
  The agreement with the data
  for various nuclear species confirms SBBN strength and
  self-consistency, making it one of the most important
  direct cosmological test supporting the hot Big Bang model
  (for recent reviews see~\cite{Schramm1998,Iocco2009,Cyburt2016}).

  Even if SBBN is able to reproduce quite well the primordial
  abundances, still some inconsistency remain:
  the so called {\it lithium problem}.
  The observation of primordial lithium in metal-poor stars gives 
  an estimate of its abundance relative to the one of Hydrogen of  
  \begin{equation}
    {\rm Li}/{\rm H}|_{\text{exp}}=(1.6\pm0.3)\times 10^{-10}\,,
  \end{equation}
  where here we reported the experimental value of Ref.~\cite{Sbordone2010}.
  Other experiments performed on different stars and stellar systems
  give quite consistent results
  (see among the others~\cite{Bonifacio1997,Bonifacio2002,Melendez2010}).
  On the other hand, the most up-to-date
  BBN prediction of $\Lis$~\cite{Cyburt2008},
  obtained  using the baryon density as measured by
  the WMAP collaboration~\cite{WMAP2009}, is
  \begin{equation}
    {\rm Li}/{\rm H}|_{\text{th}}=(5.24^{+0.71}_{-0.62})\times 10^{-10}\,.
  \end{equation}
  It is clear, by comparing the two numbers, that there is a discrepancy
  between the BBN result and the astrophysical observations. This is the
  so called ``first Lithium problem''.

  Moreover, in 2006 there was a claim of detection
  of $\Li$ in old halo stars~\cite{Asplund2006}.
  In this study, the Authors performed 
  a high resolution observations of Li absorption lines and the abundance
  of $\Li$ was found to be
  \begin{equation}
    \Li/\Lis= 5\times 10^{-2}\,,
  \end{equation}
  more than two order of magnitude larger than SBBN predictions.
  This has been called ``second lithium problem''.
  Further observation and
  analysis~\cite{Cayrel2007,Garcia2009,Steffen2009,Lind2013}
  showed that  three-dimensional
  modeling of stellar atmosphere which considers
  non local thermodynamic and includes surface convection can
  generate asymmetries in the absorption line shape
  that mimic the presence of $\Li$. 
  Therefore, the claimed ratio $\Li/\Lis$ is best interpreted
  as an upper limit,
  as confirmed by recent data on extremely low metallicity
  stars~\cite{Gonzalez2015}. On the other hand, 
  the production of $\Li$ from spallation due to 
  cosmic rays resulted to be the main contribution to
  cosmological $\Li$ abundance (see e.g.~\cite{Knauth2017}).
  However, there is not a definite conclusion of this issue yet
  and open questions
  on $\Li$ primordial abundance persist.
  
  These discrepancies can have three possible solutions~\cite{Fields2011}:
  (1) the presence of systematic errors in the astrophysical observations,
  (2) the incomplete knowledge of one or more cross-sections of the
  BBN reaction network or (3) the emergence of new physics beyond
  SM (BSM).
  Even if, as in the case of $\Li$, the first solution is the most probable,
  the situation remains still unclear and in particular
  the last possibility results to be very intriguing.
  In fact, the existence of BSM relic particles (such as dark
  matter particles), which can decay or annihilate 
  during the BBN era, can give rise to a series of SM particles which can
  interact with the BBN network changing
  light element abundances and potentially solving the lithium problem(s)
  (see for example Ref.~\cite{Jedamzik2009} for a review).
  In this sense, the increasing precision
  on the input parameters of the SBBN suggests the possibility of using BBN
  for constraining a wide range of BSM theories 
  (see the reviews of Refs.~\cite{Iocco2009,Cyburt2016}).

  In order to test these exotic scenarios, a great accuracy in the BBN
  input parameters is needed, in particular in the cross-section
  of the nuclear reactions involved in the BBN network
  (or the {\it S}-factor which is the cross-section net of 
  Coulomb barrier effect).
  In the last years, a large effort for measuring with high accuracy all the
  cross-sections of the BBN network in the BBN energy window (50-400 keV)
  was made by the experimental nuclear community.
  However, the experimental studies are very difficult, due
  to the Coulomb barrier which suppresses the cross-sections
  in the energy window of interest.
  This translates in large errors on the experimental data.
  Therefore, 
  reliable theoretical calculations result fundamental in order to reduce
  the uncertainties. 

  In this Thesis we present the theoretical study of two
  nuclear reactions that  are connected to 
  $\Li$ abundance in BBN. They are
  \begin{equation}
    \alpha+d\rightarrow\Li+\gamma\,,\label{eq:adreac}
  \end{equation}
  and 
  \begin{equation}
    {p}+\Li\rightarrow\Be+\gamma\,.\label{eq:plireac}
  \end{equation}
  The study of these reactions is also motivated by the fact that both
  were object of a recent experimental campaign at the
  deep underground facility 
  LUNA, at the Gran Sasso National Laboratory~\cite{LUNA2010}.
  In the following of this Chapter we discuss separately the
  two reactions, their connection to BBN
  and the theoretical approaches used for the calculation.
  
\section{The $\alpha+d\rightarrow \Li+\gamma$ reaction}

The $\alpha+d\rightarrow \Li+\gamma$ is considered the main reaction through
which  $\Li$ is produced during BBN~\cite{Cyburt2016}.
Therefore, it plays a fundamental
role in determining $\Li$ primordial abundance. Its role is even more
important in the case of non-standard BBN, in which there is the
possibility of having non-thermal deuteron which can interact with the $\Heq$
environment~\cite{Fields2011}. For this reason the study of the $S$-factor
in the BBN energy window becomes important.

The experimental study of this reaction is very complex not only
because of the exponential drop due to the Coulomb barrier, but also
because of the isotopic suppression of the multipole $E1$,
which is usually the main
contributor to the radiative capture reaction cross-sections.
For this reason, the experiments, which started in the 1980s with
Robertson {\it et al.}~\cite{Robertson1981}
and then continued in the 1990s~\cite{Kiener1991,Mohr1994,Cecil1996,Igamov2000},
were able only to give extrapolation or upper limits in the BBN
energy window and not to obtain direct data.
Recently, the LUNA Collaboration
started a campaign of measurements of this reaction, obtaining for the first
time some data points in the BBN window with a direct
method~\cite{LUNA2014,Trezzi2017}.

From the theoretical point of view, the study of this reaction, due to the
complexity of the problem, has been performed
using mainly cluster models
~\cite{Hammache2010,Tursunov2015,Mukh2011,Dubovichenko1998,Grassi2017},
in which the $\alpha$ particle
and the deuteron are considered as structureless particles and their interaction
is described by an {\it ad-hoc} nuclear potential. In such a way the problem
reduces to solve a simple two-body equation. Even if these models are able
to reproduce the $S$-factor behavior, the internal structure of the $d$, $\alpha$ and
$\Li$ nuclei is completely missing. Moreover, in this approach the theoretical
uncertainties are not fully controlled and only the
Authors of Ref.~\cite{Grassi2017}
give an estimate of it by comparing various potential models.
In order to recover at least the structure of $\Li$, Tursunov {\it et al.}
~\cite{Tursunov2018}
constructs the $\Li$ wave function in a three-body
model considering the lithium as an $\alpha+p+n$ system. This permitted to
have a more precise description of the final state of the reaction,
but still an
{\it ad-hoc} potential is used in order to reproduce the $\Li$ binding energy.
Furthermore, 
the scattering part is still treated considering the $\alpha+d$ as a two-body
problem. It is important to mention that the $\Li$ can be treated 
  as an {\it halo} nucleus, in which it is possible to separate
  the short-range scale of the $\alpha$ particle
  and the long-range scale of the neutron
  and proton which form the halo.
  With this assumption, it is possible to construct a phenomenological
  cluster interaction by using the so-called  halo effective field
  theory (see for example~\cite{Hammer2017}).
  The nuclear interactions constructed within this approach are
  more justifiable from a theoretical point of view  and permit also to have
  reliable estimates of the theoretical errors. However, to our knowledge,
  there are no published papers which treat the $\alpha+d$ reaction
  in such approach.

A complete study of the reaction would impose the solution of the six-body problem
both for the $\alpha+d$ scattering states and the $\Li$ bound state  
and their interaction with photon. Such approach is called an {\it ab-initio}
method.
In this approach the nuclear potential models are constructed starting from
fundamental theories (i.e. SM)  in order to reproduce exactly the interaction
among the nucleons.
 Historically, a series of phenomenological potentials
  based on the exchange of pions and heavier mesons and then fitted to reproduce
  the nucleon-nucleon (NN) data sets  have been used to study
  ground states and low-energy scattering states of light nuclei.
  Examples of these are the Argonne $v_{18}$
  (AV18)~\cite{Wiringa1995} or the charge dependent-Bonn2000 (CD-B2k)
  potentials~\cite{Machleidt2001}.
  These potentials are often used in combination with three-body forces based on the
  idea that a nucleon can be excited in a $\Delta$ particle (for example the
  Urbana IX (UIX)~\cite{Pudliner1997} potential). The use of three-body forces
  permits to reproduce the experimental energy spectra of the light nuclei.
  More recently, chiral effective field theory
  ($\chi$EFT) has provided a practical and successful scheme
to study two- and many-nucleon interactions~\cite{Epelbaum2015,Entem2015}. The
$\chi$EFT approach is based on the observation that the chiral
symmetry exhibited by quantum chromodynamics (QCD) has a noticeable impact in
the low-energy nuclear dynamics. Therefore,
the form of the strong interactions of pions among
themselves and with the nucleons is severely
constrained by the transformation properties
of the fundamental Lagrangian under chiral and
discrete symmetries~\cite{Weinberg1966,Weinberg1968,Callan1969,Weinberg1979}.
The Lagrangian terms can be
organized as an expansion in powers of $Q/\Lambda_\chi$ , where  $\Lambda_\chi\simeq1$
GeV specifies the chiral-symmetry-breaking scale and $Q$ is
the exchanged pion momentum. This scheme permits to have a control of the
theoretical error induced by the truncation of the expansion
(see for example~\cite{Epelbaum2015}).
Each Lagrangian term is associated to a
low-energy constant (LEC) which is usually determined by
fitting two- and three-body experimental data.
The nuclear potentials are derived from the $\chi$EFT Lagrangian by using
using either the time ordering perturbation theory (see for example~\cite{Baroni2016})
or the unitary transformation techniques (see for example~\cite{Epelbaum2007}).
The phenomenological and the chiral nuclear potentials may be then
used within the many-body Schr\"odinger equation
to predict other nuclear observables without any new free parameter.
Therefore, all the results obtained by using {\it ab-initio} techniques
are independent
on the experimental data that one is interested in.

The first attempt to perform an {\it ab-initio} study of 
the $\alpha+d$ radiative capture reaction was performed by Nollett
{\it et al.}~\cite{Nollet2001}
by using the Green's Function Monte Carlo (GFMC) method
(see Ref.~\cite{Carlson2015} and references therein)
with the AV18/UIX nuclear potential.
In that work, the $\Li$ bound state is described as a six-body wave function.
However, the  $\alpha+d$ scattering states are still solved in
a two-body framework.
In a more recent work, Hupin {\it et al.}~\cite{Hupin2015} have performed
for the first time
a fully {\it ab-initio} study of both the bound and the elastic
scattering states within the No-Core-Shell-Model (NCSM)
framework~\cite{Barrett2013}, using a chiral interaction. Those Authors have been able
to compute the scattering states by combining the Resonanting Group Model (RGM)
with the NCSM~\cite{Navratil2011}. In this approach,
the $A=6$ bound and scattering wave function have been 
expanded in terms of the eigenstate of the two-body and four-body Hamiltonian.
In such a way, they have been able to obtain a series of coupled channel equations that,
solved for positive energies, permitted to obtain the exact asymptotic behavior
of the scattering wave functions.
With this approach,
the Authors have been able to nicely reproduce the $\alpha+d$ elastic
scattering data. However, the calculation of the $S$-factor for the $\alpha+d$
radiative capture is still missing.

In this Thesis we make the first steps towards the calculation
of the $\alpha+d\rightarrow\Li+\gamma$ reaction by using  
another {\it ab-initio} approach, the Hyperspherical Harmonic (HH) method.
This method  was successfully used with $A\leq 4$ both for scattering
and bound states~\cite{Kievsky2008}
and it is feasible to be extended to larger nuclear systems.
In the HH method, the wave function
is expanded in HH functions and the quantum mechanical problem is then solved
using the the Rayleigh-Ritz variational principle
for the bound states and the Kohn variational principle~\cite{Khon1948}
for the scattering states.
The idea to use this approach for studying this reaction is very intriguing
since, compared to other {\it ab-initio} method,
the HH technique seems to be the best choice for studying low-energy
scattering states for nuclear reactions of astrophysical interest
~\cite{Marcucci2016,Marcucci2006}.

One of the main problems when we consider variational approaches in nuclear physics
is the enormous dimension of the basis needed to reach convergence in the expansion of the
wave function. In particular,
passing from $A=4$ to $A=6$, the dimension of the basis
grows exponentially, making hard the practical use of the HH expansion.
Therefore, one aim of this Thesis is to introduce some technical improvements
on the numerical implementation allowing for
the extension of the HH approach.
After that, we will focus on the calculation of the ground state of $\Li$
which is the final state of the reaction. The importance of this
nucleus goes beyond its relevance in the BBN. In fact, it is the first
stable nucleus beyond the $A=5$ mass gap and it is weakly bound.
Therefore, it is a good laboratory for studying nuclear force, since
its exotic structure must emerge directly from the nuclear force itself.
An important intermediate goal of this Thesis will be the study of
the Asymptotic Normalization Coefficient (ANC).
The ANC is the normalization of the tail of the $\Li$ wave function when
projected on the $\alpha+d$ cluster, and it can be extracted experimentally.
Since in the BBN energy window the nuclear reactions are almost peripheral,
only the tail of the wave functions is involved and
so the {\it S}-factor is  proportional to the square of the ANC.
It is clear then that the knowledge of the ANC is a fundamental ingredient
to determine the {\it S}-factor.
The calculation of this observable is very complex since
we need to construct in the same {\it ab-initio} framework the $\alpha$ and the $d$
wave function combining it with the $\Li$ one.
Moreover, the computation of the ANC has, from the technical point of view, a lot
in common with the calculation of the scattering states. Therefore,
it can be considered
the first step towards the construction of the HH wave function
of the $\alpha+d$ scattering states.

\section{The $p+\Li\rightarrow \Be+\gamma$ reaction}

The $p+\Li\rightarrow \Be+\gamma$ reaction is not considered to play
a direct role in the BBN since the small amount of $\Li$ present
in the early Universe environment.
However, its cross-section is
directly connected to the $\Be+\gamma\rightarrow p+\Li$ reaction,
through the detailed balance principle. The latter reaction
is supposed to become an
important contributor to $\Li$ abundance in case of presence of
high energy non-thermal photons generated by the decay of BSM particles
during the BBN~\cite{Kusukabe2006}.

The $p+\Li\rightarrow \Be+\gamma$ cross-section was extensively studied
in the past
by many experimental
groups~\cite{Switkowski1979,Tingwell1987,Cecil1992,Prior2004,He2013}.
However, large uncertainties in the {\it S}-factor at the BBN energies remain.
Furthermore, a recent work~\cite{He2013}
has pointed out the possible presence of a resonance
in the BBN energy window, with subsequent suppression at zero energy.
In order to confirm or reject such possibility,
the LUNA Collaboration has also performed
a new campaign of measurements in the Spring of 2018.

The extrapolation of the astrophysical {\it S}-factor at zero-energy
has been performed 
within the R-matrix approach in Ref.~\cite{Igamov2016}, including
somewhat by hand 
the resonance-like structure proposed in Ref.~\cite{He2013}.
On the other hand, all theoretical calculations performed within
the cluster model framework
do not reproduce the claimed resonance. The most important
theoretical studies
were performed using different approaches,
like a two-body phenomenological potential~\cite{Huang2010,Dubovichenko2011},
an optical potential~\cite{Baker1980},
a four-cluster model~\cite{Arai2002} and  the Gamow shell model~\cite{Dong2017},
obtaining all quite consistent results with each other.
However, all these studies are lacking of an estimate of the
theoretical uncertainty, especially that arising from
model dependence.

In this Thesis, we present a new theoretical study
within a cluster model of the $p+\Li\rightarrow\Be+\gamma$, using
also a two-body phenomenological potential similar to that
of Ref.~\cite{Dubovichenko2011},
but calculating not only
the astrophysical {\it S}-factor,
but also the angular distribution
of the emitted photon, for which there are also available
data~\cite{Tingwell1987}.
The main goal of this work is to give to the LUNA Collaboration,
which is finalizing the data analysis on this reaction, a reliable theoretical
calculation of the angular distribution which enters in the
determination of the efficiency of the detector.
Moreover, we also estimate a ``theoretical'' error bar, based on the different
cluster approaches presented in literature. 
Finally, we investigate on the possible presence of the
resonance structure as suggested by the data of Ref.~\cite{He2013}.

\bigskip
This thesis is divided in two parts. In the first part we discuss the
first steps towards the {\it ab-initio}
calculation of the $\alpha+d$ radiative capture.
In particular in Chapter~\ref{ch:HH} we present
the HH formalism for $A=6$ focusing on the solution of the bound state
problem. In Chapter~\ref{ch:litio} we discuss the results obtained for
the $\Li$ ground state. Moreover, in this Chapter we analyze also the
electromagnetic properties of $\Li$, namely its charge radius, its magnetic dipole
and its electric quadrupole. Finally, in Chapter~\ref{ch:ANC}
we study the $\alpha+d$ clusterization of  $\Li$ in order to
determine the ANC.
The second part of the Thesis is instead
dedicated to the phenomenological study of the $p+\Li\rightarrow \Be+\gamma$
reaction.  This study, published in Ref.~\cite{Gnech2019}, is presented in
Chapter~\ref{ch:pLi} of this Thesis.
Lastly, we conclude in Chapter~\ref{ch:conclusions}
by discussing possible future development
and applications of this work.
A series of technical details on the calculation and the implementation
of the HH method are given in the Appendices.

%% file: Chapter1_v3.tex
\chapter{The Hyperspherical Harmonic method for bound states}\label{ch:HH}

  In this chapter we  present the formalism needed to
  solve the Schr\"odinger equation for the bound state of $\Li$
  by using the HH functions.
  The HH method is a variational approach in which
  the wave function is expanded in terms of basis states constructed
  with the HH functions, transforming the problem of solving
  the Schr\"odinger equation in an eigenvalue problem.
  This chapter is organized as follows.
  We start by defining the Jacobi vectors and the hyperspherical variables
  in Section~\ref{sec:HH1} and by constructing the HH functions
  in Section~\ref{sec:HH2}. Then in Section~\ref{sec:HH3} we will define
  the basis states by including the spin and isospin quantum numbers.
  In Section~\ref{sec:HH4} we present the eigenvalue problem and we
   discuss the construction of the
  Hamiltonian matrix elements. The last section is dedicated to some
  technical details on the calculation of the potential matrix elements.
  
  In this Thesis we will consider also the wave function of ${}^4{\rm He}$.
  The construction of the ${}^4{\rm He}$ wave function follows
  the same procedure described in this chapter. However, specific
  details can be found in Ref.~\cite{Kievsky2008}.

\begin{section}{Jacobi coordinates and Hyperspherical variables}
  \label{sec:HH1}
  
  A nucleus can be considered as a generic isolated system of $A$ particles
  with spatial coordinates $\br_i$ and masses $m_i$, $i=1,\ldots ,A$. We want to
  study the system in the center-of-mass (c.m.) frame, by decoupling
  the c.m. motion in the Schr\"odinger equation. What remains are
  $N=A-1$ vectors which describe the internal spatial configurations. In order to do
  that, we introduce the Jacobi vectors $\xx_i$, $i=1,\ldots ,N$,
  which, by definition, are such that 
  the total kinetic energy operator can be written as
  \begin{equation}
    T = - \sum_{i=1}^A {\hbar^2\over 2m_i} \nabla^2_{i} =
    -{\hbar^2\over m} \sum_{i=1}^N \nabla^2_{\xx_i}
    -{\hbar^2\over 2M}  \nabla^2_{\bf X}\ ,
    \label{eq:lapl}
  \end{equation}
  where $m$ is a reference mass, $M=\sum_{i=1}^A m_i$ is the total mass
  of the system, and ${\bf X}= (1/M) \sum_{i=1}^A m_i\br_i$ is the
  c.m. coordinate. In this Thesis we consider all the
  particles to have equal mass $m=m_p$, being $m_p\simeq m_n$, where $m_p$ is
  the proton mass and $m_n$ the neutron mass. There are various
  possible choices of the Jacobi vectors. In particular we refer as
  ``standard'' set of Jacobi vectors to the following choice
  \beq
    \label{eq:jac1}
  \xx_{N-j+1} = \sqrt{2 j \over j+1 }
  \Bigl [\br_{j+1} - {\bf X}_j \Bigr ]\ ,
  \qquad j=1,\ldots,N\ ,
  \eeq
  with
  \beq
  {\bf X}_j=\frac{1}{j}\sum_{i=1}^j \br_i
  \eeq
  Explicitly, for $A=6$ this choice gives
  \begin{equation}
  \begin{aligned}
    \xx_{1p}&=\sqrt{\frac{5}{3}}\left(\br_n-\frac{\br_m+\br_l+\br_k+\br_j+\br_i}{5}
    \right)\\
    \xx_{2p}&=\sqrt{\frac{8}{5}}\left(\br_m-\frac{\br_l+\br_k+\br_j+\br_i}{4}
    \right)\\
    \xx_{3p}&=\sqrt{\frac{3}{2}}\left(\br_l-\frac{\br_k+\br_j+\br_i}{3}
    \right)\\
    \xx_{4p}&=\sqrt{\frac{4}{3}}\left(\br_k-\frac{\br_j+\br_i}{2}
    \right)\\
    \xx_{5p}&=\br_j-\br_i\,,\label{eq:jacvec}
  \end{aligned}
  \end{equation}
  where $(i,j,k,l,m,n)$ indicates a generic permutation $p$  of the particles.
  By definition the reference permutation $p=1$ is chosen to correspond to the order
  $(1,2,3,4,5,6)$ of the particles. There exists other possible choices for the
  Jacobi vectors. For example when 
  we want to describe the ${4+2}$ cluster structure of $\Li$, namely
  $\alpha+d$, a more suitable choice of the Jacobi coordinates is
  \begin{equation}
  \begin{aligned}
    \xx_{1Bp}&=\br_n-\br_m\\
    \xx_{2Bp}&=\sqrt{\frac{8}{3}}
    \left(\frac{\br_n+\br_m}{2}-\frac{\br_l+\br_k+\br_j+\br_i}{4}
    \right)\\
    \xx_{3Bp}&=\sqrt{\frac{3}{2}}\left(\br_l-\frac{\br_k+\br_j+\br_i}{3}
    \right)\\
    \xx_{4Bp}&=\sqrt{\frac{4}{3}}\left(\br_k-\frac{\br_j+\br_i}{2}
    \right)\\
    \xx_{5Bp}&=\br_j-\br_i\,,\label{eq:jacvecB}
  \end{aligned}
  \end{equation}
  which we name set $B$.
  The basis elements can be defined
  using any definition of the Jacobi coordinates. The completeness
  of the basis ensures that the expansions are completely equivalent
  using any set of Jacobi vectors. In numerical applications the expansion
  is truncated and this equivalence does not hold anymore. In this Thesis
  we will use the ``standard'' set for the expansion of the ground state
  of $\Li$, while we will consider the set $B$ when we will study
  $\alpha+d$ states.

  For a given choice of the Jacobi vectors, the hyperspherical
  coordinates are given by the hyperradius $\rho$, which is independent on the
  permutation order of the particles and is defined as
  \begin{equation}
    \rho=\sqrt{\sum_{i=1}^N x_{ip}^2}=\sqrt{2\sum_{i=1}^N(\br_i-\boldsymbol{X})^2}
    \,,\label{eq:rho}
  \end{equation}
  and by a set $\Omega_{Np}$ of angular variables.
  For the bound state calculation, we use the Zernike and Brinkman
  representation~\cite{Zernike1935,Fabre1983},
  where there are $2N$ polar angles $\hxx_{ip}=(\theta_{ip}
  ,\phi_{ip})$
  of the Jacobi vectors $\xx_{ip}$, $i=1,\dots,N$,
  and $N-1$ hyperspherical angles $\ph_{ip}$, with
  $0\leq\ph_{ip}\leq\pi/2$, given by
  \begin{equation}
    \cos \ph_{ip}=\frac{x_{ip}}{\sqrt{x_{1p}^2+\dots+x_{ip}^2}}\,,
    i=2,\dots,N\,,\label{eq:phiang}
  \end{equation}
  where $x_{ip}$ is the modulus of the Jacobi vector $\xx_{ip}$.
  Therefore we have
  \begin{equation}
    \Omega_{Np}=\{\hxx_{1p},\cdots,\hxx_{Np},
    \ph_{2p},\cdots,\ph_{Np}\}\,.\label{eq:omega}
  \end{equation}
  As for the Jacobi coordinates, there are other possible definitions
  for the hyperspherical angles $\ph_{ip}$. We explicitly discuss them
  and their utility in the next Chapters.
\end{section}
\begin{section}{Hyperspherical Harmonic Functions}\label{sec:HH2}
  The spherical harmonic functions  $Y_{\ell,m}(\theta,\phi)$ are very
  widely used in the study of quantum mechanical problem in a three-dimensional
  space. In this section we construct the generalization of the
  spherical harmonics to the $D=3N$ dimensional space, namely the
  HH functions. For simplicity we use only the ``standard''
  definitions of the Jacobi vectors and of the hyperspherical coordinates
  [Eqs.~(\ref{eq:jacvec}) and~(\ref{eq:phiang})],
  however all the results can be generalized to any set of Jacobi vector and
  hyperspherical coordinates. Moreover, the results are independent
  of the permutation of the particles, therefore we discard $p$ in the 
  notation.

  The $3N$-dimensional Laplace operator $\Delta$, which appears in the
  kinetic energy, can be written as
  \begin{equation}\label{eq:thh}
    \Delta= \sum_{i=1}^N \nabla^2_{\xx_i} =
    \biggl ( {\partial^2 \over \partial\rho^2}
    +{3N-1\over \rho} {\partial \over \partial\rho} +{\Lambda_N^2(\Omega_N)
      \over \rho^2}\biggr)\ ,
  \end{equation}
  where $\Lambda_N^2$ is the $3N$-dimensional generalized angular
  momentum operator  depending only on the
  hyperangles $\Omega_N$. 
  An homogeneous polynomial $p_{[K]}$ of
  degree $K$ in the Cartesian coordinates of $N$ Jacobi vectors,
  gives $\Delta p_{[K]}=0$. We can rewrite it in the form 
  \begin{equation}\label{eq:hp}
    p_{[K]}=\rho^K Y_{[K]}\ ,
  \end{equation}
  where $\rho$ is the hyperradius as given in Eq.~(\ref{eq:rho}) and 
  $Y_{[K]}$ depends only on the variables $\Omega_N$ defined in
  Eq.~(\ref{eq:omega}), being $p_{[K]}$ an homogeneous polynomial.
  Applying $\Delta$ to Eq.~(\ref{eq:hp}) we get
  \begin{equation}
    \begin{aligned}
    \Delta p_{[K]} &= \biggl ( {\partial^2 \over \partial\rho^2}
    +{3N-1\over \rho} {\partial \over \partial\rho} +{\Lambda_N^2(\Omega_N)
      \over \rho^2}\biggr) \rho^K Y_{[K]}(\Omega_N) \\
    \noalign{\medskip}
    &= \biggl ( \Lambda_N^2(\Omega_N) + K(K+D-2) \biggr) \rho^{K-2}
    Y_{[K]}(\Omega_N) = 0\ ,  \label{eq:equ1}
    \end{aligned}
  \end{equation}
  and dividing by $\rho^{K-2}$ we get
  \begin{equation}\label{eq:equ2}
   \biggl ( \Lambda_N^2(\Omega_N) + K(K+D-2) \biggr)
   Y_{[K]}(\Omega_{N}) = 0 \ .
  \end{equation}
  The functions that satisfies the latter equation are known as 
  Hyperspherical Harmonic functions. This is the eigenfunction of
  the generalized  angular momentum operator $\Lambda_N^2$ which
  is sometimes called the grand angular momentum operator. $K$ is called
  the grand angular quantum number while $[K]$ represents
  a set of quantum numbers as specified in the following. 

  The grand angular momentum operator
  can be written in the form ~\cite{Fabre1983}
  \begin{equation}
    \begin{aligned}
    \Lambda_i^2(\Omega_i)&={\partial^2 \over \partial \ph^2_i } +
    \biggl[3(i-2) {\rm cotan}\,\ph_i +2
      ({\rm cotan}\,\ph_i-\tan\ph_i)\biggr]
          {\partial \over \partial \ph_i }\\
          & +{{L_i}^2 \over \cos^2 \ph_i}+
          { \Lambda^2_{i-1}(\Omega_{i-1}) \over \sin^2 \ph_i}
          \ , \label{eq:thhA}
    \end{aligned}
  \end{equation}
  where $L^2_i$ is the angular momentum operator associated with
  the $i$-th Jacobi vector. In particular
  \begin{equation}\label{eq:thh1}
    \Lambda^2_{1}(\Omega_{1})= L^2_1 \ .
  \end{equation}
  The solutions of Eq.~(\ref{eq:equ2}) can be constructed by
  following a recursive procedure~\cite{Zernike1935}.
  For $N=1$ the operator reduces to Eq.~(\ref{eq:thh1}) and the
  eigenfunctions are the spherical harmonic functions
  $Y_{\ell_1 m_1} (\hat x_1)$. In this case $K=\ell_1$.
  For $N=2$ Eq.~(\ref{eq:equ2}) reduces to the form
  \begin{equation}\label{eq:aut1}
    \Lambda_2(\Omega_2) Y_{[K]}(\Omega_2)= - K(K+4) Y_{[K]}(\Omega_2)
    \ ,\quad \Omega_2\equiv \{\hxx_1,\hxx_2,\ph_2\}   \ .
  \end{equation}
  Let us look for a solution of the previous equation of the form
  \begin{equation}\label{eq:aut2}
    Y_{[K]}(\Omega_2)= 
    F(\cos 2\ph_2) (\cos\ph_2)^{\ell_2} (\sin\ph_2)^{\ell_1}
    Y_{\ell_1 m_1} (\hat x_1) Y_{\ell_2 m_2} (\hat x_2)\ ,
  \end{equation}
  where $F$ is a function to be determined. Defining $y=\cos 2\ph_2$,
  Eq.~(\ref{eq:aut1}) reduces to
  \begin{equation}\label{eq:aut3}
    (1-y^2)F''+ (\alpha - \beta y)  F' + \gamma F = 0 \ ,
  \end{equation}
  where
  \begin{equation}
  \begin{aligned}
    \alpha&= \ell_2-\ell_1\ ,\quad
    \beta=\ell_1+\ell_2+3\ , \\
    \gamma&=\frac{1}{4}\left[K(K+4) -(\ell_1+\ell_2)(\ell_1+\ell_2+4)\right] 
    \label{eq:aut3b}
    \ .
  \end{aligned}
  \end{equation}
  Eq.~(\ref{eq:aut3}) is the one satisfied by the Jacobi polynomial
  $P^{\ell_1+1/2,\ell_2+1/2}_n(z) $ provided that
  $K=2n+\ell_1+\ell_2$~\cite{AbramowitzStegun}. Therefore, a solution 
  of Eq.~(\ref{eq:aut1}) is given by
  \begin{equation}
    \begin{aligned}
    Y_{[K]}(\Omega_2)= {\cal N}^{\ell_2,\nu_2}_{n}
    (\cos\ph_2)^{\ell_2} (\sin\ph_2)^{\ell_1}
    Y_{\ell_1 m_1} (\hat x_1) Y_{\ell_2 m_2} (\hat x_2)\;
    P^{\ell_1+1/2,\ell_2+1/2}_n(\cos 2\ph_2)\ ,
    \label{eq:aut5}
    \end{aligned}
  \end{equation}
  where ${\cal N}^{\ell_2,\nu_2}_{n}$ is a normalization factor which
  will be specified later  and $\nu_2=2n+\ell_1+\ell_2+2$. For $N=2$ the symbol
  $[K]$ results to be
  \begin{equation}
    [K]=\{\ell_1,m_1,\ell_2,m_2,n\}\,.
  \end{equation}
  We can now verify that $\rho^KY_{[K]}$ is an homogeneous polynomial
  of degree $K$ in the Cartesian component $\xx_1$ and $\xx_2$. Using the
  fact that $x_1=\rho \sin\ph_2$ and $x_2=\rho \cos\ph_2$, we can write
  \begin{eqnarray}\label{eq:hop}
    \rho^K Y_{[K]}(\Omega_2) = {\cal N}^{\ell_2,\nu_2}_{n}
    \rho^{2n}
    x_1^{\ell_1} Y_{\ell_1 m_1}(\hat x_1)\;
    x_2^{\ell_2} Y_{\ell_2 m_2} (\hat x_2)\;
    P^{\ell_1+1/2,\ell_2+1/2}_n(\cos 2\ph_2)\ .
  \end{eqnarray}
  The terms $x_1^{\ell_1} Y_{\ell_1 m_1}(\hat x_1)$ and
  $x_2^{\ell_2} Y_{\ell_2 m_2}(\hat x_2)$, are homogeneous
  polynomials of degree $\ell_1$ and $\ell_2$. We remain with
  \begin{equation}
  \begin{aligned}\label{eq:hop3}
    & \rho^{2n} P^{\ell_1+1/2,\ell_2+1/2}_n(\cos 2\ph_2)=
    \rho^{2n} \sum_{m=0}^{n} a_m (\cos 2\ph_2)^m \\
    &\quad=  \sum_{m=0}^{n} a_m (\rho^2 \cos 2\ph_2)^m \rho^{2(n-m)}
    = \sum_{m=0}^{n} a_m (x_2^2-x_1^2)^m
    (x_2^2+x_1^2)^{n-m} \ ,
  \end{aligned}
  \end{equation}
  which is an homogeneous polynomial of degree $2n$. In Eq.~(\ref{eq:hop3})
  we have used $\rho^2 \cos 2\ph_2 = x_2^2-x_1^2$ and the definition of Jacobi
  polynomials~\cite{AbramowitzStegun}. In conclusion
  $\rho^K Y_{[K]}(\Omega_2)$ is a homogeneous polynomial of degree
  $K=2n+\ell_1+\ell_2$.

  We can now construct the HH functions for any $N$. Let us suppose we know
  the solution in the case $N-1$, namely the function
  $Y_{[K_{N-1}]}(\Omega_{N-1})$, which satisfies the equation
  \begin{equation}\label{eq:aut6}
    \Lambda^2_{N-1}(\Omega_{N-1}) Y_{[K_{N-1}]}(\Omega_{N-1}) =
    -K_{N-1}(K_{N-1}+D-5) Y_{[K_{N-1}]}(\Omega_{N-1})\ ,
  \end{equation}
  where $K_{N-1}$ is the corresponding grand angular quantum number.
  Analogously to Eq.~(\ref{eq:aut2}), we can look  for the eigenfunction
  of $\Lambda^2_N$ of the form
  \begin{equation}    \label{eq:aut7}
    \begin{aligned}
    Y_{[K]}(\Omega_N)= (\cos\ph_N)^{\ell_N} (\sin\ph_N)^{K_{N-1}}
    Y_{[K_{N-1}]}(\Omega_{N-1}) Y_{\ell_N m_N} (\hat x_N)
    F(\cos 2\ph_N)\ .
    \end{aligned}
  \end{equation}
  By inserting this expression for $Y_{[K]}$ in the corresponding 
  eigenvalue equation and taking into account Eq.~(\ref{eq:aut6}), 
  we get the following  solution
  \begin{equation}\label{eq:aut8}
    F(\cos 2\ph_N)= {\cal N}^{\ell_N,\nu_N}_{n_N}
    P^{\nu_{N-1},\ell_N+1/2}_{n_N}(\cos 2\ph_N)\ ,
  \end{equation}
  with
  \begin{equation}
    \begin{aligned}
    K&=2n_N+\ell_N+K_{N-1}\ ,\quad
    \nu_{N-1}=K_{N-1}+{3(N-1)\over 2}-1\ ,\\
    \nu_{N}&=K+{3N\over 2}-1\ .
    \label{eq:aut9}
    \end{aligned}
  \end{equation}
  The complete expression of the HH function can therefore be cast in
  the form~\cite{Fabre1983}
  \begin{equation}
    \begin{aligned}
    Y_{[K]}(\Omega_N) = 
      \prod_{j=1}^N Y_{\ell_j,m_j}(\hat x_j) \;
      \prod_{j=2}^N
      {\cal N}^{\ell_j,\nu_j}_{n_j}
    (\cos\ph_j)^{\ell_j} (\sin\ph_j)^{K_{j-1}}
     P^{\nu_{j-1},\ell_j+{1\over2}}_{n_j}(\cos 2\ph_j)\ ,
     \label{eq:hh}
     \end{aligned}
  \end{equation}
  and the quantum numbers  $K_j$ and $\nu_j$ are defined to be
  \begin{equation}
    K_j= \sum_{i=1}^j (\ell_i+2 n_i)\ , \quad n_1\equiv 0\ , \quad
    K\equiv K_N\ , \quad \nu_j=K_j+{3j\over 2}-1\ .
    \label{eq:go}
  \end{equation}
  The symbol $[K]$ stands for the following set of quantum numbers
  \begin{equation}\label{eq:qn1}
    [K]\equiv \{ \ell_1,\ldots,\ell_N,\ m_1,\ldots,m_N, \
    n_2,\ldots,n_N\}\ ,
  \end{equation}
  and it permits to distinguish the different HH functions having the same
  $K$ value. In total there are $3N-1$ quantum numbers which specify a HH
  functions: the $2N$ quantum numbers $\ell,m$
  associated with  the spherical harmonic functions, and the $N-1$
  quantum numbers $n$ associated with the hyperspherical polynomials.
  The normalization factors ${\cal N}$ are chosen to verify the
  orthonormal condition
  \begin{equation}\label{eq:nohh}
    \int d\Omega_N \Bigl ( Y_{[K]}(\Omega_N)\Bigr)^*
    Y_{[K']}(\Omega_N)= \delta_{[K],[K']}\ ,
  \end{equation}
  where~\cite{Fabre1983}
  \begin{equation}\label{eq:domega}
    d\Omega_N=\sin\theta_1 d\theta_1d\phi_1 \prod_{j=2}^{N}
    \sin\theta_j d\theta_jd\phi_j (\cos\ph_j)^2
    (\sin\ph_j)^{3j-4} d\ph_j
  \end{equation}
  is the surface element on the hypersphere of unit hyperradius.
  To be noticed that the integral over all the Jacobi coordinates
  can be rewritten in terms of $d\Omega_N$ and the hyperradius, namely
  \begin{equation}
    \int \prod_{i=1}^N d^3x_i=\int d\rho\,\rho^{3N-1}\int d\Omega_N\,.
  \end{equation}
  The explicit expression for ${\cal N}$ is
  \begin{equation}\label{eq:nhh}
    {\cal N}^{\ell_j,\nu_j}_{n_j}=
    \biggl[{2\nu_j\Gamma (\nu_j-n_j)n_j!\over
        \Gamma (\nu_j-n_j-\ell_j-{1\over 2})
        \Gamma (n_j+\ell_j+{3\over 2})}\biggr]^{1/2}\ .
  \end{equation}
  We need now to verify that $\rho^KY_{[K]}$ is a homogeneous polynomial
  in the Cartesian Jacobi coordinates. It is easy, using Eq.~(\ref{eq:phiang}),
  to verify that
  \begin{equation}
    \begin{aligned}
    \rho^K Y_{[K]}(\Omega_N) &\propto
      \prod_{j=1}^N x_j^{\ell_j}
      Y_{\ell_j,m_j}(\hat x_j)\\ 
     &\times\prod_{j=2}^N \rho^{2n_j}\;
      (\sin\ph_{j+1}\cdots\sin\ph_N)^{2n_j}{P}^{\nu_{j-1},\ell_j+1/2}
      _{n_j}(\cos2\ph_j)
    \ .
    \label{eq:hop4}
    \end{aligned}
  \end{equation}
  The first term, in Eq.~(\ref{eq:hop4}) is a homogeneous polynomial of degree
  $\ell_1+\cdots+\ell_N$. The second term can be rewritten as
  \begin{equation}\label{eq:hop7}
    \begin{aligned}
     (\rho\sin\ph_{j+1}&\cdots\sin\ph_N)^{2n_j}
      {P}^{\nu_{j-1},\ell_j+1/2}_{n_j}(\cos2\ph_j) \\
    &=  \rho^{2n_j}\;
    (\sin\ph_{j+1}\cdots\sin\ph_N)^{2n_j}
    \sum_{m=0}^{n_j} a_m (\cos 2\ph_j)^m \\
    &=  \sum_{m=0}^{n_j} a_m
    \biggl(x_j^2-(x_1^2+\cdots+x_{j-1}^2)\biggr)^m
    \biggl(x_1^2+\cdots+x_{j}^2\biggr)^{n_j-m} \ ,
    \end{aligned}
  \end{equation}
  where we used the following relations
  \begin{equation}\label{eq:hop5}
    \rho^2\cos2\ph_j =
        { x_j^2-(x_1^2+\cdots+x_{j-1}^2)
          \over (\sin\ph_{j+1}\cdots\sin\ph_N)^2}
        \ ,
  \end{equation}
  and
  \begin{equation}\label{eq:hop6}
    ( \rho\sin\ph_{j+1}\cdots\sin\ph_N)^{2} =
    x_1^2+\cdots+x_{j}^2\ .
  \end{equation}
  Using Eq.~(\ref{eq:hop7}),
  the second term in Eq.~(\ref{eq:hop4}) results to be a homogeneous polynomial
  of order $2\sum_{j=2,N}n_j$. Therefore, the function $\rho^KY_{[K]}$ is an
  homogeneous polynomial of order $K=\ell_1+\sum_{j=2,N} (\ell_j+2n_j)$
  in the Jacobi coordinates.

  The HH functions have several important properties. Here we report only
  the completeness of the HH basis without proof.
  By introducing the hypercoordinates  $\rho',\Omega'_N$  associated
  with the $N$ vectors $\xx_i'$, it is possible to show that~\cite{Kievsky2008}
  \begin{equation}\label{eq:pw7}
    \sum_{[K]}  Y^*_{[K]}(\Omega'_N)
    Y_{[K]}(\Omega_N)= \delta^{D-1}(\Omega_N-\Omega'_N)\ ,
  \end{equation}
  where
  \begin{equation}\label{eq:pw6}
    \delta^{D-1}(\Omega_N-\Omega'_N) = \prod_{i=1}^{N}
    \delta^2(\hat x_i-\hat x_i')
    \prod_{i=2}^{N} { \delta(\ph_i-\ph_i') \over
      (\cos\ph_i)^2   (\sin\ph_i)^{3i-4}}  \ .
  \end{equation}
  As a consequence, every continuous function $f(\Omega_N)$ can be expanded
  in terms of the HH functions as
  \begin{equation}\label{eq:pw8}
    f(\Omega_N) = \int d\Omega'_N\; \delta^{D-1}(\Omega_N-\Omega'_N) f(\Omega'_N)
    = \sum_{[K]}  a_{[K]}  Y_{[K]}(\Omega_N)\ ,
  \end{equation}
  where
  \begin{equation}\label{eq:pw9}
    a_{[K]} = \int d\Omega'_N \; Y^*_{[K]}(\Omega'_N)f(\Omega'_N)
    \ .
  \end{equation}
  We will use this property very widely in the following of this Thesis.

  In this work we will use HH functions with definite total angular momentum
  $L$. They are constructed using the following coupling
  scheme~\cite{Kievsky2008}
  \begin{equation}
    \begin{aligned}
    {\cal Y}^{KLM}_{\mu}(\Omega_N) &=& \sum_{m_1,\ldots,m_N}
    (\ell_1 m_1 \ell_2 m_2 | L_2 M_2) (L_2 M_2 \ell_3 m_3 | L_3 M_3)
    \times \\
    && \cdots \times (L_{N-1} M_{N-1} \ell_N m_N | LM)
    Y_{[K]}(\Omega_N)\ ,
    \label{eq:hh2}
    \end{aligned}
  \end{equation}
  where $Y_{[K]}(\Omega_N)$ is defined in Eq.~(\ref{eq:hh}), $(\ell_1
  m_1 \ell_2 m_2 | L_2 M_2)$ and so on are Clebsch-Gordan coefficients
  and $M_i=\sum_{j=1,i} m_j$. The symbol $\mu$ represent the complete
  set of quantum number, namely
  \begin{equation}
    \mu\equiv\{\ell_1,\dots,\ell_N,L_2,\dots,L_{N-1},n_2,\dots,n_N\}\,.
    \label{eq:qn2}
  \end{equation}
  Reintroducing the notation for the permutations, the final
  expression of the HH function for the case $A=6$ is
  \begin{equation}\label{eq:hh6}
  \begin{aligned}
    {\cal Y}^{KLM}_{\mu}(\Omega_{5p})&=
    \left(\left(\left(\left( Y_{\ell_1}(\hat x_{1p})
          Y_{\ell_2}(\hat x_{2p})\right)_{L_2} Y_{\ell_3}(\hat x_{3p}) \right)_{L_3}
        Y_{\ell_4}(\hat x_{4p}) \right)_{L_4} Y_{\ell_5}(\hat x_{5p}) \right)_{LM}
    \\
    &\times {\cal P}^{\ell_1,\ell_2,\ell_3,\ell_4,\ell_5}
         _{n_2,n_3,n_4,n_5}(\ph_{2p},\ph_{3p},\ph_{4p},\ph_{5p})\,,
  \end{aligned}
  \end{equation}
  where
  \begin{equation}\label{eq:hh6b}
  \begin{aligned}
    {\cal P}^{\ell_1,\ell_2,\ell_3.\ell_4,\ell_5}
    _{n_2,n_3,n_4,n_5}&(\ph_{2p},\ph_{3p},\ph_{4p},\ph_{5p})\\
    &={\cal N}_{n_2}^{\ell_2,\nu_2}(\cos\ph_2)^{\ell_{2}}
    (\sin\ph_{2p})^{\ell_1}P_{n_2}^{\ell_1+1/2,\ell_2+1/2}(\cos 2\ph_{2p})
       \\
       &\times{\cal N}_{n_3}^{\ell_3,\nu_3}(\cos\ph_{3p})^{\ell_3}
    (\sin\ph_{3p})^{K_2}P_{n_3}^{\nu_2,\ell_3+1/2}(\cos 2\ph_{3p})\\
       &\times{\cal N}_{n_4}^{\ell_4,\nu_4}(\cos\ph_{4p})^{\ell_4}
    (\sin\ph_{4p})^{K_3}P_{n_4}^{\nu_3,\ell_4+1/2}(\cos 2\ph_{4p})\\
       &\times{\cal N}_{n_5}^{\ell_5,\nu_5}(\cos\ph_{5p})^{\ell_5}
    (\sin\ph_{5p})^{K_4}P_{n_5}^{\nu_4,\ell_5+1/2}(\cos 2\ph_{5p})\ ,
  \end{aligned}
  \end{equation}
  and
  \begin{equation}\label{eq:qn3}
    \mu\equiv\{\ell_1,\ell_2,\ell_3,\ell_4,\ell_5,L_2,L_3,
    L_4,n_2,n_3,n_4,n_5\}\ .
  \end{equation}
\end{section}

\begin{section}{The Hyperspherical Harmonic expansion}\label{sec:HH3}
  
  In the following we will use the isospin formalism. Therefore, each
  nucleon will be described by its position and spin-isospin projection.
  Since the nucleons are Fermions,
  the wave function of the six-nucleon system has to be antisymmetric under
  the exchange of any pair of particles.
  Moreover, we require that our wave function has a well defined  total
  angular momentum $J$, $J_z$ and parity $\pi$.
  Therefore, we define a complete basis of antisymmetrical hyperangular-spin-isospin states as follows
  \begin{equation}\label{eq:hhst0}
    \Psi^{KLSTJ\pi}_\al=\sum_{p=1}^{N_p}\Phi^{KLSTJ\pi}_\al(i,j,k,l,m,n)\ ,
  \end{equation}
  where with $p$ we indicate a generic even permutation of the six particles,
  $p\equiv(i,j,k,l,m,n)$, $N_p$ being the total number of even permutation
  given by
  \begin{equation}
    N_p=\frac{6!}{2}=360\,,
  \end{equation}
  and
  \begin{equation}
  \begin{aligned}
    \Phi^{KLSTJ\pi}_\al(i,j,k,l,m,n)&=
    \Bigg \{{\cal Y}^{KLM}_{\mu}(\Omega_{5p})
    \left[\left(\left(s_is_j\right)_{S_2} s_k\right)_{S_3}
      \left(\left(s_ls_m\right)_{S_4} s_n\right)_{S_5}\right]_S\Bigg\}_{J,J_z}\\
    &\otimes\left[\left(\left(t_it_j\right)_{T_2} t_k\right)_{T_3}
      \left(\left(t_lt_m\right)_{T_4} t_n\right)_{T_5}\right]_{T,T_z}\,.
    \label{eq:hhst}
  \end{aligned}
  \end{equation}
  The function ${\cal Y}^{KLM}_{\mu}(\Omega_{5p})$ is the HH function defined in
  Eq.~(\ref{eq:hh6}) and $s_i$ $(t_i)$ denotes the spin (isospin) function
  of nucleon $i$. To be noticed that the coupling scheme of 
  the spin (isospin) does not follow the one of the hyperangular part.
  This particular choice
  simplifies the calculations of the potential matrix elements.
  There are other possible choices for the coupling of the spin (isospin)
  states that can be easily connected
  through combinations of 6j- and 9j-Wigner coefficients. 
  The total orbital angular momentum $L$ of the HH function
  is coupled to the total spin $S$ to give the total angular momentum $J$, $J_z$.
  while the total isospin is given by $T$, $T_z$. 
  The index $\al$ labels the possible choice of hyperangular, spin and
  isospin quantum numbers,
  \begin{equation}\label{eq:alpha}
    \alpha\equiv\{\ell_1,\ell_2,\ell_3,\ell_4,\ell_5,
    L_2,L_3,L_4,n_2,n_3,n_4,n_5,
    S_2,S_3,S_4,S_5,T_2,T_3,T_4,T_5\}\,,
  \end{equation}
  compatible with the given values of $K$, $L$, $S$, $T$, $J$, and $\pi$.
  The parity of the state is
  $\pi=(-1)^{\ell_1+\ell_2+\ell_3+\ell_4+\ell_5}$ and we will include in our
  basis only the states such that $\pi$ corresponds to the parity of the nuclear
  state under study.

  The total wave function must be completely antisymmetric under the exchange
  of any pair of particles. Therefore we need to impose antisymmetry on
  each state $\Psi^{KLSTJ\pi}_\al$.
  For example, after the permutation of any pair, the state given in
  Eq.~(\ref{eq:hhst0}) can be rearranged so that
  \begin{equation}\label{eq:hhst1}
    \Psi^{KLSTJ\pi}_\al
    \rightarrow\sum_{p=1}^{360}\Phi^{KLSTJ\pi}_\al(j,i,k,l,m,n)\,.
  \end{equation}
  Therefore to have antisymmetry it is sufficient to have
  \begin{equation}\label{eq:ants}
   \Phi^{KLSTJ\pi}_\al(j,i,k,l,m,n)=-\Phi^{KLSTJ\pi}_\al(i,j,k,l,m,n)\,.
  \end{equation}
  Under the exchange of $i\leftrightarrow j$ the Jacobi vector $\xx_{5p}$
  [Eq.~(\ref{eq:jacvec})] changes
  its sign, whereas all the others remain the same.
  Therefore, the HH function ${\cal Y}^{KLM}_{\mu}(\Omega_{5p})$ in
  Eq.~(\ref{eq:hh6}) transforms into itself times a factor $(-1)^{\ell_5}$.
  The spin-isospin part [see Eq.~(\ref{eq:hhst})] part
  transforms into itself times a factor $(-1)^{S_2+T_2}$ for the 
  $i\leftrightarrow j$ exchange. Therefore, we obtain 
  \begin{equation}\label{eq:ants2}
    \Phi^{KLSTJ\pi}_\al(j,i,k,l,m,n)=(-)^{\ell_5+S_2+T_2}
    \Phi^{KLSTJ\pi}_\al(i,j,k,l,m,n)\,,
  \end{equation}
  In order to fulfill the condition of Eq.~(\ref{eq:ants}) we impose
  the condition
  \begin{equation}
    \ell_5+S_2+T_2=\text{odd}\,,
  \end{equation}
  on the quantum numbers of the HH states. In such a way we 
  construct only antisymmetric states.  

  \begin{subsection}{Orthogonalization of the basis}\label{sec:ortob}
  The number $M_{KLSTJ\pi}$ of antisymmetric functions  $\Psi^{KLSTJ\pi}_\al$
  with fixed $K$, $L$, $S$, $T$, $J$ and $\pi$ 
  in general is very large, due to the high
  number of possible combinations of quantum numbers $\alpha$ that fulfill
  the requirement of antisymmetry and parity. However,
  the sum over the permutations in Eq.~(\ref{eq:hhst0}) generates
  states that are linearly dependent among each other. 
  This means that our
  basis is overcomplete and we need to extract only the independent states.
  In order to do that we  use the Transformation
  Coefficients (TC) $a^{KLSTJ\pi}_{\al,\al'}(p)$ defined as
  \begin{equation}
      \Phi^{KLSTJ\pi}_\alpha(i,j,k,l,m,n)
      =\sum_{\alpha'}a^{KLSTJ\pi}_{\alpha,\alpha'}(p)
      \Phi^{KLSTJ\pi}_{\alpha'}(1,2,3,4,5,6)\,.\label{eq:tc}
  \end{equation}
  Hence, the states $\Psi^{KLSTJ\pi}_\al$ of Eq.~(\ref{eq:hhst0})
  can be written as
  \begin{equation}
    \Psi^{KLSTJ\pi}_\al=\sum_{\al'}A^{KLSTJ\pi}_{\al,\al'}
    \Phi^{KLSTJ\pi}_\alpha(1,2,3,4,5,6)\,,\label{eq:tc1}
  \end{equation}
  where
  \begin{align}
    A^{KLSTJ\pi}_{\alpha,\alpha'}=
    \sum_{p=1}^{360} a^{KLSTJ\pi}_{\alpha,\alpha'}(p)\,.
    \label{eq:tc2}
  \end{align}
  The coefficients $A^{KLSTJ\pi}_{\alpha,\alpha'}$ contain all the properties
  of our basis, and so the knowledge of all of them coincides with the
  knowledge of the entire basis. Moreover, they permit to simplify
  the determination of the independent states as well as  
  the calculation
  of the matrix elements, as we will see in the next section.
  We will discuss the explicit calculation of the TC
  $a^{KLSTJ\pi}_{\alpha,\alpha'}(p)$
  and some of their properties in Appendix~\ref{app:tc}. For now we suppose
  we are able to compute them.

  The fundamental ingredient
  to identify the independent states is the knowledge of the norm matrix elements
  \begin{equation}\label{eq:norm}
    N^{KLSTJ\pi}_{\al,\al'}=\bra \Psi^{KLSTJ\pi}_\al | \Psi^{KLSTJ\pi}_{\al'}\ket_{\Omega}\,,
  \end{equation}
  where $\bra\cdots|\cdots\ket_{\Omega}$
  denotes the spin and isospin trace and the integration
  over the hyperspherical variables. Using the orthogonality of the
  HH basis with Eq.~(\ref{eq:tc1}) we obtain
  \begin{equation}\label{eq:norm1}
    N^{KLSTJ\pi}_{\al,\al'}=\sum_{\al''}\left(A^{KLSTJ\pi}_{\alpha,\alpha''}\right)^*
    A^{KLSTJ\pi}_{\alpha'',\alpha'}\,,
  \end{equation}
  and also
  \begin{equation}\label{eq:norm2}
    \bra \Psi^{KLSTJ\pi}_\al | \Psi^{K'L'S'T'J'\pi'}_{\al'}\ket_{\Omega}=0\quad
         {\text{for}}\quad
         \{KLSTJ\pi\}\neq\{K'L'S'T'J'\pi'\}.
  \end{equation}
  Once the matrix elements
  $N^{KLSTJ\pi}_{\al,\al'}$ are evaluated, the orthogonalization procedure
  presented in Appendix~\ref{app:tc3}
  is used to find and eliminate the linearly dependent states
  among the various $\Psi^{KLSTJ\pi}_\al$ functions.
  
  The number of independent antisymmetric states
  $M'_{KLSTJ\pi}$ is noticeably smaller
  than the corresponding $M_{KLSTJ\pi}$ as can be seen in Table~\ref{tab:states},
  where we
  report the values of $M_{KLSTJ\pi}$ and $M'_{KLSTJ\pi}$ for the case
  $J=1$, $T=0$ and $\pi=+$, which corresponds to the ground state of the
  $\Li$ up to $K=10$. Observing the table it is possible to notice that
  the number of states starts to be very
  large already for $K=6$.  For the case $K=0$ and $LSTJ\pi=0101+$,
  there is no independent state due to the Pauli principle.
  The number of states with $S=3$ is very small
  compared to the others, because the spin states which can be
  constructed are only symmetric.
  \begin{sidewaystable}
    \centering
    \footnotesize
    \begin{tabular}{r@{$\quad$}r@{$\quad$}r@{$\quad$}r@{$\quad$}
        r@{$\quad$}r@{$\quad$}r@{$\quad$}r@{$\quad$}r@{$\quad$}
        r@{$\quad$}r@{$\quad$}r@{$\quad$}r@{$\quad$}
                r@{$\quad$}r@{$\quad$}rr}
      \hline
      \hline
      \multicolumn{1}{c}{$K$} & \multicolumn{2}{c}{$L=0$ $S=1$}&
      \multicolumn{2}{c}{$L=2$ $S=1$} & \multicolumn{2}{c}{$L=2$ $S=2$}&
      \multicolumn{2}{c}{$L=2$ $S=3$} & \multicolumn{2}{c}{$L=1$ $S=0$}&
      \multicolumn{2}{c}{$L=1$ $S=1$} & \multicolumn{2}{c}{$L=1$ $S=2$}\\
      &$M_{K0101+}$ &$M'_{K0101+}$ &$M_{K2101+}$ &$M'_{K2101+}$ &$M_{K2201+}$&
      $M'_{K2201+}$ &$M_{K2301+}$ &$M'_{K2301+}$&
      $M_{K1001+}$ &$M'_{K1001+}$ &$M_{K1101+}$ &$M'_{K1101+}$ &$M_{K1201+}$&
      $M'_{K1201+}$ \\
      \hline
      0 &      21&   0 &        &     &        &    &       &    &        &    &        &     &        &     \\
      2 &     306&   1 &     327&    1&     177&   0&     34&   0&     124&   1&     222&    0&     122&   0 \\
      4 &   2,325&   7 &   4,662&   12&   2,562&   4&    504&   1&   1,744&   5&   3,132&    7&   1,732&   4 \\
      6 &  12,480&  34 &  34,065&   90&  18,815&  42&  3,730&   9&  12,536&  36&  22,533&   56&  12,483&  31 \\
      8 &  52,893& 144 & 172,500&  442&  95,500& 227& 19,000&  46&  62,540& 170& 112,470&  291&  62,370& 160 \\
     10 & 187,842& 509 & 684,885& 1535$^*$& 379,635& 804$^*$& 75,670& 145& 245,204& 615& 441,087& 1077& 244,737& 597 \\
      \hline                                               
      \hline                                               
    \end{tabular}
    \caption{\label{tab:states}
      Number of six-nucleon antisymmetrical hyperspherical-spin-isospin
      state for the case $J=1$, $T=0$ and $\pi=+$ as function of the grandangular
      quantum number $K$, the total spin $S$ and the total angular momentum $L$.
      $M_{KLSTJ\pi}$ is the total number of antisymmetric states while
      $M'_{KLSTJ\pi}$ is the number of independent antisymmetric states.
      With the symbol $^*$ we indicate
      that there are other independent states that must be included to determine the complete
      basis.}
  \end{sidewaystable}
  In Figure~\ref{fig:mk0101} we plotted both $M_{K0101+}$ and $M'_{K0101+}$ as function
  of $K$. As can be seen from the figure, the number of independent states
  is $\sim2$ orders of magnitude less than the total number of antisymmetric states.
  It is also interesting to notice that the ratio $M_{K0101+}/M'_{K0101+}$ is almost
  constant varying $K$. Moreover, from the figure it is clear
  the exponential growing of
  $M_{K0101+}$ and $M'_{K0101+}$ as function of $K$.
  A similar  behavior was found for all the combinations $LST$ considered
  in Table~\ref{tab:states}.
  \begin{figure}[h]
    \centering
    \includegraphics[scale=.8]{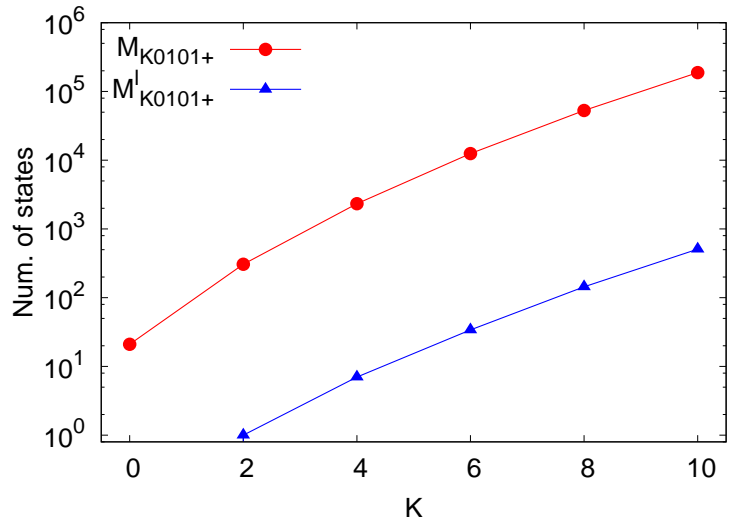}
    \caption{Number of states for the case $LSTJ\pi=0101+$ as function
      of the grandangular quantum number K. The red circles are the total
      number of antisymmetric states $M_{KLSTJ\pi}$
      while the blue triangles are the
      number of linearly independent antisymmetric states
      $M'_{KLSTJ\pi}$. To be noticed
      that the y-axis is in logarithmic scale. The lines are added to guide
      the eyes.}
    \label{fig:mk0101}
  \end{figure}
\end{subsection}
\end{section}

\begin{section}{The variational wave function}\label{sec:HH4}
  The final form of the six-nucleons bound state wave function can be
  written as
  \begin{equation}\label{eq:wf}
    \Psi_6^{J\pi}=\sum_l\sum_{KLST,\alpha}c^{KLST}_{l,\alpha}
    f_l(\rho)\Psi^{KLSTJ\pi}_\al\,,
  \end{equation}
  where the sum is restricted only on the linear independent antisymmetric
  states $\alpha$ and
 $c^{KLST}_{l,\alpha}$ are variational coefficients to be
  determined.
  The hyperradial functions $f_l(\rho)$ must be chosen in order to
  satisfy the following criteria: (i) they must constitute a complete basis;
  (ii) they must satisfy the boundary condition
  $f(\rho)\rightarrow0$ for $\rho\rightarrow \infty$;
  (iii) their form should be simple but flexible enough to well describe
  the dependence on the hyperradius.
  The choice we made is
  \begin{equation}
    f_l(\rho)=\gamma^{D/2} \sqrt{\frac{l!}{(l+D-1)!}}\,\,\, 
    L^{(D-1)}_l(\gamma\rho)\,\,e^{-\frac{\gamma}{2}\rho} \ ,
    \label{eq:fllag}
  \end{equation}
  where $L^{(D-1)}_l(\gamma\rho)$ are Laguerre polynomials~\cite{AbramowitzStegun}
  and $\gamma$ is a non-variational
  parameter that must be determined.
  A typical range for $\gamma$ is $3.5-5.5$ fm$^{-1}$.
  With such definition, the function
  $f_l(\rho)$ are orthonormal, i.e.
  \begin{equation}
    \int_0^\infty d\rho\,\rho^{D-1} f_l(\rho)f_{l'}(\rho)=\delta_{l,l'}\,.
  \end{equation}

  The coefficients $c^{KLST}_{l,\alpha}$ are determined using
  the Rayleigh-Ritz variational principle. Casting the wave function
  in the bra-ket notation, Eq.~(\ref{eq:wf}) becomes
  \begin{equation}
    |\Psi_6^{J\pi}\ket=\sum_{\xi}c_{\xi}|\Psi_\xi\ket\ ,
  \end{equation}
  where $\xi=\{l,KLST\alpha\}$ and
    $|\Psi_\xi\ket=f_l(\rho)\Psi^{KLSTJ\pi}_\al$. Using the 
  Rayleigh-Ritz variational principle, we have
  \begin{equation}
    \langle\delta_c \Psi^{J\pi}_6\,|\,H-E\,|\Psi^{J\pi}_6\rangle
    =0 \ ,
    \label{eq:rrvar}
  \end{equation}
  where $\delta_c \Psi^{J\pi}_6$ indicates the variation of 
  the wave function for arbitrary infinitesimal 
  changes of the linear coefficients $c_\xi$. By performing the variation,
  the problem is reduced to a generalized eigenvalue problem,
  where we have to determine the coefficients $c_\xi$ and
  the energy $E$, i.e.
  \begin{equation}
    \sum_{ \xi'}\,\langle\Psi_\xi\,|\,H-E\,|\, \Psi_{\xi'}\,\rangle 
    \,c_{\xi'}=0
    \ .
    \label{eq:gepb}
  \end{equation}
  Therefore, all the problem stands in calculating the matrix
  elements of the Hamiltonian $H$ with respect to the basis states.
  In nuclear physics, the Hamiltonian $H$ is usually given by
  \begin{equation}
    H=T+\sum_{i<j}V_{ij}+\sum_{i<j<k}V_{ijk}\,,
  \end{equation}
  where $T$ is the kinetic energy, $V_{ij}$ the two-body potential and
  $V_{ijk}$ the three-body potential. 
  In Sections~\ref{sec:nke} and~\ref{sec:ve}
  we will discuss the calculation of the kinetic energy and the potential
  matrix elements, respectively. In this work we will not consider
  the three-body potential.
  However, the formalism introduced in Section~\ref{sec:ve}
  can be easily generalized to this case.
 
  Due to the high number of states needed in the expansion to reach convergence,
  the generalized eigenvalue problem is then solved
  by using the preconditioned Lanczos method~\cite{Cullum1981},  whose
  main features are presented in Appendix~\ref{app:lanczos}.

  \begin{subsection}{Norm and kinetic matrix elements}\label{sec:nke}
    The first matrix element we need to compute is the norm
    that multiply the energy $E$,
    \begin{equation}
      N_{\xi,\xi'}=\langle\Psi_\xi\,|\, \Psi_{\xi'}\rangle\,,
    \end{equation}
    that explicitly translates into
    \begin{equation}
      N_{\xi,\xi'}=\int_0^{\infty}d\rho\,\rho^{14}\int
      d\Omega_5\left[f_l(\rho)\Psi^{KLSTJ\pi}_\al(\Omega_5)\right]^*
      f_{l'}(\rho)\Psi^{K'L'S'T'J\pi}_{\al'}(\Omega_5)\,.
    \end{equation}
    Using the results of Eq.~(\ref{eq:norm})
    and the
    orthonormal properties of the Laguerre polynomials, we get
    \begin{equation}
      N_{\xi,\xi'}=N^{KLSTJ\pi}_{\al,\al'}\delta_{KK'}
      \delta_{LL'}\delta_{SS'}\delta_{TT'}\delta_{ll'}\ .
    \end{equation}
    In the case of the kinetic energy matrix elements, we have
    \begin{equation}\label{eq:kinmx}
      T_{\xi,\xi'}=\int_0^{\infty}d\rho\,\rho^{14}\int
      d\Omega_5\left[f_l(\rho)\Psi^{KLSTJ\pi}_\al(\Omega_5)\right]^*
      \,{\hat T}\,f_{l'}(\rho)\Psi^{K'L'S'T'J\pi}_{\al'}(\Omega_5)\,,
    \end{equation}
    where
    \begin{equation}
      {\hat T}=-\frac{\hbar^2}{m}
      \biggl ( {\partial^2 \over \partial\rho^2}
      +{14\over \rho} {\partial \over \partial\rho} +{K(K+13)
        \over \rho^2}\biggr)\ .
    \end{equation}
    The integrals in Eq.~(\ref{eq:kinmx}) can be factorized
    in the hyperradial and hyperangular part, where the last one
    reduce to Eq.~(\ref{eq:norm}). The final result is
    \begin{equation}\label{eq:kinmxf}
      T_{\xi,\xi'}=T_{l,l'}N^{KLSTJ\pi}_{\al,\al'}\delta_{KK'}
      \delta_{LL'}\delta_{SS'}\delta_{TT'}\ ,
    \end{equation}
    where
    \begin{equation}\label{eq:kinmxf2}
      T_{l,l'}=-\frac{\hbar^2}{m}\int_0^{\infty}d\rho\,\rho^{14}f_l(\rho)
      \biggl ( {\partial^2 \over \partial\rho^2}
      +{14\over \rho} {\partial \over \partial\rho} +{K(K+13)
        \over \rho^2}\biggr)f_{l'}(\rho)\ ,
    \end{equation}
    which can be  analytically evaluated as shown in Appendix~\ref{app:kinene}.
  \end{subsection}
  \begin{subsection}{Potential matrix element}\label{sec:ve}
    In this section we  discuss the calculation of the potential matrix
    elements, focusing on non-local two-body potentials.  The formulas
    we are presenting here can be easily extended to the calculation of
    local two-body potentials and three-body potentials.

    We  need to compute the matrix element
    \begin{equation}\label{eq:vij}
      V^{KLST,K'L'S'T',J\pi}_{l\al,l'\al'}
      =\left\langle f_l(\rho) \Psi^{KLSTJ\pi}_\al
      \left|\frac{1}{2}\sum_{i,j=1}^AV_{ij}\right|
      f_{l'}(\rho')\Psi^{K'L'S'T'J\pi}_{\al'}
      \right\rangle_{\Omega,\rho,\rho'}\,,
    \end{equation}
    where $\Psi^{KLSTJ\pi}_\al$ is defined in Eq.~(\ref{eq:hhst0}),
    $f_l(\rho)$ is defined in Eq.~(\ref{eq:fllag}) and
    $\bra\cdots|\cdots\ket_{\Omega,\rho,\rho'}$ denotes the spin and isospin trace and the integration
    over the hyperspherical and hyperradial variables.
    The states in the bra and in the ket are totally antisymmetric with
    respect to the exchange of any pair of particles, so
    it is possible to 
    rewrite the matrix elements in
    Eq.~(\ref{eq:vij}) as
    \begin{equation}\label{eq:vij2}
      V^{KLST,K'L'S'T',J\pi}_{l\al,l'\al'}
      =\frac{A(A-1)}{2}\left\bra f_l(\rho)\Psi^{KLSTJ\pi}_\al\left|V_{12}\right|
      f_{l'}(\rho')\Psi^{K'L'S'T'J\pi}_{\al'}\right\ket_{\Omega,\rho,\rho'}
    \end{equation}
    In order to compute this matrix element it results convenient to use the
    $jj$-coupling scheme in which the basis state $\alpha$ results
    \begin{equation}
      \Psi_{\al}^{KLSTJ\pi} =
      \sum_\nu B^{KLSTJ\pi}_{\al,\nu} \; \Xi^{KTJ\pi}_{\nu}(1,2,3,4,5,6)
      \label{eq:PSI3jj}\,,
    \end{equation}
    where the new transformation coefficients
    $B^{KLSTJ\pi}_{\al,\nu}$ are related to the coefficients
    $A^{KLSTJ\pi}_{\al,\al'}$ via 6j- and 9j-Wigner coefficients
    (see Appendix~\ref{app:jjcoup}).
    The explicit expression for $\Xi^{KTJ\pi}_{\nu}(1,2,3,4,5,6)$ is
    given by
    \begin{equation}
    \begin{aligned}
      \Xi^{KTJ\pi}_{\nu}(1,2,3,4,5,6)&={\cal P}^{\ell_1,\ell_2,\ell_3,\ell_4,\ell_5}
      _{n_2,n_3,n_4,n_5}(\ph_2,\ph_3,\ph_4,\ph_5)\\
      &\times\bigg\{\bigg[\big(Y_{\ell_5}(\hxx_5)
        (s_1s_2)_{S_2}\big)_{j_1}\big(Y_{\ell_4}(\hxx_4)s_3\big)_{j_2}
        \bigg]_{j_{12}}\\
      &\times\bigg[\big[\big(\big(Y_{\ell_1}(\hxx_1)Y_{\ell_2}(\hxx_2)\big)_{L_2}
          Y_{\ell_3}(\hxx_3)\big]_{L_3}
        \big((s_4s_5)_{S_4}s_6\big)_{S_5}\bigg]_{j_3}
      \bigg\}_{JJ_z}\\
          &\otimes\left[\left(\left(t_1t_2\right)_{T_2} t_3\right)_{T_3}
      \left(\left(t_4t_5\right)_{T_4} t_6\right)_{T_5}\right]_{T,T_z}\,.\label{eq:PHIjj}
    \end{aligned}
    \end{equation}
    The index $\nu$ labels all possible choices of the quantum numbers
    \begin{align}
      \nu\equiv\{n_5,\ell_5,S_2,j_1,n_4,&\ell_4,j_2,j_{12},\ell_1,\ell_2,\ell_3,
      L_2,L_3,n_2,n_3,S_4,S_5,j_3,T_2,T_3,T_4,T_5\}\,,
      \label{eq:nu}
    \end{align}
    which are compatible with $K,T,J$ and $\pi$.
    To be noticed that in Eq.~(\ref{eq:PHIjj}) the $jj$-coupling scheme is used
    only for the first three particles and all the rest remains in $LS$ scheme.
    This particular mixed
    coupling scheme permits to increase the performances of the
    code which computes the potential matrix elements.
    Even if in this work we use only two-body forces, this particular
    coupling  scheme results to be very advantageous also
    when the three-nucleons interaction is included.
    
    The matrix element defined in Eq.~(\ref{eq:vij2}) can be rewritten using
    the coefficients $B^{KLSTJ\pi}_{\al,\nu}$ as
    \begin{equation}\label{eq:mxaap}
      V^{KLST,K'L'S'T',J\pi}_{l\al,l'\al'}=15
      \sum_\nu\sum_{\nu'}B^{KLSTJ\pi}_{\al,\nu}B^{K'L'S'T'J\pi}_{\al',\nu'}
      v^{KT,K'T',J\pi}_{l\nu,l'\nu'}\,,
    \end{equation}
    where
    \begin{equation}\label{eq:v12jj}
      v^{KT,K'T',J\pi}_{l\nu,l\nu'}=
      \bra f_l(\rho)\Xi^{KTJ\pi}_{\nu}(1,2,3,4,5,6)|V_{12}|
      f_l'(\rho')\Xi^{K'T'J\pi}_{\nu'}(1,2,3,4,5,6)\ket_{\Omega,\rho,\rho'}\,.
    \end{equation}
    The quantum numbers $\nu$ can be decomposed as $\nu=\{\nu_x,\nu_y,T_3\}$ where
    \begin{equation}
      \nu_x=\{j_1,n_4,\ell_4,j_2,j_{12},\ell_1,\ell_2,\ell_3,L_2,L_3,n_2,n_3,
      S_4,S_5,j_3,T_2,T_4,T_5\}\,,
    \end{equation}
    and
    \begin{equation}
      \nu_y=\{n_5,\ell_5,S_2\}\,.
    \end{equation}
    Since we consider only parity conserving forces, the potential 
    does not change the quantum numbers $j_1$ and $T_2$. Moreover,
    the potential acts only on the pair 1,2 without changing the quantum
    numbers which involves other particles.
    Therefore, we have that in the matrix element $\nu_x=\nu_x'$ and
    \begin{equation}\label{eq:v12jj2}
      v^{KT,K'T',J\pi}_{l\nu,l'\nu'}
      =\delta_{\nu_x,\nu_x'}\sum_{T_{2z}}C_{T_3,T;T_3',T'}^{T_2,T_5;T_{2z}}
      v^{K,K',j_1}_{l\nu_y,l'\nu_y'}(T_{2z})\,,
    \end{equation}
    where
    \begin{equation}
    \begin{aligned}
      C_{T_3,T;T_3',T'}^{T_2,T_5;T_{2z}}&=
      \sum_{T_{3z}T'_{3z}}( T_3 T_{3z}, T_5 T_z-T_{3z}|T T_z)
      ( T'_3 T'_{3z}, T_5 T_z-T'_{3z}|T T_z)\\
      &\times( T_2 T_{2z}, T_5 T_{3z}-T_{2z}|T_3 T_{3z})
      ( T_2 T_{2z}, T_5 T'_{3z}-T_{2z}|T'_3 T'_{3z})\,.
    \end{aligned}
    \end{equation}
    The potential term depends only on the quantum numbers
    $K$, $K'$, $j_1$, $\nu_y$, $\nu'_y$  and the indexes of the Laguerre
    polynomials $l$ and $l'$.
    Explicitly the potential term reads
    \begin{equation}
    \begin{aligned}\label{eq:mxpint}
      &v^{K,K',j_1}_{l\nu_y,l'\nu_y'}(T_{2z})=
      {\cal N}_{n_5}^{\ell_5,\nu_5}{\cal N}_{n'_5}^{\ell'_5,\nu'_5}
      \int_0^\infty d\rho_5\,(\rho_5)^{11}\int_0^\infty dx_5\, (x_5)^2
      \int_0^\infty dx'_5\, (x'_5)^2
      \\
      &\qquad\qquad f_l(\rho)(\cos \ph_5)^{\ell_5}
      (\sin \ph_5)^{K_4}P_{n_5}^{\nu_4,\ell_5+1/2}(\cos 2\ph_5)
      v^{T_2T_{2z}}_{\ell_5S_2,\ell_5'S_2';j_1}(x_5,x_5')\delta_{S_2,S_2'}\\
      &\qquad\qquad f_{l'}(\rho')(\cos \ph'_5)^{\ell'_5}
      (\sin \ph'_5)^{K_4}P_{n'_5}^{\nu_4,\ell'_5+1/2}(\cos 2\ph'_5)\,,
    \end{aligned}
    \end{equation}
    where $\rho_5^2=x_1^2+x_2^2+x_3^2+x_4^2$,
    $\rho^2=\rho_5^2+x_5^2$, $(\rho')^2=\rho_5^2+(x_5')^2$ and
    $\cos \ph_5=x_5/\rho$, $\cos \ph_5'=x'_5/\rho'$.
    Here $v^{T_2T_{2z}}_{\ell S,\ell'S',j}(r,r')$ is the
    non-local two nucleon potential acting between
    two-body states ${}^{2S+1}(\ell)_j$
    and ${}^{2S'+1}(\ell')_j$ with isospin of the pair $T_2,T_{2z}$.
    The three integrals are easily computed
    numerically with high accuracy with standard quadrature techniques.
    The final form of the potential matrix element given in Eq.~(\ref{eq:mxaap})
    is
    \begin{equation}
    \begin{aligned}\label{eq:mxpfinal}
      V^{KLST,K'L'S'T',J\pi}_{l\al,l'\al'}&=15    
      \sum_\nu\sum_{\nu'}B^{KLSTJ\pi}_{\al,\nu}\;B^{K'L'S'T'J\pi}_{\al',\nu'}
      \\
      &\times
      \sum_{T_{2z}}C_{T_3,T;T_3',T'}^{T_2,T_5;T_{2z}}\;
      v^{K,K',j_1}_{l\nu_y,l'\nu'_y}(T_{2z})\delta_{\nu_x,\nu_x'}\,.
    \end{aligned}
    \end{equation}
  \end{subsection}
\end{section}
  
  \begin{section}{Technical details of the calculation}
    The biggest computational challenge for applying the HH
    formalism to the $A=6$ system is the calculation and the storage of 
      the potential matrix elements because of 
      the high number of basis states needed to reach convergence.
      In this section we present the main feature of the algorithm
      we use to compute
      the potential matrix elements exploiting the advantages of using the TC.
      The calculations are performed using a single
      node with 48 Intel Xenon
          8160 CPUs @2.10 GHz.
      
      Before starting to discuss the algorithm, let us give an idea of
      the dimension of the problem of computing the potential matrix elements
      in this formalism.
      We start from Eq.~(\ref{eq:mxpfinal}).
      The number of
      operations needed to compute the potential matrix element
      in Eq.~(\ref{eq:mxpfinal}) is given by $\sim N_\nu\times N_{\nu'}$
      where $N_\nu$($N_{\nu'}$) is the total number of TC in $jj$-coupling.
      Once fixed the values of the quantum number
      $\gamma=\{K,L,S,T\}$ and $\gamma'=\{K',L',S',T'\}$ the number of
      matrix elements we need to compute is $N_\gamma\times N_{\gamma'}$, where
      $N_{\gamma}\equiv M'_{KLSTJ\pi}$
      which is the total number of independent states
      for given $\gamma$ as defined in Section~\ref{sec:ortob}.
      Therefore, fixed $\gamma$ and $\gamma'$, the total number of operations
      in the computation is given by 
      \begin{equation}\label{eq:nop}
        N_{op}^{\gamma,\gamma'}\sim N_{\gamma}\times N_{\gamma'}\times N_\nu\times N_{\nu'}\,.
      \end{equation}
      If we consider $\gamma=\gamma'=\{12,2,1,0\}$ which is one of the worst cases,
      we have
      $N_{\gamma}= N_{\gamma'}\sim 10^3$ and
      $N_\nu=N_{\nu'}\sim 2.3 \times 10^6$ then 
      $N_{op}^{\gamma,\gamma'}\sim 5\times10^{18}$.
      Let us suppose we are in an ideal case in which the time required
      for any of these operations is the typical clock time of a computer,
      $10^{-9}$ s, and that we are able to use in parallel $10^3$ CPUs.
      The total time required for doing all these operations is
      \begin{equation}
        T^{\gamma,\gamma'}_{op}\sim 58\,\,{\rm days}\,,
      \end{equation}
      which is a time too long for any practical purpose, especially if we
      need to repeat these operations for all the possible combinations
      $\gamma,\gamma'$ and all the potential models we want to study.
      For this reason we introduce the coefficients $D$, as follows.

      As it can be seen from Eq.~(\ref{eq:mxpint}), the potential integrals
      $v^{K,K',j_1}_{l\nu_y,l'\nu'_y}(T_{2z})$ depend only on the index of the
      Laguerre polynomials and the
      quantum numbers $T_{2z}$, $K$, $K'$, $j_1$, $\nu_y$ and $\nu_y'$.
      Therefore, Eq.~(\ref{eq:mxpfinal}) can be rewritten in a more convenient
      form as
      \begin{align}\label{eq:mxpmodified}
        V^{KLST,K'L'S'T';J\pi}_{l\al,l'\al'}&=15    
        \sum_{\nu_y,\nu_y'}\sum_{T_3,T_3'}D^{KLST,K'L'S'T';J\pi}_{\al,\nu_yT_3;
          \al',\nu_y'T_3'}
        \sum_{T_{2z}}C_{T_3,T;T_3',T'}^{T_2,T_5;T_{2z}}
        v^{K,K',j_1}_{l\nu_y,l'\nu'_y}(T_{2z})\,,
      \end{align}
      where we denote
      $D^{KLST,K'L'S'T';J\pi}_{\al,\nu_yT_3;\al',\nu_y'T_3'}$ the $D$ coefficient
      and its expression can be easily derived
      comparing Eq.~(\ref{eq:mxpfinal}) with Eq.~(\ref{eq:mxpmodified}).
      Explicitly they are given by
      \begin{align}\label{eq:DDD}
        D^{KLST,K'L'S'T',J\pi}_{\al,\nu_yT_3;\al',\nu_y'T_3'}=\sum_{\nu_x\nu'_x}
        B^{KLSTJ\pi}_{\al,\nu_yT_3\nu_x}B^{K'L'S'T'J\pi}_{\al',\nu_y'T_3'\nu'_x}\delta_{\nu_x\nu'_x}\,.
      \end{align}
      In this way the only parts which depends
      on the nuclear interaction in Eq.~(\ref{eq:mxpmodified})
      are the potential integrals $v^{K,K'}_{l\nu_y,l'\nu'_y}(T_{2z})$,
      while the coefficients defined in
      Eq.~(\ref{eq:DDD}) do not.
      Therefore, we can compute and store the $D$ coefficients
      only once for all the potential models we want to consider. However, 
      the time required for the calculation of the $D$ coefficients
      is still of the order of $50$ days.

      In order to improve the
      time consumption, we noticed that for all the possible states
      $\alpha$ and $\alpha'$ with fixed $\gamma$ and $\gamma'$,
      the states $\nu$ and $\nu'$ to be coupled are always the same.
      Therefore, the determination of the couple of states $\nu,\nu'$ that
      fulfill the $\delta_{\nu_x\nu_x'}$,
      which in general requires $N_\nu\times N_{\nu'}$ operations,
      is performed only once for all the combinations $\alpha,\alpha'$ and
      requires a typical time of 10--20 minutes for $10\leq K,K'\leq14$
      using a single node with 48 CPUs working in parallel.
      The number of operations which remain to be done in Eq.~(\ref{eq:DDD})
      is then
      equal to the number of states $\nu_x$ ($N_{\nu_x}$). In such a way 
      $N_{op}^{\gamma,\gamma'}$       reduces to
      \begin{equation}
        N_{op}^{\gamma,\gamma'}\sim \underbrace{N_\gamma\times
          N_{\gamma'}\times N_V \times N_{\nu_x}}_{D\,{\rm op.}}
        +\underbrace{N_{\nu}\times N_{\nu'}}_{\delta_{\nu_x,\nu_x'}\,{\rm op.}}\,,
      \end{equation}
      where $N_V$ is the number of combinations $\nu_y,\nu_y'$ permitted
      by the potential and is typically $<200$.
      The $N^{\gamma,\gamma'}_{op}$ in this case
      is then orders of magnitude smaller than the value
      given in Eq.~(\ref{eq:nop}).
      Empirically we found that $N_V\times N_{\nu_x}\sim 10N_\nu$ and 
      considering the optimal computational
      situation described before, with 1000 CPUs working in parallel,
      in the case $\gamma=\gamma'=\{12,2,1,0\}$
      we obtain
      \begin{equation}
        T_{op}^{\gamma,\gamma'}\sim 23\,\,{\rm s}\,,
      \end{equation}
      which is an improvement of $\sim5$ orders of magnitude.
      Actually, in a realistic situation, the typical time required for the
      computation of Eq.~(\ref{eq:DDD}), namely 
      to perform the $N_V\times N_{\nu_x}$ operations,
      is $T_D\sim 0.1$ s. Therefore, for $\gamma=\gamma'=\{12,2,1,0\}$
      \begin{equation}
        T_{op}^{\gamma,\gamma'}\sim N_\gamma\times N_{\gamma'}
        \times T_D\sim 1\,\, {\rm day}\,,
      \end{equation}
      where in this case we used  only 48 CPUs on a single node.
      In Figure~\ref{fig:ttime} we report the total time needed to compute
      the $D$ coefficients when $L=L'=0$, $S=S'=1$ and $T=T'=0$
      up to given $K=K'=K_{max}$, namely
      \begin{equation}\label{eq:ttime}
        T_D=
        \sum_{K=2}^{K_{max}}\sum_{K'=2}^{K_{max}}T_{op}^{K010,K'010}\,.
      \end{equation}
      divided by the number of CPUs ($N_{CPU}$) used in the computation.
      In particular, the blue triangles give $T_D$ by using first
      the pre-identification of the pair of states $\nu,\nu'$ to fulfill the
      $\delta_{\nu_x,\nu_x'}$ condition as discussed before,
      while the red dots correspond
      to the time spent without pre-identification.
      As it is clear from the figure,
      the computational time increases exponentially by increasing
      the values of $K_{max}$,  since it is
      proportional to the number of independent states which grows
      exponentially as well (see Figure~\ref{fig:mk0101}).
      However, by using the pre-identification, not only $T_D$ results
      to be well reduced, but also as function of $K_{max}$ it has
      a minor slope compared to the case without pre-identification.
      The exponential growth limits the maximum value of $K_{max}$
      we can use at present. However, we expect to have a great improvement
      by using a larger number of CPUs distributing the
      calculation on several nodes.
      \begin{figure}[h]
        \centering
        \includegraphics[scale=.8]{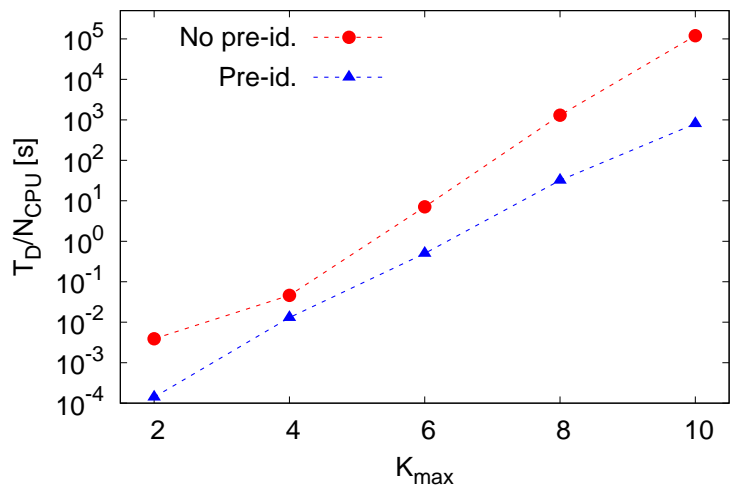}
        \caption{ Total time  needed to compute
          the $D$ coefficients [Eq.~(\ref{eq:ttime})] as function of $K=K'=K_{max}$
          divided by $N_{CPU}$, the number of CPUs used in the computation.
          These calculations are performed for fixed values of the other
          quantum numbers, in particular $L=L'=0$, $S=S'=1$ and $T=T'=0$.
          The red dots are the time spent without pre-identification,
          while the blue triangles are the time spent with pre-identification.
          The dashed lines are added to guide the eyes.
          The calculations were performed on a single node with 48 Intel Xenon
          8160 CPUs @ 2.10 GHz (i.e. $N_{CPU}=48$).}
        \label{fig:ttime}
      \end{figure}

      As regarding the storage, the total memory required fixed $\gamma$ and
      $\gamma'$ is given by the number of $D$ coefficient for each
      $\alpha,\alpha'$ combination, namely
      \begin{equation}
        M_{\gamma,\gamma'}[{\rm GB}]\sim 3\times\frac{8\times N_\gamma\times N_{\gamma'}
          \times N_V}{1024^3}\,,
      \end{equation}
      where the factor $3$ is an empirical factor
      which takes care of the additional information needed
      in the files to save the coefficients $D$.
      For example, when $\gamma=\gamma'=\{12,2,1,0\}$,
      the size of the file is only 2.2 GB.
      The total memory we used to store all the $D$ coefficients used for
      computing the $\Li$ ground state in this work is $\sim 100$ GB.
      Once computed the $D$ coefficients, the time required for the
      calculation of all the potential matrix elements is of the order of a
      couple of hours. Indeed we need only to compute the sum over the
      combinations $\nu_y,\nu_y'$ allowed by the potential ($N_V$), which
      are very few.
      
      Typically, in the
      {\it ab-initio} methods, the potential matrix elements are computed
      and stored for each potential model. This requires large amount of
      time and disk space, limiting the number of models that can be used 
      and tested. On the other hand, by  using our approach, we are able
      to save only the $D$ coefficients by eliminating the dependence on
      the potential models.
      In this sense, the HH method, in the context of
      the {\it ab-initio} approaches, can act as an efficient and precise method
      for testing various potential models up to $A=6$.

  \end{section}

%% file: Chapter2_v3.tex
\begin{chapter}{The ${}^6$Li ground state within the Hyperspherical Harmonic basis}\label{ch:litio}

In this chapter we present the results obtained for the $\Li$ ground state
computed as discussed in the previous chapter. 
The potential models we use and their major features are briefly introduced in
Section~\ref{sec:nupot}.
Section~\ref{sec:statesel} is dedicated to present the selection of the
HH basis states in order to obtain the best description of the $\Li$ wave function.
Finally in Section~\ref{sec:lithres} we present the
convergence of the HH expansion for the binding energy (BE) and the study
of the electromagnetic structure of the $\Li$ ground state.

\begin{section}{The nuclear potential}\label{sec:nupot}

In all {\it ab-initio} approaches the goal is to derive
the properties of nuclei from fundamental theories.
In order to do that the nuclear potential that appears in the Hamiltonian
must be connected directly to the QCD Lagrangian.
As we already discussed in Chapter~\ref{ch:intro}, 
the $\chi$EFT approach permits to construct an effective Lagrangian
  constrained by chiral symmetry and to organize it as an expansion
  in power of the exchanged pion momentum $Q$.
 From the Lagrangian, by performing a non-relativistic expansion,
it is then possible to  derive the nuclear potential which can be organized 
as well in a expansion in power of $Q$.
In particular, modern $\chi$EFT potentials start
from order $Q^0$, called leading-order (LO),
and then continues with the next-to-leading-order
(NLO) of order $Q^2$, the next-to-next-to-leading-order (N2LO or NNLO) of order $Q^3$,
the next-to-next-to-next-to-leading-order (N3LO) of order $Q^4$, and in some cases
they also take into account
next-to-next-to-next-to-next-to-leading-order (N4LO) terms of order $Q^5$.
In the last years, several nuclear potentials have been constructed starting from
the $\chi$EFT Lagrangian  providing
a practical and systematic scheme to derive two
and many-nucleon interactions~\cite{Epelbaum2015,Entem2015}.
Moreover, since $\chi$EFT is a low-energy theory, it is needed to introduce a cut-off
$\Lambda_C$ in order to regularize the potential.
Each term of the potential is multiplied by a LEC which is determined by reproducing
$NN$ and three-nucleon data.
The values of the LECs clearly will depend on the cut-off $\Lambda_C$
but the observables should not. Therefore, a proof of the validity of the chiral expansion
is that, by increasing the chiral
order of the expansion, the values of the observables are less and less dependent
on the cut-off.
In the future we plan to check this fact on $A=6$ observables.

However, the solution of the $A=6$ Schr\"odinger equation with these potentials
is very difficult because of the ``hard'' repulsive core due to the Pauli principle
among the quarks confined in the nucleons, which appears
for internucleon distance less than 1 fm.
In the Schr\"odinger equation the ``hard'' core gives rise to short-range
correlations. This requires to construct the wave function with
a large set of basis states, in order to reproduce
accurately all these correlations.
The required basis size is so large for $A=6$ that results very hard
to use ``bare'' potentials and reach an acceptable convergence.
Only recently, using a softer version of the NNLO potential
(NNLO$_{opt}$~\cite{Ekstrom2013})
and a huge computational power, an acceptable convergence has been obtained
(see Ref.~\cite{Forssen2018}).
However, the short-range correlations can be eliminated
decoupling high-momentum (short-range) from low-momentum (long-range) physics,
by using the Similarity Renormalization Group (SRG) unitary transformation
~\cite{Bogner2007}.
This transformation suppresses the off-diagonal matrix elements,
the ones that connect high-momentum to low-momentum physics, moving
the Hamiltonian towards a band-diagonal form. Let us give an example of
how the SRG transformation works as reported in Ref.~\cite{Bogner2007}.
The initial Hamiltonian $H=T_{rel}+V$ in the cm reference frame,
where $T_{rel}$ is the relative kinetic energy, is transformed by the unitary
operator $U(s)$ according to
\begin{equation}\label{eq:hs}
  H_s=U(s)HU^\dag(s)=T_{rel}+V_s\,,
\end{equation}
where $s$ is the flow parameter and $V_s$ is the evolved potential.
$H_s$ evolves following the relation
\begin{equation}
  \frac{dH_s}{ds}=\left[\eta(s),H_s\right]\,,
\end{equation}
with
\begin{equation}
  \eta(s)=\frac{dU(s)}{ds}U^\dag(s)=-\eta^\dag(s)\,.
\end{equation}
The selection of $\eta(s)$ determines the transformation.
Following Ref.~\cite{Bogner2007}, we take
\begin{equation}\label{eq:etas}
  \eta(s)=\left[T_{rel},H_s\right]\,.
\end{equation}
With this choice $T_{rel}$ results independent on $s$ in Eq.~(\ref{eq:hs}).
For example, using Eq.~(\ref{eq:etas}), the two-body potential evolves as
\begin{equation}\label{eq:vkk}
  \frac{dV_s(k,k')}{ds}=-(k^2-{k'}^{2})^2V_s(k,k')+\frac{2}{\pi}
  \int_0^\infty q^2dq(k^2+{k'}^{2}-2q^2)V_s(k,q)V_s(q,k')\,,
\end{equation}
where $V_s(k,k')$ is the non-local two-body potential, $k$ is the relative momentum
of the two particles and
with normalization so that $1=\frac{2}{\pi}\int|q\ket q^2 \bra q | dq$
in units where $\hbar^2/m=1$. From Eq.~(\ref{eq:vkk}),
if we do not consider the integral part, we obtain
\begin{equation}
  V_s(k,k')\propto {\rm e}^{-(k^2-{k'}^2)^2s}\,,
\end{equation}
from which it is clear that
the off-diagonal matrix elements are exponentially suppressed
when $s$ increases. The parameter $\Lambda=s^{-1/4}$ provides a
measure of the spread of off-diagonal strength.
The suppression of off-diagonal matrix elements decrease the size
of the HH basis needed for convergence.
However, the SRG transformation
on two-body forces ($V_{ij}^\Lambda=V_s$)
induces three-body forces ($V_{ijk}^\Lambda(ind)$),
the transformation of the three-body forces ($V_{ijk}^\Lambda$) induces four-body forces
($V_{ijkl}^\Lambda(ind)$) and so on~\cite{Bogner2007},
therefore the Hamiltonian we need to diagonalize becomes
\begin{equation}
  H_\Lambda=T_{rel}+\sum_{i<j}V^\Lambda_{ij}+
  \sum_{i<j<k}V^\Lambda_{ijk}(ind)+\sum_{i<j<k}V^\Lambda_{ijk}+\dots\,,
\end{equation}
where the dots stands for $n$-body forces with $N>3$. It has been shown 
that $(n+1)$-body forces are less important than the $n$-body ones~\cite{Jurgenson2011}.
Therefore we can safely discard the $n$-body forces with $N>4$.
 Moreover, in our calculation, we will not include
  three-body forces but, in order to minimize their possible effects,
  we will use values of $\Lambda$ such that
  our theoretical results will be close
  to the experimental values. In particular, we choose values of $\Lambda$ so that
  the calculated binding energy of ${}^3$He, ${}^3$H and ${}^4$He (without the inclusion
  of bare and induced three-nucleon forces) are close to their experimental values.
However, by performing this approximations we pay the price of
losing the unitarity properties of our solution.
In this Thesis, we will consider the N3LO500 chiral potential~\cite{Entem2003}
(where N3LO is the order of chiral expansion and 500 indicates that $\Lambda_C=500$ MeV)
SRG evolved with three different values of the flow parameter $\Lambda=1.2$,
1.5 and 1.8 fm$^{-1}$, which we will indicate with SRG1.2, SRG1.5 and SRG1.8,
respectively.

The wave function we  evaluate with these potentials is not the
``bare'' wave function. Because of the SRG unitary transformation we have that
\begin{equation}
  |\psi(s)\ket=U(s)|\psi_{\text{bare}}\ket\,,
\end{equation}
where $|\psi(s)\ket$ is the calculated wave function and
$|\psi_{\text{bare}}\ket$ is the ``bare'' wave function. The mean value
of an operator $O$ (we named it ``bare'') can be written as
\begin{equation}
  \bra O \ket=\bra\psi_{\text{bare}}|O|\psi_{\text{bare}}\ket
  = \bra\psi(s)|O(s)|\psi(s)\ket\,,
\end{equation}
where $O(s)=U(s)OU(s)^\dag$ is the SRG evolved operator.
This transformation is not trivial and up to
now  available only for few simple
operators (see for example Ref.~\cite{Schuster2014}).
However, it is expected from the previous discussion
that the operator $U$ would mainly take into account the short-range correlations
between nucleons. It has been argued that long-range operators would not be
affected by it, and therefore $O\approx O(s)$~\cite{Stetcu2005}. 
For example, square radius and electric quadrupole moment
operators can be considered as long-range operators.
Therefore, for these we will assume that
\begin{equation}\label{eq:srgapprox}
  \bra O \ket
  \approx \bra\psi(s)|O|\psi(s)\ket\,,
\end{equation}
where we use the ``bare'' operator instead of the evolved one.
The more this approximation is valid, the less will be the
dependence on the parameter $s$. 

In our study of $\Li$, we will use also the chiral potential NNLO$_{sat}$
~\cite{Ekstrom2015}. For this potential,
the LECs of the two- and three-body interactions
are fitted together to reproduce
the NN scattering data and also some $A>2$ nuclear properties,
like the binding energies and charge radii of carbon and oxygen isotopes.
The chiral potential constructed in such a way results ``softer''
than the standard chiral potentials, making possible to
perform reliable calculations without using the SRG evolution.
Even if the NNLO$_{sat}$ requires
three-body forces to reproduce the physical values of the observables,
we will not include them in our calculation.
Indeed, we will use this potential only as a test
for possible future application of the HH method with other ``bare''   
chiral potentials. In the following, we will indicate
  with NNLO$_{sat}$(NN) the
NNLO$_{sat}$ interaction used without considering three-body forces.
\end{section}
\begin{section}{Selection of the states}\label{sec:statesel}

The main difficulty of the HH method is the selection of
a subset of basis states which permits to have the best description
of the nuclear states we consider. Indeed, although the number of
independent states is much smaller than the degeneracy of the
basis, a brute force approach of the method, that is the inclusion
of all the HH-states having $K\leq K_M$ and then increasing $K_M$
until convergence, would be destined to fail. Moreover, as discussed previously,
it is very difficult to find all the linearly independent states already
for values of $K\geq10$, because of the loss of precision in the
orthogonalization procedure. For this reason a good
selection of a restricted and effective subset of basis state
is fundamental. Up to now we are limited to values $K_M\leq14$
which permits to reach a reasonable convergence only for the SRG evolved
and the NNLO$_{sat}$(NN) potentials.

It is convenient to separate the HH functions into classes taking
into account their properties and the fact that the convergence rate
of each class results to be rather different.
The first selection can be done considering the quantity 
$\ell_{\text{sum}}=\ell_1+\ell_2+\ell_3+\ell_4+\ell_5$. Indeed, the HH states
with large $\ell_{\text{sum}}$ are less important
because of the centrifugal barrier. The SRG potential and the
NNLO$_{sat}$(NN) are rather soft
and so, in our calculation, we can consider
only states with $\ell_{\text{sum}}\leq4$.
The second selection can be done considering the number of particles
correlated by the HH functions. The nuclear potential favors the two-body
correlations, therefore the HH states which depend only on the
coordinate of pair of particles would give important contributions,
namely the HH states with only $n_5$ and $\ell_5$ not zero.
For technical reasons related to the construction of the TC, we are not
able to work with classes defined selecting particular values of $n_2-n_5$.
Therefore we use only the criteria on the $\ell_i$ for the class definition.
Moreover, we can divide the $\Li$ ground state in
$LST$ components. The components allowed by the total
spin of $\Li$ ground state $J^\pi=1^+$  are give in Table~\ref{tab:LSTgs}.
Being the $\Li$ ground state an almost pure $T=0$ state, we do
not consider other isospin states.
\begin{table}
  \centering
  \begin{tabular}{cccc}
    \hline
    \hline
    $L$ & $S$ & $T$& ${}^{2S+1}L_J$\\
    \hline
    0 & 1 & 0 & ${}^3S_1$\\
    2 & 1 & 0 & ${}^3D_1$\\
    2 & 2 & 0 & ${}^5D_1$\\
    2 & 3 & 0 & ${}^7D_1$\\
    1 & 0 & 0 & ${}^1P_1$\\
    1 & 1 & 0 & ${}^3P_1$\\
    1 & 2 & 0 & ${}^5P_1$\\
    3 & 2 & 0 & ${}^5F_1$\\
    3 & 3 & 0 & ${}^7F_1$\\
    4 & 3 & 0 & ${}^7G_1$\\
    \hline
    \hline
  \end{tabular}
  \caption{\label{tab:LSTgs}$LST$ components 
    of the ground state wave function of $\Li$.
    In the spectroscopic notation the isospin is neglected since
    we consider $T=0$ states only.}
\end{table}

To study the $\Li$ ground state, we find very convenient to chose the
classes as described below.
\begin{itemize}
\item[a.] Class C1. In this class we include the HH states such that
  $\ell_{\text{sum}}=0$, which belong only to the wave component ${}^3S_1$.
  This class represents the main component of the $\Li$ wave function,
  and in order to obtain  convergence we include
  states up to $K_{1M}=14$.
\item[b.] Class C2. In this class we include the HH states such that
  $\ell_5=2$ and $\sum_{i=1,4}\ell_i=0$. This class contains channels belonging
  to all the $D$ waves and  its contribution is fundamental
  to obtain a bound $\Li$.
  For this class we include states up to $K_{2M}=12$.
\item[c.] Class C3. This class includes all the HH states that
  belong to ${}^3S_1$ component with $\ell_{\text{sum}}=2$.
  This class contains only many-body correlations. Therefore, its 
  impact on the binding energy is less significant.
  For this class we include states up to $K_{3M}=10$.
\item[d.] Class C4. This class includes all the HH states that
  belong to the $D$ components and are not included in class C2.
  As class C3, this class contains HH states with $\ell_{\text{sum}}=2$
  and only many-body correlations, therefore
  we expect a similar convergence.
  For this class we include states up to $K_{4M}=10$.
\item[e.] Class C5. This class includes all the
  independent HH states which
  belong to the $P$ components up to $K_{5M}=8$
  (states with $\ell_{sum}=2$ only appears).
  We stop to $K_{5M}=8$ since the contribution of the $P$
  waves to the binding energy is quite tiny.
\item[f.]Class C6. This class includes all the
  independent HH states which
  belong to the $F$ and $G$  components up to $K_{5M}=8$
  (states with $\ell_{sum}=4$ only appears).
  We stop to $K_{6M}=8$ since the contribution of the $F$ and $G$
  waves to the binding energy is very tiny.
\end{itemize}  

The convergence is studied as follows. First, only the states
of class C1 with $K\leq K_{1}$ are included in the expansion and
the convergence of the BE is  studied as the value of $K_1$ is increased
up to $K_{1M}$.
Once a satisfactory convergence for the first class is reached, the states
of the second class with $K\leq K_2$ are added to the expansion
keeping all the states of the class C1 with $K\leq K_{1M}$. The
procedure is then repeated for each new class.
Our complete calculation includes about 7000 HH states.
\end{section}

\begin{section}{Results for the $\Li$ ground state}\label{sec:lithres}
In this section we report the results obtained for the ground state
of $\Li$. The calculation are performed using the N3LO500-SRG$\Lambda$
with $\Lambda=1.2$, $1.5$, $1.8$ fm$^{-1}$. The Coulomb interaction
is included as ``bare'' (i.e. not SRG evolved). Moreover, we use also
the NNLO$_{sat}$(NN) considering only the two-body forces. In this work we have
$\hbar^2/m=41.47$ MeV fm$^2$ for all the potentials. Moreover,
we use $\gamma=4$ fm$^{-1}$ in the hyperradial functions
[see Eq.~(\ref{eq:fllag})] which is the
optimal value for the parameter $\gamma$ in order to reach convergence
on the third decimal digit with a number of Laguerre polynomials $l_{max}=16$.
For all the considered  models,
when the angular momentum of the pair $j$ is large the  NN interaction
becomes very weak.
Therefore, all the interactions for $j>6$ are discarded, since their
effects are negligible as it was already shown in Ref.~\cite{Viviani2005} for
the $\alpha$ particle.
In Appendix~\ref{app:validation} we discuss with more detail
the selection of the parameters $\gamma$ and $l_{max}$ and also the
validation of our approach by comparing our results with the results
obtained with the non-symmetrized HH (NSHH) method in Ref.~\cite{Gattobigio2011}, using
a test potential.

This section is divided in two parts. In the first one, Section~\ref{sec:convHH},
we discuss
the convergence of the HH expansion in terms of the various classes.
In the second part, Section~\ref{sec:emobs},
we examine the electromagnetic static properties of $\Li$.

\begin{subsection}{Convergence of the HH expansion}\label{sec:convHH}

We study the convergence as explained in the previous section, and the
results presented are arranged accordingly. For example in Table~\ref{tab:Kconv},
the BE reported in a row with a given set of values $K_1,\dots,K_6$
has been obtained by including in the expansion all the HH functions
of class C$i$ with $K\leq K_{i}$ $i=1,\dots,6$.
\begin{table}
  \centering
  \begin{tabular}{cccccccccc}
    \hline
    \hline
    &  &  &  &  &  &\multicolumn{3}{c}{N3LO500-SRG$\Lambda$} & \\
    \cline{7-9}
    $K_1$ & $K_2$ & $K_3$ & $K_4$ & $K_5$ & $K_6$ & $1.2$ fm$^{-1}$
    &$1.5$ fm$^{-1}$ &$1.8$ fm$^{-1}$ & NNLO$_{sat}$(NN)\\
    \hline
    $\phz$2  &    &    &    &  &   & 24.779 & 22.315 & 17.946 & $\phz9.188$\\
    $\phz$4  &    &    &    &  &   & 28.606 & 26.779 & 22.656 & 13.712\\
    $\phz$6  &    &    &    &  &   & 29.714 & 28.395 & 24.646 & 16.407\\
    $\phz$8  &    &    &    &  &   & 30.030 & 28.937 & 25.425 & 17.760\\
    10 &    &     &     &   &  & 30.150 & 29.159 & 25.781 & 18.398\\
    12 &    &     &     &   &  & 30.195 & 29.254 & 25.948 & 18.736\\
    14 &    &     &     &   &  & 30.213 & 29.295 & 26.031 & 18.931\\
       &    &     &     &   &  &        &        &        &       \\
    14 & $\phz$2  &    &    &  &   & 30.263 & 29.362 & 26.108 & 18.997\\
    14 & $\phz$4  &    &    &  &   & 30.900 & 30.481 & 27.619 & 20.828\\
    14 & $\phz$6  &    &    &  &   & 31.318 & 31.626 & 29.819 & 24.557\\
    14 & $\phz$8  &    &    &  &   & 31.413 & 32.006 & 30.827 & 26.843\\
    14 & 10 &    &    &   & & 31.437 & 32.122 & 31.195 & 27.880\\
    14 & 12 &    &    &   & & 31.444 & 32.167 & 31.352 & 28.361\\
       &    &    &    &   &   &     &        &        &       \\
    14 & 12 & $\phz$6  &    &  &  & 31.445 & 32.168 & 31.354 & 28.381\\
    14 & 12 & $\phz$8  &    &  &  & 31.477 & 32.210 & 31.396 & 28.425\\
    14 & 12 & 10 &    &  &   & 31.493 & 32.233 & 31.422 & 28.459\\
       &    &    &    &  &   &        &        &        &       \\
    14 & 12 & 10 & $\phz$4  & &  & 31.501 & 32.245 & 31.437 & 28.467\\
    14 & 12 & 10 & $\phz$6  & &  & 31.550 & 32.329 & 31.548 & 28.588\\
    14 & 12 & 10 & $\phz$8  & &  & 31.577 & 32.389 & 31.642 & 28.738\\
    14 & 12 & 10 & 10 & &   & 31.586 & 32.412 & 31.689 & 28.832\\
       &    &    &    & &  &        &        &        &       \\
    14 & 12 & 10 & 10 & 2 &  & 31.658 & 32.533 & 31.836 & 29.079\\
    14 & 12 & 10 & 10 & 4 &  & 31.710 & 32.631 & 31.970 & 29.394\\
    14 & 12 & 10 & 10 & 6 &  & 31.728 & 32.677 & 32.047 & 29.629\\
    14 & 12 & 10 & 10 & 8 &  & 31.735 & 32.699 & 32.093 & 29.771\\
    &    &    &    & &  &        &        &        &       \\
    14 & 12 & 10 & 10 & 8 & 4  & 31.736 & 32.703 & 32.101 & 29.790\\
    14 & 12 & 10 & 10 & 8 & 6  & 31.746 & 32.733 & 32.161 & 29.896\\
    14 & 12 & 10 & 10 & 8 & 8  & 31.750 & 32.751 & 32.209 & 30.008\\    
    \hline
    \hline
  \end{tabular}
  \caption{\label{tab:Kconv}Convergence of $\Li$ binding energies (MeV)
    corresponding to the inclusion in the wave function of the different
    classes C1--C6 in which the HH basis has been divided. The SRG-evolved
    potential N3LO500-SRG$\Lambda$ correspond to $\Lambda=1.2$, $1.5$ and $1.8$
    fm$^{-1}$. We show also the results for the NNLO$_{sat}$(NN) potential model.}
\end{table}

We can now analyze the results in Table~\ref{tab:Kconv}.
For the SRG evolved potentials we observe that  classes C1 and C2
are the most important and have the  slowest convergence. Indeed the largest
values of $K$ must be reached. It is evident that
increasing the value of the SRG parameter $\Lambda$, the convergence
becomes slower. This is due to the ``hardness'' of the potential that is
enhanced when $\Lambda$ is large. Moreover,
class C2 becomes less and less significant when $\Lambda$ becomes smaller.
This
effect is generated by the SRG evolution, which reduces the correlations
between the $S$ wave (class C1) and the $D$ waves (class C2), when $\Lambda$
decreases. Even if classes C1 and C2
are the slowest converging classes, they give
$98\%$ of the BE.
The contribution of classes C3, C4  is very small for all the
values of the flow parameters $\Lambda$ and also the convergence is much faster.
It is very interesting to observe that for both classes the
contribution to the BE depends less on the value of $\Lambda$
compared to that of classes C1 and C2. This gives
an indication that the many-body correlations are not very important 
for the SRG evolved potentials.
We find also that classes C5, which corresponds to
$P$ waves, and C6, which corresponds to $F$ and $G$ waves,
give a very small contributions to the ground state of the $\Li$.
Moreover, also the convergence
is very fast: in order to obtain the same convergence of the other classes,
we can stop for $K_{5M}=K_{6M}=8$. 

For the NNLO$_{sat}$(NN), even if class C1 gives still
the main contribution to the BE,
its convergence results faster than the convergence of
class C2, which indicates the prominence of the $D$ waves in the case
of this potential. 
Indeed, the contribution of the $D$ waves results
almost 1/3 of the total BE. Together,  classes C1 and C2 give 94$\%$
of the total BE.
It is clear, comparing the columns of Table~\ref{tab:Kconv} for these
classes with the ones of the SRG cases,
that we are not able to reach the same convergence of the SRG potentials.
Also classes C3 and C4  result to have a slower convergence
compare to the SRG potentials. However their contribution to
the BE does not increase dramatically as in the case of class C2.
This indicates that also
for NNLO$_{sat}$(NN) many-body correlations are much less
important. The $P$ waves (class C5) instead result to be more
significant for the NNLO$_{sat}$(NN) than for the SRG potentials.
Indeed, not only its contribution to
the BE is almost $1$ MeV, but also the convergence rate is slower compared to the
SRG case, indicating a stronger correlation of the $P$ waves
with the $S$ and $D$ waves.
Also in the case of class C6 the contribution to the BE is larger compared
to SRG potentials
which indicates a relevant presence of $F$ and $G$ waves for this potential.

Let us comment about the convergence rate of the expansion as  function
of the maximum grand angular quantum number $K_{iM}$ of the various classes of
HH states included in our expansion.
As shown in various studies~\cite{Zakhareyev1969,Schneider1972,Demin1977,Fabre1983},
the convergence of the HH functions
towards the exact BE depends primarily on the form of the potential.
For the chiral potentials, it was observed empirically that
the convergence rate has an exponential behavior as $K_{iM}$ increases.
We expect that the same rate of the convergence is obtained also for the
SRG evolved potentials as
already observed for example in Ref.~\cite{Jurgenson2011}.

To study the convergence behavior, we indicate with $B(K_1,K_2,K_3,K_4,K_5,K_6)$
the BE obtained by including in the expansion all the HH states of
class C1 with $K\leq K_{1}$, all the HH states of class C2 having $K\leq K_{2}$
and so on. Let us define
\begin{align}
  \Delta_1(K)&=B(K,K_{2M},0,K_{4M},K_{5M},K_{6M})-B(K-2,K_{2M},0,K_{4M},K_{5M},K_{6M})\,,\label{eq:d1}\\
  \Delta_2(K)&=B(K_{1M},K,K_{3M},0,K_{5M},K_{6M})-B(K_{1M},K-2,K_{3M},0,K_{5M},K_{6M})\,,\label{eq:d2}\\
  \Delta_3(K)&=B(K_{1M},K_{2M},K,K_{4M},K_{5M},K_{6M})\nonumber\\
  &\qquad\qquad\qquad\qquad\qquad
  -B(K_{1M},K_{2M},K-2,K_{4M},K_{5M},K_{6M})\,,\label{eq:d3}\\
  \Delta_4(K)&=B(K_{1M},K_{2M},K_{3M},K,K_{5M},K_{6M})\nonumber\\
  &\qquad\qquad\qquad\qquad\qquad
  -B(K_{1M},K_{2M},K_{3M},K-2,K_{5M},K_{6M})\,,\label{eq:d4}\\
  \Delta_5(K)&=B(K_{1M},K_{2M},K_{3M},K_{4M},K,K_{6M})\nonumber\\
  &\qquad\qquad\qquad\qquad\qquad
  -B(K_{1M},K_{2M},K_{3M},K_{4M},K-2,K_{6M})\,,\label{eq:d5}\\
  \Delta_6(K)&=B(K_{1M},K_{2M},K_{3M},K_{4M},K_{5M},K)\nonumber\\
  &\qquad\qquad\qquad\qquad\qquad
  -B(K_{1M},K_{2M},K_{3M},K_{4M},K_{5M},K-2)\,,\label{eq:d6}
\end{align}
where with $K_{iM}$ we indicate that, for the class C$i$, we are including all the
HH states up to the maximum $K$ considered in this work.
With these definitions, we can compute the ``missing'' energy for each class
due to the truncation of the expansion up to a given $K_{iM}$,
by taking care of the modification of the convergence of a
class C$i$ due to the inclusion of the other classes.
To be noticed that for $\Delta_1$($\Delta_2$) we put $K_3=0$($K_4=0$).
This is because the HH states included in class C3(C4) cannot be added to
the basis without adding before class C1(C2) due to the orthogonalization
procedure. For example, we cannot add the HH states of class C3 with $K_3=6$
without adding before the HH states of class C1 with $K_1=6$. Therefore, to have a clear
convergence pattern for class C1(C2), we studied it without adding class C3(C4).
The changes in the convergence pattern of  class C1(C2) due to the coupling
with the class C3(C4) are in any case negligible, since class
C3(C4) gives a very small contribution to the total BE.

In Figure~\ref{fig:Kconv} we plot the values of $\Delta_1$, $\Delta_2$ and $\Delta_5$ for
the four potential models considered. By inspecting the figure, we can see
a clear exponential
decreasing behavior of the $\Delta_i$ as function of $K$,
even if  the values of $K$ we used are rather small. 
In particular, we can assume for each class that
\begin{equation}\label{eq:fitb}
  B_i(K)=B_i(\infty)+a_i\,{\rm e}^{-b_iK}\,,
\end{equation}
where $B_i(\infty)$ is the asymptotic BE of the class C$i$ for $K\rightarrow\infty$,
while $a_i$ and $b_i$ are  parameters which depend on the potential and on
the class of the HH functions we are studying. In particular, the
parameter $b_i$ indicates the  convergence rate of the class C$i$.
From Eq.~(\ref{eq:fitb}) we obtain
\begin{equation}\label{eq:deltai}
  \Delta_i(K)=a_i{\rm e}^{-b_i K}\left(1-{\rm e}^{2b_i}\right)\,,
\end{equation}
which is used  for fitting  $\Delta_i(K)$. The results of the fits
are the dashed lines in Figure~\ref{fig:Kconv}.
By observing Figs.~\ref{fig:Kconv}a--\ref{fig:Kconv}c
in which we study the SRG evolved potentials,
it is clear that the convergence rate
diminishes by increasing the values of $\Lambda$.
It is also interesting to observe that for $\Lambda=1.2$ fm$^{-1}$ we have
$\Delta_1(K)>\Delta_2(K)$, for $\Lambda=1.5$ fm$^{-1}$ we have
$\Delta_1(K)\approx\Delta_2(K)$ while for $\Lambda=1.8$ fm$^{-1}$ we have
$\Delta_1(K)<\Delta_2(K)$, which confirms the increasing importance of the
tensor term of the potential which correlates the $S$ and $D$ waves
by increasing $\Lambda$.
Moreover, for all the values of the flow parameter $\Lambda$, we find
$\Delta_5(K)<\Delta_1(K),\Delta_2(K)$, confirming the small contribution
of the $P$ waves to the BE.
For the NNLO$_{sat}$(NN) (Fig.~\ref{fig:Kconv}d)
we find a situation very similar to the case
of $\Lambda=1.8$ fm$^{-1}$ in which we have $\Delta_1(K)<\Delta_2(K)$,
even if in this case the convergence is much slower. Also the $P$
waves (class C5) result to have a more important role in this case,
as demonstrate by the fact
that from the fit $\Delta_5(K)\approx\Delta_1(K)$ when $K=14$.
\begin{figure}[h]
  \centering
  \includegraphics[width=\linewidth]{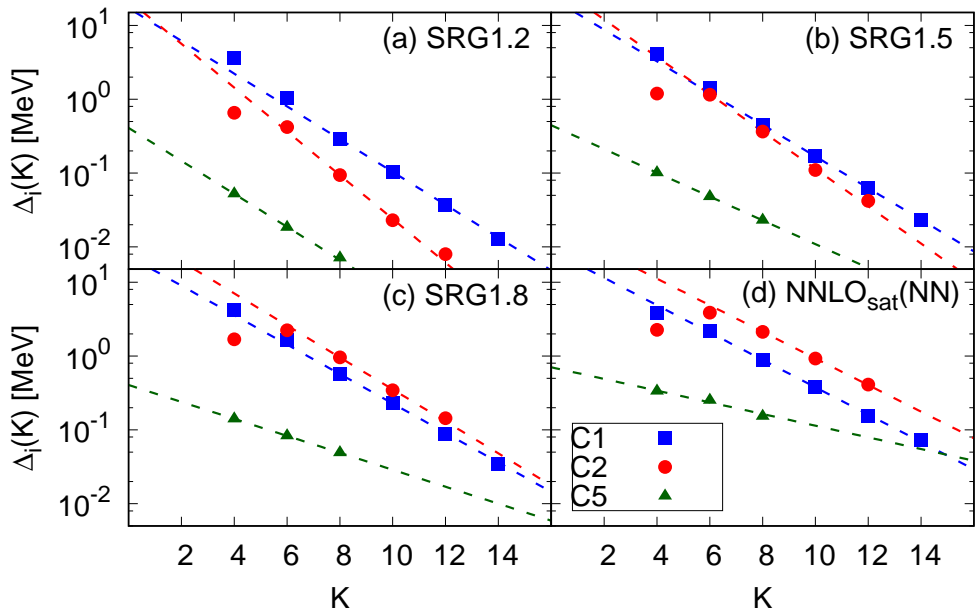}
  \caption{ Binding energy differences for the $\Li$ for the
  classes C1 (squares), C2 (circles), and C5 (triangles) as function
  of the grand angular value $K$ for the four potential models considered
  (see the text for more details).
  The dashed lines are the fits obtained using Eq.~(\ref{eq:deltai}).}
  \label{fig:Kconv}
\end{figure}

This effects can be seen also by comparing the values of $b_i$ obtained by
the fits and reported in Table~\ref{tab:missingbe}. 
For classes C1, C2 and C5 in the case of the SRG evolved potentials,
the values of $b_i$ decrease when $\Lambda$ grows. This indicates
a more and more repulsive core of the potential when $\Lambda$ increases.
Even smaller values of $b_i$ are obtained for the NNLO$_{sat}$(NN) potential since it has
stronger short-range correlations.

For the classes C3, C4 and C6, the calculated values of $\Delta_i(K)$ are not
enough to perform a fit and so we extract the parameter $b_i$ by using
only the last two values of $\Delta_i$, namely
\begin{equation}\label{eq:deltai2}
  \frac{\Delta_i(K_{iM}-2)}{\Delta_i(K_{iM})}={\rm e}^{2b_i}\,.
\end{equation}
This formula gives only a rough estimate of the convergence rate.
Therefore, we cannot compare the values of $b_i$ for different potentials.
The obtained values of $b_i$ are reported in Table~\ref{tab:missingbe} as well,
except for the class C6 in the cases of NNLO$_{sat}$(NN) potential
for which the extrapolation was not possible.
We will take care of it by considering larger errors on the extrapolations.

Before discussing the calculation of the ``missing'' energy, we
want to underline that Eq.~(\ref{eq:deltai})
represents the asymptotic behavior of the convergence pattern when $K$ is large,
while we are using value of $\Delta_i$ computed for not so large values of $K$.
For this reason, for the final fit of class C1 and C2
we used only  $\Delta_i(K)$ with $K\geq8$.
Indeed, in Figure~\ref{fig:Kconv} it is possible to observe that, for $K\leq4$, 
$\Delta_i$ deviates from the fit. This is usual for the convergence of HH
states, as already observed in the case of the $\alpha$ particle
in Ref.~\cite{Viviani2005},
and is due to the fact that for small values of $K$  the number of states
are not enough to give a good description of the wave function.

The ``missing'' BE due to the truncation of the expansion for each class
to finite values of $K = K_{iM}$ can be defined as in Ref.~\cite{Viviani2005}
\begin{equation}\label{eq:dbi}
  \left(\Delta B\right)_i=\sum_{K= K_{iM}+2,K_{iM}+4,\dots}\Delta_i(K)\,,
\end{equation}
and by using Eq.~(\ref{eq:deltai}) we obtain
\begin{equation}
  \left(\Delta B\right)_i=\Delta_i(K_{iM})\frac{1}{{\rm e}^{2b_i}-1}\,.
\end{equation}
The ``total missing'' BE is then computed as
\begin{equation}\label{eq:dbt}
  \left(\Delta B\right)_T=\sum_{i=1,6}\Delta_i(K_{iM})
  \frac{1}{{\rm e}^{2b_i}-1}\,.
\end{equation}
In Table~\ref{tab:missingbe} we summarize the ``missing'' BE of each class and
the ``total missing'' BE. By inspecting the table we observe
that the ``total missing'' BE is less than $1\%$ of the total BE for the SRG evolved
potential and of the order of $3\%$ for the NNLO$_{sat}$(NN). This confirms the high accuracy
of the computed binding energies. The estimate of the ``total missing'' BE
suffers from the fact that the extrapolation is not really done for large $K$,
in particular for the class C3, C4 and C6.

As regarding the errors on the ``missing'' BE ($\delta(\Delta B)_i$),
in the case of the class C1, C2 and C5 we propagate the errors
on $b_i$ evaluated in the fits, while for the class C3, C4 and C6 we consider a
conservative error of  $\delta(\Delta B)_i/(\Delta B)_i=0.5$.
For the class C6,
in the case of the SRG1.8, for which we are not able
to extrapolate a reliable value of $b_i$,
and NNLO$_{sat}$(NN) potentials, for which we are not able to estimate $b_i$,
we consider a more conservative error of
$\delta(\Delta B)_i=(\Delta B)_i$.  The error
on the ``total missing'' BE is then computed as
\begin{equation}
  \delta(\Delta B)_T=\sqrt{\sum_{i=1,6}\left(\delta(\Delta B)_i\right)^2}
\end{equation}
For all the potential considered, the relative
error $\delta(\Delta B)_T/(\Delta B)_T$ is of the order
of $\sim15-30\%$.
\begin{table}
  \centering
  \begin{tabular}{cccccccc}
    \hline
    \hline
    &  &\multicolumn{3}{c}{SRG1.2} &\multicolumn{3}{c}{SRG1.5} \\
    \hline
    $i$ & $K_{iM}$ &  $\Delta_i(K_{iM})$ &  $b_i$ & $(\Delta B)_i$ &
    $\Delta_i(K_{iM})$ &  $b_i$ & $(\Delta B)_i$ \\
    \hline
    1 & 14 & 0.013 & 0.51 & 0.007(0) & 0.023 & 0.49 & 0.014(0) \\ 
    2 & 12 & 0.008 & 0.68 & 0.003(1) & 0.042 & 0.58 & 0.019(0) \\
    3 & 10 & 0.015 & 0.37 & 0.014(7) & 0.022 & 0.32 & 0.024(12) \\
    4 & 10 & 0.008 & 0.60 & 0.004(2) & 0.022 & 0.49 & 0.013(6) \\
    5 &  8 & 0.007 & 0.52 & 0.004(0) & 0.023 & 0.37 & 0.021(0) \\
    6 &  8 & 0.004 & 0.44 & 0.003(1) & 0.018 & 0.26 & 0.026(13) \\
    $(\Delta B)_T$ & & &  & 0.034(7) &  & & 0.117(19)\\
    \hline
    &  &\multicolumn{3}{c}{SRG1.8} &\multicolumn{3}{c}{NNLO$_{sat}$(NN)} \\
    \hline
    $i$ & $K_{iM}$ &  $\Delta_i(K_{iM})$ &  $b_i$ & $(\Delta B)_i$ &
    $\Delta_i(K_{iM})$ &  $b_i$ & $(\Delta B)_i$ \\
    \hline
    1 & 14 & 0.035 & 0.46 & 0.023(0) & 0.074 & 0.43 & 0.05(0) \\ 
    2 & 12 & 0.144 & 0.50 & 0.084(11)& 0.411 & 0.42 & 0.32(1) \\
    3 & 10 & 0.024 & 0.30 & 0.029(15)& 0.031 & 0.17 & 0.07(4) \\
    4 & 10 & 0.045 & 0.38 & 0.039(20)& 0.093 & 0.25 & 0.14(7) \\
    5 &  8 & 0.049 & 0.26 & 0.070(1) & 0.153 & 0.18 & 0.35(14) \\
    6 &  8 & 0.048 & 0.11 & 0.19(5)   & 0.112 & --   & -- \\
    $(\Delta B)_T$ & & &  & 0.43(9)  &  & & 0.93(20)\\
    \hline
    \hline
  \end{tabular}
  \caption{\label{tab:missingbe}Increments of the $\Li$ BE $\Delta_i(K_{iM})$,
    computed using Eqs.~(\ref{eq:d1})--(\ref{eq:d5}) for the various classes
    $i=1,\dots,6$ and the N3LO500-SRG$\Lambda$ and NNLO$_{sat}$(NN) potentials.
    The coefficients $b_i$ are
    fitted on the $\Delta_i(K)$ for the classes $i=1,2,5$ and computed as in
    Eq.~(\ref{eq:deltai2}) for the classes $i=3,4$ and $6$. $(\Delta B)_i$ is computed  as in
    Eq.~(\ref{eq:dbi}) and it represents the ``missing'' BE of each class due to
    the truncation of the expansion up to a given $K_{iM}$. Finally, the
    ``total missing'' BE $(\Delta B)_T$ is computed from Eq.~(\ref{eq:dbt}).
    Between the parenthesis we report the errors. With $(0)$ we indicate that the
    errors are smaller than the precision of the digits reported in the table.}
\end{table}

In Table~\ref{tab:summary} we report the computed BE, the
extrapolated ``exact'' BE and a series of other observables associated with the
wave function, such as the mean values of the kinetic $\bra K \ket$
and the potential $\bra V \ket$ energy,
the percentage of $P$, $D$, $F$ and $G$ states, as well as the percentage of
the $S=0,1,2,3$ total spin components. As we already observed, the percentage
of $D$ states increases when $\Lambda$ increases and it is almost $8.5\%$
in the case of the NNLO$_{sat}$(NN). Moreover, for every potential model, the
$\Li$ ground state turns out to be an almost pure ($>93\%$) $S=1$ state.

In order to determine if the computed $\Li$ wave function describes a bound nucleus,
we need to compare the ``exact'' BE with the energy of the $\alpha+d$ system
$B_{\alpha+d}=B_\alpha+B_d$ where $B_\alpha$ and $B_d$ are the BE of
the $\alpha$-particle and the deuteron computed with the same potentials
(see Chapter~\ref{ch:ANC} for details). For all the interactions considered,
we have found $\Li$ bound since $B_{extr.}-B_{\alpha+d}>0$.
However, the experimental value $(B-B_{\alpha+d})_{exp}=1.4743$ MeV
~\cite{Tilley2002}
is not very well reproduced by any potential model, very likely because
we are not including  three-nucleon forces.
By comparing the values reported in Table~\ref{tab:Kconv} with
$B_{\alpha+d}$, it is clear that for the SRG evolved potentials
the $S$ and $D$ states
are sufficient to construct a $\Li$ bound of more than 1 MeV.
 For the case of the NNLO$_{sat}$(NN) the $P$ waves seem to play a more important
role in the determination of the $\Li$ binding energy. However, since
for the NNLO$_{sat}$ potential two- and three-body forces were fitted together,
we cannot discuss the role of the different partial waves in determining
if the $\Li$ is bound or not, using only  the NN part of the interaction.
\begin{table}
  \centering
  \begin{tabular}{lrrrrr}
    \hline
    \hline
    & SRG1.2 & SRG1.5 & SRG1.8 & NNLO$_{sat}$(NN) & Exp.\\
    \hline
     $B$ & 31.75 &  32.75 & 32.21 & 30.01 & 31.99\\
    $B_{extr.}$ & 31.78(1) & 32.87(2) &  32.64(9) & 30.93(20) & \\
    $B_{c}$ & 3.00(1) & 2.46(2) & 2.02(9) & 2.11(20) & 1.47\\
    $\bra K \ket$ & 68.67 & 77.68 & 82.55 & 89.62 & \\
    $\bra V \ket$ & $-100.42$ & $-110.43$ & $-114.76$ & $-119.63$ & \\ 
    $P_D$ & 3.03 & 4.68 & 6.61 & 8.81  & \\
    $P_P$ & 0.76 & 1.23 & 1.51 & 2.64  & \\
    $P_F$ & 0.01 & 0.03 & 0.05 & 0.11  & \\
    $P_G$ & 0.01 & 0.02 & 0.04 & 0.05  & \\
    $P_{S=0}$ & 0.65 & 1.02 & 1.22 & 2.03  & \\
    $P_{S=1}$ & 98.53& 97.32& 96.06& 93.64 & \\ 
    $P_{S=2}$ & 0.51 & 0.99 & 1.59 & 2.60  & \\
    $P_{S=3}$ & 0.31 & 0.66 & 1.12 & 1.73  & \\
    \hline
    \hline
  \end{tabular}
  \caption{\label{tab:summary} The $\Li$ binding energy $(B)$,
    the extrapolated binding energy $(B_{extr.})$ and  the separation
    energy $B_c=B_{extr.}-B_\alpha-B_d$ with the error computed as in
    Eq.~(\ref{eq:dbt})
    (errors on $B_\alpha$ and $B_d$ are not relevant - see Chapter~\ref{ch:ANC}),
    the kinetic $(\bra K \ket)$ and the potential
    $(\bra V \ket)$ energy for various potential
    models used in this work. All these quantities are given in units of MeV.
    Moreover, we report also the percentage of components corresponding
    to the different waves and
    of $S=0,1,2,3$ total spin states of $\Li$.
    In the last column we report the available experimental values.}
\end{table}
In Table~\ref{tab:comparison} we compare our results with those of
Ref.~\cite{Jurgenson2011}, obtained using the NCSM.
As it can be observed by inspecting the table, the results obtained with
the same N3LO-SRG$\Lambda$ potentials in Ref.~\cite{Jurgenson2011},  
seem to be systematically
larger, especially for the case $\Lambda=1.8$ fm$^{-1}$.
The explanation can be found in the fact that
the SRG evolution of the N3LO500 potential is 
slightly different and that the Coulomb potential is included
in the SRG evolution~\cite{Navratil2019}. We can exclude that  our underbinding is related
to errors in our expansion, since we validated it by comparing
the results  with the ones in Ref.~\cite{Gattobigio2011}
and obtaining exactly the same values with a test potential
(see Appendix~\ref{app:validation}).
In any case, the present results are in a reasonable agreement with those
of Ref.~\cite{Jurgenson2011}, especially considering that they are obtained
using two completely different computational techniques.

Finally, we note that no potential model, since we are not including the three-nucleon interactions,
is able to reproduce the experimental value of the BE,
$B_{exp}=31.99$ MeV. However, in the case
of SRG1.2 and SRG1.8,
the probable cancellation between the induced and the ``pure'' three-body
forces, drives the calculated BE closer to the experimental one.

\begin{table}[h]
  \centering
  \begin{tabular}{lrr}
    \hline
    \hline
    & This work & Ref.~\cite{Jurgenson2011}  \\
    \hline
    SRG1.2 & 31.78(1)  &  31.85(5)\\
    SRG1.5 & 32.87(2) &  33.00(5)\\
    SRG1.8 & 32.64(9) &   32.8(1)\\
    NNLO$_{sat}$(NN) & $30.93(20)$ &  -- \\
    \hline
    \hline
  \end{tabular}
  \caption{\label{tab:comparison}Comparison of the $\Li$ binding energy
    calculated here with the extrapolated values of Ref.~\cite{Jurgenson2011},
    obtained with the NCSM basis up to $N_{max}=10$. This corresponds 
    to a smaller basis compared to the one used in this work.}
\end{table}
\end{subsection}

\begin{subsection}{Electromagnetic static properties}\label{sec:emobs}
In this section we  report of the calculated values of the
charge radius, magnetic dipole moment and
electric quadrupole moment
of $\Li$. The study of these observables allows for a better understanding of
the structure of $\Li$.
For the SRG evolved potentials,
we use the approximation given in Eq.~(\ref{eq:srgapprox}).
Moreover, we discuss the convergence of these observables
as function of $K$. In this section with $K$  we indicate the fact that for each class
we include all the HH states with $K_i<K$. When $K>K_{iM}$ for a given class C$i$,
this means that
we include HH states of this class up to $K_{iM}$.

\subsubsection{Charge radius}
The mean square (ms) charge radius of a nucleus is given by~\cite{Friar1997}
\begin{equation}\label{eq:rcdef}
  \bra r_c^2 \ket=\bra r_p^2 \ket+\bra R_p^2 \ket + \frac{N}{Z}\bra R_n^2 \ket
  +\frac{3\hbar^2}{2m_p^2c^2}\,,
\end{equation}
where $\bra R_p^2 \ket$ and $\bra R_n^2 \ket$ are the ms charge
radii of proton and neutron respectively, and $\frac{3\hbar^2}{2m_p^2c^2}$
is the Darwin-Foldy relativistic correction~\cite{Foldy1950}.
The values used in this work for these
three terms are summarized in Table~\ref{tab:par}. 
The expression in Eq.~(\ref{eq:rcdef}) does not include the contribution
of spin-orbit terms and meson-exchange currents.
\begin{table} 
  \centering
  \begin{tabular}{lr}
    \hline
    \hline
    $\sqrt{\bra R_p^2 \ket}$ [fm] & 0.8751 \\
    $\bra R_n^2 \ket$ [fm$^2$] & -0.1161  \\
    $\hbar c$ [MeV fm] & 197.327  \\
    $m_pc^2$ [MeV] & 938.272 \\
    $\mu_p$ [$\mu_N$] & 2.7928  \\
    $\mu_n$ [$\mu_N$] &-1.9130  \\
    \hline
    \hline
  \end{tabular}
  \caption{\label{tab:par}
    Recommended values from evaluation of experimental data
    of the various quantities appearing in Eq.~(\ref{eq:rcdef})
    and~(\ref{eq:mudef})
    as obtained from Ref.~\cite{PDG2018}.
    We have indicated with $\mu_N$  the nuclear magneton.}
\end{table}
 The ms value of the proton point radius $\bra r_p^2 \ket$ is computed as
\begin{equation}
  \bra r_p^2 \ket= \bra \Psi_{\Li} | {\hat r_p}^2 |\Psi_{\Li}\ket\,,
\end{equation}
where the operator ${\hat r_p}^2$ is defined as
\begin{equation}
  {\hat r_p}^2=\frac{1}{Z}\sum_{i=1}^A(\br_i-\boldsymbol{R_{\text{cm}}})^2\left(\frac{1+\tau_z(i)}{2}\right)\,,
\end{equation}
with $\boldsymbol{R_{\text{cm}}}$ the cm coordinate of the nucleus and
$Z$ the proton number.
Our wave function of  $\Li$ is  a pure state of isospin $T=0$. Therefore,
by using the Wigner-Eckart theorem,
the term proportional to $\tau_z(i)$ does not contribute. Indeed
\begin{equation}\label{eq:tz0}
  \bra T=0\,\, T_z=0|\tau_z(i)|T=0\,\, T_z=0\ket=( 0 0, 1 0| 0 0 )
  \frac{\bra T=0||\tau_z(i)||T=0\ket}{\sqrt{3}}=0\,,
\end{equation}
where $( 0 0, 1 0| 0 0 )$ is the Clebsch-Gordan coefficient and
with $\bra \cdots || \cdots || \cdots \ket$ we indicate the reduced matrix element.
The remaining part can be rewritten in terms of the hyperspherical
variables. Using Eq.~(\ref{eq:rho}) we obtain
\begin{equation}\label{eq:r_prho}
  {\hat r_p}^2=\frac{1}{6}\sum_{i=1}^6(\br_i-\boldsymbol{R_{\text{cm}}})^2=\frac{\rho^2}{12}\,.
\end{equation}
The explicit remaining calculation is discussed in Appendix~\ref{app:charge_radius}.
In Figure~\ref{fig:radius} we plot the values of the
root mean square charge radius $r_c=\sqrt{\bra r_c^2\ket}$
as function of $K$. From the figure we can observe for the SRG evolved potentials
an exponential behavior as $K$ increases. In order to extrapolate
the  full converged value, we fit our results with
\begin{equation}\label{eq:fitr}
  r_c(K)=r_c(\infty)+a{\rm e}^{-bK}\,,
\end{equation}
where $r_c(\infty)$ is the extrapolated value for $K\rightarrow\infty$.
From the fit we exclude the values of $r_c$ obtained
for $K=2$ and $K=14$, since they do not follow the exponential behavior,
as it can be seen in Figure~\ref{fig:radius}. This is due to the fact that
for $K=2$ there are not enough states to well define the structure of $\Li$, and
among the states with $K=14$ we are not considering channels with  
$D$ states which are fundamental for describing properly the $\Li$ radius.
  \begin{figure} 
    \centering
    \includegraphics[scale=0.85]{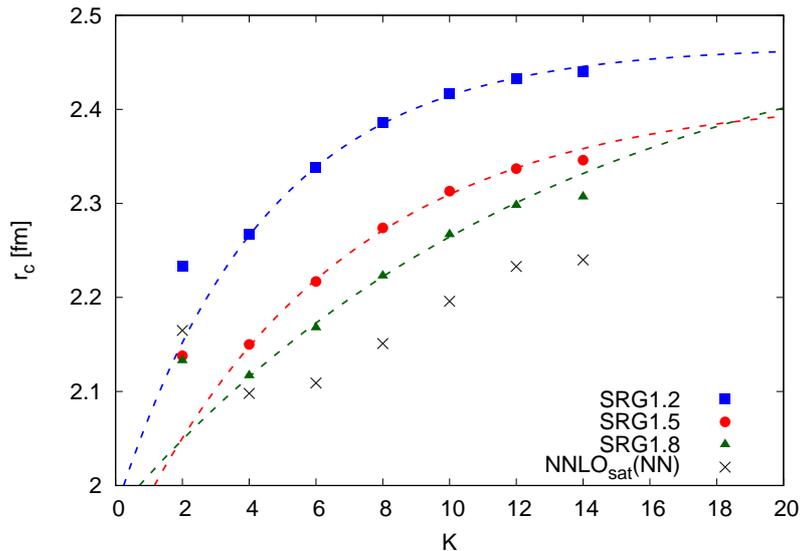}
    \caption{$\Li$ charge radius in fm as function of $K$
      for the potential models
      SRG1.2 (blue), SRG1.5 (red), SRG1.8 (green) and NNLO$_{sat}$(NN) (black).
      The exponential fit performed on the SRG evolved potentials
      using Eq.~(\ref{eq:fitr}) is also
      shown (dashed lines).}
    \label{fig:radius}
  \end{figure}
In Table~\ref{tab:radius} we report the value of the charge radius for
the various values of $K$, and the extrapolated one for the SRG evolved potentials
in the last row.
The error reported in parentheses is due to our fitting procedure.
\begin{table}[h] 
  \centering
  \begin{tabular}{rcccc}
    \hline
    \hline
    $K$ & SRG$1.2$ & SRG$1.5$ & SRG$1.8$ & NNLO$_{sat}$(NN)\\
    \hline
     2  & 2.233 &  2.138  &  2.133  &    2.165 \\
     4  & 2.267 &  2.150  &  2.117  &    2.098 \\
     6  & 2.338 &  2.217  &  2.168  &    2.109 \\
     8  & 2.386 &  2.274  &  2.223  &    2.151 \\
     10 & 2.417 &  2.313  &  2.267  &    2.196 \\
     12 & 2.433 &  2.337  &  2.298  &    2.233 \\
     14 & 2.440 &  2.346  &  2.307  &	 2.240 \\
    \hline
    $r_c(\infty)$ & 2.47(1)& 2.42(2) & 2.52(10) & -- \\
    \hline
    \hline
  \end{tabular}
  \caption{\label{tab:radius} Charge radii as function of $K$ for
    the various potential models considered. In the last row
    we report the extrapolated values for the
    SRG evolved potentials with the errors (between parenthesis)
    obtained from the fits of Eq.~(\ref{eq:fitr}). All the values are given in fm.}
\end{table}
As it can be seen from Table~\ref{tab:radius} and Figure~\ref{fig:radius}, the
convergence is quite slow. Indeed, the HH basis is a ``compact''
basis and it is not able to describe perfectly the tail of the wave function,
which in the case of $\Li$ has a $\alpha+d$ dominant structure,  
very hard to be reproduced.
We will treat this point with more details in Chapter~\ref{ch:ANC}.
By looking at the results obtained with the SRG potentials, it is clear that
the convergence rate is faster for the smallest values of the parameter $\Lambda$,
since in these cases the correlations between the nucleons are
reduced, favoring the convergence.

 As regarding the NNLO$_{sat}$(NN) potential, the convergence
  is even slower than the SRG evolved potentials
  and does not show a clear exponential behavior. Therefore,
  in Table~\ref{tab:radius} we report only our calculations
  without the extrapolation.
  Since the NNLO$_{sat}$(NN) potential is less attractive than
  the SRG evolved ones,
  we can speculate that $r_c(\infty)$ will be larger.
  We also expect that by including three-body forces, since they are attractive,
this will induce a more
compact structure for $\Li$ by reducing the charge radius.

\subsubsection{The magnetic dipole moment}

The magnetic dipole moment operator for the $A=6$ case is defined as
\begin{equation}\label{eq:mudef}
\begin{aligned}
  {\hat \mu_z}&=\mu_p\sum_{i=1}^6\sigma_z(i)
  \left(\frac{1+\tau_z(i)}{2}\right)
  +\mu_n\sum_{i=1}^6\sigma_z(i)
  \left(\frac{1-\tau_z(i)}{2}\right)\\
  &+\sum_{i=1}^6\ell_z(i) \left(\frac{1+\tau_z(i)}{2}\right)\,.
\end{aligned}
\end{equation}
where $\mu_p$ and $\mu_n$ are the proton and neutron intrinsic magnetic moment,
given in Table~\ref{tab:par}, $\sigma_z(i)$ are the Pauli matrices and
$\ell_z(i)=(\boldsymbol{r}_i\times\boldsymbol{p}_i)_z$ is the orbital
angular momentum of a single particle. We consider the operator
projected along the $z$-axis since
measurements are performed polarizing the nucleus along such axis.
As for the charge radius case,
the terms proportional to $\tau_z(i)$ do not contribute.
Defining the total spin as
\begin{align}
  {\hat S_z}=\frac{1}{2}\sum_{i=1}^6\sigma_z(i)\,\label{eq:sz0},
\end{align}
and the total angular momentum as
\begin{align}
  {\hat L_z}&=\sum_{i=1}^6\ell_z(i)\,,\label{eq:lz0}
\end{align}
we can rewrite the magnetic dipole moment as
\begin{equation}
  {\hat \mu_z}=(\mu_p+\mu_n){\hat S_z}+\frac{1}{2}{\hat L_z}\,.
\end{equation}

By definition, the mean value of the
magnetic dipole moment of $\Li$ is given by
\begin{equation}
  \bra \mu_z\ket=\bra \Psi_{\Li}(J_z=+1)|{\hat \mu_z}|\Psi_{\Li}(J_z=+1)\ket\,,
\end{equation}
where the wave function of $\Li$ is in the maximum total angular momentum projection.
Analogously we can define the mean values
of $\hat{S_z}$ and $\hat{L_z}$,  which explicitly reads
\begin{align}
  \bra S_z\ket&=
  P_{01}+\frac{1}{2}P_{11}+\frac{3}{2}P_{12}
  -\frac{1}{2}P_{21}+\frac{1}{2}P_{22}+2P_{23}
  -P_{32}+\frac{1}{2}P_{33}-\frac{3}{2}P_{43}\,,\label{eq:sz}\\
  \bra L_z\ket&=P_{10}+\frac{1}{2}P_{11}+\frac{1}{2}P_{12}
  +\frac{3}{2}P_{21}+\frac{1}{2}P_{22}-P_{23}
  +2P_{32}+\frac{1}{2}P_{33}+\frac{5}{2}P_{43}\,.\label{eq:lz}
\end{align}
where $P_{LS}$ is the percentage of the ${}^{2S+1}L$ partial wave component
in $\Li$ wave function. The derivation of these formulas is reported in
Appendix~\ref{app:mdm}.

In Table~\ref{tab:mdp1} and Figure~\ref{fig:mdp}
we report the results obtained for the various potentials
from $K=2$ to $K=14$. As it can be seen
by inspecting the table, the values of the magnetic dipole
moment are practically at convergence. This is due to the fact
that the percentages of the different wave components are very
stable starting already from $K=6$.
For the SRG potentials, the value of the magnetic dipole moment decreases by
increasing the value of $\Lambda$. Indeed, when $\Lambda$ increases
the correlations induced by the nuclear potential
are stronger, generating a larger amount
of $^3D_1$ component in the wave function, which is
the only relevant negative component in Eq.~(\ref{eq:sz}). In the case of the
NNLO$_{sat}$(NN) potentials, this effect is even more important, giving a smaller value of the
magnetic dipole moment.
\begin{table} 
  \centering
  \begin{tabular}{rcccc}
    \hline
    \hline
    $K$ & SRG$1.2$ & SRG$1.5$ & SRG$1.8$ & NNLO$_{sat}$(NN)\\
    \hline
    2   & 0.872  &  0.868  &   0.866 &	0.862\\
    4   & 0.864  &  0.858  &   0.852 & 	0.846\\
    6   & 0.864  &  0.857  &   0.850 & 	0.840\\
    8   & 0.864  &  0.857  &   0.850 & 	0.840\\
    10  & 0.864  &  0.858  &   0.851 & 	0.843\\
    12  & 0.864  &  0.858  &   0.852 & 	0.845\\
    14  & 0.865  &  0.858  &   0.852 &  0.845\\
    \hline
    \hline
  \end{tabular}
  \caption{\label{tab:mdp1} Values of the magnetic dipole moment as function of $K$ for
    the various potential models considered.
    All values are given in units of $\mu_N$.}
\end{table}
  \begin{figure}[h] 
    \centering
    \includegraphics[scale=.85]{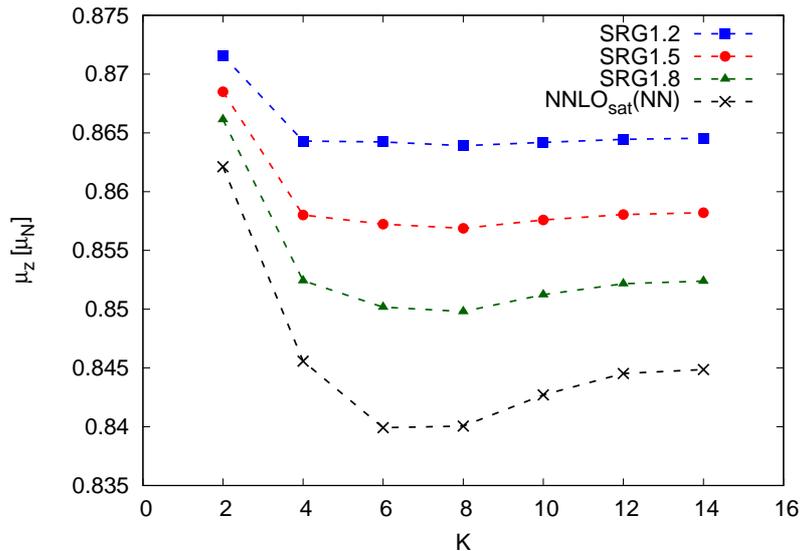}
    \caption{Values of the magnetic dipole moment as function of $K$,
      for three different SRG potential
      1.2(blue), 1.5(red), 1.8(green) and the NNLO$_{sat}$(NN) (black).
      }
    \label{fig:mdp}
  \end{figure}

If we consider $\Li$ to be formed as a $\alpha+d$ cluster, 
we can expect that
\begin{equation}
  \mu_z({\Li})\approx\mu_z(d)\,,
\end{equation}
because the $\alpha$-particle has no magnetic dipole moment. However, the internal
structure of $\Li$ plays a fundamental role decreasing the value of the magnetic
dipole moment compared to the deuteron one, as it can be observed comparing the experimental
values (last row of Table~\ref{tab:mdp2}).
For this reason, a good potential model should be able to reproduce
the differences between the magnetic dipole moments of $d$ and $\Li$.
In Table~\ref{tab:mdp2} we compare the results of the magnetic dipole moment of
$\Li$ and $d$ evaluated with 
SRG potentials for the three values of $\Lambda$ and the one computed with
NNLO$_{sat}$(NN). In all the cases, 
the $\Li$ magnetic dipole moment is reduced compared to the $d$ ones,
showing that the potential models are going in the right direction, even
if they are not able to reproduce the experimental value.
Obviously this is partially due to the
fact we are not considering the evolved operator in the SRG case and also
that we are not including  three-body forces in both the SRG evolved and
NNLO$_{sat}$(NN) potentials. However, as it was shown in Refs.~\cite{Carlson1998,
Schiavilla2019} for $\Ht$ and $\Het$,
the magnetic dipole moment receives important contributions
from two-body electromagnetic currents.
We expect that similar corrections are necessary
to reproduce the experimental value of $\mu_z(\Li)$.
\begin{table}[h] 
  \centering
  \begin{tabular}{lcc}
    \hline
    \hline
    & $\mu_z(d)$ & $\mu_z(\Li)$ \\
    \hline
    SRG1.2  & 0.872 & 0.865  \\
    SRG1.5  & 0.868 & 0.858  \\
    SRG1.8  & 0.865 & 0.852  \\
    NNLO$_{sat}$(NN) & 0.860 & 0.845 \\
    \hline
    Exp. &  0.857  & 0.822     \\
    \hline
    \hline
  \end{tabular}
  \caption{\label{tab:mdp2} Values of the magnetic dipole moment of
    $\Li$ and $d$ evaluated with the various considered potential models.
    In the last row the experimental values are listed~\cite{Stone2014}.
    All the values are given in units of $\mu_N$.}
\end{table}

\subsubsection{Electric quadrupole moment}\label{sec:eqm}

The electric quadrupole moment operator is defined as
\begin{equation}\label{eq:qdef}
  {\hat Q}=\sum_{i=1}^6(3z_i^2-r_i^2)\left(\frac{1+\tau_z(i)}{2}\right)\,.
\end{equation}
Therefore, we can write the mean value as
\begin{equation}\label{eq:defquad}
  \bra Q \ket=\bra \Psi_{\Li}(J_z=+1)|{\hat Q}|\Psi_{\Li}(J_z=+1)\ket\,,
\end{equation}
where the wave function of $\Li$ is evaluated in the maximum projection of
the total angular momentum $J_z=+1$ since the measurements are performed by polarizing
the nuclei along the $z$-axis.
Again, in Eq.~(\ref{eq:defquad}) the term proportional to
$\tau_z(i)$ does not give any contribution as shown in Eq.~(\ref{eq:tz0}).
By taking advantage of the antisymmetry of the wave function,
the calculation
of the mean value of ${\hat Q}$ is reduced to the calculation of the mean value of the
operator
\begin{equation}\label{eq:qzz6}
  {\hat Q}(6)=3\times(3z_6^2-r_6^2)\,,
\end{equation}
where with 6 we indicate that it acts only on the sixth particle.
In order to write in a simple form the ${\hat Q}(6)$ operator
in terms of the HH states,
we define a new set of Jacobi vectors, hereafter denoted as ``Q'' set, such that
\begin{equation}\label{eq:xxq_def}
  \xx_1^Q=\xx_5\,,\qquad\xx_2^Q=\xx_2\,,\qquad\xx_3^Q=\xx_3\,,\qquad
  \xx_4^Q=\xx_4\,,\qquad\xx_5^Q=\xx_1\,,\qquad
\end{equation}
and the hyperangular variables $\ph^Q$ as
\begin{equation}
  \cos \ph_i^Q=\frac{x_i^Q}{\sqrt{(x_1^Q)^2+\dots+(x_i^Q)^2}}\,,\qquad
  i=2,\dots,5\,,\label{eq:phiangq}
\end{equation}
while the definition of $\rho$ does not change.
Using these hyperangular variables the operator in
Eq.~(\ref{eq:qzz6}) can be rewritten as
\begin{equation}
  {\hat Q}(6)=2\sqrt{5\pi}\rho^2\cos \ph_5^Q\, Y_{20}(\hat{x}_5^Q)\,.
\end{equation}
The use of the ``Q'' set of  hyperspherical variables permits to reduce drastically the
number of HH states coupled by the electric quadrupole moment
operator in Eq.~(\ref{eq:defquad}), reducing in this way the computational time.
The explicit calculation of the electric quadrupole moment in terms of the
HH states is given in Appendix~\ref{app:eqm}.

The study of this observable is crucial for understanding the
accuracy of the wave function we computed.
Indeed, from the experiment, we know that
the electric quadrupole moment of $\Li$ is very small and negative. For this reason
it is challenging for all the potential models
to reproduce this value.
In Table~\ref{tab:quadr} and in Figure~\ref{fig:quadr}
we report the values of the electric quadrupole
moment as function of $K$ for the potential models considered in this Thesis.
As it can be observed by inspecting the table and the figure,
the trend is irregular up to $K=8$, and this is due to the cancellations
among the contributions coming from different sets of HH states.
After $K=8$, the behavior becomes stable
and the convergence very fast since the cancellations are smaller.
\begin{table}[h]
  \centering
  \begin{tabular}{rcccc}
    \hline
    \hline
    $K$ & SRG$1.2$ & SRG$1.5$ & SRG$1.8$ & NNLO$_{sat}$(NN) \\
    \hline
    2  &  $-0.188$ &   $-0.162$ &   $-0.173$ & $-0.115$\\ 
    4  &  $-0.212$ &   $-0.132$ &   $-0.091$ & $+0.065$\\
    6  &  $-0.170$ &   $-0.066$ &   $-0.016$ & $+0.136$\\
    8  &  $-0.184$ &   $-0.094$ &   $-0.052$ & $+0.086$\\
    10 &  $-0.191$ &   $-0.101$ &   $-0.055$ & $+0.079$\\
    12 &  $-0.191$ &   $-0.101$ &   $-0.055$ & $+0.069$\\
    14 &  $-0.191$ &   $-0.101$ &   $-0.055$ & $+0.068$\\
    \hline
    \hline
  \end{tabular}
  \caption{\label{tab:quadr}
    Values of the electric quadrupole moment varying $K$ for
    SRG potentials with $\Lambda=1.2$, $1.5$ and $1.8$ fm$^{-1}$ and
    for the NNLO$_{sat}$(NN) model.
    All the values are given in unit of $e$ fm$^2$.}
\end{table}
\begin{figure}[h]
  \centering
  \includegraphics[scale=0.85]{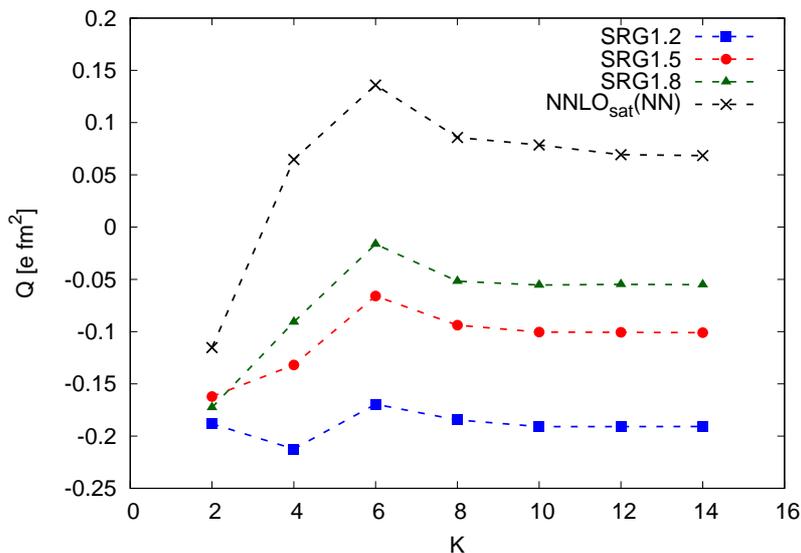}
  \caption{Values of the electric quadrupole moment
    for the potential models 
    SRG1.2 (blue), SRG1.5 (red), SRG1.8 (green) and
    NNLO$_{sat}$(NN) (black) as function of $K$.}
  \label{fig:quadr}
\end{figure}

The values obtained for the SRG potentials
seem to be quite dependent on the value of the
parameter $\Lambda$.
This is due to the fact we are using the bare operator
in Eq.~(\ref{eq:defquad})
and not the one transformed by the SRG unitary transformation.
Therefore, for the electric quadrupole moment the approximation
of Eq.~(\ref{eq:srgapprox}) is not good at all.
However, all the SRG models are able to reproduce a small and
negative value for the electric quadrupole moment. In particular, for
$\Lambda=1.5$  and 1.8 fm$^{-1}$ we obtain a value
quite close to the experimental measurement $-0.0806(6)\,e$ fm.
This is not the case for
NNLO$_{sat}$(NN) potential, for which the electric quadrupole moment results to be
positive even if very small. In order to understand why we have these
differences between the various models, we report in Table~\ref{tab:quadr_wave} 
the partial wave contributions to this observable.
As it can be seen, we have large differences only in the contribution
coming from $S-D$ matrix element. In particular the value of this matrix element
increases when $\Lambda$ increases and it becomes positive with the NNLO$_{sat}$(NN).
Therefore, the value of the quadrupole moment seems directly connected to the
strength of the tensor term in the nuclear potential: the more is important, the
larger is the value of the electric quadrupole moment.
Obviously, we need to consider also that three-body forces are not included. However
we do not expect that they will be significant.
On the other hand, the corrections of two-body currents could be more significant and  necessary to obtain a reliable prediction of the
electric quadrupole moment.

\begin{table}[h]
  \centering
  \begin{tabular}{lccccc}
    \hline
    \hline
     & $S-D$ & $D-D$ & $P-P$ & $P-D$ & remaining \\
    \hline
    SRG1.2 & $-0.187$ & $-0.023$ & $0.009$ & $0.009$ & $<0.001$ \\
    SRG1.5 & $-0.102$ & $-0.023$ & $0.014$ & $0.010$ & $<0.001$\\
    SRG1.8 & $-0.058$ & $-0.024$ & $0.016$ & $0.010$ & $\phantom{<}0.001$\\
    NNLO$_{sat}$(NN) & $\phantom{-}0.049$ & $-0.018$ & $0.023$ & $0.011$ & $\phantom{<}0.003$\\
    \hline
    \hline
  \end{tabular}
  \caption{\label{tab:quadr_wave}
    Partial wave contributions to the electric quadrupole moment of $\Li$.
    All the values are given in unit of $e$ fm$^2$.}
\end{table}

\end{subsection}
\begin{subsection}{Comparison with literature}

In order to compare the electromagnetic static properties of $\Li$
calculated here with the results obtained using other techniques,
we report in Table~\ref{tab:radius_comp} our N3LO500-SRG1.5 results compared
with the values of
Ref.~\cite{Forseen2009}, calculated using the
CD-B2k potential~\cite{Machleidt2001}
evolved with SRG $\Lambda=1.5$ fm$^{-1}$.
Also the experimental values are listed. Even if the starting interactions
are different, the SRG evolution makes the CD-B2k and the N3LO500
potentials very similar, and therefore we can
qualitatively compare the results.
As regarding the charge radius,
our result is in very good agreement with the result
reported in Ref.~\cite{Forseen2009}.
The experimental value however is underestimated. The reason can be found in the
fact that we are using the bare operator and not the evolved one
[see Eq.~(\ref{eq:srgapprox})]
and we also do not include three-nucleon forces. 
Comparing our calculation for the magnetic dipole moment at $\Lambda=1.5$ fm$^{-1}$
with the result reported in Ref.~\cite{Forseen2009}
(see Table~\ref{tab:radius_comp}),
we see that our result is larger
and not compatible within the error bar. This is probably due to the different
potential used in Ref.~\cite{Forseen2009}, the CD-B2k, which presumably
induces  larger $D$ wave components in the wave function.
The electric quadrupole moment result, instead, is quite close
to the one reported in Ref.~\cite{Forseen2009}, taking into account
the relative uncertainty.
This is also very close to the experimental value, but, as already discussed, this is just
a chance, due to the particular choice of the evolution parameter $\Lambda$.
\begin{table}[h]
  \centering
  \begin{tabular}{lccc}
    \hline
    \hline
    & This work(SRG$1.5$) & Ref.~\cite{Forseen2009} &  Exp. \\
    \hline
    $r_c$ [fm] & 2.42(2) & 2.40(6) & 2.540(28)~\cite{Puchalski2013} \\
    $\mu_z$ [$\mu_N$] & 0.858 & 0.843(6) & 0.822~\cite{Stone2014}\\
    $Q$ [$e$ fm$^2$] & -0.101 & -0.066(40)& -0.0806(6)~\cite{Stone2014}\\
    \hline
    \hline
  \end{tabular}
  \caption{Values of the $\Li$ charge radius, $r_c$, magnetic dipole
    moment, $\mu_z$, and electric quadrupole moment, $Q$, obtained in this
    work for the N3LO500-SRG1.5, compared
    with the results of Ref.~\cite{Forseen2009}, obtained with the CDB2k-SRG1.5,
    and with the experimental values.}\label{tab:radius_comp}
\end{table}
\end{subsection}
\end{section}
\end{chapter}

%% file: Chapter3_v3.tex
\chapter{The $\Li$ Asymptotic Normalization Coefficient}\label{ch:ANC}

In this chapter we study the $\alpha+d$ clusterization of $\Li$ with the
goal of determining the ANC, which is the
normalization of the long-range tail of the $\Li$ wave function.
This quantity  plays a fundamental role in the determination of the
$\alpha+d\rightarrow\Li+\gamma$ radiative capture cross section
since this reaction is peripheral
at energies relevant for BBN, and therefore
the astrophysical factor is practically
determined by the square of the ANC (see for example~\cite{Mukham2016}).

This chapter is organized as follows. In the first section we briefly
discuss the calculation of the deuteron and $\alpha$-particle wave functions.
In Section~\ref{sec:overlap}
we formally define the $\alpha+d$ cluster form factor (CFF)
and the ANC,
discussing the calculation and the results. In Section~\ref{sec:eqcff}
we introduce a different approach for computing the CFF from a
differential equation. To perform this calculation
we introduce a new technique called
``projection'' method, which we use to compute the terms of the equation.
In the last Section we discuss the results obtained by
computing the CFF with the equation derived in Section~\ref{sec:eqcff},
and we compare these results with the one obtained in Section~\ref{sec:overlap}.

\section{The deuteron and the $\alpha$-particle wave functions}\label{sec:adbound}
The first ingredients needed to compute the $\alpha+d$ clusterization
are the deuteron and the $\alpha$-particle wave functions.
Both of them are obtained by solving the two-bodies and four-bodies Schr\"odinger
equation respectively, by using the variational
approach presented in Chapter~\ref{ch:HH}. In this section we discuss only the main
features and the results. 

The variational deuteron wave function is expressed in term of
spherical harmonics, namely
\begin{equation}\label{eq:psid}
  \Psi_d(\xx_{1})=\sum_{\ell_d=0,2}u_{\ell_d}(x_{1})\left[Y_{\ell_d}
    (\hxx_{1})(\sg_1\sg_2)_{S_d}\right]_{J_d J_{dz}}(\tp_1\tp_2)_{T_dT_{dz}}\,,
\end{equation}
where $\xx_1=\br_2-\br_1$. The fact that the deuteron is a $J_d^\pi=1^+$
states implies that $\ell_d=0,2$ and that the spin $S_d=1$. By imposing
antisymmetry we obtain $T_d=0$. The radial wave functions $u_{\ell_d}(x_1)$
is expanded as
\begin{equation}
  u_{\ell_d}(x_1)=\sum_{l_d=0}^{l_d^{max}}c_{\ell_d,l_d}f_{l_d}(x_1)\,,
\end{equation}
where $c_{\ell_d,l_d}$ are the variational parameters determined
by using the Rayleigh-Ritz variational principle, and
\begin{equation}
  f_{l_d}(x_1)=\gamma_d^{\frac{3}{2}}\sqrt{\frac{l_d!}{(l_d+2)!}}\,
    L^{(2)}_{l_\al}(\gamma_d x_1)e^{-\frac{\gamma_d x_1}{2}}\,.
\end{equation}
In this work we use $\gamma_d=4$ fm$^{-1}$ and $l_d^{max}=40$.
The potential models N3LO500-SRG$\Lambda$ and
NNLO$_{sat}$(NN) are equivalent in the two-body sector and they must reproduce
almost exactly the experimental binding energy of the deuteron ($B_d$).
In Table~\ref{tab:deut} we report the results for the deuteron obtained with the four
potentials used in this Thesis, and we compare them with the experimental value.
As it can be seen, all the potentials are able to reproduce at the third decimal digit
the deuteron binding energy.
\begin{table}[h!]
\centering
\begin{tabular}{lc}
  \hline
  \hline
  Potential & $B_d$ (MeV)\\
  \hline
  SRG1.2 & $2.2248$\\
  SRG1.5 & $2.2248$\\
  SRG1.8 & $2.2248$\\
  NNLO$_{sat}$(NN) & $2.2249$\\
  \hline
  Exp. & $2.224575$\\
  \hline
  \hline
\end{tabular}
\caption{\label{tab:deut}
  Binding energy of the deuteron computed with the N3LO500-SRG$\Lambda$ and
  the NNLO$_{sat}$(NN) potential. In the last row we report for comparison  the
  experimental binding energy.}
\end{table}

The $\alpha$-particle variational wave function is written as
\begin{equation}
\begin{aligned}\label{eq:psialpha}
  \Psi_\alpha(\xx_{1},\xx_{2},\xx_{3})
  &=\sum_{l_\alpha=0}^{l_\alpha^{max}}\sum_{K_\al}
  \sum_{\beta_\al}a_{\beta_\al,l_\al}
  f_{l_\al}(\rho)
  \sum_{p=1}^{12}
  \Big\{{\cal Y}^{K_\al,L_\al}_{\beta_\al}(\Omega_{3p})\\
   &\otimes \left[\left((\sg_i\sg_j)_{S_{2\al}}\sg_k\right)_{S_{3\al}}
      \sg_l\right]_{S_{\al}}
  \Big\}_{J_\al J_{\al z}}
  \left[\left((\tp_i\tp_j)_{T_{2\al}}\tp_k\right)_{T_{3\al}}
    \tp_l\right]_{T_{\al}T_{\al z}}\,,
\end{aligned}
\end{equation}
where $\xx_i$ are the Jacobi vectors as defined in Eq.~(\ref{eq:jac1}) by using $N=3$
and $a_{\beta_\al,l_\al}$ are the variational parameters. The hyperradial
function are defined to be,
\begin{equation}
  f_{l_\al}(\rho)=\gamma_\al^{\frac{9}{2}}\sqrt{\frac{l_\al!}{(l_\al+8)!}}\,
    L^{(8)}_{l_\al}(\gamma_\al\rho)e^{-\frac{\gamma_\al\rho}{2}}\,,
\end{equation}
where in our calculation we use $\gamma_\al=4$ fm$^{-1}$ and
$l_\al^{max}=16$. The expression of the HH function
${\cal Y}^{K_\al,L_\al}_{\beta_\al}(\Omega_{3p})$
can be easily derived from
Eq.~(\ref{eq:hh}) and Eq.~(\ref{eq:hh2}) by substituting $A=4$ and $N=3$.
Moreover, with $\beta_\al$ we indicate all the quantum numbers
needed to define the basis, namely
\begin{equation}
  \beta_\al
  \equiv\{\ell_{1\al},\ell_{2\al},\ell_{3\al},L_{2\al},L_\al,n_{2\al},n_{3\al},
  S_{2\al},S_{3\al},S_{\al},T_{2\al},T_{3\al},T_{\al}\}\,.
\end{equation}
The $\alpha$-particle is a $J_\al^\pi=0^+$ state, therefore $L_\al=S_\al$.
Moreover we consider only $T_\alpha=0$, since it is mainly a isospin $0$ state, and
all the other $T_\alpha$ contributions can be safely neglected~\cite{Viviani2005}.
In Eq.~(\ref{eq:psialpha}), the permutation $p=1$
corresponds to the order $(1,2,3,4)$ of the particles. This fact will be
exploited later in this chapter.
We limit our basis states so that $\ell_{sum}^\al=\ell_{1\al}+
\ell_{2\al}+\ell_{3\al}\leq2$. States with $\ell_{sum}^\al\geq4$ do not give any
significant contribution. Unlike the $\Li$, being the number of basis states
much smaller, we do not perform any other pre-selection.
In Table~\ref{tab:alpha}
we report the convergence of the binding energy of the $\alpha$-particle
ground state ($B_\alpha$) for the four potentials we used.
As it can be seen comparing the binding energy at $K_\al=10$,
with the extrapolated value,
it is clear that we reached the convergence to 1 keV
for the SRG evolved potentials and to $10$ keV for NNLO$_{sat}$(NN) potential.
For the SRG1.5 and SRG1.8 potentials, the binding energy is very close
to the experimental one, presumably
due to the cancellation between the pure three-body forces
and the induced ones. For the NNLO$_{sat}$(NN) we obtain an underbinding of
$\sim1.7$ MeV, due to the fact that we do not include the three-body forces.
\begin{table}[h!]
\centering
\begin{tabular}{lcccc}
  \hline
  \hline
      &\multicolumn{4}{c}{$B_\alpha$ (MeV)}\\
  $K_\al$ & SRG1.2 & SRG1.5 & SRG1.8 & NNLO$_{sat}$(NN)\\
  \hline
  0 & 25.524 & 25.896 & 24.122 & 17.906\\
  2 & 26.352 & 27.450 & 26.490 & 21.459\\
  4 & 26.543 & 27.999 & 28.178 & 25.609\\
  6 & 26.555 & 28.172 & 28.364 & 26.391\\
  8 & 26.556 & 28.179 & 28.396 & 26.562\\
  10 & 26.557 & 28.180 & 28.402 & 26.597\\
 $\infty$ & 26.557 & 28.181 & 28.403 & 26.606\\
 \hline
 Exp.& \multicolumn{4}{c}{28.30}\\
 \hline
 \hline
\end{tabular}
\caption{\label{tab:alpha}
  Convergence of the binding energy of the
  $\alpha$-particle ground state computed with the N3LO500-SRG$\Lambda$ and
  the NNLO$_{sat}$(NN) potentials as function of $K_\al$.
  We also report the extrapolated
  values at $K_\al\rightarrow\infty$.
  In the last row we report the experimental binding energy.}
\end{table}

\section{The $\alpha+d$ cluster form factor}\label{sec:overlap}
This section is dedicated to study the $\alpha+d$ clusterization of the $\Li$
wave function. We begin by defining the cluster wave function as
\begin{equation}\label{eq:clusterw}
  \Psi^{(L)}_{\alpha+d}=N_{\alpha+d}{\cal A}
  \left[\left(\Psi_\alpha\times\Psi_d\right)_S
    Y_L(\hat r)\right]_J\,,
\end{equation}
where the symbol ${\cal A}$ is the antisymmetrization operator,
$\Psi_\alpha$ and $\Psi_d$ are the wave functions of the $\alpha$-particle
and the deuteron calculated
in the previous section and $\br$ the distance between the $\alpha$ and $d$
c.m. position. The quantum numbers
$L$, $S$ and $J$ are determined by considering that
$J_\alpha^\pi=0^+$ and $J_d^\pi=1^+$ and so $S=1$.
Moreover, since we are studying
the $\Li$ ground state, we consider only the case $J^\pi=1^+$,
therefore only $L=0$ ($S$-wave) and $L=2$ ($D$-wave) are possible.
The normalization  $N_{\alpha+d}$ is chosen in order to have
\begin{equation}
  \int_{\Omega_S} d\Omega_S\, |\Psi^{(L)}_{\alpha+d}|=1\,,
\end{equation}
where we are integrating over the hyperspherical surface $\Omega_S$ with an
hyperradius $\rho\rightarrow\infty$. This condition corresponds of imposing
unit flux at infinite distance. Since we are integrating on a surface with
infinite radius,
this integral receives contributions only from the asymptotic regions
in which the six particles are clusterized in $4+2$, which corresponds
to $\frac{6!}{4!\,2!}=15$ regions. Therefore we obtain
\begin{equation}
  N_{\alpha+d}=\frac{1}{\sqrt{15}}\,.
\end{equation}
We need also to normalize properly the bound state wave function. For a generic
wave function of $A$ nucleons, $\Psi_A$, the normalization is given by
\begin{align}
  \bra\Psi_A|\Psi_A\ket&=\int \prod_{i=1}^Ad\br_i\,
  \delta\left(\brr^A_{\text{c.m.}}-\frac{1}{A}\sum_{j=1}^A\br_j
  \right)|\Psi_A|^2=1\,\label{eq:norma},
\end{align}
where  $\brr_{\text{c.m.}}^A$
is a generic position of the c.m. of the $A$ particles.
This result follows since the $\Psi_A$ do not depend on $\brr_{\text{c.m.}}^A$. 
We note that from the solution of the eigenvalue problem 
our  wave functions are normalized as
\begin{equation}
  \begin{aligned}
  \int& d\xx_1\, |\Psi_d(\xx_1)|^2=1\,,\quad  \int d\xx_1d\xx_2d\xx_3\,
  |\Psi_\alpha(\xx_1,\xx_2,\xx_3)|^2=1\,,\\
  \int& d\xx_1d\xx_2d\xx_3d\xx_4d\xx_5\,
  |\Psi_{\Li}(\xx_1,\xx_2,\xx_3,\xx_4,\xx_5)|^2=1\,,
  \end{aligned}
\end{equation}
where $\xx_i$ are the Jacobi coordinates as defined in Eq.~(\ref{eq:jac1})
by using $N=1$ in the case of the deuteron,
$N=3$ in the case of the $\alpha$-particle and $N=5$ for the $\Li$.
By using Eq.~(\ref{eq:norma}) the norms of $d$, $\alpha$ and $\Li$ wave functions
result
\begin{align}
  \bra\Psi_d|\Psi_d\ket&=1\,,\label{eq:normd1}\\
  \bra\Psi_\al|\Psi_\al\ket&=
  \left(\frac{1}{\sqrt{2}}\right)^3\,,\label{eq:norma1}\\
  \bra\Psi_{\Li}|\Psi_{\Li}\ket&=\left(\frac{\sqrt{3}}{4}\right)^3\,.\label{eq:normli1}
\end{align}

We can now define the $\alpha+d$ cluster form factor (CFF) as
\begin{equation}\label{eq:overlap_df}
  \frac{f_{L}(r)}{r}=\frac{\bra \Psi_{\Li}|\Psi^{(L)}_{\al+d}\ket_{r}}
  {\left[\bra\Psi_d|\Psi_d\ket\bra\Psi_\al|\Psi_\al\ket\bra\Psi_{\Li}|\Psi_{\Li}\ket\right]^{1/2} }
\,,   
\end{equation}
where $f_L(r)$ is the so called reduced CFF.
In Eq.~(\ref{eq:overlap_df}) $\bra\Psi_i|\Psi_i\ket$
represents the normalization of the wave functions as defined
in Eqs.~(\ref{eq:normd1})--(\ref{eq:normli1}),
$\Psi_{\alpha+d}^{(L)}$
the cluster wave function defined in Eq.~(\ref{eq:clusterw}),
and  $\bra\cdots|\cdots\ket_{r}$
the traces over the spin and isospin and integration
over the internal variables except the intercluster distance $r$, namely
\begin{equation}\label{eq:intV}
  \int_{r}=\int  d\br_1d\br_2d\br_3d\br_4d\br_5
  d\br_6\,
  \delta\left(\brr^{\Li}_{\text{c.m.}}-\frac{1}{6}\sum_{i=1}^6\br_i\right)
  \delta\left(r-|\brr_{\text{c.m.}}^d-\brr_{\text{c.m.}}^\alpha|)\right)\,.
\end{equation}
We want to underline that by using this definition, the CFF
is not sensitive to the choice of the internal coordinates.
Considering the results of Eqs.~(\ref{eq:normd1})-(\ref{eq:normli1})
and expressing the integral of Eq.~(\ref{eq:intV}) in terms of
the Jacobi vectors of set ``B'' given in Eq.~(\ref{eq:jacvecB}),
the $\alpha+d$ CFF reduces to
\begin{equation}\label{eq:overlap_df2}
  \frac{f_{L}(r)}{r}= \frac{1}{\sqrt{15}}
    \left(\frac{\sqrt{6}}{4}\right)^{\frac{3}{2}}
    \int \prod_{i=1}^5d\xx_{iB}\,
    \delta\left(r-\sqrt{\frac{8}{3}}x_{2B}\right)
    \Psi_{\Li}^\dag {\cal A}
    \left[\left(\Psi_\alpha\times\Psi_d\right)_1
      Y_L(\hat r)\right]_1\,.   
\end{equation}
If we indicate with $\Psi_\alpha(1,2,3,4)$
the wave function of the $\alpha$-particle  constructed as the sum over
the 12 even permutations of the particles $(1,2,3,4)$ and 
with $\Psi_d(5,6)$ the wave function of the deuteron constructed with
the particles $(5,6)$, we can rewrite
Eq.~(\ref{eq:overlap_df2}) as
\begin{equation}
\begin{aligned}\label{eq:overlap_df3}
  \frac{f_{L}(r)}{r}= \sqrt{15}
    \left(\frac{\sqrt{6}}{4}\right)^{\frac{3}{2}}&
    \int \prod_{i=1}^5d\xx_{iB}\,
    \delta\left(r-\sqrt{\frac{8}{3}}x_{2B}\right)\\
    &\times\Psi_{\Li}^\dag
    \left[\left(\Psi_\alpha(1,2,3,4)\times\Psi_d(5,6)\right)_1
      Y_L(\hat r)\right]_1\,,
\end{aligned}
\end{equation}
where we used the antisymmetry of the $\Li$ wave function to eliminate the
antisymmetrization operator $\cal{A}$, and we
have multiplied for a factor 15 to take care 
of the fact that the initial function $\Psi_{\alpha+d}^{(L)}$ contains all the 
$4+2$ partitions of the six particles.
The explicit form of Eq.~(\ref{eq:overlap_df3}) in terms of the HH
functions is derived in Appendix~\ref{app:cff_cal}.
From the CFF, it is possible to calculate the so-called
spectroscopic factor, which is defined as
\begin{equation}
  {\cal S}_L=\int_0^\infty dr\,f_L(r)^2\,,
\end{equation}
and it can be interpreted
as the percentage of $\alpha+d$ clusterization in the $\Li$ wave function.

From the CFF, it is possible  to derive also the ANC.
In the asymptotic regions where the $\Li$ is completely clusterized in $\alpha+d$, the
wave function can be written as
\begin{equation}
  \overline{\Psi}_{\Li}=
  \Psi^{(0)}_{\alpha+d}\frac{\overline{f}_0(r)}{r}
  +\Psi^{(2)}_{\alpha+d}\frac{\overline{f}_2(r)}{r}\,,
  \quad r\rightarrow\infty\,.
\end{equation}
In these regions, the Schr\"odinger equation of the six particles can be written as
\begin{align}
  \left[H_\alpha+H_d
    -\frac{\hbar^2}{2m}\nabla^2_{x_{2B}}+\sum_{i\in \alpha}\sum_{j\in d}
    V_{ij}\right]\overline{\Psi}_{\Li}=-B_{\Li}\overline{\Psi}_{\Li}\,,\label{eq:schcff}
\end{align}
where $H_\alpha$ and $H_d$ are the Hamiltonian of the $\alpha$-particle and
the deuteron which, acting on $\overline{\Psi}_{\Li}$,
gives the binding
energies $B_\alpha$ and $B_d$.
If we consider the limit in which the intercluster distance goes to infinity, 
the intercluster interaction reduces to the long range Coulomb interactions between
protons, and thus Eq.~(\ref{eq:schcff}) becomes
\begin{equation}
  \Big[-\frac{\hbar^2}{2m}\nabla^2_{x_{2B}}+
    \sum_{i\in\alpha}\sum_{j\in d}
    \frac{e^2}{r_{ij}}\Big(\frac{1+\tau_z(i)}{2}\Big)
    \Big(\frac{1+\tau_z(j)}{2}\Big)+B_c\Big]
  \overline{\Psi}_{\Li}(r\rightarrow\infty)=0
  \,,
\end{equation}
where $B_c=B_{\Li}-B_\alpha-B_d$ is the cluster binding energy and
$e$ the electric charge.
Moreover, in this limit, the Coulomb potential reduces to
\begin{equation}
  \sum_{i\in\alpha}\sum_{j\in d}
  \frac{e^2}{r_{ij}}\Big(\frac{1+\tau_z(i)}{2}\Big)
  \Big(\frac{1+\tau_z(j)}{2}\Big)\overline{\Psi}_{\Li}(r\rightarrow\infty)
  \xrightarrow[]{r_\rightarrow\infty}
  \frac{2e^2}{r}\overline{\Psi}_{\Li}(r\rightarrow\infty)\,.
\end{equation}
Therefore, Eq.~(\ref{eq:schcff}) results
\begin{equation}\label{eq:rinfty}
  \Big[-\frac{\hbar^2}{2\mu}\Big(\frac{d^2}{dr^2}
    +\frac{L(L+1)}{r^2}\Big)+
    \frac{2e^2}{r}+B_c\Big]\overline{f}_L(r)=0\,,\qquad{\text{with}}\,\, L=0,2
  \,,
\end{equation}
where $\mu=\frac{4}{3}m$ is the reduced mass of the $\alpha+d$ system.
The solution of this equation is the Whittaker function, i.e. 
\begin{equation}
  \overline{f}_L(r)=W_{-\eta,L+1/2}(2k r)\,.
\end{equation}
with $k=\sqrt{\frac{2\mu B_c}{\hbar^2}}$ and $\eta=\frac{2e^2 \mu}{\hbar^2k}$.
Therefore, the asymptotic form of the wave function results to be
\begin{align}\label{eq:psiinfty}
  \overline{\Psi}_{\Li}(r\rightarrow\infty)=
  C_0\frac{W_{-\eta,1/2}(2k r)}{r}\Psi_{\alpha+d}^{(0)}
  +C_2\frac{W_{-\eta,5/2}(2k r)}{r}\Psi_{\alpha+d}^{(2)}\,,
\end{align}
where the coefficients $C_0$ and $C_2$ are the so called ANCs for
the $S$- and the $D$-wave, respectively.
By comparing the overlap given in Eq.~(\ref{eq:overlap_df}) with the
asymptotic behavior given in Eq.~(\ref{eq:psiinfty}) we can identify
\begin{equation}
  \lim_{r\rightarrow\infty}f_L(r)=C_L W_{-\eta,L+1/2}(2kr)\,.
\end{equation}
Therefore, by defining the ratio
\begin{equation}\label{eq:clr}
  C_L(r)=\frac{f_L(r)}
  {W_{-\eta,L+1/2}(2k r)}\,,
\end{equation}
we get the ANC as
\begin{equation}
  C_L=\lim_{r\rightarrow\infty}C_L(r)\,.
\end{equation}

\subsection{Results for the $\alpha+d$ cluster form factor and ANCs}\label{sec:res1}
In this section we discuss the results obtained for the $\alpha+d$ CFF
and ANC. In particular we focus on the convergence of the CFF and the
ANC as function of the maximum $K$ used in the expansion of the $\Li$
wave function.
In our calculation we use $l_d^{max}=40$ for the calculation of the
deuteron, and $l_\alpha^{max}=16$ and $K_{\al}=8$ for the
$\alpha$-particle. We use the  $\alpha$-particle computed only up to $K_{\al}=8$,
since the number of HH states used in the expansion
is large enough to have good convergence in
the binding energy (see Table~\ref{tab:alpha}), but at the same time it
is small enough to be easily managed in the computation of the CFF.

\begin{figure}
  \centering
  \includegraphics[scale=1]{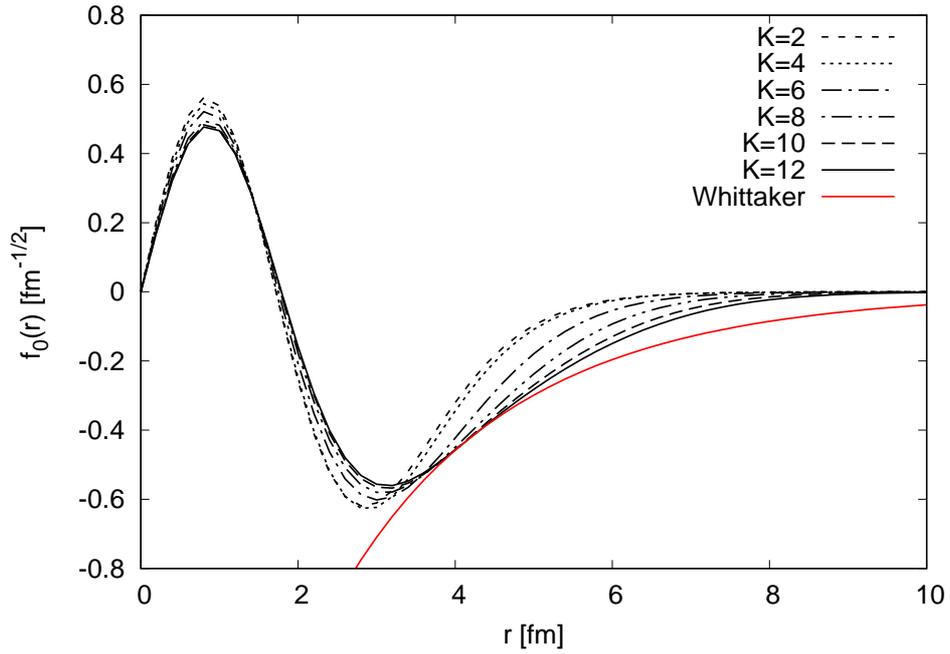}
  \caption{\label{fig:over_conv0}
    $S$-wave component of the reduced CFF for different values of
    $K$ used in the expansion of the $\Li$ wave
    function. The full red line represents the correct asymptotic behavior
    of the CFF given by the Whittaker function. These results are
    obtained with N3LO500-SRG1.5 potential.}
\end{figure}

\begin{figure}
  \centering
  \includegraphics[scale=1]{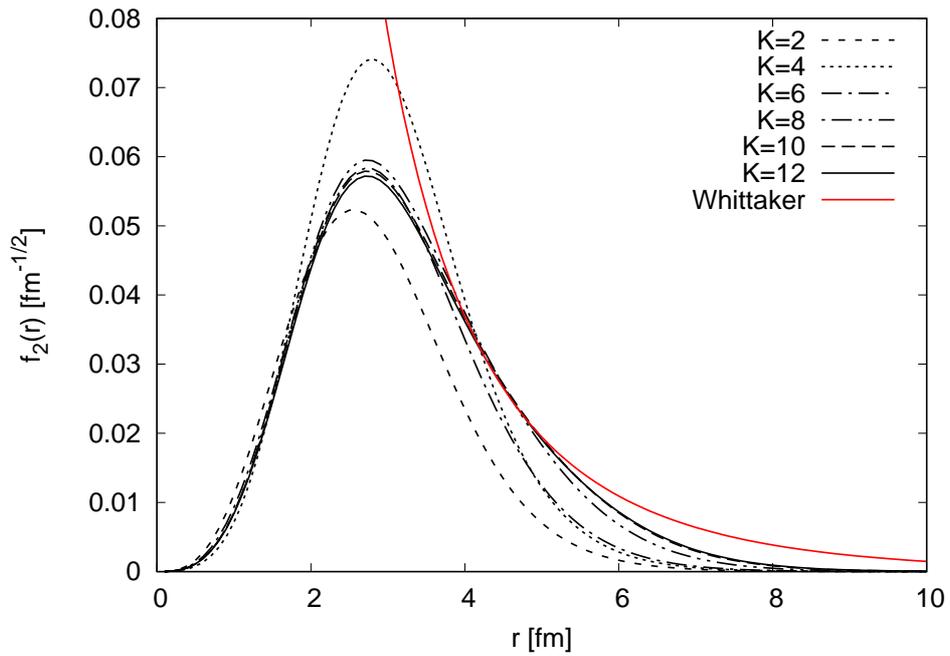}
  \caption{\label{fig:over_conv2}
  The same as Figure~\ref{fig:over_conv0}, for the $D$-wave component.}
\end{figure}

In Figure~\ref{fig:over_conv0} we plot the $S$-wave
component of the reduced CFF obtained with the N3LO500-SRG1.5 potential.
From the figure, it is clear that
the tail of the reduced CFF has not the correct behavior of the Whittaker
function (full red line) as predicted in Eq.~(\ref{eq:psiinfty}). This is
due to the limited number of HH states used in the expansion
of the  $\Li$ wave function, which are not enough to reproduce the
correct asymptotic behavior of the reduced CFF. However, it is also clear
that the HH states are slowly constructing the correct asymptotic
slope when $K$ increases. On the other hand, for the
short range part ($r<4$ fm) the convergence is fast.  Similar comments
apply also to the $D$-wave component given in Figure~\ref{fig:over_conv2}.

\begin{figure}
  \centering
  \includegraphics[scale=1]{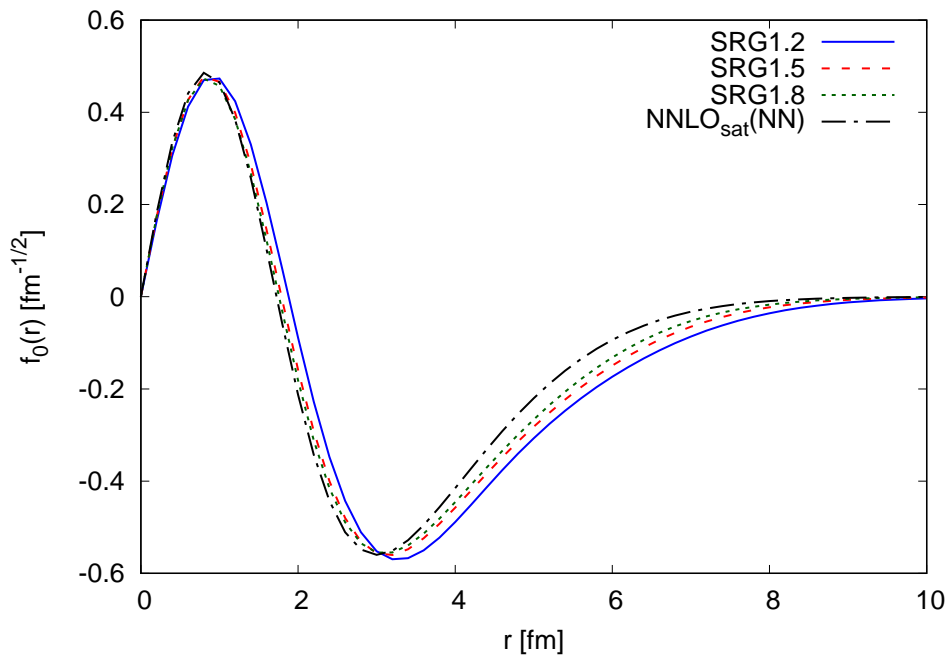}
  \caption{\label{fig:over_comp0} Comparison of the $S$-wave component of the
    reduced CFF calculated with  SRG1.2(blue), SRG1.5(red), SRG1.8(green)
    and NNLO$_{sat}$(NN)(black) potentials. The CFF are obtained
    using the $\Li$ wave function computed with $K=12$.}
\end{figure}

\begin{figure}
  \centering
  \includegraphics[scale=1]{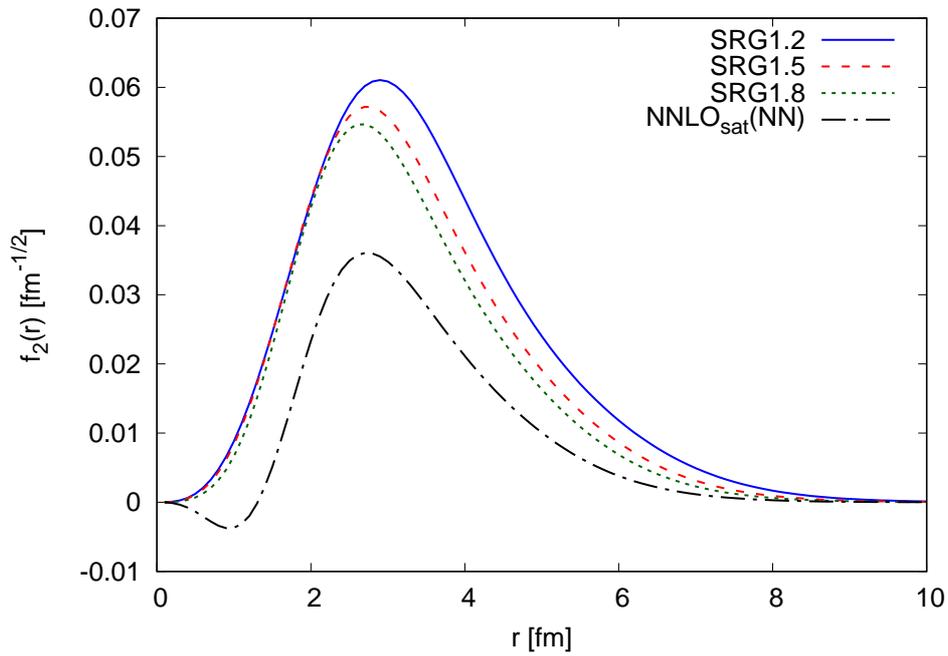}
  \caption{\label{fig:over_comp2} The same as Figure~\ref{fig:over_comp2}
  for the $D$-wave component.}
\end{figure}

In Figures~\ref{fig:over_comp0} and~\ref{fig:over_comp2}
we compare the reduced CFF calculated with $K=12$ using
the various potentials, for $S$- and $D$-wave component, respectively.
By inspecting the figures, it is clear that 
the ``harder'' is the potential, the worse the tail of the CFF is constructed.
In fact, for ``hard'' potentials, the HH states are mainly used to construct the
short range correlations, and not the tail.
As regarding the short range part of the $S$-wave component, it is possible
to note the presence of a node which does not depend on the particular
potential we use, but is strictly related to the symmetry properties of the
$\alpha+d$ clusterization. In particular, the presence of a node reflects
the fact that a lower bound state in the $\alpha+d$ system exists.
However, this bound state results totally symmetric and therefore it
is forbidden by the Pauli principle. The results obtained for the
$S$-wave  CFF are consistent with the one reported in
Refs.~\cite{Forest1996,Nollet2001,Navratil2004,Navratil2011}. 
For the $D$-wave, the shapes of the CFF are very similar for the
SRG evolved potentials but not for the NNLO$_{sat}$(NN).
In particular, in the NNLO$_{sat}$(NN) case, a node, which is not present
in the SRG cases, appears.
A similar behavior was found in the GFMC
calculation of the CFF,
done with the AV18/UIX potential in Refs.~\cite{Forest1996,Nollet2001}.
The presence of this node can be related to the strength of
the tensor term in the potential, as already discussed in Ref.~\cite{Kukulin1995}.
This hypothesis is confirmed also by the fact that the node
disappears in the case of the
SRG evolved potentials, where the impact of tensor forces is reduced.
The results  for the SRG potentials are perfectly consistent with the
ones reported in Refs.~\cite{Navratil2004,Navratil2011},
obtained for SRG evolved potentials as well.

\begin{figure}
  \centering
  \includegraphics[scale=1]{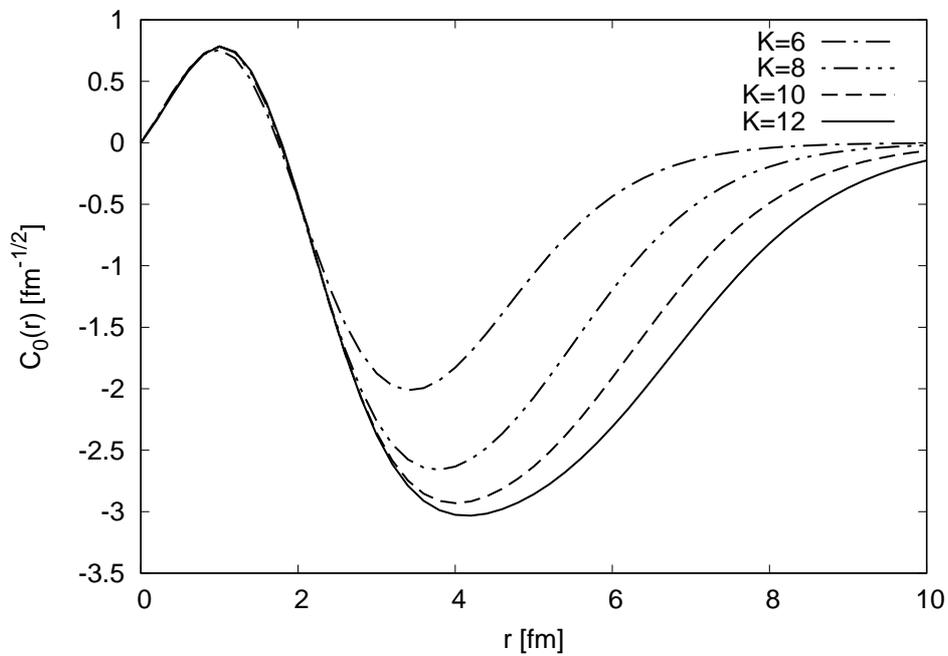}
  \caption{\label{fig:C0r} Function $C_0(r)$ as defined in Eq.~(\ref{eq:clr})
    computed for the N3LO500-SRG1.5 by varying $K$ in the expansion of
    $\Li$ wave function. The result for $K=2$ and 4 are not present since
    $B_c<0$.}
\end{figure}

\begin{figure}
  \centering
  \includegraphics[scale=0.7]{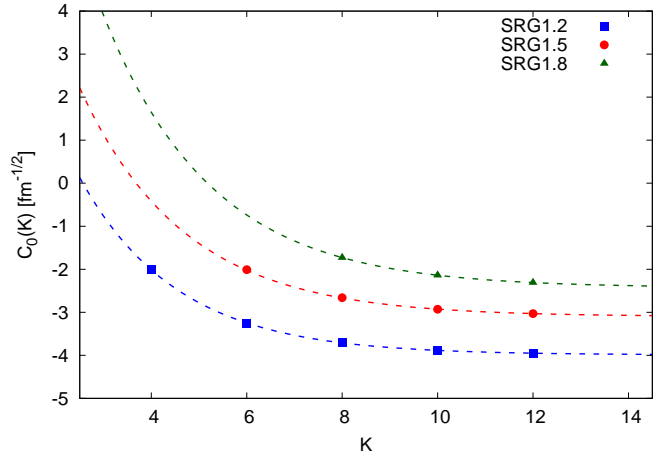}
  \caption{\label{fig:cl0} Values of $C_0$ as function of $K$
    for the SRG potential models,
    SRG1.2 (blue), SRG1.5 (red) and  SRG1.8 (green).
    We report also the fit performed
    using Eq.~(\ref{eq:fitcl}) (dashed lines).}
\end{figure}

By using Eq.~(\ref{eq:clr}), we can extrapolate the ANC. In Figure~\ref{fig:C0r} we
plot the ratio $C_0(r)$ for the SRG1.5 potential, as function of the maximum
value of $K$ in the expansion of $\Li$. The ratio $C_0(r)$ shows a sort of
``plateau'' around the minimum, from which we can try to estimate the ANC. It is
nice to observe that by increasing $K$, the ``plateau'' is enlarging and
slowly converging. Similar results are obtained for $C_2(r)$.
In Table~\ref{tab:cl_coeff} we report the values of the ANCs
estimated by the ``plateau'' for all the potentials considered, as function of $K$
($C_L(K)$).
Moreover, we indicate also the energy $B_c$ computed for each
$K$, since the Whittaker function depends on it. Obviously, when $B_c<0$ the
Whittaker function is not defined and so the ANC, since the
$\Li$ results not bound. From the values of $C_L(K)$ we can extrapolate the
asymptotic ANC $C_L(\infty)$ by fitting
\begin{equation}\label{eq:fitcl}
  C_L(K)=C_L(\infty)+A{\rm e}^{-bK}\,.
\end{equation}
on the calculated $C_L(K)$.
In the case of the NNLO$_{sat}$(NN) we cannot give an extrapolation, since the
available $C_L(K)$ are not enough for performing a fit. As regarding the
theoretical errors we consider a conservative error on $C_L(\infty)$ defined as
\begin{equation}\label{eq:fiterr}
  \Delta C_L=0.5\times|C_L(12)-C_L(\infty)|\,.
\end{equation}
In Figure~\ref{fig:cl0} we plot the values of
$C_0$ as function of $K$ and the fit performed with Eq.~(\ref{eq:fitcl}).
A similar result is obtained for $C_2$.
By comparing Figures~\ref{fig:over_comp0} and~\ref{fig:over_comp2} with
the results in Table~\ref{tab:cl_coeff},
the differences in the tails of the reduced CFF for the various potentials
are not enough to explain the numerical differences
in the ANCs. The main reason of the differences between
the ANCs must be found in the different binding energy $B_c$ used to compute
the Whittaker functions. However, these estimates of the ANC are rather uncertain,
due to the lack of well established ``plateau'', as shown in Figure~\ref{fig:C0r}.
Therefore, to this extrapolation ``error'' we should add also a systematic error of uncertain
magnitude.
\begin{table}
  \centering
  \begin{tabular}{r|ccc||ccc}
    \hline
    \hline
     & \multicolumn{3}{c||}{SRG1.2} & \multicolumn{3}{c}{SRG1.5}\\
    $K$ & $B_c$ & $C_0$ & $C_2$& $B_c$ & $C_0$ & $C_2$ \\
    \hline
    2 & $-3.736$ & $-$ & $-$ &  $-5.359$ & $-$ & $-$\\ 
    4 & $\m0.891$ & $-2.01$ & $0.030$ & $-1.615$ & $-$ & $-$\\
    6 & $\m2.359$ & $-3.25$ & $0.075$ & $\m1.071$ & $-2.01$ & $0.025$\\
    8 & $\m2.766$ & $-3.70$ & $0.096$ & $\m1.929$ & $-2.66$ & $0.050$\\
    10& $\m2.909$ & $-3.89$ & $0.103$ & $\m2.222$ & $-2.93$ & $0.059$\\
    12& $\m2.955$ & $-3.95$ & $0.104$ & $\m2.323$ & $-3.03$ & $0.061$\\
    $\infty$ &$\m3.00(1)$ & $-3.99(2)$ & $0.107(2)$ & $\m2.46(2)$ & $-3.10(4)$ & $0.063(1)$\\
    \hline
    \hline
     & \multicolumn{3}{c||}{SRG1.8} & \multicolumn{3}{c}{NNLO$_{sat}$(NN)}\\
    $K$ & $B_c$ & $C_0$ & $C_2$& $B_c$ & $C_0$ & $C_2$ \\
    \hline
    2 & $-12.108$   & $-$     & $-$ &  $-18.74$ & $-$ & $-$\\ 
    4 & $-5.105$    & $-$     & $-$ &  $-10.99$ & $-$ & $-$\\
    6 & $-0.911$    & $-$     & $-$ &  $-4.16$ & $-$ & $-$\\
    8 & $\m0.740$ & $-1.73$ & $0.018$ & $-0.74$ & $-$ & $-$\\
    10& $\m1.332$ & $-2.14$ & $0.030$ & $\m0.61$ & $-1.59$ & $0.010$\\
    12& $\m1.551$ & $-2.31$ & $0.034$ & $\m1.15$ & $-1.93$ & $0.017$\\
    $\infty$ &$\m2.02(9)$ & $-2.43(6)$ & $0.036(1)$ & $2.11(20)$ & n.a. & n.a.\\
    \hline
    \hline
  \end{tabular}
  \caption{\label{tab:cl_coeff} Values of the binding energy $B_c$ in MeV
    and the ANC $C_0$ and $C_2$ in fm$^{-1/2}$ as function of $K$
    for the various potential models considered. In the rows indicated with
    ``$\infty$'' we report the extrapolated values with the errors
    (between parenthesis) obtained with Eq.~(\ref{eq:fiterr}).
    However, these estimates of the ANC are rather uncertain
    due to the lack of well established ``plateau'' as shown in
    Fig.~\ref{fig:C0r}. So to this extrapolation ``error''
    we should add also a systematic error of uncertain magnitude.}
\end{table}

From the CFF it is also possible to compute the spectroscopic
factor ${\cal S}_L$. 
In Table~\ref{tab:scal} we report the  values of the
spectroscopic factors obtained for the various potentials. We do not
report the full convergence pattern obtained by varying $K$, but only
the converged digits when $K=12$.
For this quantity the convergence issues are less
problematic, being an integral quantity.
Independently on the potential used,
it results that $\Li$ is clusterized in an $\alpha+d$ system for
more than $80\%$. The differences between the values obtained with
the SRG evolved potentials
can be due to the fact that the induced and proper three-body forces
are not included. Moreover, in the table we compare our results
with the calculation of Refs.~\cite{Forest1996,Nollet2001,Navratil2004}.
The values are similar, even if obtained with different potential
models. We want to underline that our calculation of the spectroscopic
factor is also in line with the experimental
estimate of Ref.~\cite{Robertson1981}, based on an optical potential analysis
of the data.
\begin{table}
  \centering
  \begin{tabular}{llccc}
    \hline
    \hline
    Method & Potential & ${\cal S}_0$ & ${\cal S}_2$ & ${\cal S}_0+{\cal S}_2$\\
    \hline
    \multirow{4}{*}{HH (This work)}& SRG1.2 & 0.909  & 0.008 & 0.917\\
                                   & SRG1.5 & 0.868  & 0.007 & 0.875\\
                                   & SRG1.8 & 0.840  & 0.006 & 0.846\\
                                   & NNLO$_{sat}$(NN)& 0.805 & 0.002 & 0.807\\
    \hline
    GFMC (Ref.~\cite{Forest1996})  & AV18/UIX & $0.82$ & $0.021$ & 0.84\\
    GFMC (Ref.~\cite{Nollet2001})  & AV18/UIX & $-$ & $-$ & 0.87(5)\\
    NCSM (Ref.~\cite{Navratil2004})& CD-B2k   & 0.822 & 0.006 & 0.828\\
    Exp. (Ref.~\cite{Robertson1981}) &          & $-$ & $-$ & 0.85(4)\\ 
    \hline
    \hline
  \end{tabular}
  \caption{\label{tab:scal} Values of the spectroscopic factors
    for the various potential models considered. In the last rows 
    we compare our results with the most recent values in literature and
    the experimental one.}
\end{table}

\section{Equation for the cluster form factor}\label{sec:eqcff}

Since the procedure adopted so far results somewhat unsatisfactory, due to
the difficult identification of the ``plateau'',
in this section we follow another procedure,
based on Ref.~\cite{Timofeyuk1998}.
With this approach, we can extrapolate the ANC with greater accuracy.

The $\Li$  wave function 
$\Psi_{\Li}(J^\pi=1^+)$ can be written in terms of
a complete basis of  4+2 cluster
wave functions, namely
\begin{equation}\label{eq:adnm}
  \Psi_{\Li}=\sum_{LSnm}I_{LSnm}(r)
  \frac{1}{\sqrt{15}}{\cal A}
  \left[\left(\Psi_4^{n}
    \times\Psi_2^{m}
    \right)_S Y_L(\hat r)\right]_1
    \,,
\end{equation}
where  $\Psi_4^{n}$($\Psi_2^{m}$) is a four(two)-body
cluster wave function,  $n(m)$  indicates that it is an eigenstate  
of the four(two)-body Hamiltonian with eigenvalue $E_n(E_m)$, namely
$H_4\Psi_4^n=E_n\Psi_4^n(H_2\Psi_2^m=E_m\Psi_2^m)$.
The function $I_{LSnm}(r)$ is  the  overlap integral, and it is defined as
\begin{equation}
  I_{LSnm}(r)=\frac{\bra \frac{1}{\sqrt{15}}{\cal A}
  \left[\left(\Psi_4^{n}
    \times\Psi_2^{m}
    \right)_S Y_L(\hat r)\right]_1|\Psi_{\Li}\ket_{r}}
    {\left[\bra\Psi_2^{m}|\Psi_2^{m}\ket
        \bra\Psi_4^{n}|\Psi_4^{n}\ket
        \bra\Psi_{\Li}|\Psi_{\Li}\ket\right]^{1/2}}
        \,,
\end{equation}
where $\bra\cdots|\cdots\ket_{r}$ indicates that we are using the integral over the
internal coordinates as defined in Eq.~(\ref{eq:intV}), and we have normalized
the overlap with the norm of the wave function as defined in
Eq.~(\ref{eq:norma}).
In such a way, if we have $\Psi_4^{n}=\Psi_\alpha$ and
$\Psi_2^{m}=\Psi_d$, we obtain exactly the definition of the $\alpha+d$
CFF, namely
\begin{equation}
  I_{LS\alpha d}(r)=\frac{f_L(r)}{r}\,,
\end{equation}
of Eq.~(\ref{eq:overlap_df}).
The wave function $\Psi_{\Li}$ is the solution of the six-body Hamiltonian
\begin{equation}\label{eq:h6}
  H_6\Psi_{\Li}=-B_{\Li}\Psi_{\Li}\,,
\end{equation}
where the Hamiltonian $H_6$ can be rewritten in terms of the Hamiltonian of the
four-body system, the Hamiltonian of the two-body systems and the Hamiltonian
describing the relative motion of the two clusters, i.e.
\begin{equation}\label{eq:h6ad}
  H_6=H_4+H_2
  -\frac{\hbar^2}{2\mu}\nabla^2_r+
  \sum_{i\in\al}\sum_{j\in d}V_{ij}\,.
\end{equation}
Inserting Eq.~(\ref{eq:adnm}) in Eq.~(\ref{eq:h6}), multiplying
the left-hand-side by the $\alpha+d$ cluster function $\bra\Psi_{\al+d}^{(L)}|$
defined in Eq.~(\ref{eq:clusterw}), and
integrating over all the internal variables except the intercluster distance $r$,
we obtain the differential equation for the $\alpha+d$ reduced CFF $f_L(r)$, i.e.
\begin{equation}\label{eq:fr}
  \left[-\frac{\hbar^2}{2\mu}\left(\frac{d^2}{dr^2}
    -\frac{L(L+1)}{r^2}\right)
    +\frac{2e^2}{r}+B_c\right]f_L(r)+g_L(r)=0\,,
\end{equation}
where $B_c=B_{\Li}-B_\alpha-B_d$.
The function $g_L(r)$ is a source term which results to be
\begin{equation}\label{eq:GL}
    \frac{g_L(r)}{r}=\frac{\bra\Psi_{\Li}|
    \left(\sum_{i\in\al}\sum_{j\in d}V_{ij}-\frac{2e^2}{r}\right)
    |\Psi_{\al+d}^{(L)}\ket_r}{\left[\bra\Psi_d|\Psi_d\ket
        \bra\Psi_\al|\Psi_\al\ket
        \bra\Psi_{\Li}|\Psi_{\Li}\ket\right]^{1/2}}
    \,.
\end{equation}
When $r\rightarrow\infty$ the source term
$g_L(r)\rightarrow 0$ and we obtain exactly Eq.~(\ref{eq:rinfty}) and so
the asymptotic solution coincides with the Whittaker function.
In this form, Eq.~(\ref{eq:fr}) can be solved with high accuracy,
using the Numerov method, thus obtaining the correct trend of the
$f_L(r)$ for $r\rightarrow\infty$ and a more accurate value for the ANC.

  \subsection{The projection of the asymptotic states}\label{sec:pjmethod}

  The calculation of the source term $g_L(r)$ is quite involved, due to the
  presence of the cluster wave function $|\Psi_{\alpha+d}^{(L)}\ket$
  which does not
  allow us to use the properties of the HH states in the computation.
  In order to restore these
  properties and simplify the calculation, in this section we introduce
  a new numerical approach called ``projection'' method. 

  Any wave function can be expanded in terms of
  states given by products of HH functions, spin-isospin states, and functions of the
  hyperradius, since they form a complete basis.
  Denoting these states
  by $|HH_\mu\ket$, let us expand a generic antisymmetric
  cluster wave function $|\Psi_{A+B}\ket$ as follows
  \begin{equation}
    |\Psi_{A+B}\ket=\sum_{\mu}c_\mu |HH_\mu\ket\,,
  \end{equation}
  where $c_\mu$ are coefficients defined by projecting
  the scattering wave function on the $|HH_\mu\ket$ states, namely
  \begin{equation}
    c_\mu=\bra HH_\mu | \Psi_{A+B}\ket\,,
  \end{equation}
  and the states $|HH_\mu\ket$ are normalized such that
  $\bra HH_{\mu'}|HH_\mu\ket=\delta_{\mu'\mu}$.
  By expressing our cluster wave function in such a way, it is then possible
  to compute the source term in Eq.~(\ref{eq:GL}), taking advantage of
  the HH properties and in particular of the TC.
  This method can be applied in different situations but
  in this section we will focus only on $A=6$ considering the
  $\alpha+d$ cluster wave function given in Eq.~(\ref{eq:clusterw}), namely
  \begin{equation}
  \begin{aligned}
    \Psi_{A+B}=\Psi^{LSJ,\xi}_{\alpha+d}&=\frac{1}{\sqrt{15}}{\cal A}
    \left\{\left[\left(\Psi_\alpha\times\Psi_d\right)_S
      Y_L(\hat r)\right]_JF(r,\xi)\right\}\\
    &=\frac{1}{\sqrt{15}}\sum_{p=1}^{360}
    \left[\left(\Psi_\alpha(i,j,k,l)\times\Psi_d(m,n)\right)_S
      Y_L(\hat r)\right]_JF(r,\xi)\,,
  \end{aligned}
  \end{equation}
  where $F(r,\xi)$ is a generic intercluster
  function which depends on $r$ and on other possible quantum numbers
  designed by $\xi$.
  Let us define the generic cluster wave function $\Psi^{LSJ,\xi}_{\alpha+d}$
  in the reference permutation $p=1$ (1,2,3,4,5,6), which reads
  \begin{equation}
    \Psi^{LSJ,\xi}_{\alpha+d}(p=1)=
    \left\{\left[\Psi_\alpha(p=1)\times\Psi_d(p=1)\right]_S Y_L(\hat r)
    \right\}_J F(r,\xi)\,,
  \end{equation}
  With  $\Psi_\alpha(p=1)$ we indicate the term of the $\alpha$ wave function
  given in Eq.~(\ref{eq:psialpha}) corresponding to the
  reference permutation $p=1$ of the particles (1,2,3,4) only,
  while with  $\Psi_d(p=1)$ we 
  indicate the wave function of the deuteron constructed with the particles 5,6.

  As basis for our expansion we will consider the HH functions constructed with
  the set ``B'' of the Jacobi coordinates as defined in Eq.~(\ref{eq:jacvecB})
  since they are the  natural set for describing $\alpha+d$ clusterization.
  For the hyperradial part we consider the Laguerre
  polynomials $f_\lbb(\rho)$ as defined in Eq.~(\ref{eq:fllag}).
  Explicitly, the basis in which we expand is given by
  \begin{equation}
  \begin{aligned}\label{eq:pjad}
    \Phi^{\kb\,\lb\,\sbb\,\tb}_{\lbb,\bb}(p=1)&=f_{\lbb}(\rho)\,
    \left\{{\cal Y}^{\kb\,\lb}_{\bar\mu}(\Omega_B)
    \left[\left((\sg_1\sg_2)_{\bar S_2}\sg_3\right)_{\bar S_3}
      \left((\sg_4\sg_5)_{\bar S_4}\sg_6\right)_{\bar S_5}\right]_{\bar S}
    \right\}_{J}\\
    &\times\left[\left((\tp_1\tp_2)_{\bar T_2}\tp_3\right)_{\bar T_3}
      \left((\tp_4\tp_5)_{\bar T_4}\tp_6\right)_{\bar T_5}\right]_{\bar T}\,,
  \end{aligned}
  \end{equation}
  where ${\cal Y}^{\kb\,\lb}_{\bar\mu}(\Omega_B)$
  is the HH function defined in Appendix~\ref{app:cff_cal},
  and $\bb$ is the complete set of quantum numbers
  which defines the basis state as given in Eq.~(\ref{eq:alpha}). From now on we indicate
  with overlined indices the quantum numbers which refers to the ``projection''
  basis states.
  Being this a complete basis on the
  spatial-spin and isospin space, we can rewrite our cluster wave function
  as
  \begin{equation}\label{eq:cbar0}
    \Psi^{LSJ,\xi}_{\alpha+d}(p=1)=\sum_{\lbb=0}^{\lbb_{max}}
    \sum_{\kb=0}^{\kb_{max}}
    \sum_{\lb\,\sbb\,\tb,\bb}c^{\kb\,\lb\,\sbb\,\tb}_{\lbb,\bb}(LSJ,\xi)
    \Phi^{\kb\,\lb\,\sbb\,\tb}_{\lbb,\bb}(p=1)\,.
  \end{equation}
  The equivalence holds exactly only when $\kb_{max}\rightarrow\infty$ and
  $\lbb_{max}\rightarrow\infty$. Obviously this is not the case in practical
  applications, since
  we are limited in the dimension of the basis we can use.
  However, we can check the quality of our approximation by looking to the
  convergence of the observables when we increase $\kb_{max}$ and $\lbb_{max}$.
  The coefficients of the expansion are defined as
  \begin{equation}\label{eq:cbar}
    c^{\kb\,\lb\,\sbb\,\tb}_{\lbb,\bb}(LSJ,\xi)
    =\bra \Phi^{\kb\,\lb\,\sbb\,\tb}_{\lbb,\bb}(p=1)|
    \Psi^{LSJ,\xi}_{\alpha+d}(p=1)\ket_{\Omega_B,\rho}\,,
  \end{equation}
  where $\bra\cdots|\cdots\ket_{\Omega_B,\rho}$ represents the integration over all
  the hypercoordinates ``B'' and the trace over spins and isospins. Their explicit
  expression and the details of the calculations are given
  in Appendix~\ref{app:coeffcbar}. We note here that this calculation is rather easy,
  as both wave functions are constructed with the same permutation. Therefore, the
  angular integration and the spin-isospin traces can be performed analytically (most
  of them are just $\delta$-functions).
  
  Before concluding this section,
  we need to remember that the  cluster function must be
  totally antisymmetric, namely
  \begin{equation}
    \Psi^{LSJ,\xi}_{\alpha+d}=\frac{1}{\sqrt{15}}
    \sum_{p=1}^{360}\Psi^{LSJ,\xi}_{\alpha+d}(p)\,.
  \end{equation}

  Let us expand $\Psi^{LSJ}_{\alpha+d}(p)$ in terms of the basis states
  $\Phi^{\kb\,\lb\,\sbb\,\tb}_{\lbb,\bb}(p)$
  constructed with the same permutation.
  The expansion coefficients in this case are nothing else than the coefficients of
  Eq.~(\ref{eq:cbar0}), as it is easy to prove, redefining the numberings of the
  particles in order to obtain the reference permutation $(1,2,3,4,5,6)$. Thus,
  \begin{equation}
    \Psi^{LSJ,\xi}_{\alpha+d}=\frac{1}{\sqrt{15}}
    \sum_{\lbb=0}^{\lbb_{max}}
    \sum_{\kb=0}^{\kb_{max}}
    \sum_{\lb\,\sbb\,\tb\,\bb}c^{\kb\,\lb\,\sbb\,\tb}_{\lbb,\bb}(LSJ,\xi)
    \sum_{p=1}^{360}\Phi^{\kb\,\lb\,\sbb\,\tb}_{\lbb,\bb}(p)\,,
  \end{equation}
  where thanks to the fact that we are using the HH formalism, we can
  impose antisymmetrization by selecting only the states $\bb$ such that
  $(-1)^{\bar\ell_5+\bar S_2+\bar T_2}=-1$.
  Therefore, expressing the sum over the permutations through the TC, we have
  \begin{equation}\label{eq:antisyad}
    \Psi^{LSJ,\xi}_{\alpha+d}=\frac{1}{\sqrt{15}}
    \sum_{\lbb=0}^{\lbb_{max}}
    \sum_{\kb=0}^{\kb_{max}}
    \sum_{\lb\,\sbb\,\tb,\bb}c^{\kb\,\lb\,\sbb\,\tb}_{\lbb,\bb}(LSJ,\xi)
    \sum_{\bb'}
    A^{\kb\,\lb\,\sbb\,\tb,B}_{\bb,\bb'}
    \Phi^{\kb\,\lb\,\sbb\,\tb}_{\lbb,\bb'}(p=1)\,,
  \end{equation}
  where  $A^{\kb\,\lb\,\sbb\,\tb,B}_{\bb,\bb'}$
  are the TC computed in the set of coordinates ``B''
  (see Appendix~\ref{app:transform} for more details).
  We are now ready for computing the source term. 
 
  \subsection{Calculation of the source term}

  The source term given in Eq.~(\ref{eq:GL})
  can be  divided  in a  Coulombian term, defined as
  \begin{equation}
  \begin{aligned}
    \frac{g^C_L(r)}{r}=-\frac{2e^2}{r}&
    \sqrt{15}
    \left(\frac{\sqrt{6}}{4}\right)^{\frac{3}{2}}
    \int\prod_{i=1}^5d\xx_{iB}\,
    \delta\left(r-\sqrt{\frac{8}{3}}x_{2B}\right)\\
    &\times\Psi_{\Li}^\dag
    \left[\left(\Psi_\alpha(1,2,3,4)\times\Psi_d(5,6)\right)_1
      Y_L(\hat r)\right]_1\,, 
  \end{aligned}
  \end{equation}
  and a potential term, defined as
  \begin{equation}
  \begin{aligned}\label{eq:GVL}
    \frac{g^V_L(r)}{r}&= \sqrt{15}
    \left(\frac{\sqrt{6}}{4}\right)^{\frac{3}{2}}
    \int \prod_{i=1}^5d\xx_{iB}\,
    \delta\left(r-\sqrt{\frac{8}{3}}x_{2B}\right)\\
    &\times\Psi_{\Li}^\dag
    \left(\sum_{i\in\alpha}\sum_{j\in d}V_{ij}\right)
    \left[\left(\Psi_\alpha(1,2,3,4)\times\Psi_d(5,6)\right)_1
      Y_L(\hat r)\right]_1\,,
  \end{aligned}
  \end{equation}
  where the symbols and the arguments used to derive these expressions
  are the same as used in Eq.~(\ref{eq:overlap_df3}). In this section, in order to
  simplify the notations, we consider only local potentials. The generalization
  to the non-local ones is straightforward.

  As regarding the calculation of the Coulomb part, we can repeat the 
  procedure followed for the $\alpha+d$ CFF in Section~\ref{sec:overlap}.
  The computation of the potential part is more involved 
  due to the presence of the $\delta$-function.
  In order to overcome this difficulty,
  we can expand  $g_L^V(r)/r$ in terms of the Laguerre polynomials, i.e.
  \begin{equation}\label{eq:glvexp}
    \frac{g_L^V(r)}{r}=\sum_{n=0}^{N_{\text{max}}}C_n^{L}f_n(r)\,,
  \end{equation}
  where
  \begin{equation}
    f_n(r)=\gamma^{\frac{3}{2}}\sqrt{\frac{n!}{(n+2)!}}
      L_n^{(2)}(\gamma r)\mathrm{e}^{-\frac{\gamma r}{2}}\,.
  \end{equation}
  The parameter $\gamma$ is chosen to optimize the expansion.
  The coefficients $C_n^{L}$ are given by
  \begin{equation}\label{eq:cngl}
    C_n^L=\int dr \,r^2 \frac{g_L^V(r)}{r}f_n(r)\,,
  \end{equation}
  and, substituting Eq.~(\ref{eq:GVL}) in Eq.~(\ref{eq:cngl}), we obtain
  \begin{equation}
  \begin{aligned}\label{eq:cnglx}
    C_n^L&=\sqrt{15}
    \left(\frac{\sqrt{6}}{4}\right)^{\frac{3}{2}}
    \int \prod_{i=1,5}d\xx_{iB}
    \,\Psi_{\Li}^\dag
    \left(\sum_{i\in\alpha}\sum_{j\in d}V_{ij}\right)\\
    &\times\left[\left(\Psi_\alpha(1,2,3,4)\times\Psi_d(5,6)\right)_1
      Y_L(\hat r)\right]_1f_n(r)
    \Big|_{\br=\sqrt{\frac{8}{3}}{\boldsymbol{x}}_{2B}}\,.
  \end{aligned}
  \end{equation}
  We can now define the cluster wave function as
  \begin{equation}\label{eq:clustern}
    \Phi_{L,n}=\left[\left(\Psi_\alpha(1,2,3,4)\times\Psi_d(5,6)\right)_1
      Y_L(\hat r)\right]_1f_n(r)\,.
  \end{equation}
  which can be expanded by using the ``projection'' method.
  Let us remember that $\Psi_\alpha(1,2,3,4)$ is built by
  summing over the 12 even permutations of $(1,2,3,4)$. Therefore
  Eq.~(\ref{eq:clustern}), in terms of the ``projection'' basis states given
  in Eq.~(\ref{eq:pjad}) reads
  \begin{equation}\label{eq:clustern1}
    \Phi_{L,n}=    \sum_{\lbb=0}^{\lbb_{max}}
    \sum_{\kb=0}^{\kb_{max}}
    \sum_{\lb\,\sbb\,\tb,\bb}c^{\kb\,\lb\,\sbb\,\tb}_{\lbb,\bb}(L,n)
    \sum_{p_\alpha=1}^{12}\Phi^{\kb\,\lb\,\sbb\,\tb}_{\lbb,\bb}(p_\al)
    \,,
  \end{equation}
  where with  $p_\alpha$ we indicate the even permutation of $(i,j,k,l)$ of the
  six particles $(i,j,k,l,5,6)$.
  To be noticed that in this case, contrarily to Eq.~(\ref{eq:antisyad}),
  since we are not
  summing over all the 360 permutations of the six-particles, the states $\bb$
  are not necessarily antisymmetric.
  Now let us change the numberings of the particles so that the potential acts
  always on particles (1,2). 
  In this way Eq.~(\ref{eq:cnglx}) reduces to
  \begin{equation}
  \begin{aligned}
    C_n^L&=4\sqrt{15}
    \left(\frac{\sqrt{6}}{4}\right)^{\frac{3}{2}}\sum_{\lbb=0}^{\lbb_{max}}
    \sum_{\kb=0}^{\kb_{max}}
    \sum_{\lb\,\sbb\,\tb,\bb}c_{\lbb,\bb}^{\kb\,\lb\,\sbb\,\tb}(L,n)\\
    &\times\int \prod_{i=1}^5 d\xx_{iB}
    \Psi_{\Li}^\dag(\rho,\Omega_B) V_{1,2}(\xx_{5B})
    f_\lbb(\rho)
    \sum_{p^*}\Phi_{\bb}^{\kb\,\lb\,\sbb\,\tb}(p^*)\,,
  \end{aligned}
  \end{equation}
  where $p^*$ are $12\times8=96$ even permutations deriving from this operation
  of reordering
  (the 96 permutations derive from the $\sum_{p_\alpha}$ and $\sum_{ij}$). 
  In order to compute this integral it is convenient to return to the standard
  set of hypercoordinates. Therefore, we  rewrite 
  the sum over permutation $p^*$ by using the TC, which in $jj$-coupling results in
  \begin{equation}
    \sum_{p^*}\Phi_{\bb}^{\kb\,\lb\,\sbb\,\tb}(\rho,\Omega_{Bp^*})=
    \sum_{\bar\nu} B^{\kb\,\lb\,\sbb\,\tb,B*\rightarrow 1}_{\bb, \bar \nu}
    \Xi_{\bar \nu}^{\kb\,\lb\,\sbb\,\tb}(\rho,\Omega_5)\,,
  \end{equation}
  where the coefficients $B^{\kb\,\lb\,\sbb\,\tb,B*\rightarrow 1}_{\bar \nu}$ are computed
  summing only on the permutations $p^*$ and transforming
  the HH functions written in
  terms of set ``B'' in terms of the standard one (see
  Appendix~\ref{app:transform}), while the functions
  $\Xi_{\bar \nu}^{\kb\,\lb\,\sbb\,\tb}(\rho,\Omega_5)$ are defined in Eq.~(\ref{eq:PHIjj}).
  Finally, the coefficients $C_n^L$ are given by
  \begin{equation}
  \begin{aligned}
    C_n^L&=4\sqrt{15}
    \left(\frac{\sqrt{6}}{4}\right)^{\frac{3}{2}}\sum_{\lbb=0}^{\lbb_{max}}
    \sum_{\kb=0}^{\kb_{max}}
    \sum_{\lb\,\sbb\,\tb,\bb}c_{\lbb,\bb}^{\kb\,\lb\,\sbb\,\tb}(L,n)\\
    &\times\sum_{KLST}\sum_{l,\alpha} a^{KLST}_{l,\alpha}
    \sum_{\nu,\bar \nu}B^{KLST}_{\alpha,\nu}B^{\kb\,\lb\,\sbb\,\tb,B*\rightarrow1}_{\bb,\bar \nu}
    v^{KT,\bar K\bar T}_{l\nu,\lbb\bar\nu}\,,
  \end{aligned}
  \end{equation}
  where $a^{KLST}_{l,\alpha}$ are the variational coefficients of the $\Li$
  wave function and $\nu$ is defined in Eq.~(\ref{eq:nu}). Here we have used
  the expansion of $\Li$ wave function in terms of the standard set of HH functions.
  The potential matrix elements
  $v^{KT,\bar K\bar T}_{l\nu,\lbb\bar\nu}$ are exactly the ones given
  in Eq.~(\ref{eq:v12jj}) and can be computed by using the same approach of
  Section~\ref{sec:ve}.
  Note that this formula
  does not depend on the locality/non-locality of the potential,
  which appears only
  in the potential matrix elements.
  
  \section{Results}
  In this section we discuss the results obtained
  for the $\alpha+d$ CFF by using Eq.~(\ref{eq:fr}). Furthermore, we compare
  the CFF obtained with the two methods, i.e. using the direct overlap
  (Section~\ref{sec:overlap}) and the equation (Section~\ref{sec:eqcff}).

  For the calculation of the source term we use the following parameters.
  We use $N_{max}=20$ for the expansion  with the Laguerre polynomials of the potential
  part of the source term [Eq.~(\ref{eq:glvexp})]. We use $\lbb_{max}=40$
  for the hyperradial part of the cluster function
  [Eq.~(\ref{eq:clustern1})]. Both these values permit to reach full convergence
  in the respective expansions.
  More difficult results to be the expansion in 
  HH states, as shown by the dependence on  $\kb$.
  In Table~\ref{tab:pjstate}
  we report the total number of HH states used in the projection
  of both $S$- and $D$-wave components of the cluster wave function for
  fixed $\kb$, $\lb$ and $\sbb$.
  \begin{table}
    [h!]
    \centering
    \begin{tabular}{ll|rr|rr|rr|rr|rr}
      \hline
      \hline
      & & \multicolumn{2}{c|}{$\kb=0$}&
      \multicolumn{2}{c|}{$\kb=2$}&
      \multicolumn{2}{c|}{$\kb=4$}&
      \multicolumn{2}{c|}{$\kb=6$}&
      \multicolumn{2}{c}{$\kb=8$}\\
      $\lb$ & $\sbb$ & $S$& $D$& $S$& $D$& $S$& $D$& $S$& $D$& $S$& $D$\\
      \hline
      0& 1 & 6 & 0 &18&0 & 60 &12 &168 & 60 &414 &228 \\
      1& 0 & 0 & 0 &4 &0 & 30 & 8 &124 & 68 &376 &292 \\
      1& 1 & 0 & 0 &6 &0 & 48 &18 &234 &162 &732 &672 \\
      1& 2 & 0 & 0 &2 &0 & 20 &10 &126 & 80 &416 &362 \\
      2& 1 & 0 & 0 &12&6 & 72 &48 &294 &252 &906 &936 \\
      2& 2 & 0 & 0 &6 &0 & 42 &12 &180 &100 &576 &452 \\
      2& 3 & 0 & 0 &2 &0 & 12 & 2 & 50 & 20 &158 &100 \\
      3& 2 & 0 & 0 &0 &0 & 10 &10 & 56 & 72 &218 &304 \\
      3& 3 & 0 & 0 &0 &0 &  2 & 2 & 12 & 16 & 50 & 72 \\
      4& 3 & 0 & 0 &0 &0 &  2 & 2 & 10 & 12 & 36 & 46 \\
      \hline
      \multicolumn{2}{l|}{Total}& 6 & 0 &50 & 6 &298 & 124 &1254 & 842 & 3882 & 3464\\ 
      \hline
      \hline
    \end{tabular}
    \caption{\label{tab:pjstate} Total number of HH states used in
      the expansion of the 
      $S$- and $D$-wave of the cluster wave functions
      for given values of $\kb$, $\lb$ ans $\sbb$.}
  \end{table}
  As regarding the
  isospin we consider only $\tb=0$ states. As can be seen
  by inspecting the table,
  the number of states grows very fast as function of $\kb$.
  This limits our calculation to $\kb_{max}=8$. However,  by considering all the
  projecting states up to $\kb_{max}=8$, we are reproducing completely the
  hyperangular-spin-isospin structure of the $\alpha$ wave function
  with our projections, since we use
  the $\alpha$-particle computed with $K_\al=8$.
  What remains are the radial part of the deuteron wave function
  and the hyperradial part of
  the $\alpha$-particle wave function, which are reproduced by the remaining 
  Jacobi polynomials and the hyperradial functions
  (see Appendix~\ref{app:overlapint}).
  Hereafter, in this section,
  with $\kb$ we indicate that we are including all the possible states $\bb$ with
  all the possible $\lb$, $\sbb$ and $\tb$ allowed such that $\kb(\bb)\leq\kb$.

\begin{figure}
  \centering
  \includegraphics[scale=1]{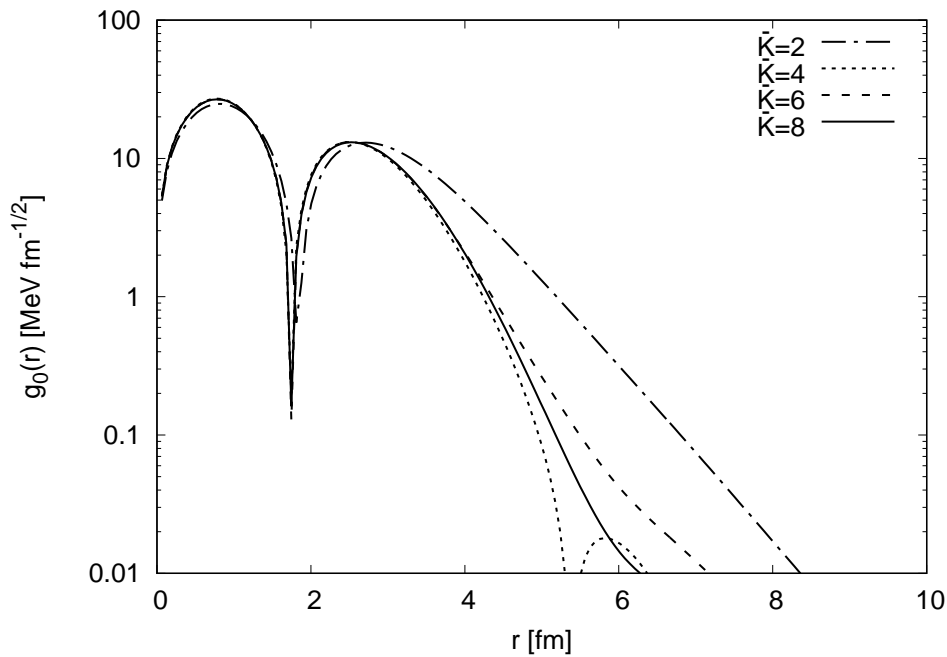}
  \caption{\label{fig:source0}
    $S$-wave component of the reduced source term $g_0(r)$  for different values of
    $\kb$ used in the projection of the $\alpha+d$ cluster wave function.
    This figure is obtained by using N3LO500-SRG1.5 potential, and
    the $\Li$ wave function computed with $K=12$.}
\end{figure}

  In Figure~\ref{fig:source0} we plot the reduced source term $g_L(r)$
  for the $S$-wave component of the CFF
  in the case of N3LO500-SRG1.5 potential for different values of $\kb$. The
  calculation shown in this plot is performed by considering
  the $\Li$ wave function computed for $K=12$.
  From the figure it is immediately
  clear that we have a nice convergence in $\kb$ for the short-range
  part ($r<3-4$ fm) but not for larger $r$.
  This effect is due to the fact that the Jacobi polynomials are not flexible enough
  in reproducing the exponential behavior of the wave function.
  This means that a larger number of states in the projection are needed
  to describe correctly $g_L(r)$ for $r>4$ fm.
  However, by inspecting the figure it results clear that
  there are problems only in a region where $g_0(r)$ is  a factor 100
  smaller than the peak. A similar convergence behavior in $\kb$
  is founded also for the other potentials studied and for $g_2(r)$.

\begin{figure}
  \centering
  \includegraphics[scale=1]{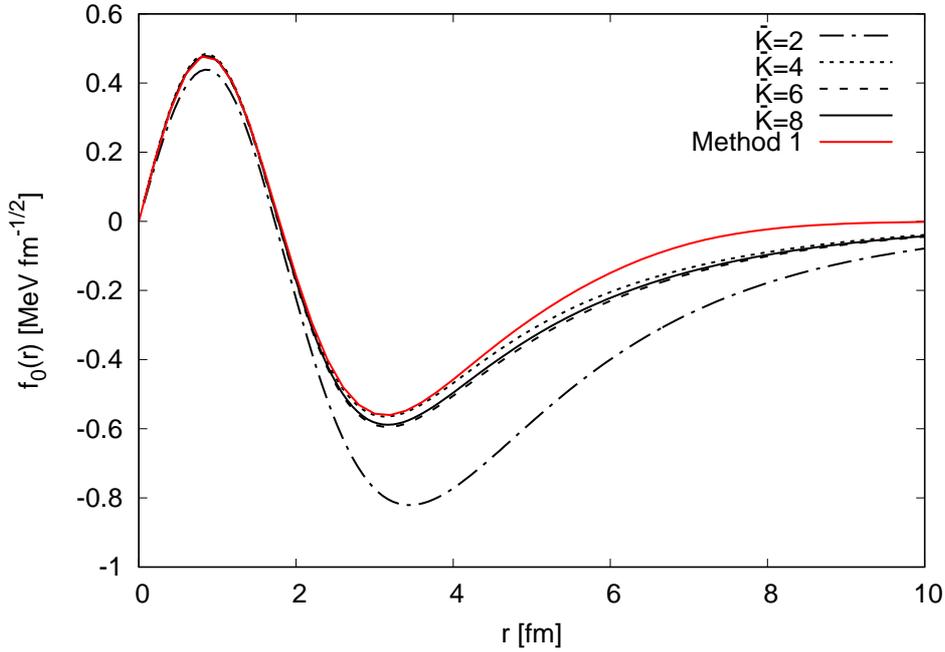}
  \caption{\label{fig:fr0_method1}
    $S$-wave component calculated via Eq.~(\ref{eq:fr})
    of the reduced $\alpha+d$ CFF $f_0(r)$ for different values of the
    $\kb$ used in the projection of the  cluster wave function. For comparison we
    report also the result obtained with the direct overlap (red line).
    These results are obtained with N3LO500-SRG1.5 potential and
    the $\Li$ wave function computed with $K=12$.}
\end{figure}

  Let us now study how these considerations on $g_L(r)$ are reflected
  on the CFF calculated via Eq.~(\ref{eq:fr}).
  We find that, even if the source term
  is exponentially decreasing and vanishing for large
  $r$, the different behavior of the tail as function of $\kb$ has an
  impact on the ANC, although it is not large. In Figure~\ref{fig:fr0_method1}
  we compare the $S$-wave
  reduced CFF computed using the N3LO500-SRG1.5 potential
  and the $\Li$ wave function with $K=12$ for various values of $\kb$.
  As already discussed for the source term, in the region $r$ below $3-4$ fm we reach
  full convergence with $\kb=8$. As regarding the tail,
  thanks to the vanishing term $g_L(r)$ for larger $r$, we obtain the Whittaker
  function which is the correct asymptotic behavior. 
  However its normalization (i.e. the ANC) depends a bit on $\kb$ as the tail of
  $g_L(r)$. In any case this effect  is quite small  in particular for the
  ``harder'' potentials, as can be seen from the good convergence in $\kb$
  in Figure~\ref{fig:fr0_method1} and in Table~\ref{tab:anckb}.
  On the other hand, the $D$-wave ANCs result to be more dependent on $\kb$.  

  \begin{table}
    \centering
    \begin{tabular}{rcccc}
      \hline
      \hline
      $\kb$ & SRG1.2 & SRG1.5 & SRG1.8 & NNLO$_{sat}$(NN)\\
      \hline
      \multicolumn{5}{c}{$C_0$ [fm$^{-1/2}$]}\\
      \hline
      2 &  $-7.86$ & $-6.34$ & $-5.22$ & $-4.77$ \\
      4 &  $-4.00$ & $-3.17$ & $-2.59$ & $-2.51$ \\
      6 &  $-4.40$ & $-3.59$ & $-2.98$ & $-2.83$ \\
      8 &  $-4.17$ & $-3.45$ & $-2.90$ & $-2.84$ \\
      \hline
      \multicolumn{5}{c}{$C_2$ [fm$^{-1/2}$]}\\
      \hline
      2 &  0.136 & 0.084 & 0.051 & 0.027 \\
      4 &  0.170 & 0.115 & 0.073 & 0.044 \\
      6 &  0.080 & 0.043 & 0.025 & 0.013 \\
      8 &  0.115 & 0.072 & 0.044 & 0.026 \\
      \hline
    \hline
  \end{tabular}
    \caption{\label{tab:anckb} Values of the $C_0$ and $C_2$ in fm$^{-1/2}$
      as function of $\kb$. All the results given in this table are obtained by
      considering the $\Li$ wave function computed at $K=12$.}
\end{table}
      
  In Table~\ref{tab:cl_coeff2} we report the value of the ANC
  computed with the projection method for $\kb=8$ as function of
  $K$ used for computing the $\Li$ wave function.
  Since we can not give a reliable extrapolation for $\kb\rightarrow\infty$
  due to the fact the convergence pattern is not clear (see Table~\ref{tab:anckb})
  we report also an error given by the half of difference between the
  ANC computed for $\kb=6$ and $\kb=8$, namely
  \begin{equation}\label{eq:dclbar0}
    \Delta C^{(0)}_L(K)=0.5\times|C_L(K,\kb=6)-C_L(K,\kb=8)|\,.
  \end{equation}
  By repeating the same procedure used in Section~\ref{sec:res1}, we fit the
  values of the ANCs obtained with $\kb=8$ for different values of $K$
  by using the exponential function given in Eq.~(\ref{eq:fitcl}).
  The extrapolated
  values for $K\rightarrow\infty$ are reported in the last row for each
  potential.
  As error due to the extrapolation on $K$, we consider a conservative error
  estimated as
  \begin{equation}
    \Delta C^{(1)}_L=0.5\times|C_L(12,\kb=8)-C_L(\infty)|\,
  \end{equation}
  and a total error on our final extrapolation of
  \begin{equation}\label{eq:dclbar}
    \Delta C_L=\sqrt{\left(\Delta C^{(0)}_L(12)\right)^2+\left(\Delta C^{(1)}_L\right)^2}\,.
  \end{equation}

  \begin{table}
    \centering
    \begin{tabular}{r|ccc||ccc}
      \hline
      \hline
     & \multicolumn{3}{c||}{SRG1.2} & \multicolumn{3}{c}{SRG1.5}\\
    $K$ & $B_c$ & $C_0$ & $C_2$& $B_c$ & $C_0$ & $C_2$ \\
    \hline
    2 & $-3.736$  & $-$ & $-$ &  $-7.634$ & $-$ & $-$\\ 
    4 & $\m0.891$ & $-3.04(3)$  & $0.041(1)$  & $-1.615$ & $-$ & $-$\\
    6 & $\m2.359$ & $-3.74(12)$ & $0.092(4)$  & $\m1.071$ & $-2.82(8)$ & $0.029(3)$\\
    8 & $\m2.766$ & $-4.08(5)$  & $0.111(13)$ & $\m1.929$ & $-3.30(1)$ & $0.063(9)$\\
    10& $\m2.909$ & $-4.10(12)$ & $0.114(16)$ & $\m2.222$ & $-3.37(7)$ & $0.070(13)$\\
    12& $\m2.955$ & $-4.17(12)$ & $0.115(18)$ & $\m2.323$ & $-3.45(7)$ & $0.072(15)$\\
    $\infty$ &$\m3.00(1)$ & $-4.19(12)$ & $0.116(18)$ & $\m2.46(2)$ & $-3.44(7)$ & $0.072(15)$\\
    \hline
    \hline
     & \multicolumn{3}{c||}{SRG1.8} & \multicolumn{3}{c}{NNLO$_{sat}$(NN)}\\
    $K$ & $B_c$ & $C_0$ & $C_2$& $B_c$ & $C_0$ & $C_2$ \\
    \hline
    2 & $-12.108$   & $-$     & $-$ &  $-18.74$ & $-$ & $-$\\ 
    4 & $-5.105$    & $-$     & $-$ &  $-10.99$ & $-$ & $-$\\
    6 & $-0.911$    & $-$     & $-$ &  $-4.16$ & $-$ & $-$\\
    8 & $\m0.740$ & $-2.61(1)$ & $0.024(4)$ & $-0.74$ & $-$ & $-$\\
    10& $\m1.332$ & $-2.80(4)$ & $0.039(7)$ & $\m0.61$ & $-2.61(1)$ & $0.015(3)$\\
    12& $\m1.551$ & $-2.90(4)$ & $0.044(10)$ & $\m1.15$ & $-2.83(1)$ & $0.026(6)$\\
    $\infty$ &$\m2.02(9)$ & $-3.01(7)$ & $0.047(10)$ & $2.11(20)$ & n.a. & n.a.\\
    \hline
    \hline
  \end{tabular}
  \caption{\label{tab:cl_coeff2} Values of the binding energy $B_c$ in MeV
    and the ANCs $C_0$ and $C_2$ in fm$^{-1/2}$ as function of $K$
    for the various potential models  computed using Eq.~(\ref{eq:fr}).
    In the row labeled with
    ``$\infty$'' we report the extrapolated values. The errors 
    (between parentheses) are obtained with Eq.~(\ref{eq:dclbar0}) for the various $K$
    and with Eq.~(\ref{eq:dclbar}) for the extrapolated values.}
\end{table}

\begin{figure}
    \centering
    \includegraphics[width=\linewidth]{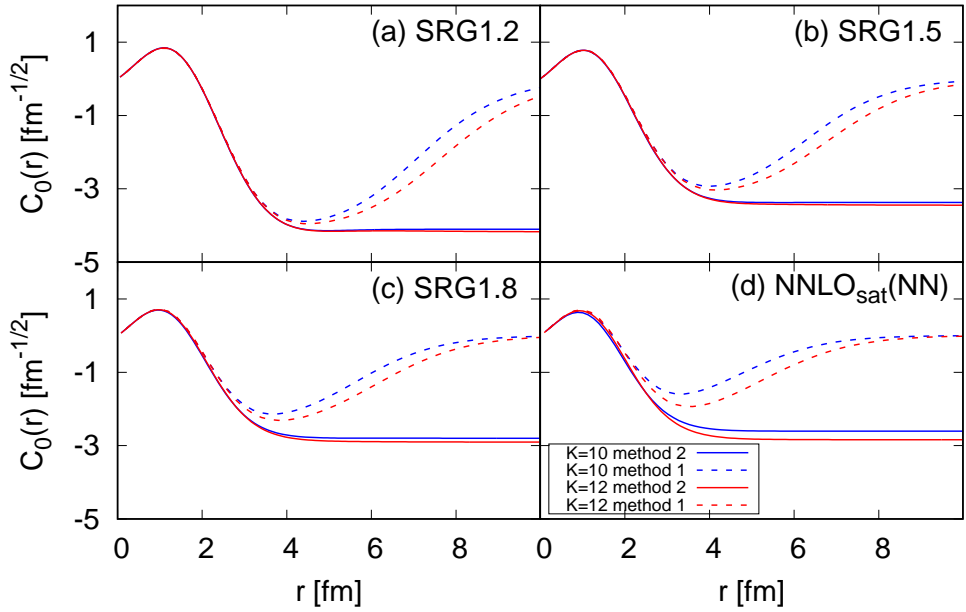}
    \caption{\label{fig:fr0_all}
      Function $C_0(r)$ computed with
      the overlap method (dashed lines) and the equation method (continuous
      lines) for all the potential considered.
      The calculations are performed with the $\Li$ wave function
      computed with $K=10$ (blue lines) and $K=12$ (red lines). 
      Results of the equation method are obtained using $\kb=8$.}
  \end{figure}

\begin{figure}
    \centering
    \includegraphics[width=\linewidth]{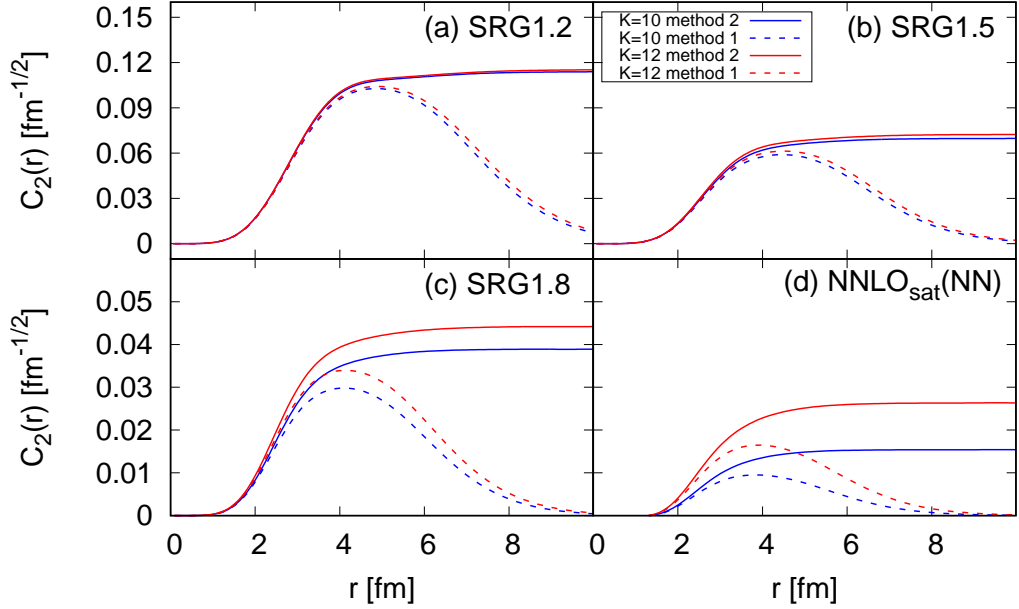}
    \caption{\label{fig:fr2_all} Same as Figure~\ref{fig:fr0_all}, but
    for the function $C_2(r)$.}
  \end{figure}

  We can now compare the ratio of the $\alpha+d$ CFF with the Whittaker function
  obtained with the two methods.
  In Figure~\ref{fig:fr0_all} we plot the  functions $C_0(r)$ defined in
  Eq.~(\ref{eq:clr})
  obtained with the two methods
  for two different values of $K$ used for the calculation of the $\Li$ wave function.
  For the equation method we consider the results obtained with $\kb=8$.
  For the SRG potentials it is clear that the short-range part of $C_0(r)$
  computed with the two methods are essentially indistinguishable.
  This is not the case of the tail, since the $C_0(r)$ computed with the
  direct overlap is not following the correct asymptotic behavior, as already
  discussed. Moreover, it is possible to observe that, when the potential becomes
  ``harder'' as in the case of SRG1.8 and NNLO$_{sat}$(NN),
  the differences between the $C_0(r)$ functions
  computed with the two methods start
  at smaller values of $r$.
  For example, in the case of NNLO$_{sat}$(NN) even the short range part does not
  completely coincide. This effect is directly related to the convergence in both
  $\kb$ and in $K$: the better is the convergence in both the expansions, the
  closer are the results of the two methods.
  In Figure~\ref{fig:fr2_all} we plot the same for $C_2(r)$.
  In this case, even if the general shapes result  similar,
  for all the potentials the $C_2(r)$ obtained with the equation method seems
  systematically larger compared to the one obtained with the direct overlap method.
  We suppose that such a difference is related to the poor description of the
  $D$-wave components in the $\Li$ wave function.
  
  In Table~\ref{tab:ANC_comp} we compare the extrapolated values of the ANC computed
  with the two methods. By inspecting the table it is evident that the
  results obtained with the equation method are systematically larger than
  the ones obtained in Section~\ref{sec:res1}. Even if the equation method
  results are affected by larger errors due to the expansion in the HH states 
  of the cluster wave function, this systematic difference cannot be
  explained simply giving the responsibility to a not complete convergence in $\kb$
  for the projection of the cluster wave function.
  Indeed, in the case of the SRG1.2 potential, the convergence
  in $\kb$ is
  almost reached and still this systematic difference persists.
  As discussed previously, the ANCs directly calculated from the overlap suffer of
  unknown systematic uncertainties, therefore we consider the ones calculated from
  the equation as more reliable.
  \begin{table}
    \centering
    \begin{tabular}{llcccc}
      \hline
      \hline
      &Model & $B_c$ [MeV] & $C_0$ [fm$^{-1/2}$] & $C_2$ [fm$^{-1/2}$] &
      $C_2/C_0$ \\
      \hline
               & SRG1.2 & 3.00(1) & $-3.99(2) $& 0.107(2) &$ -0.027(1) $\\
      Method 1 & SRG1.5 & 2.46(2) & $-3.10(4) $& 0.063(1) &$ -0.020(1) $\\
               & SRG1.8 & 2.02(9)  & $-2.43(6) $& 0.036(1) &$ -0.015(1) $\\
      \hline
               & SRG1.2 & 3.00(1) & $-4.19(12)$& 0.116(18)&$ -0.028(5) $\\
      Method 2 & SRG1.5 & 2.46(2) & $-3.44(7) $& 0.072(15)&$ -0.021(5) $\\
               & SRG1.8 & 2.02(9) & $-3.01(7) $& 0.047(10)&$ -0.016(4) $\\
      \hline
      Ref.~\cite{Nollet2001}
      & AV18/UIX   & 1.47     & $-2.26(5) $& -- &$ -0.027(1) $\\
      Ref.~\cite{Hupin2015}
      & SRG2.0(3b) & 1.49     & $-2.695   $& 0.074 & $-0.027$\\
      \hline
      Ref.~\cite{George1999} & Exp.       & 1.4743   & $-2.91(9) $& 0.077(18) & $-0.025(11)$\\
      \hline
      \hline
      \end{tabular}
    \caption{\label{tab:ANC_comp}
      Values of the extrapolated ANC $C_0$ and $C_2$ in fm$^{-1/2}$ for the
      various SRG-evolved potential models and the two methods
      used for the calculation.
      We report also the binding energy $B_c$ (MeV)
      used in the calculation of the
      ANC and the ratio $C_2/C_0$.
      For completeness we shows also the results of the {\it ab-initio} calculation
      of Refs.~\cite{Nollet2001,Hupin2015} obtained with the AV18/UIX potential and
      the N3LO500-SRG1.5 including three-body forces (3b).
      Also the experimental values of Ref.~\cite{George1999} are listed.}
  \end{table}
  For completeness in the table we report also the experimental values of Ref.~\cite{George1999}
  and the calculation of Refs.~\cite{Nollet2001,Hupin2015} obtained using
  the AV18/UIX and the N3LO500-SRG1.5 with three-body forces, respectively.
  We can not  really
  perform a comparison, since our results do not contain the contribution
  of the three-body forces, and they are not computed
  at the physical energy $B_c$. However, from a qualitative point of view, there is
  a nice agreement since all the SRG potential models used are able to reproduce
  the correct magnitude of the ANC.
  This result is quite remarkable, if we consider that the potential models
  we use are constrained only to nucleon-nucleon data and do not have any information
  on the $A=6$ nuclei.

%% file: Chapter4.tex
\chapter{The $p$+$^6$Li$\rightarrow$$^7$Be+$\gamma$ radiative capture reaction}
\label{ch:pLi}
In this chapter we present a theoretical study of the $p+\Li\rightarrow\Be+\gamma$
radiative capture reaction within a cluster model.
The cluster model approach is based on the fact that the two colliding nuclei
can be considered as structureless particles,
which interact through an {\it ad hoc} potential.  
This approach is completely different from the {\it ab-initio}
approach used in the previous chapters.
First of all, here, we are not starting from first principles,
therefore the potential
model we consider is limited to describe only the $p-\Li$ system.
Moreover, in the {\it ab-initio} approaches the full $A$-body
Hamiltonian is solved, while in this chapter we consider a simple two-body Hamiltonian
which is a very crude approximation. However, even if this is a quite simple
model, it permits to predict the $S$-factor and the angular distribution of
the emitted photons with a good accuracy.

This chapter is organized as follows. In Section~\ref{sec:theory} we introduce
the general formalism of the cluster model approach. In Section~\ref{sec:clusterp}
we present the $p-\Li$ potential model used in this calculation,  while in Section
~\ref{sec:res} we discuss our results. All the results of this
chapter have been already published in Ref.~\cite{Gnech2019}.

\section{Theoretical formalism}
\label{sec:theory}

Let us consider a generic reaction $A_1+A_2\rightarrow A_3+\gamma$.
By considering the nuclei $A_1$ and $A_2$ as structurless particles,
the scattering wave function is written as
\begin{equation}
  \begin{aligned}
  \psi_{1,2}(\br,p)=\frac{\sqrt{4\pi}}{p}\sum_{LSJJ_z}
  i^L\, \sqrt{2L+1}\,
   (J_1 M_{1}, J_2 M_{2}| S J_z)
   ( SJ_z, L 0| J J_z )
  \,\psi_{1,2}^{LSJJ_z}(\br,p)\,,
  \label{eq:psi12}
  \end{aligned}
\end{equation}
with
\begin{equation}
  \psi_{1,2}^{LSJJ_z}(\br,p)=
  R_{LSJ}(r,p)\,\Big[Y_L(\hat{r})\otimes\chi_S\Big]_{JJ_z}\,,
  \label{eq:psi12_LSJ}
\end{equation}
where $p$ is the relative momentum of the two particles, $\br$
the intercluster distance, $L$, $S$ and $J$ the total orbital,
spin and angular momentum of the system,
with $J_1,M_1$ and $J_2,M_2$ being the total
angular momenta and third components of the two nuclei.
In Eq.~(\ref{eq:psi12_LSJ}), $\chi_S=\left(\phi_{J_1}^{p}\phi^{\Li}_{J_2}
\right)_S$ is the ``spin'' state of the two clusters.
The function $R_{LSJ}(r,p)$ is the scattering wave function,
that has been
determined solving the two-body Schr\"odinger equation similarly to
what has been done in Ref.~\cite{Grassi2017}.
For the bound states of the final nucleus $A_3$ we write the wave function
as
\begin{equation}
  \psi_{3}^{J_3M_{3}}(\br)=
  u_{L_3S_3}(r)\,\Big[Y_{L_3}(\hat{r})\otimes\chi_{S_3}\Big]_{J_3M_{3}}\,,
  \label{eq:bound}
\end{equation}
where $\br$ is again the intercluster distance.
The function $u_{L_3S_3}(r)$ has also been determined as
explained above.
The total cross section for a radiative capture in a bound state with
total angular momentum $J_3$ is written as
\begin{eqnarray}
  \sigma_{J_3}(E)=\frac{32\pi^2}{(2J_1+1)(2J_2+1)}\frac{\alpha}{v_{\rm rel}}
  \frac{q}{1+q/m_3}
  \sum_{\Lambda\geq1}\sum_{LSJ}
  \Big(|E_{\Lambda}^{LSJ,J_3}|^2+|M_{\Lambda}^{LSJ,J_3}|^2\Big)
  \label{eq:sigtot}\,,
\end{eqnarray}
where $\alpha=e^2/4\pi$, $v_{\rm rel}$ is the relative velocity of the
two incoming particles, $q$ is the photon momentum and $m_3$ is the mass of
$A_3$ nucleus. Finally, $T_{\Lambda}^{LSJ,J_3}$, with $T=E/M$,
are the reduced matrix element of the electromagnetic
operator and $\Lambda$ is the multipole order.
Using the Wigner-Eckart theorem, they are defined as
\begin{equation}
  T^{LSJ,J_3}_\Lambda=\bra \psi_{1,2}^{LSJJ_z}(\br,p)|
  T_{\Lambda\lambda}|\psi_{3}^{J_3M_{3}}(\br)\ket\frac{\sqrt{2J_3+1}}
  {( J_3 M_{3}, \Lambda \lambda | J J_z)}\,,
  \label{eq:tlambda}
\end{equation}
where $\lambda=\pm1$ is
the photon polarization.
In our calculation we include only the electric multipoles,
which are typically much larger than the magnetic ones. Then, in 
the long-wavelength approximation~\cite{Walecka}, by using
Eqs.~(\ref{eq:psi12_LSJ}) and~(\ref{eq:bound}), it results
\begin{equation}
  \begin{aligned}
  E_{\Lambda}^{LSJ,J_3}&=(-1)^{2J_f+\Lambda+L+S-J}
  \hat{J}\hat{J_3}\hat{L_3}\hat{\Lambda}
  ( L_3 0, \Lambda 0 | L 0)
  \begin{Bmatrix}
      J   & L   & S \\
      L_3 & J_3 &\Lambda
    \end{Bmatrix}\\
  &\times\frac{Z_e^{(\Lambda)}}{(2\Lambda+1)!!}
  \sqrt{\frac{\Lambda+1}{\Lambda}}\frac{q^\Lambda}{\sqrt{4\pi}p}
  \int_0^\infty dr\, r^2\,u_{L_3S_3}(r)r^\Lambda
  R_{LSJ}(r,p)\delta_{S,S_3}\label{eq:eleop}
  \,.
  \end{aligned}
\end{equation}
Here we have defined $\hat{x}=\sqrt{2x+1}$ and
\begin{equation}
  Z_e^{(\Lambda)}=Z_1\left(\frac{m_2}{m_1+m_2}\right)^\Lambda+
  Z_2\left(-\frac{m_1}{m_1+m_2}\right)^\Lambda\,
  \label{eq:zeff}
\end{equation}
is the effective charge, in which $Z_1(Z_2)$ is the charge
and $m_1(m_2)$ is the mass of the $A_1(A_2)$ nucleus.
Given the radial wave functions $u_{L_3S_3}(r)$
and $R_{LSJ}(r,p)$, the one-dimensional integral of
Eq.~(\ref{eq:eleop}) is simple and it can be calculated easily with standard
numerical techniques.
The astrophysical $S$-factor is then defined as
\begin{equation}
  S_{J_3}(E)=E\exp(2\pi\eta)\sigma_{J_3}(E)\,,
\label{eq:sfactor}
\end{equation}
where $\sigma_{J_3}(E)$ is the total cross section of
Eq.~(\ref{eq:sigtot}) and
$\eta$ will be discussed below [see Eq.~(\ref{etabar})]. 

The other observable of interest is the
photon angular distribution, which can be written as
\begin{equation}
  \sigma_{J_3}(E,\theta)=\sigma_0(E)\sum_{k\geq0}a_k^{J_3}(E) P_k(\cos\theta)\,,
  \label{eq:sigma_theta}
\end{equation}
where $\sigma_0(E)$ is a kinematic factor defined as
\begin{equation}
  \sigma_0(E)=\frac{16\pi^2}{(2J_1+1)(2J_2+1)}\frac{\alpha}{v_{\rm rel}}
  \frac{q}{1+q/m_3}\,,\label{eq:sig0}
\end{equation}
and $P_k(\cos\theta)$ are the Legendre polynomials.
Note that we have defined the $z$-axis as the direction of the
incoming protons in the lab, and the $xy$ plane as the plane where
the photon is emitted. Since the initial particles are unpolarized and the
final polarizations are not measured, the cross section depends only on
$\theta$, the angle between $\bm{p}$ and $\bm{q}$.
The coefficients $a_k$ are given by
\begin{equation}
  \begin{aligned}
  a_k^{J_3}(E)&=\sum_{LL'SJJ'\Lambda\Lambda'}(-)^{J+J'+J_3+L'+\Lambda+S+1}
  \,i^{L+L'+\Lambda+\Lambda'}
  \hat{L}\hat{L'}\hat{\Lambda}\hat{\Lambda'}\hat{J}\hat{J'}\\
  &\times( L 0, L' 0|k 0)   \begin{Bmatrix}
    L   & L'   & k \\
    J' & J & S
  \end{Bmatrix}
  \begin{Bmatrix}
      J'   & J   & k \\
      \Lambda & \Lambda' & J_3
    \end{Bmatrix}
  \sum_{\lambda=\pm1}( \Lambda' -\lambda, \Lambda \lambda |k 0)\\
  &\times\left(\lambda M_{\Lambda'}^{L'SJ',J_3}+E_{\Lambda'}^{L'SJ',J_3}\right)
  \left(\lambda M_{\Lambda}^{LSJ,J_3}+E_{\Lambda}^{LSJ,J_3}\right)^*\,.
  \label{eq:ak}
  \end{aligned}
  \end{equation}
The photon angular distribution can be casted in the final form
\begin{equation}
  \sigma_{J_3}(E,\theta)
  =\sigma_{J_3}(E)\left(1+\sum_{k\geq1}A^{J_3}_k(E) P_k(\cos\theta)\right)\,,
  \label{eq:angulardistr}
\end{equation}
where $\sigma_{J_3}(E)$ is defined in Eq.~(\ref{eq:sigtot}),
and $A^{J_3}_k(E)=a_k^{J_3}(E)/a_0^{J_3}(E)$. All these formulas
are derived explicitly in Appendix~\ref{app:sig_formula}.

\section{The $p-\Li$ potential model}
\label{sec:clusterp}

In our study we consider as clusters the two colliding nuclei,
$p$ $(J^\pi=1/2^+)$ and $\Li$ $(J^\pi=1^+)$.
The potential model we build is tuned to reproduce the $\Be$ properties and the
elastic scattering phase shifts.
Following Ref.~\cite{Dubovichenko2011}, we consider a $p-\,\Li$ potential of the form
\begin{equation}
  V(r)=-V_0\exp{(-a_0r^2)}\,,
\label{eq:vr}
\end{equation}
where $V_0$ and $a_0$ are two parameters to be selected independently for
phase-shifts of each different $L,S,J$ waves. We also add a point-like
Coulomb interaction, i.e.
\begin{equation}
  V(r)=\alpha \frac{Z_1Z_2}{r} \ ,
\label{eq:pointCoul}
\end{equation}
where $\alpha=1.439975$ MeV fm. All the other coefficients
entering the two-body Schr\"odinger equation
solved in this framework are given for completeness
in Table~\ref{tab:params}.
\begin{table}[h]
\begin{center}
  \begin{tabular}{lc}
    \hline
    \hline
    $m_p$    & $1.00727647$ u~\cite{PDG2018}\\
    $m_{\Li}$& $6.01347746$ u~\cite{Tilley2002}\\
    $\hbar c$& $197.3269788$ MeV fm~\cite{PDG2018}\\
    \hline
    \hline
\end{tabular}
\caption{ \label{tab:params}
  Values of the parameters used in the Schr\"odinger equation obtained.
  Note that we have used $1$ u $=931.4940954$ MeV~\cite{PDG2018}.}
\end{center}
\end{table}
All the following results have been
obtained using the Numerov algorithm to solve the Schr\"odinger equation
and then 
tested using the R-matrix method
(see Ref.~\cite{Descouvemont2010} and references therein).

\begin{figure}[ht]
  \begin{center}
    \includegraphics[scale=0.8]{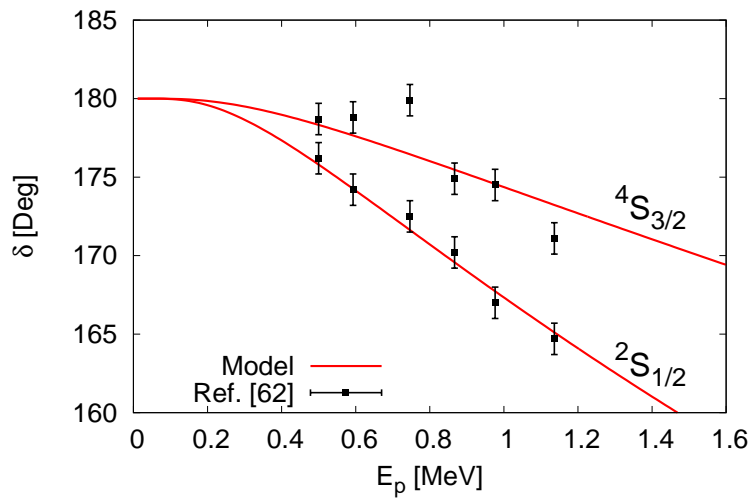}
    \caption{\label{fig:phase-shifts}
      Phase shifts for the ${}^2S_{1/2}$ and the ${}^4S_{3/2}$ partial waves
      as function of the proton energy.
      The data are taken from Ref.~\cite{Dubovichenko2011}. The full red line are 
      the calculated phase shifts with the potential parameters given in
      Table~\ref{tab:chim}.}
  \end{center}
\end{figure}

The parameters of the intercluster potential given in Eq.~(\ref{eq:vr})
are chosen in order to reproduce
the elastic scattering phase shifts, which are derived from partial wave
analysis of the experimental elastic scattering data of Ref.~\cite{Dubovichenko2011}.
In Table~\ref{tab:par-wave}
we report all possible partial waves
up to orbital angular momentum $L=2$ that are needed,
both for the doublet $S=1/2$ and quartet $S=3/2$ states, $S$
being the sum of the proton and $\Li$ spins, 1/2 and 1 respectively.
\begin{table}[h]
\begin{center}
  \begin{tabular}{l|cc}
    \hline
    \hline
& $S=1/2$ & $S=3/2$\\
\hline
$L=0$ & ${}^2S_{1/2}$ & ${}^4S_{3/2}$ \\
$L=1$ & ${}^2P_{1/2}$, ${}^2P_{3/2}$ & ${}^4P_{1/2}$, ${}^4P_{3/2}$, ${}^4P_{5/2}$\\
$L=2$ & ${}^2D_{3/2}$, ${}^2D_{5/2}$ & ${}^4D_{1/2}$, ${}^4D_{3/2}$, ${}^4D_{5/2}$
${}^4D_{7/2}$\\
\hline
\hline
\end{tabular}
\caption{ \label{tab:par-wave}
  Partial waves of the $p(J^\pi=1/2^+)-\, \Li(J^\pi=1^+)$ system up to $L=2$.
We indicate with $S$ the total spin.}
\end{center}
\end{table}
While the value of $a_0$ has been fixed and kept as in Ref.~\cite{Dubovichenko2011},
the values of $V_0$ has been obtained minimizing the $\chi^2$ function,
defined as
\begin{equation}
  \chi^2=\sum_i\frac{\left(\delta^i_{\rm{EXP}}(E)
    -\delta^i_{\rm{TH}}(V_0,E)\right)^2}
      {(\Delta\delta^i_{\rm{EXP}})^2}\ .
\label{eq:chi2}
\end{equation}
Here $\delta^i_{\rm{EXP}}(E)$ are the experimental phase shifts
and $\delta^i_{\rm{TH}}(V_0,E)$ are the calculated ones.
The minimization has been performed using the COBYLA algorithm~\cite{Powell1998}.
The values of $V_0$ and $a_0$ for the various partial waves and the
corresponding $\chi^2$/datum are listed in
Table~\ref{tab:chim}. To be noticed that the phase shift for the ${}^2P$ wave
is given by $\delta_{{}^2P}=\delta_{{}^2P_{1/2}}+\delta_{{}^2P_{3/2}}$
as defined in Ref.~\cite{Dubovichenko2011}.
The data set we used from Ref.~\cite{Dubovichenko2011} has no information
  on the $D$ waves. Therefore, in order to evaluate the ${}^2D$ and the ${}^4D$ waves,
  we use the same potential parameters used for the ${}^2S_{1/2}$ and ${}^4S_{3/2}$
  respectively, changing only the angular momentum $L$ in the Schr\"odinger equation.
  Moreover, we impose that the $D$ waves have the same radial part regardless
  of $J$.
In Figure~\ref{fig:phase-shifts}
we report the
experimental values and the calculated phase shifts for the $S$ waves.
As we can see from the figure, a nice agreement is found for the
$S$-wave phase shifts, especially for ${}^2S_{1/2}$.
\begin{table}[ht]
  \begin{center}
    \begin{tabular}{lccc}
      \hline
      \hline
        wave &  $V_0$ (MeV) & $a_0$ (fm$^{-2}$)& $\chi^2/$datum\\ 
        \hline
        ${}^2S_{1/2}$ & 124.63 & 0.15 & 0.4\\
        ${}^4S_{3/2}$ & 141.72 & 0.15 & 3.6\\
        ${}^2P$ & 67.44  & 0.1 & 1.9\\
        \hline
        \hline
    \end{tabular} 
    \caption{Values for the parameters of the
      Gaussian potential and $\chi^2/$datum for the different
      partial waves.}\label{tab:chim}
  \end{center}
\end{table}

The $p-\,\Li$ potential of Eq.~(\ref{eq:vr}) is used
also in order to describe the $\Be$ nucleus. In this case
we need to reproduce the binding energies of the two bound states,
the ground state (GS) $J^\pi=3/2^-$ with $B=5.6068$ MeV
and the first excited state (FES) $J^\pi=1/2^-$ with $B=5.1767$
MeV~\cite{Tilley2002}.
We fixed again the parameter $a_0$ as in Ref.~\cite{Dubovichenko2011},
while in order to obtain $V_0$ we impose 
that the calculated binding energies reproduce the experimental ones up to the
sixth digit. Moreover, we have evaluated also the ANC, defined in this
case as
\begin{equation}
    {\rm{ANC}}=\frac{u_{LS}(r)}{\sqrt{2k}W_{L+1/2}(2kr,\eta)}\,,
\label{eq:anc}
\end{equation}
where $u_{LS}(r)$ is the radial part of the wave function
[see Eq.~(\ref{eq:bound})],
$r$ is the intercluster
distance, $k=\sqrt{\frac{2\mu B}{\hbar^2}}$ with
$\mu=\frac{m_pm_{\Li}}{m_p+m_{\Li}}$, $B$ the binding
energy of the bound state, and
$W_{L+1/2}(2kr,\eta)$
is the Whittaker function~\cite{AbramowitzStegun}, with $\eta$ defined as
\begin{equation}
    \eta=1.439975\times Z_pZ_{\Li}\frac{\mu}{2k\hbar^2}\,.
    \label{etabar}
\end{equation}
Moreover, in this case we divided the ANC for $\sqrt{2k}$ in order to have it
dimensionless.
In Table~\ref{tab:bspar}
we report the values for $V_0$ and $a_0$, 
and the calculated values of the binding energies and ANCs for both the GS
and FES. Note that, to our knowledge, there are no experimental data
for the ANCs.
\begin{table}[h]
  \begin{center}
    \begin{tabular}{lcccc}
      \hline 
      \hline
      $J^\pi$ &  $V_0$ (MeV) & $a_0$ (fm$^{-2}$)& $B$ (MeV) & ANC\\ 
        \hline
        $3/2^-$ & 254.6876510 & 0.25 & 5.606800 & 2.654\\
        $1/2^-$ & 252.7976803 & 0.25 & 5.176700 & 2.528\\
        \hline
        \hline
    \end{tabular} 
    \caption{Values for the parameters of the
      Gaussian potential of Eq.~(\ref{eq:vr})
      and the calculated binding energy and ANC for 
      both the GS and the FES.
      }\label{tab:bspar}
  \end{center}
\end{table}
\section{Results}
\label{sec:res}

In this section we compare our theoretical predictions for the
astrophysical $S$-factor and the
angular distribution of the emitted photon
with the available experimental data. In the last part,
we also discuss the possibility of introducing in our model the resonance
proposed in Ref.~\cite{He2013}.

Before discussing the results, we note that in the $p-\,\Li$ reaction
the open 
${}^3{\rm He}-\,^4{\rm He}$ channel should in principle
be included. However, we do not consider
this channel in our work. This can be done, because
the experimental phase shifts of Ref.~\cite{Dubovichenko2011} used to fit
our potential were obtained considering only the $p-\,\Li$ channel.
Therefore the ${}^3{\rm He}-\,^4{\rm He}$ channel
results to be hidden in the experimental phase shifts that we reproduce
with our potential. On the other hand, for the $\Be$ bound states,
the ${}^3{\rm He}-\,^4{\rm He}$
component has to be considered, and this is
done phenomenologically, introducing in our calculation the
spectroscopic factors, as explained in the next subsection.

\subsection{The astrophysical $S$-factor}
\label{subsec:sfactor}

The main contribution to the  
$p+\Li\rightarrow\Be+\gamma$ radiative capture reaction $S$-factor
comes from the electric dipole ($E1$) transition.
The structure of the electric operator in the long wavelength
approximation implies a series of selection rules due to the presence of
the Wigner-6j coefficient as shown in
Eq.~(\ref{eq:eleop}).
Therefore, the only waves allowed by the $E1$ transition operator
up to $L=2$ are ${}^2S_{1/2}$, ${}^2D_{3/2}$ and ${}^2D_{5/2}$
for the GS, and  
${}^2S_{1/2}$ and ${}^2D_{3/2}$ for the FES.
From the calculation, it turns out that up to energies
of about 400 keV, the contribution
of the ${}^2D$ waves is very small. However, for higher values of the energy,
this contribution becomes significant.

\begin{figure}[t]
  \begin{center}
    \includegraphics[scale=0.8]{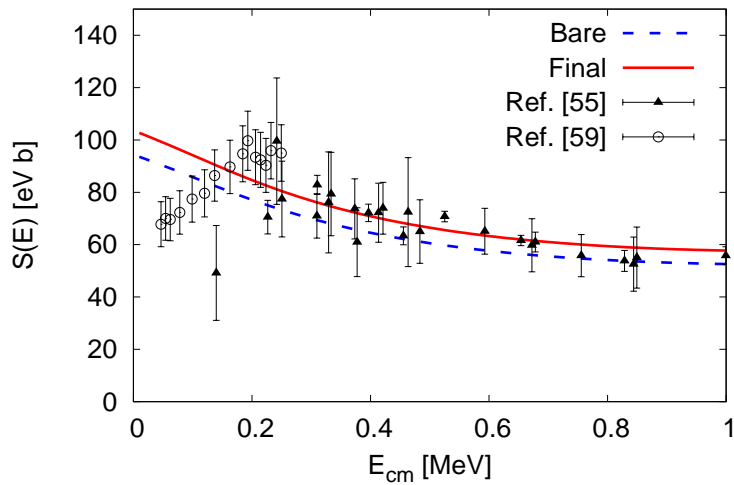}
    \caption{\label{fig:sftot}
      Total astrophysical
      $S$-factor for the $\Li(p,\gamma)\Be$ radiative capture reaction.
      The (blue) dashed line is the bare calculation, while
      the (red) full line is the obtained including
      the spectroscopic factors ${\cal{S}}_0$ and ${\cal S}_1$
      of Table~\ref{tab:spectro}.
      The data are taken from
      Refs.~\cite{Switkowski1979} and~\cite{He2013}.}
  \end{center}
\end{figure}

In Figure~\ref{fig:sftot} we compare our results for the
astrophysical $S$-factor with the experimental
data of Ref.~\cite{He2013} and~\cite{Switkowski1979}. The calculation is
performed summing up the
contributions to both the GS and the FES.
Since the data of Ref.~\cite{He2013} are still under debate, 
in discussing the results of Figure~\ref{fig:sftot} we will
consider only the data of Ref.~\cite{Switkowski1979}. By inspection of the
figure, we can conclude that our calculated (bare) $S$-factor is
systematically lower than the data.
The reason can be simply traced back
to the fact that in our model we do not take into account
the internal structure of $\Li$ and $\Be$.
In order to overcome this limitation,
we introduce the spectroscopic factor $\cal{S}$,
for both bound states of $\Be$,
so that the total cross section can be rewritten as
\begin{equation}
  \sigma(E)={\cal{S}}_0^2\sigma^{\rm{bare}}_0(E)+
        {\cal{S}}_1^2\sigma^{\rm{bare}}_1(E)\,.
\end{equation}
Here $\sigma^{\rm{bare}}_0(E)\,[\sigma^{\rm{bare}}_1(E)]$ 
and ${\cal{S}}_0({\cal{S}}_1)$ are the calculated bare cross section
and spectroscopic factor
for the transition to the  GS (FES) of $\Be$.

In order to determine the two spectroscopic factors ${\cal{S}}_0$
and ${\cal{S}}_1$, we proceed as follows: we notice that
in Ref.~\cite{Switkowski1979} there are two sets of data,
which corresponds to the radiative
capture to GS and FES, and the total $S$-factor
is given by multiplying the
data for the relative branching ratio (BR).
Therefore, we divide the two data sets for the corresponding BR and
we fit the spectroscopic factors, calculating the $S$-factor for GS and FES
captures separately.
In such a way we are able to reproduce not only the total $S$-factor but also the
experimental BR for the FES radiative capture of $\sim39\%$~\cite{Switkowski1979},
defined as ${\cal{S}}_1^2\sigma^{\rm{bare}}_1(E)/\sigma(E)$.
The values of the spectroscopic
factors and the $\chi^2/N$, where $N$ is the number of independent data,
are given in Table~\ref{tab:spectro}.
We used the data of Ref.~\cite{Switkowski1979} only, before
($\chi^2_0/N$) and after ($\chi^2_{\cal S}/N$) adding
the spectroscopic factors.

From the values of the $\chi^2_0/N$ given
in Table~\ref{tab:spectro}, it is possible to conclude that the 
description of the radiative capture reaction to the GS
using the bare wave function 
is quite accurate, while this is not the case for the FES.
\begin{table}[h]
  \begin{center}
    \begin{tabular}{lccc}
      \hline
      \hline 
      $J^\pi$ & ${\cal S}$ & $\chi_0^2/N$ & $\chi^2_{\cal_S}/N$\\ 
      \hline
      $3/2^-$ & 1.003 & 0.064  & 0.064 \\
      $1/2^-$ & 1.131 & 2.096  & 0.219 \\
      \hline
      \hline
    \end{tabular} 
    \caption{Spectroscopic
      factors ${\cal S}$ and $\chi^2/N$ obtained by fitting
      the data of Ref.~\cite{Switkowski1979}, before
      ($\chi^2_0/N$) and after ($\chi^2_{\cal S}/N$) adding
      the spectroscopic factors themselves. With $J^\pi=3/2^-$ and $J^\pi=1/2^-$
      we indicate the GS and FES of $\Be$.}\label{tab:spectro}
  \end{center}
\end{table}

In order to extrapolate the astrophysical $S$-factor at zero energy, we
perform a polynomial fit of our
calculated points up to second order, i.e. we rewrite the $S$-factor $S(E)$
in the energy range between $0$ and $300$ keV as
\begin{equation}
  S(E)=S(0)+S_1(0)E+S_2(0)E^2\,.\label{eq:sfexp}
\end{equation}
In Table~\ref{tab:fit} we report
the values obtained for $S(0)$, $S_1(0)$, and $S_2(0)$
obtained for our model compared with other
phenomenological models
of Refs.~\cite{Dubovichenko2011,Huang2010,Baker1980}.

\begin{table}[h]
  \begin{center}
    \begin{tabular}{lcccccc}
      \hline
      \hline
      & This work & Ref.~\cite{Dubovichenko2011} & Ref.~\cite{Huang2010} &
      Ref.~\cite{Baker1980} \\ 
      \hline
      $S(0)$ [eV b]          & $\m103.9$&$\m106$&$\m98.5$&$\m108$\\
      $S_1(0)$ [eV b/MeV]    & $ -105.1$&$ -215$&$ -71.5$&$ -130$\\
      $S_2(0)$ [eV b/MeV$^2$]& $\m45.0$ &$\m312$&$\m32.5$&$\m81.7$ \\
      \hline
      \hline
    \end{tabular} 
    \caption{Astrophysical factor at zero energy $S(0)$ and the coefficients
      $S_1(0)$ and $S_2(0)$ obtained from the polynomial fit of the $S$-factor
      as defined in Eq.~(\ref{eq:sfexp}).
      For Refs.~\cite{Dubovichenko2011,Huang2010,Baker1980}
      we report the expansion obtained from the fit of digitalized curves of the
      total $S$-factor.}
    \label{tab:fit}
  \end{center}
\end{table}
In particular, we can conclude that 
all the results for $S(0)$ are within $3\%$.
As regarding the shape
of the $S$-factor, determined by $S_1(0)$ and $S_2(0)$, our results
are quite in agreement with those of Ref.~\cite{Huang2010} and~\cite{Baker1980}.
On the other hand, the results obtained in Ref.~\cite{Dubovichenko2011}
with an approach similar to ours, give a higher value for $S_1(0)$ and $S_2(0)$.
The origin of this discrepancy is still unknown.
The results of Ref.~\cite{Arai2002}, although obtained with
a more sophisticated model than the one presented here, are consistent
with ours, while those of Ref.~\cite{Dong2017}
show a different energy dependence.
All the theoretical calculations, except the studies
of Refs.~\cite{Cecil1992,Prior2004},
agree in a negative slope in the $S$-factor at low energies,
and none of them predict a resonance structure, as suggested instead
by the data of Ref.~\cite{He2013}.

\begin{figure}[h]
  \begin{center}
    \includegraphics[scale=0.8]{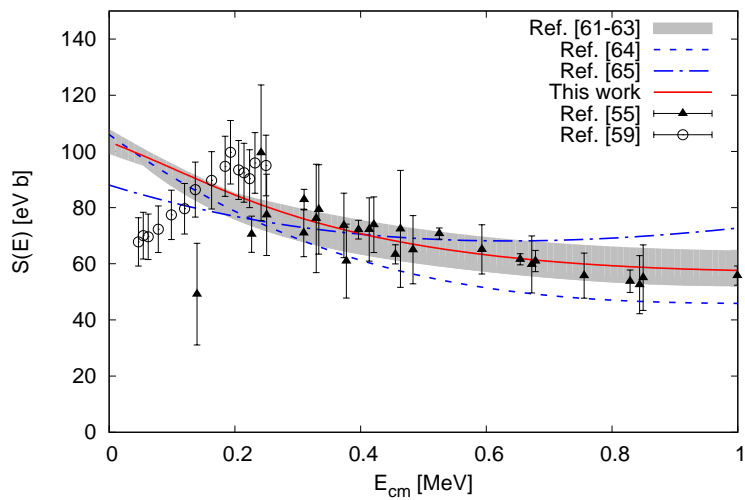}
    \caption{\label{fig:sfcomparison}
      Comparison between our predictions
      (full red line) and the studies available in the
      literature. We show as a (gray) band
      the calculated astrophysical $S$-factors of
      Refs.~\cite{Dubovichenko2011,Huang2010,Baker1980}, obtained with
      phenomenological potentials.
      For completeness we also report
      the results of Ref.~\cite{Arai2002} (blue dashed line) and of
      Ref.~\cite{Dong2017} (blue dot-dashed line).}
  \end{center}
\end{figure}

In order to estimate the theoretical uncertainty
arising from a calculation performed in the phenomenological
two-body cluster approach, we have reported in Figure~\ref{fig:sfcomparison}
in a (gray) band all the results
available in literature.
As we can conclude
by inspection of the figure, the theoretical error
which can be estimated by the width of the band is quite
significant,
of the same order of
the experimental errors on the data.
If we take into account all the results obtained
with the phenomenological potentials of
Refs.~\cite{Huang2010,Dubovichenko2011,Baker1980}
and the results of the present work, 
we obtain for the $S$-factor at zero energy the value
\begin{equation}
  S(0)= (103.5\pm 4.5)\ {\rm eV\ b}\,.
  \label{eq:s0}
\end{equation}
We remark that also the value for $S(0)$ obtained in Ref.~\cite{Descouvemont2010}
is within this range.
  
\subsection{Angular distribution of photons}
\label{subsec:angular}

We present in this section the photon angular
distribution results obtained within the framework
outlined in Sec.~\ref{sec:theory}, and we compare our results
with the data of Ref.~\cite{Tingwell1987}.
This provides a further check on our model.

By using Eq.~(\ref{eq:angulardistr}),
we have found that the main contribution to the $A_k^{J_3}(E)$ coefficients
comes from the interference of the
$E1$ multipole generated by the ${}^2S_{1/2}$ wave
with the $E1$ generated
by the ${}^2D$ waves and with the $E2$ generated by the ${}^2P$ waves.
Note that for the ${}^2P$ and ${}^2D$ waves, we do not have a complete
set of data for the phase shifts for all  possible
total angular momentum $J^\pi$ values. 
Therefore we use the same radial function for
the ${}^2D_{3/2}$ and ${}^2D_{5/2}$ waves,
and also for the ${}^2P_{1/2}$ and ${}^2P_{3/2}$ waves.
The relative phases for these waves, being arbitrary, are fixed in order to have the best
description of the data of Ref.~\cite{Tingwell1987}.
   
\begin{figure}[h]
  \begin{center}
    \includegraphics[scale=0.8]{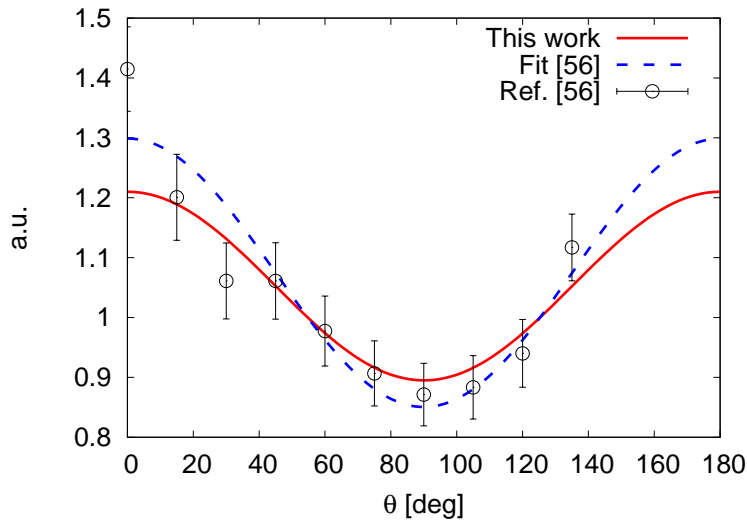}
    \caption{\label{fig:angdist0}
      Photon angular distribution for the radiative capture to the
      GS for $E_p=0.5$ MeV. Our calculation (full red line) is compared
      to the fit (dashed blue line) and the data of Ref.~\cite{Tingwell1987}.}
  \end{center}
\end{figure}

\begin{figure}[h]
  \begin{center}
    \includegraphics[scale=0.8]{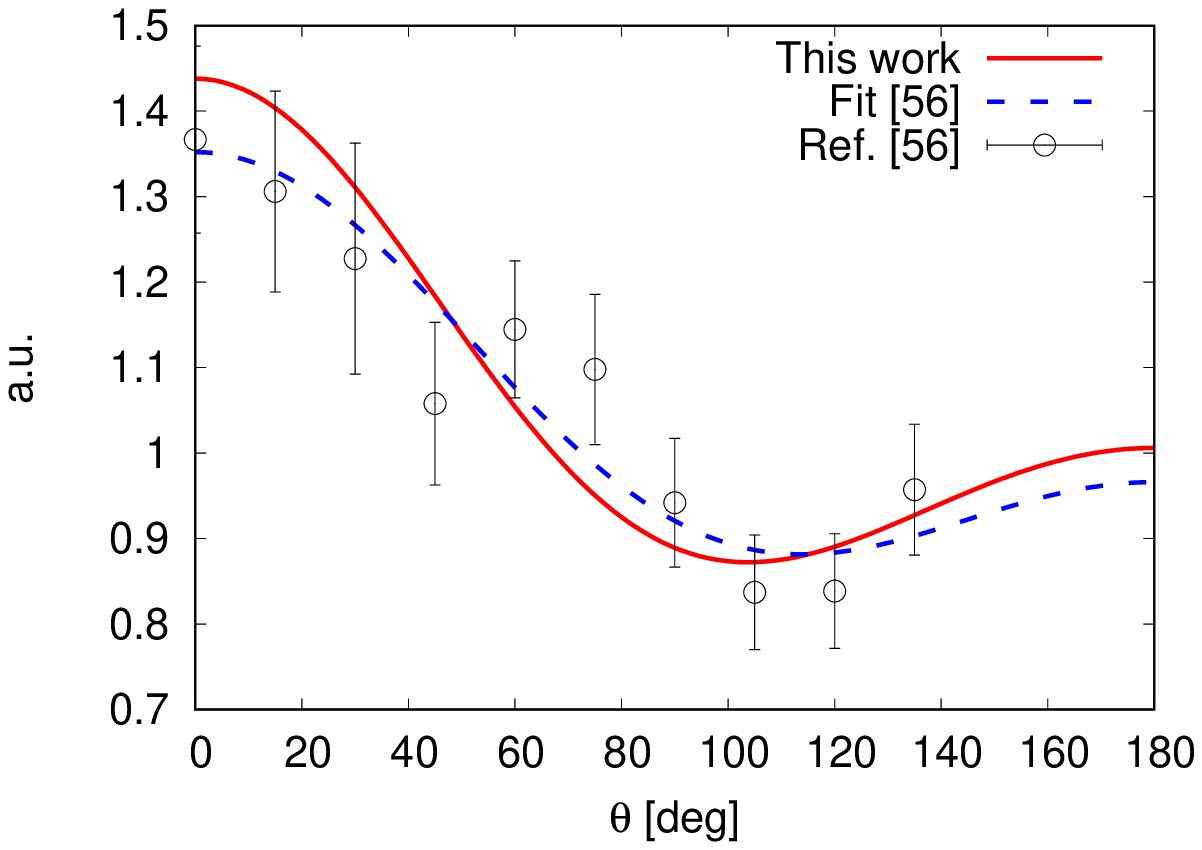}
    \caption{\label{fig:angdist1}
      The same as Figure~\ref{fig:angdist0} for the FES.}
  \end{center}
\end{figure}

The results for the $A_k^{J_3}$ coefficients for various incident
proton energies $(E_p)$ are reported in
Table~\ref{tab:t32},  where
they are compared with the values fitted on the
experimental data of Ref.~\cite{Tingwell1987}.
\begin{table}[h]
  \begin{center}
    \begin{tabular}{c|cc|cc}
      \hline
      \hline
      &\multicolumn{2}{c}{$J_3=3/2$} &\multicolumn{2}{c}{$J_3=1/2$}\\
      \hline
      $k$ & This work & Fit~\cite{Tingwell1987}& This work & Fit~\cite{Tingwell1987}\\ 
      \hline
      \multicolumn{5}{c}{$E_p=500$ keV}\\
      \hline
      1 &0.000   &   -             &0.188   &$0.193\pm0.055$      \\
      2 &0.210   & $0.299\pm0.045$ &0.222   &$0.159\pm0.074$      \\
      3 &0.000   &   -             &0.028   &      -              \\
      \hline
      $\chi^2/N$ & 0.97 & 0.90 & 0.62 & 0.78 \\
      \hline 
      \multicolumn{5}{c}{$E_p=800$ keV}\\
      \hline
      1 &0.000   &   -            &0.237   &$0.283\pm0.042$      \\
      2 &0.303   & $0.390\pm0.031$&0.322   &$0.257\pm0.051$      \\
      3 &0.000   &   -            &0.056   &     -               \\
      \hline
      $\chi^2/N$ & 0.87 & 1.17 & 1.37 & 0.76 \\
      \hline 
      \multicolumn{5}{c}{$E_p=1000$ keV}\\
      \hline
      1 &0.000   &   -            &0.262   &$0.205\pm0.043$      \\
      2 &0.354   & $0.368\pm0.036$&0.376   &$0.281\pm0.054$      \\    
      3 &0.000   &   -            &0.078   &     -               \\
      \hline
      $\chi^2/N$ & 1.03 & 1.21 & 1.61 & 1.67 \\
      \hline
      \hline
    \end{tabular} 
    \caption{\label{tab:t32}
      Values of the coefficients $A_k^{3/2}(E)$ and $A_k^{1/2}(E)$
      for three proton energies
      compared with the fit to the data of Ref.~\cite{Tingwell1987}.
      The $\chi^2/N$ is also reported.}
  \end{center}
\end{table}
In Figures~\ref{fig:angdist0} and~\ref{fig:angdist1}
we report the calculated angular distribution of the emitted photon for
the capture to the GS  and to the FES, respectively, for protons of
laboratory energy $E_p=0.5$ MeV. 
The data of Ref.~\cite{Tingwell1987} are also shown.
The theoretical values are in agreement with the fitted data for the GS.
In particular, the $A_1^{3/2}$ coefficient,
obtained using Eq.~(\ref{eq:ak}), results to be
\begin{equation}
  A_1^{3/2}\propto E_1^{0\frac{1}{2}\frac{1}{2},\frac{3}{2}}
  \left(E_2^{1\frac{1}{2}\frac{1}{2},\frac{3}{2}}
  -E_2^{1\frac{1}{2}\frac{3}{2},\frac{3}{2}}\right)+\ldots \ ,
  \label{eq:a1-32}
\end{equation}
where the dots indicate the interferences between the $E1$ generated by
the $D$ waves, which
give a negligible contribution.
If now we suppose that
$E_2^{1\frac{1}{2}\frac{1}{2},\frac{3}{2}}\simeq E_2^{1\frac{1}{2}\frac{3}{2},\frac{3}{2}}$,
the value for $A_1^{3/2}$ goes to zero, explaining why we
do not need this coefficient to reproduce the data.
In our case the values of the coefficients $A_1^{3/2}$
are exactly zero, because we use the same radial function for
different $J^\pi$. The same happens also for $A_3^{3/2}$.
As regarding to the capture to the FES, our calculation shows
also a nice agreement to the values obtained by fit to the data
of Ref.~\cite{Tingwell1987}.
In this case, there is no cancellation as in Eq.~(\ref{eq:a1-32}), and
therefore the values of the
coefficients $A_k^{1/2}$ are strongly dependent on the ${}^2P$ and ${}^2D$
waves, which are very uncertain in our model.

\begin{figure}[h!]
  \begin{center}
    \includegraphics[width=.7\textwidth]{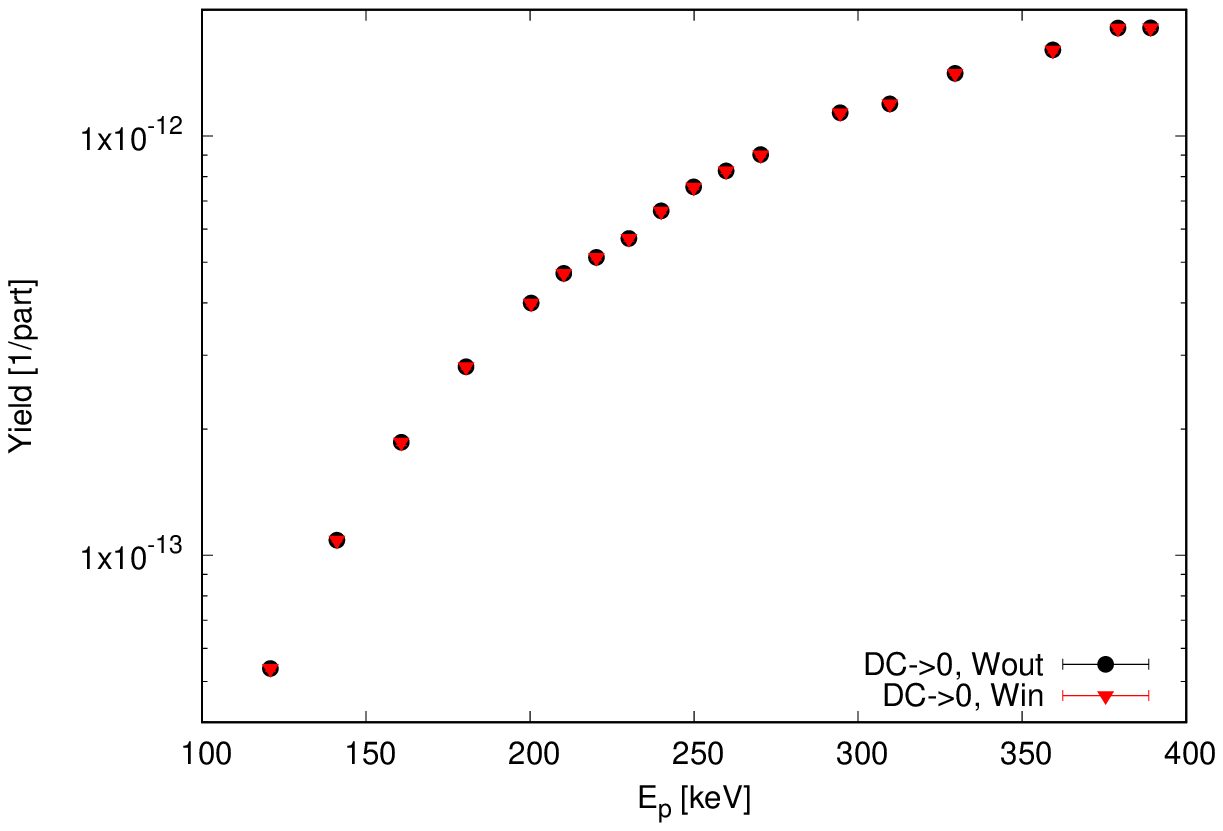}\\
    \includegraphics[width=.7\textwidth]{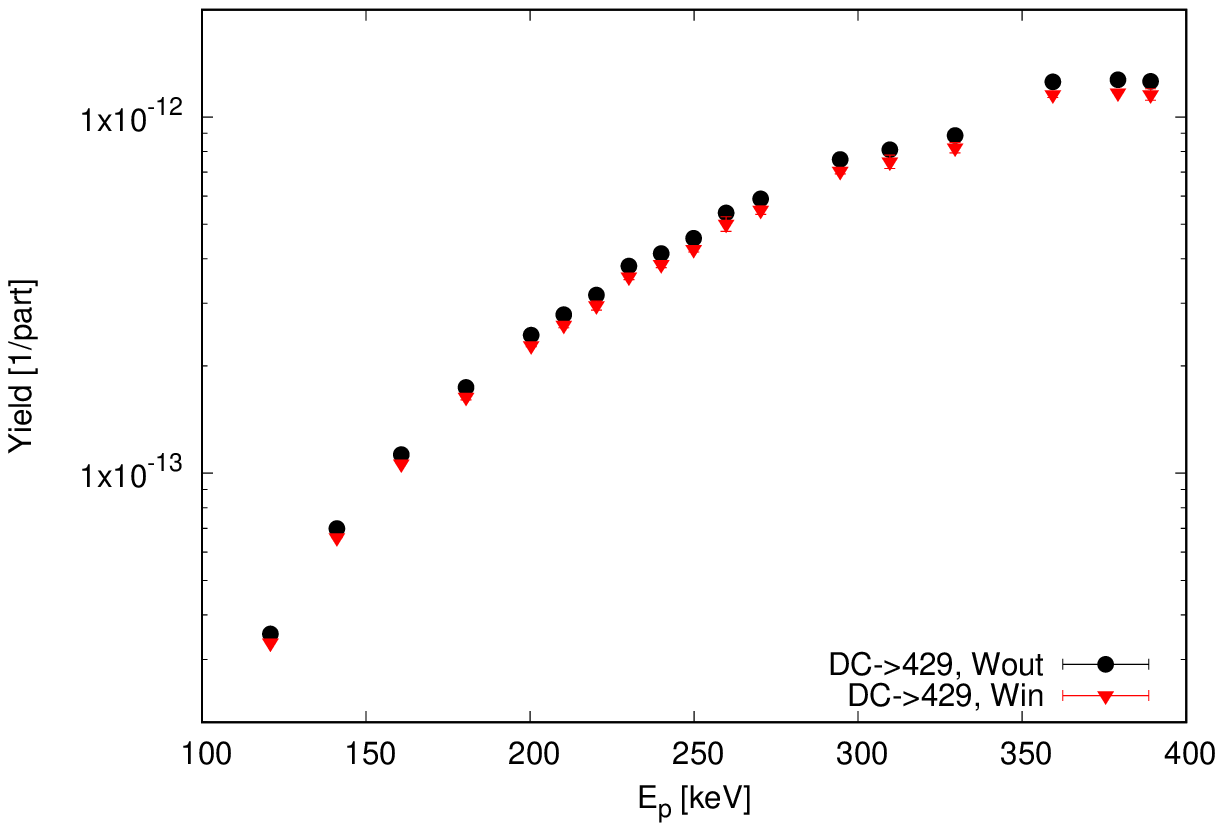}
    \caption{\label{fig:ce} Experimental yields for the radiative capture
      reaction to the
      GS (top) and the FES (bottom) as function of the incoming proton energy.
      The triangular red dots are the yields obtained 
      without considering the photon angular distribution while the circular black
      dots are obtained considering the photon angular distribution computed in
      this work. Figures courtesy of the LUNA Collaboration~\cite{Depalo2019}.
    }
  \end{center}
\end{figure}

The photon angular distribution has a noticeable impact on the experimental
measurements of the $S$-factor. Many experiments are done measuring
only the photon emitted in a small part of the total solid angle ($\Omega_d$).
Therefore, the rough data must
be corrected by a factor related to the angular distribution which enters
in the determination of the efficiency of the detector.
The factor of correction is then given by
\begin{eqnarray}
  \varepsilon(E,\Omega_d)\propto\int_{\Omega_d} d\cos\theta\, d\phi
  \left(1+\sum_{k\geq1}A_k(E)P_k(\cos\theta)\right)\,.
\label{eq:cea}
\end{eqnarray}
where $\Omega_d$ is the solid angle of the detector on which the
angular distribution is integrated.
The LUNA experiment, in order to minimize the effect of the
angular distributions, aligned the center of the
photon detector to an angle $\theta_0\simeq 55^\circ$ respect to the beam axis.
In this way the contribution of $A_2^{J_3}(E)$ is almost negligible  since
$P_2(\cos\theta_0)\simeq0$ and so the integrals
between angles close to $\theta_0$ is quite small.
However, the contribution which comes from the
$A_1^{J_3}(E)$ coefficient cannot be neglected.
Using our calculations of the coefficient $A_k^{J_3}(E)$, which are not known
experimentally, the LUNA Collaboration  was able to
estimate the impact of the angular distribution
of the photon to the measurement of the $p+\Li\rightarrow\Be+\gamma$
cross section. 
In Figure~\ref{fig:ce} we show the yields, namely the number of photon measured
normalized on the number of incoming protons, for the capture to the
GS and the FES as function of the incoming proton energy.
By inspecting the figure
we can conclude that the correction given by the photon
angular distribution is negligible for the capture to the GS (less than $0.2\%$)
~\cite{Depalo2019}.
This can be traced back to the fact that the
coefficient $A_1^{1/2}(E)\simeq0$ and the coefficient $A_2^{1/2}(E)$
does not significantly contribute, because of the geometry of the detector.
On the other hand, for the FES, the
corrections are of the order of $\sim 6-9\%$~\cite{Depalo2019}, which translates in
an appreciable modification of the experimental yield, as can be seen in the figure.
As before, this effect is mainly due to the $A_1^{3/2}(E)$, coefficient
which is not 0 for the FES.
 
  Before concluding the discussion, we want to remark that the poor knowledge on
  the phase shifts on the $P$ and $D$ waves introduce large uncertainties.
  These were estimated by comparing our angular distributions with the available
  data and then used in the data analysis performed by the LUNA Collaboration.

\subsection{The ``He''-resonance}
\label{subsec:heres}

In a recent work~\cite{He2013}, He {\it et al.} considered the possibility
of introducing a resonance-like structure
in the $\Li(p,\gamma)\Be$ $S$-factor data at low energies, and they estimated
the energy and width in the proton decay channel to
be $E_R=195$ keV and $\Gamma_p=50$ keV, respectively. The total
angular momentum of the resonance
reported in Ref.~\cite{He2013} was conjectured to be either
$J^\pi=1/2^+$ or $J^\pi=3/2^+$. In this section
we give for granted the existence
of this resonance, and we
explore the effects of introducing such a resonance
structure in our model. The comparison with the available data
will tell us whether this assumption is valid or not.

\begin{figure}[h]
  \begin{center}
    \includegraphics[scale=0.8]{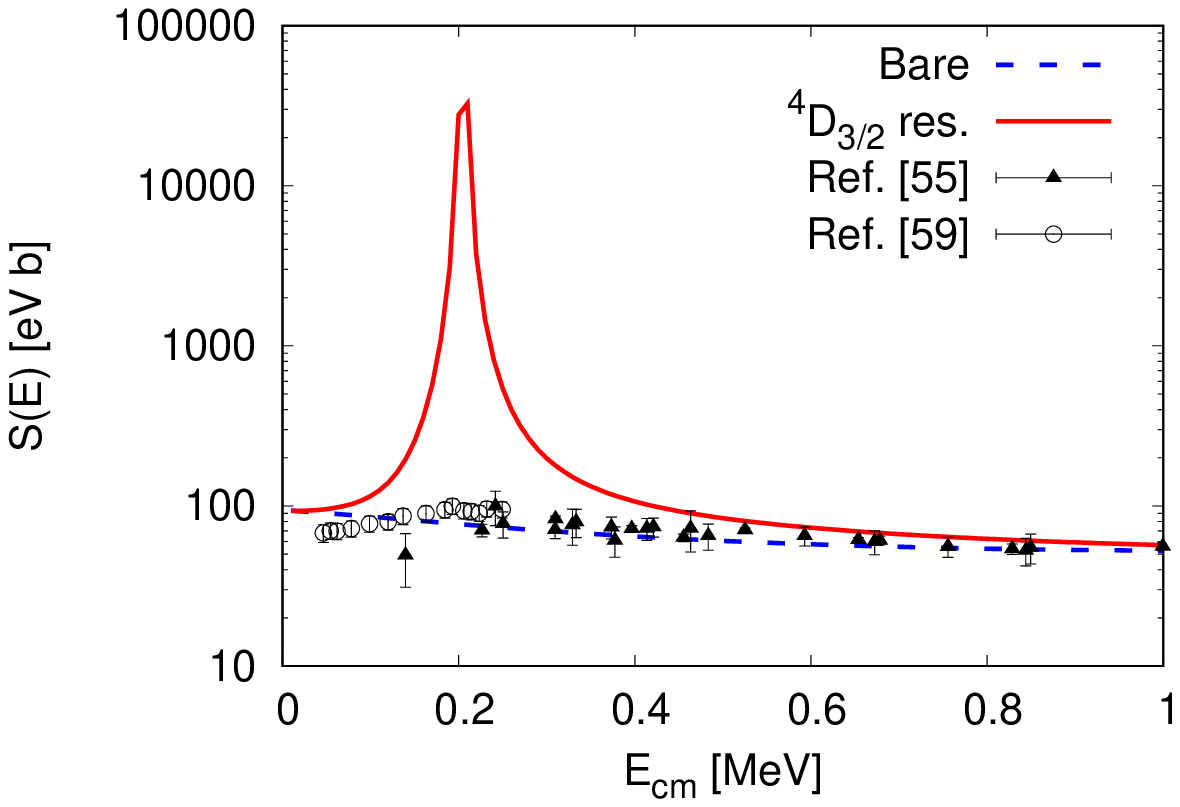}
    \caption{\label{fig:sfresd}
      Bare astrophysical $S$-factor (blue dot-dashed line)
      to which it is summed a resonance structure in
      the ${}^2D_{3/2}$ wave (full red line).
      See text for more details.}
  \end{center}
\end{figure}

The first step of our study consists in constructing
the nuclear potentials in order to obtain 
$190$ keV $<E_R<$ $200$ keV and to reproduce
the width of the resonance in the $S$-factor data.
In a first calculation, we consider to introduce the resonance in the
partial wave of spin 1/2. In particular we use the wave ${}^2S_{1/2}$
for $J^\pi=1/2^+$ and ${}^2D_{3/2}$ for $J^\pi=3/2^+$.
In both cases, we were not able to find values for the parameters $V_0$ and $a_0$
[see Eq.~(\ref{eq:vr})] that give a consistent description of all the available data. 
For the ${}^2S_{1/2}$ the introduction of such a resonance is completely
inconsistent with the experimental phase shifts.
For the ${}^2D_{3/2}$ we do not have experimental constrains on
the experimental phase shifts, but we were not able to obtain
the strength of the resonance as given in Ref.~\cite{He2013}. 
The best result obtained adding the resonance in the ${}^2D_{3/2}$ wave is given
in Figure~\ref{fig:sfresd}.

\begin{figure}[h]
  \begin{center}
    \includegraphics[scale=0.8]{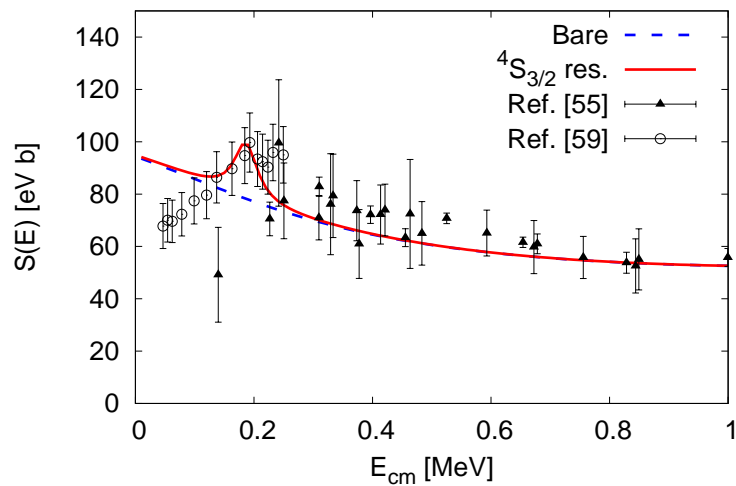}
    \caption{\label{fig:sfres}
      Bare astrophysical $S$-factor (blue dot-dashed line)
      to which it is summed a resonance structure in
      the ${}^4S_{3/2}$ wave (full red line). See text for more details.}
  \end{center}
\end{figure}

\begin{figure}[h]
  \begin{center}
    \includegraphics[scale=0.8]{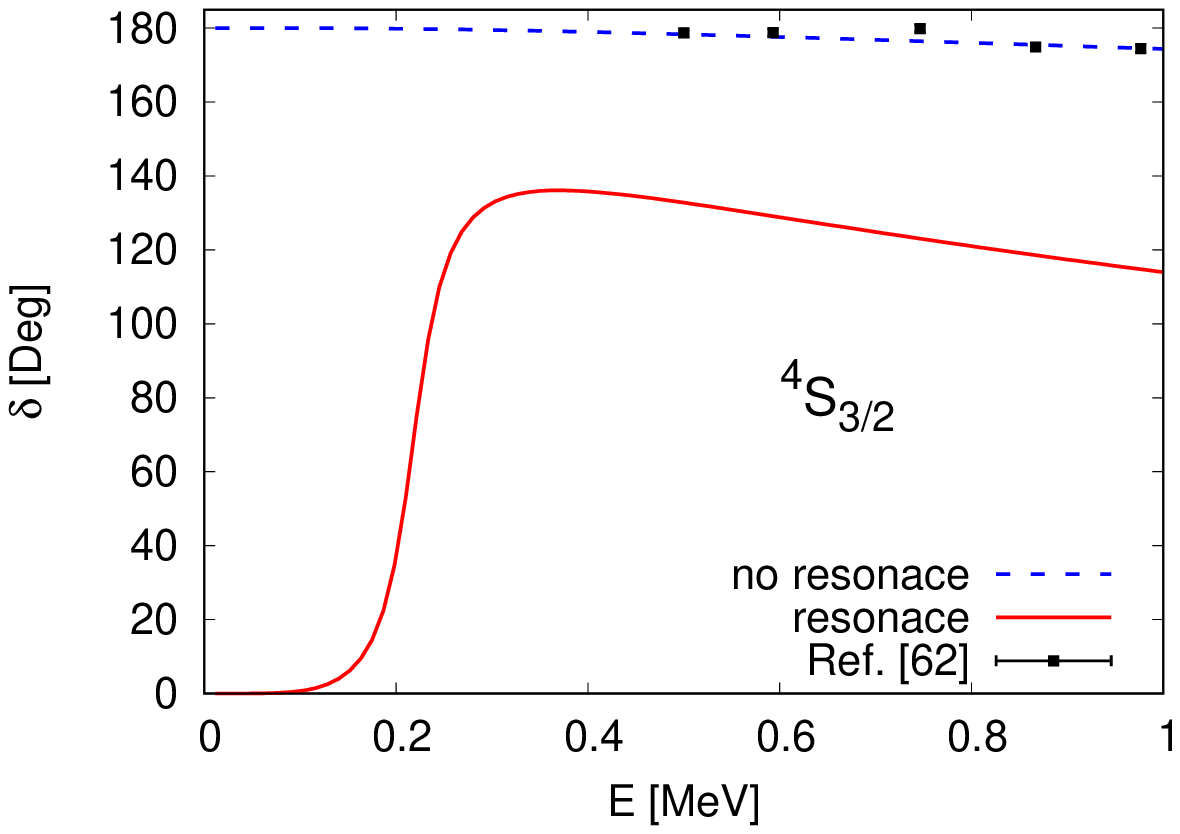}
    \caption{\label{fig:phsres}
      Elastic scattering phase shifts for the ${}^4S_{3/2}$ wave
      calculated in the
      case of resonance (red full line) and no resonance (blue dashed line),
      compared with the data of Ref.~\cite{Dubovichenko2011}.}
  \end{center}
\end{figure}

In a second calculation, we considered the GS of $\Be$ to be a mixed
state of spin $1/2$ and $3/2$. In this way the $E1$ operator can couple
the scattering wave ${}^4S_{3/2}$ to the ${}^4P_{3/2}$ component of the GS. Therefore,
we can introduce the $J^\pi=3/2^+$ resonance in the ${}^4S_{3/2}$ partial
wave. In this calculation, we use the ${}^2P_{3/2}$ radial wave function for
the ${}^4P_{3/2}$ component of the GS.
We select as potential parameters for the
${}^4S_{3/2}$ component
$V_0=438.7$ MeV and
$a_0=0.2$ fm$^{-2}$.
With this potential we get a resonance energy of $E_R=197$ keV and a width
of the resonance $\Gamma\sim15$ keV. The difference in the width
compared to the value reported by Ref.~\cite{He2013} is mainly due to the
fact that we do not include interferences with the ${}^3{\rm He}-\,^4{\rm He}$
channel.
Then we rewrite the total cross section as
\begin{equation}
  \sigma(E)={\cal{S}}_0^2\sigma^{\rm{bare}}_0(E)+
        {\cal{S}}^2_1\sigma^{\rm{bare}}_1(E)
        +{\cal{S}}_{\rm{res}}^2\sigma^{\rm{bare}}_{\rm{res}}(E)\,,
\label{eq:sigmae-he}
\end{equation}
where ${\cal{S}}_{\rm{res}}$ is the spectroscopic factor of the ${}^4P_{3/2}$
wave component in the GS and $\sigma^{\rm{bare}}_{\rm{res}}$ is
the calculated capture reaction cross section in the resonance wave.
The result obtained imposing ${\cal{S}}_0={\cal{S}}_1\sim1$ and
${\cal{S}}_{\rm{res}}\sim0.011$ is in good agreement
with both the data set of Refs.~\cite{Switkowski1979} and~\cite{He2013}
and it is shown in Figure~\ref{fig:sfres}. 
The small value of ${\cal{S}}_{\rm{res}}$ reflects
the small percentage of spin 3/2 component in the $\Be$ GS.
To be noticed that our results are also consistent with the
R-matrix fit reported in Ref.~\cite{He2013}. 
However, using the potential model which describes the resonance
in the $S$-factor data, we
were not able to reproduce the ${}^4S_{3/2}$ elastic phase shifts data.
Indeed, as shown in Figure~\ref{fig:phsres}, the ${}^4S_{3/2}$ phase shift
is badly underpredicted.
Therefore we can conclude that by including the resonance structure
in the  ${}^4S_{3/2}$ wave, we obtain a nice description of the $S$-factor
data, but we destroy the agreement between theory and experiment for the
elastic phase shifts. This puts under question the
real existence of the resonance structure proposed in Ref.~\cite{He2013}.

%% file: Conclusions_v2.tex
\chapter{Conclusions}\label{ch:conclusions}
In this Thesis we studied two reactions related to the
production of $\Li$ in the BBN: the 
$\alpha+d\rightarrow\Li+\gamma$ and the $p+\Li\rightarrow\Be+\gamma$ radiative
capture reactions.

For the first reaction we used a so-called {\it ab-initio} approach, in which
we solved the full six-body quantum mechanical problem. In particular, we considered  
a variational approach, in which we expanded our wave functions in terms of the
HH functions. Since solving the full six-body problem is very complex, in this work
we moved only the first steps towards the analysis of the $\alpha+d$ radiative capture reaction.
In particular, we  studied in details the  $\Li$ ground state, which is the
final state of the reaction. As first we analyzed the convergence rate
of the binding energy as function of the number of HH states
used in the wave function expansion.
In order to reach reasonable convergence,
we considered chiral nuclear potentials evolved
using the SRG unitary transformation~\cite{Bogner2007},
which are usually considered in {\it ab-initio} variational approaches.
We were able to obtain convergence rates 
even better than the ones obtained
using the NCSM reported in Ref.~\cite{Jurgenson2011}.
On the other hand,
in our calculation we did not consider three-body forces which are fundamental
to give an accurate description of the nuclear structure.
However, in order to have models which give values of the observables close to
the experimental one,
we have selected SRG transformed interaction for which presumably
the effects of proper three-body forces and the ones
induced by the SRG evolution cancel out.

Then we focused on $\Li$ electromagnetic static structure. Since we did not considered
SRG evolved operators and three-body forces, we were not able to reproduce exactly
the experimental values. However, we were able to obtain a lot of information
on the $\Li$ structure.
In particular, we noticed a strong dependence of the
electromagnetic observables on the strength of tensor forces in the nuclear
potentials. Moreover, we found indication that
two-body electromagnetic currents are presumably needed to explain the 
small and negative electric quadrupole moment found in the experiments.

Finally, we considered the $\alpha+d$ clusterization of $\Li$ which plays a
crucial role in the determination of the astrophysical {\it S}-factor.
Indeed, by defining the
overlap between the $\Li$ and the $\alpha+d$  wave functions
in relative $S$ and $D$ wave, we were able to
obtain the ANCs, which appear squared in the formula for the calculation
of the radiative capture cross section. To compute the ANCs we used two different
approaches, one computing it directly from the overlap of the wave functions,
the other from a differential equation. The results obtained with
the two approaches were not compatible within the error bars. However, 
we considered the second approach more reliable, since in the first one
there are systematic uncertainties which cannot be properly controlled.
The  obtained values well reproduce the correct magnitude of the
experimental values of the ANCs. The result is quite remarkable, if we consider
that the potential models  we used  have no parameters constrained
on this particular nuclear system.

The most important  result of this work is the extension of the
HH approach to nuclei up to $A=6$ mass number. In the Thesis we have presented
not only the theoretical aspect but also some of the algorithmic
improvements needed to perform calculations at $A=6$.
In perspective, using more refined computing codes, we believe possible to
increase the HH basis compared to the one used in this work. Therefore we
expect that it will be possible for the first time with an {\it ab-initio}
variational approach to
reach good convergence with not-SRG evolved potentials at $A=6$.
Moreover, in this work we established the first steps towards the
{\it ab-initio} study calculation
of the $\alpha+d$ scattering. In solving scattering problems, the main difficulty
is to compute the potential matrix elements between scattering states and pure
HH states. In this work, we presented and validated an approach that permits
to project scattering states
on the HH basis making easy then to compute the potential matrix elements.
Therefore,
this Thesis can be considered as the starting point
of the application of the HH approach to study
a wide range of nuclear phenomena that appear at $A=6$.

To study of the  $p+\Li\rightarrow\Be+\gamma$
we used a phenomenological approach in which
we treated $p$ and $\Li$ as structureless particles.
In this approach the parameters of the
phenomenological nuclear potential are fitted  in order
to reproduce the experimental binding energies and scattering phase shifts.
The theoretical $S$-factor is able to reproduce the experimental data
once we introduced the spectroscopic factor to take care of the internal
structure of the colliding nuclei.
By comparing different phenomenological approaches presented in the literature, we
estimated also a theoretical error for the $S$-factor.

The main goal of this study was the determination
of the photon angular distribution,
which is an unknown input for determining the efficiency of
experimental setups.
We validated our theoretical calculations obtained with our model
by comparing with the available data. Then, they were
used to determine the efficiency of the LUNA experimental setup and determine
the final experimental yields.

Finally, we have studied the possibility of introducing in our model the
resonance-like structure proposed in Ref.~\cite{He2013}. Even if, by introducing it
in the ${}^4S_{3/2}$ partial wave, we were able to obtain a nice description
of the $S$-factor data as given in Ref.~\cite{He2013}, we completely lost the possibility
to reproduce the ${}^4S_{3/2}$ experimental scattering phase shifts. 
We can conclude therefore that the presence of
a resonant structure cannot be accepted in our theoretical
framework.

%% file: trans_coeff.tex
\chapter{The Transformation Coefficients}\label{app:tc}

  In this appendix we present the calculation of the 
  TC as defined in Eq.~(\ref{eq:tc}). We follow
  the method presented in Refs.~\cite{Viviani1998,Dohet2019},
  based on the recursion relations, generalizing the formulas for the $A=6$ case.
  This appendix is divided in three sections. In 
  Section~\ref{app:tc1} we introduce the various terms needed to construct
  the TC. In Section~\ref{app:tc2} we discuss the recursion procedure.
  Finally in Section~\ref{app:tc3} we present the othogonalization procedure.

    \begin{section}{Construction of the Transformation Coefficients}
      \label{app:tc1}
    The TC are defined in Chapter~\ref{ch:HH} as the coefficients
    $a^{KLSTJ\pi}_{\alpha,\alpha'}(p)$ which
    transforms a state $\alpha$,
    written  in a generic permutation $p$,  in a sum of
    states $\al'$ written in the reference permutation, namely
    \begin{equation}
      \Phi^{KLSTJ\pi}_\alpha(i,j,k,l,m,n)
      =\sum_{\alpha'}a^{KLSTJ\pi}_{\alpha,\alpha'}(p)
      \Phi^{KLSTJ\pi}_{\alpha'}(1,2,3,4,5,6)\,,\label{eq:tc_app}
    \end{equation}
    where $\alpha$ is defined in Eq.~(\ref{eq:alpha}) and
    $\Phi^{KLSTJ\pi}_\alpha(i,j,k,l,m,n)$ is defined in Eq.~(\ref{eq:hhst}).
    By construction, the functions $\Phi^{KLSTJ\pi}_\alpha$ are the product
    of a HH, a spin and an isospin function.
    We can define coefficients that transform each part
    from a permutation to another.
    Therefore, for the HH part we define the coefficients
    $c^{KL,p}_{\mu;\mu'}$ such that
    \begin{equation}
      {\cal Y}^{KLM,p}_{\mu}
      =\sum_{\mu'}c^{KL,p}_{\mu;\mu'}
      {\cal Y}^{KLM,1}_{\mu'}\,,\label{eq:cc0}
    \end{equation}
    where ${\cal Y}^{KLM,p}_{\mu'}$ is defined in Eq.~(\ref{eq:hh6}) and
    the apex $p$ indicates that the HH function depends on $\Omega_{5p}$.
    We will discuss the calculation of this coefficients in Section~\ref{app:tc2}.
    
    As regarding the spin(isospin) part, we define similar coefficients
    $c^{S,p}_{\sigma;\sigma'}$($c^{T,p}_{\tau;\tau'}$), such that
    \begin{equation}\label{eq:sp_sig}
      {\cal S}^{SS_z,p}_{\sigma}=\sum_{\sigma'}
      c^{S,p}_{\sigma;\sigma'}{\cal S}^{SS_z,1}_{\sigma'}\,,
    \end{equation}
    where $\sigma=\{S_2,S_3,S_4,S_5,S\}$($\tau=\{T_2,T_3,T_4,T_5,T\}$), and
    \begin{equation}
      {\cal S}^{SS_z,p}_{\sigma}
      =\left[\left(\left(s_is_j\right)_{S_2} s_k\right)_{S_3}
        \left(\left(s_ls_m\right)_{S_4} s_n\right)_{S_5}\right]_{S,S_z}\,,
    \end{equation}
    and analogously for the isospin. By using Eq.~(\ref{eq:sp_sig}) it is
    possible to compute the coefficients as
    \begin{equation}
      c^{S,p}_{\sigma;\sigma'}=\left({\cal S}^{SS_z,1}_{\sigma'}\right)^\dag
      {\cal S}^{SS_z,p}_{\sigma}\,,
    \end{equation}
    which can be written explicitly as a sum of products of Clebsch-Gordan
    coefficients that, for the specific used coupling scheme, reads
    \begin{equation}
    \begin{aligned}
    &c^{S,p}_{\sigma;\sigma'}=\sum_{m_1,m_2,m_3,\atop m_4,m_5,m_6}
    \sum_{m_i,m_j,m_k,\atop m_l,m_n,m_m}
    ( S_3M_3,S_5M_5|S  M_6 ) ( S_3'M_3',S_5'M_5'|S M_6 )\\
    &\times( S_2M_2,\frac{1}{2}m_3|S_3M_3 ) ( S_2'M_2',\frac{1}{2}m_k |S_3'M_3')
    ( S_4M_4,\frac{1}{2}m_6|S_5M_5 ) ( S_4'M_4',\frac{1}{2}m_n |S_5'M_5')\\
    &\times( \frac{1}{2}m_1,\frac{1}{2}m_2|S_2M_2 ) ( \frac{1}{2}m_i,\frac{1}{2}m_j|S_2'M_2' )
    ( \frac{1}{2}m_4,\frac{1}{2}m_5|S_4M_4 ) ( \frac{1}{2}m_l,\frac{1}{2}m_m|S_4'M_4' )\,,
    \end{aligned}
    \end{equation}
    where $M_i=m_i+M_{i-1}$ with $M_0=0$.
    This formula is valid also for the isospin part, substituting $S$, $\sigma$ 
    and $\sigma'$ with $T$, $\tau$ and $\tau'$, respectively.
    By definition, a state $\alpha$ in a generic permutation $p$
    can be written as
    \begin{equation}\label{eq:tc_app2}
      \Phi^{KLSTJ\pi}_\alpha(i,j,k,l,m,n)=\left[{\cal Y}^{KLM,p}_{\mu}
        \,{\cal S}^{SS_z,p}_\sigma\right]_J\, {\cal T}^{TT_z,p}_{\tau}\,,
    \end{equation}
    where, by definition, $\alpha=\{\mu,\sigma,\tau\}$.
    Using Eqs.~(\ref{eq:cc0}) and~(\ref{eq:sp_sig}), Eq.~(\ref{eq:tc_app2})
    reads
    \begin{equation}\label{eq:tc_app3}
      \Phi^{KLSTJ\pi}_\alpha(i,j,k,l,m,n)=\sum_{\mu'}\sum_{\sigma'}
      \sum_{\tau'}c^{KL,p}_{\mu;\mu'}c^{S,p}_{\sigma;\sigma'}c^{T,p}_{\tau;\tau'}
      \left[{\cal Y}^{KLM,1}_{\mu'}\,
        {\cal S}^{SS_z,1}_{\sigma'}\right]_{JJ_z}\, {\cal T}^{TT_z,1}_{\tau'}\,.
    \end{equation}
    Comparing Eq.~(\ref{eq:tc_app}) with Eq.~(\ref{eq:tc_app3}), it is easy
    to derive the final expression for the TC in terms of the coefficients $c$,
    i.e.
    \begin{equation}
      a^{KLSTJ\pi}_{\alpha,\alpha'}(p)
      =c^{KL,p}_{\mu;\mu'}c^{S,p}_{\sigma;\sigma'}c^{T,p}_{\tau;\tau'}\,.      
    \end{equation}

    Once constructed the TC for each permutation, we can evaluate the
    coefficients $A^{KLSTJ\pi}_{\alpha,\alpha'}$ as given in Eq.~(\ref{eq:tc2}).
  \end{section}

  \begin{section}{The recursion procedure}\label{app:tc2}

    In this section we discuss the calculation of the coefficients
    $c^{KL,p}_{\mu\mu'}$. 
    In order to make the notation easier
    we rewrite the generic HH function, defined in
    Eq.~(\ref{eq:hh6}), as
    \begin{equation}
    \begin{aligned}\label{eq:hh6app}
    {\cal Y}^{KLM}_{\mu}(\Omega_{5p})&=
    \left[\left(\left(\left(Y_{\ell_1}(\hxx_{1p})
      Y_{\ell_2}(\hxx_{2p})\right)_{L_2}Y_{\ell_3}(\hxx_{3p})
      \right)_{L_3}Y_{\ell_4}(\hxx_{4p})\right)_{L_4}
      Y_{\ell_5}(\hxx_{5p})\right]_{LM}\\
    &\times\prod_{j=2}^5{\cal N}_{n_j,K_{j-1}}^{\ell_j,\nu_j}
    (1+y_{jp})^{\frac{\ell_j}{2}}
    (1-y_{jp})^{\frac{K_{j-1}}{2}}P_{n_j}^{\nu_{j-1},\ell_j+1/2}(y_{jp})\,,
    \end{aligned}
    \end{equation}
    where
    \begin{equation}
      y_{jp}=\cos 2\varphi_{jp}\,,
    \end{equation}
    and the hyperangles $\ph_{jp}$ are defined in Eq.~(\ref{eq:phiang}).
    The $P_{n_j}^{\nu_{j-1},\ell_j+1/2}(y_{jp})$ are the Jacobi polynomials, and
    \begin{equation}\label{eq:norm_def}
      {\cal N}_{n_j,K_{j-1}}^{\ell_j,\nu_j}=
      \left(\frac{1}{2}\right)^{\frac{(\ell_j+K_{j-1})}{2}}
      \left[\frac{2\nu_j\Gamma(\nu_j-n_j)n_j!}
      {\Gamma(\nu_j-n_j-\ell_j-1/2)\Gamma(n_j+\ell_j+3/2)}\right]^{1/2}\,,
    \end{equation}
    are normalization factors [see Eq.~(\ref{eq:nhh})],
    where $\nu_j$ and $K_j$ are defined in
    Eq.~(\ref{eq:go}). The index $\mu$ in Eq.~(\ref{eq:hh6app})
    has been introduced in Eq.~(\ref{eq:qn3}).

  Let us assume to know the coefficients $c^{KL,p}_{\mu;\mu'}$ for a given $K$, so that
  \begin{equation}
    {\cal Y}^{KLM,p}_{\lambda, n_2, n_3, n_4, n_5}
    =\sum_{\mu'}c^{KL,p}_{\lambda, n_2, n_3, n_4, n_5;\mu'}
    {\cal Y}^{KLM,1}_{\mu'}\,,\label{eq:aa0}
  \end{equation}
  where for convenience, we have defined
  \begin{eqnarray}\label{eq:lambda}
    \lambda=\{\ell_1,\ell_2,\ell_3,\ell_4,\ell_5,L_2,L_3,L_4\}\,.
  \end{eqnarray}
  The recurrence is applied separately on the quantum number $n_2$, $n_3$,
  $n_4$ and $n_5$, once fixed the quantum numbers $\lambda$.

  In the following we will use the quantities
    \begin{align}
    a_{n_i}&=\frac{{\cal N}_{n_2,\dots,n_i+1,0,\dots,0}^{\ell_1,\dots,\ell_5}}
    {{\cal N}_{n_2,\dots,n_i,0,\dots,0}^{\ell_1,\dots,\ell_5}}
    \frac{(2n_i+\al_i+\be_i+1)(\al_i^2-\be_i^2)}{2(n_i+1)(n_i+\al_i+\be_i+1)(2n_i+\al_i+\be_i)}\,,\\
    b_{n_i}&=\frac{{\cal N}_{n_2,\dots,n_i+1,0,\dots,0}^{\ell_1,\dots,\ell_5}}
    {{\cal N}_{n_2,\dots,n_i,0,\dots,0}^{l_1,\dots,l_5}}
    \frac{(2n_i+\al_i+\be_i+1)(2n_i+\al_i+\be_i+2)}{2(n_i+1)(n_i+\al_i+\be_i+1)}\,,\\
    c_{n_i}&=-\frac{{\cal N}_{n_2,\dots,n_i+1,0,\dots,0}^{\ell_1,\dots,\ell_5}}
    {{\cal N}_{n_2,\dots,n_i,0,\dots,0}^{\ell_1,\dots,\ell_5}}
    \frac{(n_i+\al_i)(n_i+\be_i)(2n_i+\al_i+\be_i+2)}{(n_i+1)(n_i+\al_i+\be_i+1)(2n_i+\al_i+\be_i)}\,,
    \end{align}
    which are defined for $i=2,\dots,5$.
  Here we have defined $\alpha_i=\nu_{i-1}$, $\beta_i=\ell_i+1/2$ and
  \begin{equation}\label{eq:norm_coeff}
    {\cal N}_{n_1,\dots,n_5}^{\ell_1,\dots,\ell_N}=
    \prod_{j=2}^5{\cal N}_{n_j,K_{j-1}}^{\ell_j,\nu_j}\,.
  \end{equation}
  If we consider the function
  ${\cal Y}^{KLM,p}_{\lambda n_2 n_3 n_4 n_5+1}$,
  the Jacobi polynomial $P_{n_5+1}(y_{5p})$ entering the expression can be
  written in terms of $P_{n_5}(y_{5p})$ and $P_{n_5-1}(y_{5p})$ by using
  the Jacobi polynomial recursion relation. Explicitly we obtain
  \begin{equation}
    {\cal Y}^{K+2LM,p}_{\lambda, n_2, n_3, n_4, n_5+1}=
    (a_{n_5}+b_{n_5}y_{5p}){\cal Y}^{KLM,p}_{\lambda, n_2, n_3, n_4, n_5}
    +c_{n_5}{\cal Y}^{K-2LM,p}_{\lambda, n_2, n_3, n_4, n_5-1}\,.\label{eq:rec5}
  \end{equation}
  Now, $y_{5p}=2\frac{(x_{5p})^2}{\rho^2}-1$, and
  \begin{equation}\label{eq:xlp2}
    (x_{lp})^2=\sum_{i,j}\Gamma_{i,j}^p(l)\,\xx_i\cdot\xx_j\qquad l,i,j=1,\dots,5\,,
  \end{equation}
  $\xx_i$ being the Jacobi vectors evaluated in the 
  permutation 1 and in this case $l=5$.
  The coefficients $\Gamma_{i,j}^p(l)$ are numerical coefficients
  which depend on the permutation.
  Using the orthonormalization properties of the HH
  functions, one gets
  \begin{equation}
    c^{K+2L,p}_{\mu;\mu'}=\int d\Omega_5
    \left[{\cal Y}^{K+2LM}_{\mu}(\Omega_5)\right]^\dag
         {\cal Y}^{K+2LM}_{\mu'}(\Omega_{5p})\,,\label{eq:aa}
  \end{equation}
  where
  \begin{equation}
    \int d\Omega_5 =\frac{1}{2^{21}}\int \prod_{i=1}^5d\hat{x}_i
    \prod_{j=2}^5dy_j(1+y_j)^\frac{1}{2}(1-y_j)^\frac{3j-5}{2}\,.
  \end{equation}
  If now we insert Eq.~(\ref{eq:rec5}) in the integral~(\ref{eq:aa}), only
  the term containing the factor $b_{n_5}(x_{5p})^2$ gives a non-vanishing
  contribution. Indeed only this term is of 
  order $K+2$ in Eq.~(\ref{eq:rec5}). Finally, by using
  Eq.~(\ref{eq:aa0}), the recursion relation is obtained
  \begin{equation}\label{eq:n5f}
    c^{K+2L,p}_{\lambda, n_2, n_3, n_4, n_5+1;\mu'}=
    2b_{n_5}\sum_{\mu''}c^{KL,p}_{\lambda, n_2, n_3, n_4, n_5;\mu''}
    \times\sum_{i,j}\Gamma_{i,j}^p(5)I_{i,j}^{\mu',\mu''}\,,
  \end{equation}
  where
  \begin{equation}\label{eq:Iij}
    I_{i,j}^{\mu',\mu''}=\int d\Omega_5
    \left[{\cal Y}^{K+2LM}_{\mu'}(\Omega_5)
        \right]^\dag\frac{{\boldsymbol x}_i
    \cdot{\boldsymbol x}_j}{\rho^2}{\cal Y}^{KLM}_{\mu''}(\Omega_5)\,.
  \end{equation}
  The explicit expressions for the integrals $I_{i,j}^{\mu',\mu''}$ for
  $A=6$ are given in Section~\ref{app:int}.

  Let us now consider the HH function ${\cal Y}^{KLM}_{\lambda, n_2, n_3, n_4+1, 0}$.
  By proceeding as in the previous case we have
  \begin{equation}
    {\cal Y}^{K+2LM,p}_{\lambda,n_2,n_3,n_4+1,0}=
    (1-y_{5p})(a_{n_4}+b_{n_4}y_{4p}){\cal Y}^{KLM,p}_{\lambda,n_2,n_3,n_4,0}
    +c_{n_4}(1-y_{5p})^2{\cal Y}^{K-2LM,p}_{\lambda,n_2,n_3,n_4-1,0}\,.
    \label{eq:n40}
  \end{equation}
  The first term of Eq.~(\ref{eq:n40}) can be written as
  \begin{equation}
  \begin{aligned}
    (1-y_{5p})(a_{n_4}&+b_{n_4}y_{4p}){\cal Y}^{KLM,p}_{\lambda,n_2,n_3,n_4,0}
    =\\
    &\Big[4b_{n_4}\frac{(x_{4p})^2}{\rho^2}+
      2(b_{n_4}-a_{n_4})\frac{(x_{5p})^2}{\rho^2}\Big]
    {\cal Y}^{KLM,p}_{\lambda,n_2,n_3,n_4,0}+{\cal O}(K)\,,
  \end{aligned}
  \end{equation}
  where with ${\cal{O}}(K)$ we group all the terms of order smaller than $K+2$,
  which, therefore, do not contribute to the integral in Eq.~(\ref{eq:aa}).
  The square of the moduli of the Jacobi vectors
  $(x_{5p})^2$ and $(x_{4p})^2$ can be expressed using
  Eq.~(\ref{eq:xlp2}).
  The second term in Eq.~(\ref{eq:n40}) can be rewritten as
  \begin{equation}
  \begin{aligned}\label{eq:n41}
    (1-y_{5p})^2{\cal Y}^{K-2LM,p}_{\lambda,n_2,n_3,n_4-1,0}&=
    A_{n_2,n_3,n_4-1}^{(4)}{\cal Y}^{K+2LM,p}_{\lambda,n_2,n_3,n_4-1,2}
    +B_{n_2,n_3,n_4-1}^{(4)}{\cal Y}^{KLM,p}_{\lambda,n_2,n_3,n_4-1,1}\\
    &+C_{n_2,n_3,n_4-1}^{(4)}{\cal Y}^{K-2LM,p}_{\lambda,n_2,n_3,n_4-1,0}
    \,,
  \end{aligned}
  \end{equation}
  where we expressed the factor $(1-y_{5p})^2$ in terms of the Jacobi polynomials
  of order 0, 1 and 2. Above, we have defined
  \begin{align}
    A^{(i)}_{n_2,\dots,n_i}
    &=\frac{{\cal N}_{n_2,\dots,n_i,0,\dots,0}^{\ell_1,\dots,\ell_5}}
    {{\cal N}_{n_2,\dots,n_i,2,0,\dots,0}^{\ell_1,\dots,\ell_5}}
    \frac{8}{(\al_i+\be_i+4)(\al_i+\be_i+3)}\,,\\
    B^{(i)}_{n_2,\dots,n_i}&=
    -\frac{{\cal N}_{n_2,\dots,n_i,0,\dots,0}^{\ell_1,\dots,\ell_5}}
    {{\cal N}_{n_2,\dots,n_i,1,0,\dots,0}^{\ell_1,\dots,\ell_5}}
    \frac{8(\al_i+2)}{(\al_i+\be_i+4)(\al_i+\be_i+2)}\,,\\
    C^{(i)}_{n_2,\dots,n_i}&=\frac{(\al_i+1)(\al_i+2)}
    {(\al_i+\be_i+3)(\al_i+\be_i+2)}\,,
  \end{align}
    where $\alpha_i=\nu_{i-1}$ and $\beta_i=\ell_i+1/2$.
    When the expression in Eq.~(\ref{eq:n41}) is inserted in Eq.~(\ref{eq:aa}),
    only the term proportional to $A_{n_2,n_3,n_4-1}^{(4)}$ survives, having order
    $K+2$. In conclusion, writing the integral with respect to $\Omega_5$
    in terms of the integral $I_{i,j}^{\mu',\mu''}$,
    the recursion formula for $n_4$ results
    \begin{equation}
    \begin{aligned}\label{eq:n4f}
    c^{K+2L,p}_{\lambda,n_2,n_3,n_4+1,0;\mu'}&=
    c_{n_4}A^{(4)}_{n_2,n_3,n_4-1}c^{K+2L,p}_{\lambda,n_2,n_3,n_4-1,2;\mu'}
    +\sum_{\mu''}c^{KL,p}_{\lambda,n_2,n_3,n_4,0;\mu''}\\
    &\times\sum_{i,j}\left[4b_{n_4}\Gamma_{i,j}^p(4)+
      2(b_{n_4}-a_{n_4})\Gamma_{i,j}^p(5)\right]I_{i,j}^{\mu',\mu''}\,.
    \end{aligned}
    \end{equation}
  In the case of the HH function ${\cal Y}^{K,L,M,p}_{\lambda, n_2, n_3+1, 0, 0}$,
  using the recursion of the Jacobi polynomials on the index $n_3$ we get
  \begin{equation}
  \begin{aligned}
    {\cal Y}^{K+2LM,p}_{\lambda,n_2,n_3+1,0,0}&=
    (1-y_{5p})(1-y_{4p})(a_{n_3}+b_{n_3}y_{3p})
    {\cal Y}^{KLM,p}_{\lambda,n_2,n_3,0,0}\\
    &+c_{n_3}(1-y_{5p})^2(1-y_{4p})^2
    {\cal Y}^{K-2LM,p}_{\lambda,n_2,n_3-1,0,0}\,.
    \label{eq:n30}
  \end{aligned}
  \end{equation}
  The first term of Eq.~(\ref{eq:n30}) can be written as
  \begin{align}
    (1-y_{5p})&(1-y_{4p})(a_{n_3}+b_{n_3}y_{3p})
    {\cal Y}^{KLM,p}_{\lambda,n_2,n_3,0,0}=\nonumber\\
    &\left[8b_{n_3}\frac{(x_{3p})^2}{\rho^2}+
    4(b_{n_3}-a_{n_3})\left(\frac{(x_{4p})^2}{\rho^2}+
    \frac{(x_{5p})^2}{\rho^2}\right)\right]
         {\cal Y}^{KLM,p}_{\lambda,n_2,n_3,0,0}
    +{\cal O}(K)\,,
  \end{align}
  where the terms ${\cal O}(K)$ do not give contributions when inserted
  in Eq.~(\ref{eq:aa}). The second term of Eq.~(\ref{eq:n30}) is rewritten as
  \begin{equation}
  \begin{aligned}
    (1-&y_{5p})^2(1-y_{4p})^2
    {\cal Y}^{K-2LM,p}_{\lambda,n_2,n_3-1,0,0}=
    A^{(3)}_{n_2,n_3-1}{\cal Y}^{K+2LM,p}_{\lambda,n_2,n_3-1,2,0}\\
    &+B^{(3)}_{n_2,n_3-1}\eta^{(4)}_{n_2,n_3-1,1}
    {\cal Y}^{K+2LM,p}_{\lambda,n_2,n_3-1,1,1}
    +C^{(3)}_{n_2,n_3-1}A^{(4)}_{n_2,n_3-1,0}{\cal Y}^{K+2LM,p}_{\lambda,n_2,n_3-1,0,2}+{\cal O}(K)
    \,,
  \end{aligned}
  \end{equation}
  where we define
  \begin{align}
    \eta^{(i)}_{n_2,\dots,n_i}
    &=\frac{{\cal N}_{n_2,\dots,n_i,0,\dots,0}^{\ell_1,\dots,\ell_5}}
    {{\cal N}_{n_2,\dots,n_i,1,0,\dots,0}^{\ell_1,\dots,\ell_5}}
    \frac{2}{(\al_i+\be_i+2)}\,,\\
    \gm^{(i)}_{n_2,\dots,n_i}&=\frac{2(\al_i+1)}{(\al_i+\be_i+2)}\,,\
  \end{align}
  with $\alpha_i=\nu_i$ and $\beta_i=\ell_{i+1}+1/2$.
  Inserting these results in Eq.~(\ref{eq:aa}), we obtain the recursion
  formula for $n_3$, which reads
  \begin{equation}
  \begin{aligned}\label{eq:n3f}
    c^{K+2L,p}_{\lambda,n_2,n_3+1,0,0;\mu'}&=
    c_{n_3}\big[A^{(3)}_{n_2,n_3-1}c^{K+2L,p}_{\lambda,n_2,n_3-1,2,0;\mu'}
    +B^{(3)}_{n_2,n_3-1}\eta^{(4)}_{n_2,n_3-1,1}
    c^{K+2L,p}_{\lambda,n_2,n_3-1,1,1;\mu'}\\
    &+C^{(3)}_{n_2,n_3-1}A^{(4)}_{n_2,n_3-1,0}
    c^{K+2L,p}_{\lambda,n_2,n_3-1,0,2;\mu'}\big]
    +\sum_{\mu''}c^{KL,p}_{\lambda,n_2,n_3,0,0;\mu''}\\&\times
    \sum_{i,j}\big[8b_{n_3}\Gamma_{i,j}^p(3)
    +4(b_{n_3}-a_{n_3})(\Gamma_{i,j}^p(4)+\Gamma_{i,j}^p(5))\big]
    I_{i,j}^{\mu',\mu''}\,.
  \end{aligned}
  \end{equation}
  
  The last recursion is done on the index $n_2$. Imposing $n_3,n_4,n_5=0$,
  we rewrite the HH function ${\cal Y}^{K+2LM,p}_{\lambda,n_2+1,0,0,0}$
  using the recursion relation on the Jacobi polynomials, i.e.
  \begin{equation}
  \begin{aligned}
    {\cal Y}^{K+2LM,p}_{\lambda,n_2+1,0,0,0}&=
    (1-y_{5p})(1-y_{4p})(1-y_{3p})(a_{n_2}+b_{n_2}y_{2p})
    {\cal Y}^{KLM,p}_{\lambda,n_2,0,0,0}\\
    &+c_{n_2}(1-y_{5p})^2(1-y_{4p})^2(1-y_{3p})^2
    {\cal Y}^{K-2LM,p}_{\lambda,n_2-1,0,0,0}\,.
    \label{eq:n20}
  \end{aligned}
  \end{equation}
  The first term of Eq.~(\ref{eq:n20}) can be expressed as
  \begin{equation}
  \begin{aligned}
    (1-y_{5p})&(1-y_{4p})(1-y_{3p})(a_{n_2}+b_{n_2}y_{2p})
    {\cal Y}^{KLM,p}_{\lambda,n_2,0,0,0}=\\
   & 8b_{n_2}\left(\frac{(x_{2p})^2}{\rho^2}
      -\frac{(x_{1p})^2}{\rho^2}\right)+
    8a_{n_2}\left(\frac{(x_{2p})^2}{\rho^2}+
      \frac{(x_{1p})^2}{\rho^2}\right){\cal Y}^{KLM,p}_{\lambda,n_2,0,0,0}
    +{\cal O}(K)\,,
  \end{aligned}
  \end{equation}
  where also in this case the ${\cal O}(K)$ terms do not give contributions
  in the integrals of Eq.~(\ref{eq:aa}).
  We can rewrite the factor $(1-y_{5p})^2(1-y_{4p})^2(1-y_{3p})^2$
  of the first term in Eq.~(\ref{eq:n20}) using the Jacobi polynomial of order
  0, 1, and 2. Here we report the final result, which reads
  \begin{equation}
  \begin{aligned}
    (1-y_{5p})^2&(1-y_{4p})^2(1-y_{3p})^2
    {\cal Y}^{K-2LM,p}_{\lambda,n_2,n_3-1,0,0}=
    A^{(2)}_{n_2-1}{\cal Y}^{K+2LM,p}_{\lambda,n_2-1,2,0,0}\\
    &+B^{(2)}_{n_2-1}\Big[\eta^{(3)}_{n_2-1,1}
      {\cal Y}^{K+2LM,p}_{\lambda,n_2-1,1,1,0}
      +\gamma^{(3)}_{n_2-1,1}\eta^{(4)}_{n_2-1,1,0}
      {\cal Y}^{K+2LM,p}_{\lambda,n_2-1,1,0,1}\Big]\\
      &+C^{(2)}_{n_2-1}\Big[
        A^{(3)}_{n_2,0}{\cal Y}^{K+2LM,p}_{\lambda,n_2-1,0,2,0}
        +B^{(3)}_{n_2-1,0}\eta^{(4)}_{n_2-1,0,1}
        {\cal Y}^{K+2LM,p}_{\lambda,n_2-1,0,1,1}\\
        &+C^{(3)}_{n_2-1,0}A^{(4)}_{n_2-1,0,0}
        {\cal Y}^{K+2LM,p}_{\lambda,n_2-1,0,0,2}\Big]
        +{\cal O}(K)
    \,,
  \end{aligned}
  \end{equation}
    where again the terms ${\cal O}(K)$ can be discarded in Eq.~(\ref{eq:aa}).
    Inserting these result in Eq.~(\ref{eq:aa}) we get the final expression
    for the recursion on $n_2$
    \begin{equation}
    \begin{aligned}\label{eq:n2f}
    &c^{K+2L,p}_{\lambda,n_2+1,0,0,0;\mu'}=
    c_{n_3}\Big[A^{(2)}_{n_2-1}c^{K+2L,p}_{\lambda,n_2,2,0,0;\mu'}
    +B^{(2)}_{n_2-1}\Big(\eta^{(3)}_{n_2-1,1}
      c^{K+2L,p}_{\lambda,n_2-1,1,1,0;\mu'}\\
      &\qquad\qquad+\gamma^{(3)}_{n_2-1,1}\eta^{(4)}_{n_2-1,1,0}
      c^{K+2L,p}_{\lambda,n_2-1,1,0,1;\mu'}\Big)
    +C^{(2)}_{n_2-1}\Big(A^{(3)}_{n_2-1,0}
    c^{K+2L,p}_{\lambda,n_2-1,0,2,0;\mu'}\\
    &\qquad\qquad+B^{(3)}_{n_2-1,0}\eta^{(4)}_{n_2-1,0,1}
    c^{K+2L,p}_{\lambda,n_2-1,0,1,1;\mu'}
    +C^{(3)}_{n_2-1,0}A^{(4)}_{n_2-1,0,0}
    c^{K+2L,p}_{\lambda,n_2-1,0,0,2;\mu'}\Big)\Big]\\
    &\qquad\qquad+\sum_{\mu''}c^{KL,p}_{\lambda,n_2,0,0,0;\mu''}
    \sum_{i,j}\big[8(b_{n_2}+a_{n_2})\Gamma_{i,j}^p(2)
      +8(b_{n_2}-a_{n_2})\Gamma_{i,j}^p(1)\big]
    I_{i,j}^{\mu',\mu''}.
    \end{aligned}
    \end{equation}

  Eqs.~(\ref{eq:n5f}), (\ref{eq:n4f}), (\ref{eq:n3f}) and~(\ref{eq:n2f})
  are the recurrence relations giving the coefficients at order $K+2$ in
  terms of grandangular momentum quantum number $K$.
  It is possible to notice that writing the recursion in this way there
  are no inconsistency.  Indeed, the coefficients
  $c^{K+2L,p}_{\lambda,n_2,n_3,n_4,n_5+1;\mu'}$ depend only on the
  coefficients with grandangular momentum $K$. The coefficients
  $c^{K+2L,p}_{\lambda,n_2,n_3,n_4+1,0;\mu'}$ depend on the coefficients
  at order $K$ and only on the order $K+2$ coefficients
  evaluated in the recursion on $n_5+1$, and so on for all the
  other recursions. Moreover, the four relations already involve
  only coefficients
  with the same quantum numbers $\lambda$. Therefore, the algorithm is well
  suited in our case, since we need to compute TC up to different $K$ for
  different values of $\lambda$.

  In order to complete the discussion we compute the starting
  case of the recursion
  by using an approach similar to the one presented in Ref.~\cite{Efros1995}.
  Fixed $L$ and the minimum
  valid value of $K$ for this $L$, we generate $N$ sets of Jacobi
  coordinates in a random way, where $N$ is the number of states $\mu$
  allowed by $K$ and $L$. For any choice $j$ of the Jacobi coordinates,
  the following relation must be verified
  \begin{equation}
    \sum_{\mu'}c^{KL,p}_{\mu;\mu'}{\cal Y}^{KLM}_{\mu'}
    (\Omega_5(j))={\cal Y}^{KLM}_{\mu}
    (\Omega_{5p}(j))\,,
  \end{equation}
  where the $c^{K,L,p}_{\mu,\mu'}$ coefficients are unknown. If we
  consider the $N$ sets of Jacobi coordinates, we have $N$ independent
  equations for $N$ unknowns which can be solved with standard
  algebraic techniques.
  In principle, this method can be used to find the transition coefficients
  at any order $K$, but the numerical precision of the calculation decrease
  rapidly as the order $K$ increases. We have used this method
  to compute the starting TC corresponding to
  $n_2=n_3=n_4=n_5=0$ so $K=\ell_1+\ell_2+\ell_3+\ell_4+\ell_5$. Since
  this value is always small, the number $N$ of HH functions is not critical.
  Moreover, we used this approach for selected cases of
  $K>\ell_1+\ell_2+\ell_3+\ell_4+\ell_5$ as benchmark
  for the recursion method in order to test the previous formulas and
  to debug the code.
  
  \begin{subsection}{The integrals $I_{i,j}^{\mu,\mu'}$}\label{app:int}
    In this section, we present the detailed calculation for the integrals
    $I_{i,j}^{\mu,\mu'}$ defined in Eq.~(\ref{eq:Iij}). The integrals
    $I_{i,j}^{\mu,\mu'}$  involve only functions constructed with the Jacobi
    polynomials in the permutation $p=1$.
    In the case $A=6$, there are 15 different terms $(\xx_i\cdot\xx_j)/\rho^2$,
    which have the following expression in terms of the hyperangular variables
    $\Omega_5\equiv\{\hxx_1,\hxx_2,\hxx_3,\hxx_4,\hxx_5,y_2,y_3,y_4,y_5\}$:
      \begin{align}
     \frac{x_1^2}{\rho^2}&=\frac{(1-y_5)(1-y_4)(1-y_3)(1-y_2)}{16}\,,\\
     \frac{\xx_1\cdot\xx_2}{\rho^2}
    &=-\frac{4\pi}{\sqrt{3}}\frac{(1-y_5)(1-y_4)(1-y_3)\sqrt{(1-y_2^2)}}{16}
    \left[Y_1(\hxx_1)Y_1(\hxx_2)\right]_{00}\,,\\
    \frac{\xx_1\cdot\xx_3}{\rho^2}
    &=-\frac{4\pi}{\sqrt{3}}\frac{(1-y_5)(1-y_4)\sqrt{(1-y_3^2)(1-y_2)}}{8\sqrt{2}}
    \left[Y_1(\hxx_1)Y_1(\hxx_3)\right]_{00}\,,\\
    \frac{\xx_1\cdot\xx_4}{\rho^2}
    &=-\frac{4\pi}{\sqrt{3}}\frac{(1-y_5)\sqrt{(1-y_4^2)(1-y_3)(1-y_2)}}{8}
    \left[Y_1(\hxx_1)Y_1(\hxx_4)\right]_{00}\,,\\
   \frac{\xx_1\cdot\xx_5}{\rho^2}
    &=-\frac{4\pi}{\sqrt{3}}\frac{\sqrt{(1-y_5^2)(1-y_4)(1-y_3)(1-y_2)}}{4\sqrt{2}}
    \left[Y_1(\hxx_1)Y_1(\hxx_5)\right]_{00}\,,\\
    \frac{x_2^2}{\rho^2}&=\frac{(1-y_5)(1-y_4)(1-y_3)(1+y_2)}{16}\,,\\
     \frac{\xx_2\cdot\xx_3}{\rho^2}
    &=-\frac{4\pi}{\sqrt{3}}\frac{(1-y_5)(1-y_4)\sqrt{(1-y_3^2)(1+y_2)}}{8\sqrt{2}}
    \left[Y_1(\hxx_2)Y_1(\hxx_3)\right]_{00}\,,\\
    \frac{\xx_2\cdot\xx_4}{\rho^2}
    &=-\frac{4\pi}{\sqrt{3}}\frac{(1-y_5)\sqrt{(1-y_4^2)(1-y_3)(1+y_2)}}{8}
    \left[Y_1(\hxx_2)Y_1(\hxx_4)\right]_{00}\,,\\
    \frac{\xx_2\cdot\xx_4}{\rho^2}
    &=-\frac{4\pi}{\sqrt{3}}\frac{\sqrt{(1-y_5^2)(1-y_4)(1-y_3)(1+y_2)}}{8}
    \left[Y_1(\hxx_2)Y_1(\hxx_5)\right]_{00}\,,\\
    \frac{x_3^2}{\rho^2}&=\frac{(1-y_5)(1-y_4)(1+y_3)}{8}\,,\\
    \frac{\xx_3\cdot\xx_4}{\rho^2}
    &=-\frac{4\pi}{\sqrt{3}}\frac{(1-y_5)\sqrt{(1-y_4^2)(1+y_3)}}{4\sqrt{2}}
    \left[Y_1(\hxx_3)Y_1(\hxx_4)\right]_{00}\,,\\
    \frac{\xx_3\cdot\xx_5}{\rho^2}
    &=-\frac{4\pi}{\sqrt{3}}\frac{\sqrt{(1-y_5^2)(1-y_4)(1+y_3)}}{4}
    \left[Y_1(\hxx_3)Y_1(\hxx_5)\right]_{00}\,,\\
    \frac{x_4^2}{\rho^2}&=\frac{(1-y_5)(1+y_4)}{4}\,,\\
    \frac{\xx_4\cdot\xx_5}{\rho^2}
    &=-\frac{4\pi}{\sqrt{3}}\frac{\sqrt{(1-y_5^2)(1+y_4)}}{2\sqrt{2}}
    \left[Y_1(\hxx_4)Y_1(\hxx_5)\right]_{00}\,,\\
    \frac{x_5^2}{\rho^2}&=\frac{(1+y_5)}{2}\,.
  \end{align}
      By using these relations the integrals of Eq.~(\ref{eq:Iij}) can be
      factorized in products of five integrals of the kind
      \begin{equation}
        {\cal J}_j\left[\ell_j',m_j',\ell,m,\ell''_j,m''_j\right]=
        \int d\hat{x}_j\,\left[Y_{\ell'_jm'_j}(\hxx_j)\right]^*
        Y_{\ell m}(\hxx_j)Y_{\ell''_jm''_j}(\hxx_j)\,,
      \end{equation}
      where $j$ goes from 1 to 5 and $\ell=0$ or 1, and four integrals of the
      kind
      \begin{equation}
        \cali_j(2a,2b)=\int_{-1}^{1} dy_j (1-y_j)^a(1+y_j)^b
        P_{n_j}^{\nu_j,\ell_j+\frac{1}{2}}(y_j)
        P_{n_j'}^{\nu_j',\ell_j'+\frac{1}{2}}(y_j)\,,
      \end{equation}
      where $j$ goes from 2 to 5.

      The integrals type ${\cal J}_j$ can be easily calculated in terms of
      Wigner 3j coefficients. The integrals $\cali_j$ can be evaluated
      by using Gauss integration formula. Let us define the polynomial
      \begin{equation}
        P_N(x)=(1-x)^{\text{int}(a)}(1+x)^{\text{int}(b)}
        P_{n}^{\nu,l_+\frac{1}{2}}(x)
        P_{n'}^{\nu',l'+\frac{1}{2}}(x)\,,
      \end{equation}
      where $\text{int}(a)$ stands for integer part of $a$ and $N=\text{int}(a)
      +\text{int}(b)+n+n'$ is the degree of the polynomial.
      We can distinguish four case:
      \begin{itemize}
      \item $a$ and $b$ integers. In this case the integrand polynomial
        has degree $N=a+b+n+n'$, which can be evaluated using Gauss-Legendre
        quadrature, i.e.
        \begin{equation}
          \cali_j(2a,2b)=\int_{-1}^{1}dx P_N(x)=\sum_{i=1}^{N_G}\omega_i^{(N_G)}
          P_N(x_i^{(N_G)})\,
        \end{equation}
        where $\omega_i^{(N_G)}$ and $x_i^{(N_G)}$ are the
        Gauss-Legendre weights and points.
      \item $a$ half-integer and $b$ integer. We can write the integral as
        \begin{equation}
          \cali_j(2a,2b)=\int_{-1}^{1}dx \sqrt{1-x}P_N(x)
          =2^{\frac{3}{2}}\sum_{i=1}^{2{N_G}+1}
          \tilde{\omega}_i^{(2{N_G}+1)}P_N(2\tilde{x}_i^{(2{N_G}+1)}-1)\,,
        \end{equation}
        where
        \begin{align}
          \tilde{\omega}_i^{(2{N_G}+1)}&=2\epsilon_i^2\omega_i^{(2{N_G}+1)}\,
          ,\label{eq:wtilde}\\ 
          \tilde{x}_i^{(2{N_G}+1)}&=1-\epsilon_i^2\,,\label{eq:xtilde} 
        \end{align}
        $\omega_i^{(2{N_G}+1)}$ being the $2{N_G}+1$ Gauss-Legendre weights
        and $\epsilon_i$ the zeros of the $P_{2{N_G}+1}(x)$
        Legendre polynomials.
      \item $a$ integer and $b$ half-integer. We can write the integral as
        \begin{equation}
          \cali_j(2a,2b)=\int_{-1}^{1}dx \sqrt{1+x}P_N(x)
          =2^{\frac{3}{2}}\sum_{i=1}^{2{N_G}+1}
          \tilde{\omega}_i^{(2{N_G}+1)}P_{N}(1-2\tilde{x}_i^{(2{N_G}+1)})\,,
        \end{equation}
        where $\tilde{\omega}_i^{(2{N_G}+1)}$
        and $\tilde{x}_i^{(2{N_G}+1)}$ are defined in
        Eqs.~(\ref{eq:wtilde}) and~(\ref{eq:xtilde}).
      \item $a$ half-integer and $b$ half-integer. We can write the integral as
        \begin{equation}
          \cali_j(2a,2b)=\int_{-1}^{1}dx \sqrt{1-x^2}P_N(x)=\sum_{i=1}^{{N_G}}
          \omega_i^{(N_G)}P_N(x_i^{{(N_G)}})\,,
        \end{equation}
        where $\omega_i^{(N_G)}$ and $x_i^{(N_G)}$ are the Gauss-Tchebishev
        weights and points.
  \end{itemize}
      The Gauss abscissa and coefficients have the property to give exact results
      when $N_G>N=\text{int}(a)+\text{int}(b)+n+n'$. The precision is limited
      only by the internal precision of the computer.

      The final expression of the integrals $I_{i,j}^{\mu,\mu'}$ in terms of
      ${\cal I}_j$ and the Wigner 3j coefficients are
        \begin{align}
    I_{1,1}^{\mu,\mu'}&=\frac{N_{21}}{16}
    {\cal N}_{\mu}{\cal N}_{\mu'}
    \cali_2(3+K_1+K_1',3+\ell_2+\ell_2')
    \,\cali_3(6+K_2+K_2',1+\ell_3+\ell_3')\nonumber\\
    &\times\cali_4(9+K_3+K_3',1+\ell_4+\ell_4')\,
    \cali_5(12+K_4+K_4',1+\ell_5+\ell_5')\nonumber\\
    &\times\prod_{i=1}^5\delta_{\ell_i\ell_i'}
    \prod_{i=2}^4\delta_{L_iL_i'}\,,\\
    I_{1,2}^{\mu,\mu'}&=\frac{N_{21}}{16}{\cal N}_{\mu}{\cal N}_{\mu'}
    \cali_2(2+K_1+K_1',2+\ell_2+\ell_2')\,
    \cali_3(6+K_2+K_2',1+\ell_3+\ell_3')\nonumber\\
    &\times\cali_4(9+K_3+K_3',1+\ell_4+\ell_4')\,
    \cali_5(12+K_4+K_4',1+\ell_5+\ell_5')\nonumber\\
    &\times(-1)^{1+L_2}\hat{l}_1\hat{\ell}_1'\hat{\ell}_2\hat{\ell}_2'
    \begin{Bmatrix}
     \ell_1' & \ell_1 & 1 \\
     \ell_2  & \ell_2'& L_2
   \end{Bmatrix}
    \begin{pmatrix}
     1 & \ell_1'& \ell_1 \\
     0 & 0   & 0
   \end{pmatrix}
    \begin{pmatrix}
     1 & \ell_2'& \ell_2 \\
     0 & 0   & 0
   \end{pmatrix}\nonumber\\
    &\times\delta_{\ell_3\ell_3'}\delta_{\ell_4\ell_4'}\delta_{\ell_5\ell_5'}
    \prod_{i=2}^4\delta_{L_iL_i'}\,,\\
    I_{1,3}^{\mu,\mu'}&=\frac{N_{21}}{8\sqrt{2}}
    {\cal N}_{\mu}{\cal N}_{\mu'}
    \cali_2(2+K_1+K_1',1+\ell_2+\ell_2')\,
    \cali_3(5+K_2+K_2',2+\ell_3+\ell_3')\nonumber\\
    &\times\cali_4(9+K_3+K_3',1+\ell_4+\ell_4')\,
    \cali_5(12+K_4+K_4',1+\ell_5+\ell_5')\nonumber\\
    &\times(-1)^{1+L_3+l_2}
    \hat{\ell}_1\hat{\ell}_1'\hat{\ell}_3\hat{\ell}_3'\hat{L}_2\hat{L}_2'\nonumber\\
    &\times
    \begin{Bmatrix}
     \ell_3 & \ell_3' & 1 \\
     L_2'& L_2  & L_3
   \end{Bmatrix}
    \begin{Bmatrix}
     L_2 & L_2' & 1 \\
     \ell_1'& \ell_1  & \ell_2
   \end{Bmatrix}
    \begin{pmatrix}
     1 & \ell_1'& \ell_1 \\
     0 & 0   & 0
   \end{pmatrix}
    \begin{pmatrix}
     1 & \ell_3'& \ell_3 \\
     0 & 0   & 0
   \end{pmatrix}\nonumber\\
    &\times\delta_{\ell_2\ell_2'}\delta_{\ell_4\ell_4'}\delta_{\ell_5\ell_5'}
    \delta_{L_3L_3'}\delta_{L_4L_4'}\,,\\
    I_{1,4}^{\mu,\mu'}&=\frac{N_{21}}{8}{\cal N}_{\mu}{\cal N}_{\mu'}
    \cali_2(2+K_1+K_1',1+\ell_2+\ell_2')\,
    \cali_3(5+K_2+K_2',1+\ell_3+\ell_3')\nonumber\\
    &\times\cali_4(8+K_3+K_3',2+\ell_4+\ell_4')\,
    \cali_5(12+K_4+K_4',1+\ell_5+\ell_5')\nonumber\\
    &\times(-1)^{1+L_4+L_2'+L_2+\ell_3+\ell_2}
    \hat{\ell}_1\hat{\ell}_1'\hat{\ell}_4\hat{\ell}_4'\hat{L}_2\hat{L}_2'
    \hat{L}_3\hat{L}_3'\nonumber\\
    &\times
    \begin{Bmatrix}
     \ell_4 & \ell_4' & 1 \\
     L_3'& L_3  & L_4
   \end{Bmatrix}
    \begin{Bmatrix}
     L_3 & L_3' & 1 \\
     L_2'& L_2  & \ell_3
   \end{Bmatrix}
    \begin{Bmatrix}
     L_2 & L_2' & 1 \\
     \ell_1'& \ell_1  & \ell_2
   \end{Bmatrix}\nonumber\\
    &\times    \begin{pmatrix}
     1 & \ell_1'& \ell_1 \\
     0 & 0   & 0
   \end{pmatrix}
    \begin{pmatrix}
     1 & \ell_4'& \ell_4 \\
     0 & 0   & 0
   \end{pmatrix}
    \delta_{\ell_2\ell_2'}\delta_{\ell_3\ell_3'}\delta_{\ell_5\ell_5'}
    \delta_{L_4L_4'}\,,\\
    I_{1,5}^{\mu,\mu'}&=\frac{N_{21}}{4\sqrt{2}}
    {\cal N}_{\mu}{\cal N}_{\mu'}
    \cali_2(2+K_1+K_1',1+\ell_2+\ell_2')\,
    \cali_3(5+K_2+K_2',1+\ell_3+\ell_3')\nonumber\\
    &\times\cali_4(8+K_3+K_3',1+\ell_4+\ell_4')\,
    \cali_5(11+K_4+K_4',2+\ell_5+\ell_5')\nonumber\\
    &\times(-1)^{1+L+L_3'+L_3+L_2'+L_2+\ell_4+\ell_3+\ell_2}
    \hat{\ell}_1\hat{\ell}_1'\hat{\ell}_5\hat{\ell}_5'\hat{L}_2\hat{L}_2'
    \hat{L}_3\hat{L}_3'\hat{L}_4\hat{L}_4'\nonumber\\
    &\times
    \begin{Bmatrix}
     \ell_5 & \ell_5' & 1 \\
     L_4'& L_4  & L
   \end{Bmatrix}
    \begin{Bmatrix}
     L_4 & L_4' & 1 \\
     L_3'& L_3  & \ell_4
   \end{Bmatrix}
    \begin{Bmatrix}
     L_3 & L_3' & 1 \\
     L_2'& L_2  & \ell_3
   \end{Bmatrix}
    \begin{Bmatrix}
     L_2 & L_2' & 1 \\
     \ell_1'& \ell_1  & \ell_2
   \end{Bmatrix}\nonumber\\
    &\times    \begin{pmatrix}
     1 & \ell_1'& \ell_1 \\
     0 & 0   & 0
   \end{pmatrix}
    \begin{pmatrix}
     1 & \ell_5'& \ell_5 \\
     0 & 0   & 0
   \end{pmatrix}
    \delta_{\ell_2\ell_2'}\delta_{\ell_3\ell_3'}\delta_{\ell_4\ell_4'}
    \,,\\
    I_{2,2}^{\mu,\mu'}&=\frac{N_{21}}{16}
    {\cal N}_{\mu}{\cal N}_{\mu'}
    \cali_2(1+K_1+K_1',3+\ell_2+\ell_2')
    \,\cali_3(6+K_2+K_2',1+\ell_3+\ell_3')\nonumber\\
    &\times\cali_4(9+K_3+K_3',1+\ell_4+\ell_4')\,
    \cali_5(12+K_4+K_4',1+\ell_5+\ell_5')\nonumber\\
    &\times\prod_{i=1}^5\delta_{\ell_i\ell_i'}
    \prod_{i=2}^4\delta_{L_iL_i'}\,,\\
    I_{2,3}^{\mu,\mu'}&=\frac{N_{21}}{8\sqrt{2}}
    {\cal N}_{\mu}{\cal N}_{\mu'}
    \cali_2(1+K_1+K_1',2+\ell_2+\ell_2')\,
    \cali_3(5+K_2+K_2',2+\ell_3+\ell_3')\nonumber\\
    &\times\cali_4(9+K_3+K_3',1+\ell_4+\ell_4')\,
    \cali_5(12+K_4+K_4',1+\ell_5+\ell_5')\nonumber\\
    &\times(-1)^{1+L_3+L_2'+L_2+\ell_1}
    \hat{\ell}_2\hat{\ell}_2'\hat{\ell}_3\hat{\ell}_3'\hat{L}_2\hat{L}_2'
    \nonumber\\
    &\times
    \begin{Bmatrix}
     \ell_3 & \ell_3' & 1 \\
     L_2'& L_2  & L_3
   \end{Bmatrix}
    \begin{Bmatrix}
     L_2 & L_2' & 1 \\
     \ell_2'& \ell_2  & \ell_1
   \end{Bmatrix}
    \begin{pmatrix}
     1 & \ell_2'& \ell_2 \\
     0 & 0   & 0
   \end{pmatrix}
    \begin{pmatrix}
     1 & \ell_3'& \ell_3 \\
     0 & 0   & 0
   \end{pmatrix}\nonumber\\
    &\times
    \delta_{\ell_1\ell_1'}\delta_{\ell_4\ell_4'}\delta_{\ell_5\ell_5'}
    \delta_{L_3L_3'}\delta_{L_4L_4'}
    \,,\\
    I_{2,4}^{\mu,\mu'}&=\frac{N_{21}}{8}
    {\cal N}_{\mu}{\cal N}_{\mu'}
    \cali_2(1+K_1+K_1',2+\ell_2+\ell_2')\,
    \cali_3(5+K_2+K_2',1+\ell_3+\ell_3')\nonumber\\
    &\times\cali_4(8+K_3+K_3',2+\ell_4+\ell_4')\,
    \cali_5(12+K_4+K_4',1+\ell_5+\ell_5')\nonumber\\
    &\times(-1)^{1+L_4+\ell_3+\ell_1}
    \hat{\ell}_2\hat{\ell}_2'\hat{\ell}_4\hat{\ell}_4'
    \hat{L}_2\hat{L}_2'\hat{L}_3\hat{L}_3'
    \nonumber\\
    &\times
    \begin{Bmatrix}
     \ell_4 & \ell_4' & 1 \\
     L_3'& L_3  & L_4
   \end{Bmatrix}
    \begin{Bmatrix}
     L_3 & L_3' & 1 \\
     L_2'& L_2  & \ell_3
   \end{Bmatrix}
    \begin{Bmatrix}
     L_2 & L_2' & 1 \\
     \ell_2'& \ell_2  & \ell_1
   \end{Bmatrix}\nonumber\\
    &\times\begin{pmatrix}
     1 & \ell_2'& \ell_2 \\
     0 & 0   & 0
   \end{pmatrix}
    \begin{pmatrix}
     1 & \ell_4'& \ell_4 \\
     0 & 0   & 0
   \end{pmatrix}
    \delta_{\ell_1\ell_1'}\delta_{\ell_3\ell_3'}\delta_{\ell_5\ell_5'}
    \delta_{L_4L_4'}
    \,,\\
    I_{2,5}^{\mu,\mu'}&=\frac{N_{21}}{4\sqrt{2}}
    {\cal N}_{\mu}{\cal N}_{\mu'}
    \cali_2(1+K_1+K_1',2+\ell_2+\ell_2')\,
    \cali_3(5+K_2+K_2',1+\ell_3+\ell_3')\nonumber\\
    &\times\cali_4(8+K_3+K_3',1+\ell_4+\ell_4')\,
    \cali_5(11+K_4+K_4',2+\ell_5+\ell_5')\nonumber\\
    &\times(-1)^{L+L_3+L_3'+\ell_4+\ell_3+\ell_1}
    \hat{\ell}_2\hat{\ell}_2'\hat{\ell}_5\hat{\ell}_5'
    \hat{L}_2\hat{L}_2'\hat{L}_3\hat{L}_3'\hat{L}_4\hat{L}_4'
    \nonumber\\
    &\times
    \begin{Bmatrix}
     \ell_5 & \ell_5' & 1 \\
     L_4'& L_4  & L
   \end{Bmatrix}
    \begin{Bmatrix}
     L_4 & L_4' & 1 \\
     L_3'& L_3  & \ell_4
   \end{Bmatrix}
    \begin{Bmatrix}
     L_3 & L_3' & 1 \\
     L_2'& L_2  & \ell_3
   \end{Bmatrix}
    \begin{Bmatrix}
     L_2 & L_2' & 1 \\
     \ell_2'& \ell_2  & \ell_1
   \end{Bmatrix}\nonumber\\
    &\times\begin{pmatrix}
     1 & \ell_2'& \ell_2 \\
     0 & 0   & 0
   \end{pmatrix}
    \begin{pmatrix}
     1 & \ell_5'& \ell_5 \\
     0 & 0   & 0
   \end{pmatrix}
    \delta_{\ell_1\ell_1'}\delta_{\ell_3\ell_3'}\delta_{\ell_4\ell_4'}
    \,,\\
    I_{3,3}^{\mu,\mu'}&=\frac{N_{21}}{8}
    {\cal N}_{\mu}{\cal N}_{\mu'}
    \cali_2(1+K_1+K_1',1+\ell_2+\ell_2')
    \,\cali_3(4+K_2+K_2',3+\ell_3+\ell_3')\nonumber\\
    &\times\cali_4(9+K_3+K_3',1+\ell_4+\ell_4')\,
    \cali_5(12+K_4+K_4',1+\ell_5+\ell_5')\nonumber\\
    &\times\prod_{i=1}^5\delta_{\ell_i\ell_i'}
    \prod_{i=2}^4\delta_{L_iL_i'}\,,\\
    I_{3,4}^{\mu,\mu'}&=\frac{N_{21}}{4\sqrt{2}}
    {\cal N}_{\mu}{\cal N}_{\mu'}
    \cali_2(1+K_1+K_1',1+\ell_2+\ell_2')\,
    \cali_3(4+K_2+K_2',2+\ell_3+\ell_3')\nonumber\\
    &\times\cali_4(8+K_3+K_3',2+\ell_4+\ell_4')\,
    \cali_5(12+K_4+K_4',1+\ell_5+\ell_5')\nonumber\\
    &\times(-1)^{L_4+L_3+L_3'+L_2}
    \hat{\ell}_3\hat{\ell}_3'\hat{\ell}_4\hat{\ell}_4'
    \hat{L}_3\hat{L}_3'
    \nonumber\\
    &\times
    \begin{Bmatrix}
     \ell_4 & \ell_4' & 1 \\
     L_3'& L_3  & L_4
   \end{Bmatrix}
    \begin{Bmatrix}
     L_3 & L_3' & 1 \\
     \ell_3'& \ell_3  & L_2
   \end{Bmatrix}
   \begin{pmatrix}
     1 & l_3'& \ell_3 \\
     0 & 0   & 0
   \end{pmatrix}
    \begin{pmatrix}
     1 & \ell_4'& \ell_4 \\
     0 & 0   & 0
    \end{pmatrix}\nonumber\\
    &\times
    \delta_{\ell_1\ell_1'}\delta_{\ell_2\ell_2'}\delta_{\ell_4\ell_4'}
    \delta_{L_2L_2'}\delta_{L_4L_4'}
    \,,\\
    I_{3,5}^{\mu,\mu'}&=\frac{N_{21}}{4}
    {\cal N}_{\mu}{\cal N}_{\mu'}
    \cali_2(1+K_1+K_1',1+\ell_2+\ell_2')\,
    \cali_3(4+K_2+K_2',2+\ell_3+\ell_3')\nonumber\\
    &\times\cali_4(8+K_3+K_3',1+\ell_4+\ell_4')\,
    \cali_5(11+K_4+K_4',2+\ell_5+\ell_5')\nonumber\\
    &\times(-1)^{1+L+L_2+\ell_4}
    \hat{\ell}_3\hat{\ell}_3'\hat{\ell}_5\hat{\ell}_5'
    \hat{L}_3\hat{L}_3'\hat{L}_4\hat{L}_4'
    \nonumber\\
    &\times
    \begin{Bmatrix}
     \ell_5 & \ell_5' & 1 \\
     L_4'& L_4  & L
   \end{Bmatrix}
    \begin{Bmatrix}
     L_4 & L_4' & 1 \\
     L_3'& L_3  & \ell_4
   \end{Bmatrix}
    \begin{Bmatrix}
     L_3 & L_3' & 1 \\
     \ell_3'& \ell_3  & L_2
   \end{Bmatrix}
    \nonumber\\
    &\times
    \begin{pmatrix}
     1 & \ell_3'& \ell_3 \\
     0 & 0   & 0
   \end{pmatrix}
    \begin{pmatrix}
     1 & \ell_5'& \ell_5 \\
     0 & 0   & 0
    \end{pmatrix}
    \delta_{\ell_1\ell_1'}\delta_{\ell_2\ell_2'}\delta_{\ell_4\ell_4'}
    \delta_{L_2L_2'}
    \,,\\
    I_{4,4}^{\mu,\mu'}&=\frac{N_{21}}{4}
    {\cal N}_{\mu}{\cal N}_{\mu'}
    \cali_2(1+K_1+K_1',1+\ell_2+\ell_2')\,
    \cali_3(4+K_2+K_2',1+\ell_3+\ell_3')\nonumber\\
    &\times\cali_4(8+K_3+K_3',3+\ell_4+\ell_4')\,
    \cali_5(12+K_4+K_4',1+\ell_5+\ell_5')\nonumber\\
    &\times\prod_{i=1}^5\delta_{\ell_i\ell_i'}
    \prod_{i=2}^4\delta_{L_iL_i'}\,,\\
    I_{4,5}^{\mu,\mu'}&=\frac{N_{21}}{2\sqrt{2}}
    {\cal N}_{\mu}{\cal N}_{\mu'}
    \cali_2(1+K_1+K_1',1+\ell_2+\ell_2')\,
    \cali_3(4+K_2+K_2',1+\ell_3+\ell_3')\nonumber\\
    &\times\cali_4(7+K_3+K_3',2+\ell_4+\ell_4')\,
    \cali_5(11+K_4+K_4',2+\ell_5+\ell_5')\nonumber\\
    &\times(-1)^{L+L_4+L_4'+L_3}
    \hat{\ell}_4\hat{\ell}_4'\hat{\ell}_5\hat{\ell}_5'
    \hat{L}_4\hat{L}_4'
    \nonumber\\
    &\times
    \begin{Bmatrix}
     \ell_5 & \ell_5' & 1 \\
     L_4'& L_4  & L
   \end{Bmatrix}
    \begin{Bmatrix}
     L_4 & L_4' & 1 \\
     \ell_4'& \ell_4  & L_3
   \end{Bmatrix}
    \begin{pmatrix}
     1 & \ell_4'& \ell_4 \\
     0 & 0   & 0
   \end{pmatrix}
    \begin{pmatrix}
     1 & \ell_5'& \ell_5 \\
     0 & 0   & 0
    \end{pmatrix}\nonumber\\
    &\times
    \delta_{\ell_1\ell_1'}\delta_{\ell_2\ell_2'}\delta_{\ell_3\ell_3'}
    \delta_{L_2L_2'}\delta_{L_3L_3'}
    \,,\\
    I_{5,5}^{\mu,\mu'}&=\frac{N_{21}}{2}
    {\cal N}_{\mu}{\cal N}_{\mu'}
    \cali_2(1+K_1+K_1',1+\ell_2+\ell_2')\,
    \cali_3(4+K_2+K_2',1+\ell_3+\ell_3')\nonumber\\
    &\times\cali_4(7+K_3+K_3',1+\ell_4+\ell_4')\,
    \cali_5(10+K_4+K_4',3+\ell_5+\ell_5')\nonumber\\
    &\times\prod_{i=1}^5\delta_{\ell_i\ell_i'}
    \prod_{i=2}^4\delta_{L_iL_i'}\,,
  \end{align}
        where $N_{21}=1/2^{21}$ and
        \begin{equation}
          {\cal N}_{\mu}=\prod_{j=2}^5{\cal N}_{n_j,K_{j-1}}^{\ell_j,\nu_j}\,.
        \end{equation}  

  \end{subsection}

    \begin{subsection}{Details of the implementation}
    In this section we present some technical details of the code used
    for computing the TC. The code was written in FORTRAN90
    using quadrupole precision.
    The first part of the code computes the integrals
    $I_{i,j}^{\mu,\mu'}$ and stores them in files.
    If we define $N_{K,L}$ as the number of states $\mu$ for a give value
    of $K$ and $L$, the number of integrals to be calculated at
    each step of the recursion is
    \begin{equation}
      N_I=15N_{K+2,L}N_{K,L}\,.
    \end{equation}
    Due to the $\delta$-functions and the Wigner 3j ``triangular conditions'', 
    many of these integrals are vanishing. This 
    reduces drastically the number of integrals we actually need
    to compute.
    For example, for the case $L=2$ and $K=12$, the number of integrals
    is $N_I\approx 8.3\times10^{10}$. However the number of non-zero integrals
    is  $10,099,724$ which are 4 order of magnitude less.
    Also the amount of space needed to save the integrals on the disk
    is enormously reduced:
    the total space occupied by the integrals used in this work,
    saved in quadrupole precision, is only 3.7 GB. Moreover,
    the explicit implementation of the vanishing condition combined
    with the parallelization of the calculation on the states $\mu'$ reduced
    drastically the computing time. For example, in   
    the case of $L=2$ and $K=12$, the calculation
    of the integrals takes $\sim15$ minutes
    on a 48-core CPU with 2.10 GHz.

    In the second part of the code
    we select a state $\lambda$ [see Eq.~(\ref{eq:lambda})]
    and  the recursion procedure can start
    following the algorithm showed in Section~\ref{app:tc2}.
    The calculation of the TC
    implies two main problems we need to consider:
    (i) the high number of permutations (360) increase dramatically
    the time required for the computation, and
    (ii) at each step
    we need to store all the coefficients $c^{KL,p}_{\mu;\mu'}$
    needed then for the case $K+2$. 
    The first problem was easily overcome by parallelizing on the permutations
    which are independent from each other. The second
    problem is related to the memory needed for storing all the coefficients.
    If we want to compute the coefficients for the case $K+2$ for a given
    $\lambda$, we need to store the coefficients $c^{KL,p}_{\mu;\mu'}$ for all
    the combinations of $n_2,n_3,n_4,n_5$ and for all the  permutations.
    If we define $\overline{N}=(K-\ell_{\text{sum}})/2$, where $\ell_{\text{sum}}=
    \ell_1+\ell_2+\ell_3+\ell_4+\ell_5$, the number of combinations of
    $n_2,n_3,n_4,n_5$ is given by
    \begin{equation}\label{eq:Nn}
      N_n(K,\ell_{\text{sum}})=\frac{1}{6}(\overline{N}+1)
      (\overline{N}^2+5\overline{N}+6)\,.
    \end{equation}
    Multiplying this number for the 360 permutations, the number of
    coefficients $N_{K,L}$ of the given state $\mu$ and the 16 byte occupied by
    each coefficient, we have the total memory needed for the storage at
    each recursion step, namely
    \begin{equation}\label{eq:mem}
      M[\text{GB}]=16\times\frac{360\times N_{K,L}\times N_n(K,\ell_{\text{sum}})}
      {1024^3}\,.
    \end{equation}
    For example, if we want to compute the coefficients $c_\mu^{12\,,0,p}$ where
    the states $\mu$ have $\ell_{\text{sum}}=2$,
    we need the corresponding coefficients at $K=10$. Considering that
    $N_{10,2}=30,525$, using Eqs.~(\ref{eq:Nn}) and~(\ref{eq:mem}),
    the memory required for the storage is $\sim6$ GB which can be easily managed
    by standard supercomputers. The possibility of working only with
    the single core memory reduces enormously the computing time. 
    However, 
    as it can be seen from Eq.~(\ref{eq:Nn}),
    the memory requirement is proportional to $N_{K,L}$.
    Therefore it grows exponentially
    for increasing $K$. Hence,
    for future applications in which we plan to increase $K$, we will be
    obliged to store the coefficients on the disk reducing drastically
    the performances of the code, or implement more sophisticated codes,
    using a distributed memory over different cores.

    As regarding the precision, the main source of error was found to be
    in the zero case of the recursion. All the other loss of precision are
    due to the round-off of the computer.
    In order to have a run-time check on the precision, we use the orthonormal
    properties of the HH functions which requires that
    \begin{equation}
      C^p_{\mu}=\sum_{\mu'}\left(c^{KL,p}_{\mu;\mu'}\right)^2=1\,.
    \end{equation}
    In particular we impose that
    \begin{equation}
      |C^p_{\mu}-1|<10^{-28}\,,
    \end{equation}
    condition  fulfilled by each state $\mu$ we have considered.
  \end{subsection}
  \end{section}
    
  \begin{section}{Orthogonalization procedure}\label{app:tc3}

    The sum over the permutation present in Eq.~(\ref{eq:tc2}) generates a
    strong redundancy that must be eliminated as already discussed in
    Chapter~\ref{ch:HH}. In this section we present the procedure we used to
    individuate the independent states.    
    The orthogonalization process is the real bottleneck
    of the construction of the 6-body wave function within the HH formalism.
    Let us suppose to have already defined a set of
    states that are linear independent
    and we are checking a new state. There are two
    possibilities: or the new state is completely dependent on the orthogonal
    basis or the new state has an orthogonal component to the basis
    we have calculated so far, and we have to include it.
    However, this orthogonal component can be very small, and therefore,
    very hard to identify.
    This means that it can happen very easily to not identify some orthogonal
    component due to the numerical round-off precision of the computer.
    This problem grows exponentially
    increasing the value of the grandangular momentum $K$,
    since smaller and smaller
    components must be identified.
    
    This problem forces us to improve the precision both from the
    algorithmic and the implementation point of view.
    As already discussed, we implemented the code
    in quadrupole precision (32 digits) in order to have less round-off errors.
    As regarding the algorithm, typically the Gram-Schmidt method
    is usually used to orthogonalization of a basis. In our case, we implement
    an algorithm using the {\it LINPACK} libraries which are based on
    Gaussian elimination. The algorithm works as follow:
    (i) selected an order of the states, 
    we keep as first state the first that has non-zero norm.
    (ii) Let us suppose we have already individuated $M$ independent states
    and let us define the matrix
    \begin{equation}
      N^{KLSTJ\pi}_M=
      \begin{pmatrix}
        N_{\al_1,\al_1}&\dots&N_{\al_1,\al_M}\\
        \vdots & \ddots &\vdots\\
        N_{\al_M,\al_1}&\dots&N_{\al_M,\al_M}
      \end{pmatrix}\,,
    \end{equation}
    where
    \begin{equation}
      N_{\al_i,\al_j}=\sum_{\al''}\left(A^{KLSTJ\pi}_{\alpha,\alpha''}\right)^*
      A^{KLSTJ\pi}_{\alpha'',\alpha'}\,.
    \end{equation}
    Considering now a new state
    $\alpha_{M+1}$, let us compute the norms
    $N_{\al_{M+1},\al_i}$ with all the $M$
    independent states $\alpha_i$.
    (iii) The new norm matrix of dimensions $(M+1)\times (M+1)$ is then
    \begin{equation}\label{eq:mxnorm2}
      N^{KLSTJ\pi}_{M+1}=
      \begin{pmatrix}
        N_{\al_1,\al_1}&\dots&N_{\al_1,\al_M}&N_{\al_1,\al_{M+1}}\\
        \vdots & \ddots &\vdots&\vdots\\
        N_{\al_M,\al_1}&\dots&N_{\al_M,\al_M}&N_{\al_{M},\al_{M+1}}\\
        N_{\al_{M+1},\al_1}&\dots&N_{\al_{M+1},\al_M}&N_{\al_{M+1},\al_{M+1}}\,,
      \end{pmatrix}
    \end{equation}
    which is inverted using  {\it LINPACK} libraries, modified 
    to  work in quadrupole precision
    and in parallel.
    From this we get the condition number $R$ which tells us if the
    matrix can be inverted or not.
    (iv) If $R>R_{min}$, where $R_{min}=10^{-30}$,
    the matrix is invertible and so we put the new state in the independent
    basis, in the other case the matrix is not invertible and we discard it.
    We select $R_{min}=10^{-30}$
    in order to be conservative, compared to the precision of the computer
    (i.e. $10^{-32}$).
    The procedure is then repeated for all the states of given $KLSTJ\pi$.
    This approach permits to individuate very small orthogonal components when we
    add new states. For example for $K=12$ and $LSTJ\pi=0101+$,
    we were able to find independent states up to value of $R\sim 10^{-18}$,
    which would be impossible in double precision.

    In the implementation of the orthogonalization
    procedure it is important to consider the time consumption,
    due to the large amount of states we need to test.
    Moreover, the parallelization results not trivial because
    the algorithm can check only one state at a time.
    Indeed  at the end, this part
    results to be the most inefficient part of the code.
    For example, if we consider $L=0$ and a given $\lambda$, the
    time needed for computing all the TC for all the states
    $\alpha$  up to $K=12$ results to be
    $\sim 20$ minutes without the orthogonalization procedure.
    On the other hand,  if we want to identify the independent states through
    the orthogonalization procedure, the time spent
    is more than 10 hours. Much of this time is
    used to invert the norm matrix. In fact, while 
    much of the computation time of the norm matrix can be saved by storing
    the norm matrix element, the inversion of the norm matrix given in
    Eq.~(\ref{eq:mxnorm2}) has to be performed  every time.
    Therefore, only a preliminary identification
    of the orthogonal states can really improve the algorithm
    efficiency. On the other hand, this orthogonalization has to be done
    only once.

  \end{section}

%% file: trans_HH.tex
\chapter{Transformation of the HH states}\label{app:transform}
The calculation of some observables can be performed more
easily and precisely by changing the definitions of the Jacobi
vectors and the hypercoordinates compared to the standard ones
defined in Eqs.~(\ref{eq:jacvec}) and~(\ref{eq:phiang}).
This implies that we need to
express the HH-spin-isospin basis functions defined in Eq.~(\ref{eq:hhst0})
according to the
new coordinate definition. This transformation can be done by using
particular TC, which we discuss in this appendix.

From the computational point of view,
the most efficient way to perform this transformation is to 
compute separately the TC for the spatial, spin, and isospin parts for each permutation,
and then sum over the permutations, as already discussed in Appendix~\ref{app:tc}.
The transformation we are considering involves only the spatial coordinates, while
the spin-isospin states are the same as considered in Appendix~\ref{app:tc}, so for spin
and isospin we can proceed as already discussed.
The only part we will modify is the spatial part for which
we define new coefficients
\begin{equation}
  {\cal Y}^{KLM}_{\mu}(\Omega_{5p})
  =\sum_{\mu'}c^{KL,p\rightarrow X}_{\mu;\mu'}
  {\cal Y}^{KLM}_{\mu'}(\Omega_X)\,,\label{eq:aaX}
\end{equation}
where $\Omega_{5p}$ represents a generic permutation of the standard Jacobi
coordinates, while $\Omega_X$ represents a generic definition of the Jacobi
vectors in the reference permutation and $c^{KL,p\rightarrow X}_{\mu;\mu'}$ are the
coefficients which permit to transform the HH states defined in the
Jacobi coordinates $\Omega_{5p}$ to the ones defined in the coordinates $\Omega_X$.
It is clear that only small modification of the algorithm presented in
Appendix~\ref{app:tc} are really needed. In this Appendix we will discuss only this
small modifications for the case of the change of variables for the calculation
of the quadrupole moment and for the scattering case.

\section{From the standard variables to ``X'' variables}

In this Section we will discuss the calculation of the coefficients
$c^{KL,p\rightarrow X}_{\mu;\mu'}$  defined in Eq.~(\ref{eq:aaX}),
where the ``X'' stands for the set ``Q'' in the case of the quadrupole moment
[see Eqs.~(\ref{eq:xxq_def}] and~(\ref{eq:phiangq})] or the set ``B'' in the case of the scattering states
[see Eqs.~(\ref{eq:jacvecB}) and~(\ref{eq:phib})].
By following Appendix~\ref{app:tc}, the coefficients are obtained by using 
\begin{equation}
  c^{K,L,p\rightarrow X}_{\mu,\mu'}=\int d\Omega_X
  \left[{\cal Y}^{K,L,M}_{\mu}(\Omega_X)\right]^\dag
       {\cal Y}^{K,L,M}_{\mu'}(\Omega_{5p})\,,
\end{equation}
from which it is possible to derive the recursion relations. They turn out
to be exactly identical to the ones derived in Appendix~\ref{app:tc},  because we
are applying the recursion formula of the Jacobi polynomials. The only changes are:
\begin{itemize}
\item[1)]the integration must be done over the ``X'' variables,
  therefore the integrals $I^{\mu,\mu'}_{i,j}$ result
\begin{equation}\label{eq:IijQ}
  I_{i,j}^{\mu,\mu'}=\int d\Omega_X
  \left[{\cal Y}^{K+2LM}_{\mu}(\Omega_X)
    \right]^\dag\frac{{\boldsymbol x}^X_i
    \cdot{\boldsymbol x}^X_j}{\rho^2}{\cal Y}^{KLM}_{\mu'}(\Omega_X)\,.
\end{equation}
\item[2)]The coefficients $\Gamma_{i,j}$ of Eq.~(\ref{eq:xlp2})
  must be re-evaluated by inverting the equation
\begin{equation}\label{eq:xlp2Q}
  (x_{lp})^2=\sum_{i,j}\Gamma_{i,j}^{p\rightarrow X}(l)\xx^X_i\cdot\xx^X_j\,.
\end{equation}
\end{itemize}
All the other steps remain exactly the same.

\section{From the scattering to the standard Jacobi}
This section is dedicated to the calculation of the coefficients
$c^{KL,Bp\rightarrow 1}_{\mu,\mu'}$ defined such that
\begin{equation}
  {\cal Y}^{KLM}_{\mu}(\Omega_{Bp})
  =\sum_{\mu'}c^{KL,Bp\rightarrow 1}_{\mu;\mu'}
  {\cal Y}^{KLM}_{\mu'}(\Omega_5)\,,\label{eq:aaB}
\end{equation}
where $\Omega_{Bp}$ represents a generic permutation of
the coordinates defined in Eq.~(\ref{eq:jacvecB}), with the hyperangular
variables defined in Eq.~(\ref{eq:phib}),
while $\Omega_5$ represent the coordinate in the standard definition 
and in the reference permutation.
By following Appendix~\ref{app:tc}, the coefficients are obtained by
\begin{equation}
  c^{KL,Bp\rightarrow 1}_{\mu;\mu'}=\int d\Omega_5
  \left[{\cal Y}^{KLM}_{\mu}(\Omega_5)\right]^\dag
       {\cal Y}^{KLM}_{\mu'}(\Omega_{Bp})\,,
\end{equation}
from which we derive the new recursion relations. From the formal
point of view, the result is identical to the one given in Appendix~\ref{app:tc}.
However, it is fundamental to redefine the normalization of the recursion
factors given in Eq.~(\ref{eq:norm_coeff}) by taking care of the new definition of the quantum
number $K_i$ and $\nu_i$ [see Eqs.~(\ref{eq:Kalpha}) and~(\ref{eq:nualpha})].
By using the new definitions we obtain
\begin{equation}
  {\cal N}_{n_1,\dots,n_N}^{\ell_1,\dots,\ell_N}=
  {\cal N}_{n_2,K_1}^{\ell_4,\nu_2}{\cal N}_{n_3,K_2}^{\ell_5,\nu_2}
  {\cal N}_{n_4,K_3}^{\ell_1,\nu_4}{\cal N}_{n_5,K_4}^{\ell_4,\nu_5}\,,
\end{equation}
where ${\cal N}_{n_i,K_{i-1}}^{\ell_j,\nu_i}$ are defined in Eq.~(\ref{eq:norm_def}).
All the rest remains exactly identical to what reported in Appendix~\ref{app:tc}.

%% file: matrix_element.tex
\chapter{Useful formulas for the matrix elements}
In this Appendix we give the explicit formula for the
calculation of the kinetic energy matrix elements (Section~\ref{app:kinene})
and
the calculation of the TC in the $jj$-coupling (Section~\ref{app:jjcoup}).

\section{Kinetic energy matrix element}\label{app:kinene}
   The kinetic energy matrix element given in Eq.~(\ref{eq:kinmxf2})
   contains the integral on the hyperradius, that for a generic number
   of particle $A$ reads
    \begin{equation}\label{eq:tllp}
      T_{l,l'}=-\frac{\hbar^2}{m}\int_0^{\infty}d\rho\,\rho^{3N-1}f_l(\rho)
      \biggl ( {\partial^2 \over \partial\rho^2}
      +{3N-1\over \rho} {\partial \over \partial\rho} +{K(K+3N-2)
        \over \rho^2}\biggr)f_{l'}(\rho)\ ,
    \end{equation}
    where $N=A-1$, $D=3N$ and $f_l(\rho)$ are defined in Eq.~(\ref{eq:fllag}).
    Using the properties of the Laguerre polynomials, Eq.~(\ref{eq:tllp}) can be
    cast in the form~\cite{AbramowitzStegun}
    \begin{equation}
    \begin{aligned}
      T_{l,l'}=&+\frac{\hbar^2\gamma^2}{m}\bigg[ (l'+K(K+3N-2))I_{l,l'}^{(2)}+
        \left(l'+\frac{3N-1}{2}\right)I_{l,l'}^{(1)}\\
        &-\sqrt{l'(l'+3N-1)}I_{l,l'-1}^{(2)}-\frac{1}{4}\delta_{l,l'}\bigg]\,,
    \end{aligned}
    \end{equation}
    where
    \begin{align}
      I^{(1)}_{l,l'}&=\frac{N_lN_{l'}}{\gamma^{D}}\int_{0}^{+\infty}dx\, x^{D-2}e^{-x}
      L_l^{(D-1)}(x)L_{l'}^{(D-1)}(x)\,,\label{eq:ill1}\\
      I^{(2)}_{l,l'}&=\frac{N_lN_{l'}}{\gamma^{D}}\int_{0}^{+\infty}dx\, x^{D-3}e^{-x}
      L_l^{(D-1)}(x)L_{l'}^{(D-1)}(x)\,,\label{eq:ill2}
    \end{align}
    and
    \begin{equation}
      N_l=\gamma^{D/2}\sqrt{\frac{l!}{(l+D-1)!}}\,.
    \end{equation}
    The integrals given in Eqs.~(\ref{eq:ill1})
    and~(\ref{eq:ill2}) can be obtained analytically.
    
    \section{$jj$-coupling scheme}\label{app:jjcoup}
    In this Section we present the connection between the so called $jj$-coupling
    scheme defined in Eq.~(\ref{eq:PHIjj})
    and the definition given in Eq.~(\ref{eq:hhst}).
    By rewriting a state $\alpha$ in the $jj$-coupling scheme
    we obtain
    \begin{equation}
      \Phi^{KLSTJ\pi}_\al(i,j,k,l,m,n)=\sum_{j_1j_2j_3j_{12}}
      C^{j_1j_2j_3j_{12},J}_{\ell_4\ell_5L_3L_4LS_2S_3S_5S}
      \Xi^{KTJ\pi}_{\nu}(i,j,k,l,m,n)\,,
    \end{equation}
    where the coefficient $C^{j_1j_2j_3j_{12},J}_{\ell_4\ell_5L_3L_4LS_2S_3S_5S}$
    explicitly reads
    \begin{equation}
    \begin{aligned}\label{eq:cjj}
      C^{j_1j_2j_3j_{12},J}_{\ell_4\ell_5L_3L_4LS_2S_3S_5S}&=
      \hat{L}_4\hat{L}\hat{S}_3\hat{S}\hat{j}_1\hat{j}_2\hat{j}_3\hat{j}_{12}\\
      &\times\sum_{\lambda}(2\lambda+1)
      \begin{Bmatrix}
        L_3 & \ell_4 & L_4 \\
        \ell_5 & L & \lambda
      \end{Bmatrix}
      \begin{Bmatrix}
        \ell_5 & \ell_4 & \lambda \\
        S_2 & 1/2 & S_3 \\
        j_1 & j_2 & j_{12}
      \end{Bmatrix}
      \begin{Bmatrix}
        \lambda & L_3 & L \\
        S_3 & S_5 & S \\
        j_{12} & j_3 & J
      \end{Bmatrix}\,,
    \end{aligned}
    \end{equation}
    and $\Xi^{KTJ\pi}_{\nu}(i,j,k,l,m,n)$ is defined in Eq.~(\ref{eq:PHIjj})
    for the reference permutation (1,2,3,4,5,6).
    In Eq.~(\ref{eq:cjj}) we have used the notation $\hat{l}=\sqrt{2l+1}$.
    Note that  these coefficients do not depend on
    the permutation.
    In such a way the state $\Psi_{\al}^{KLSTJ\pi}$ of Eq.~(\ref{eq:PSI3jj})
    results
    \begin{equation}
    \begin{aligned}
      \Psi_{\al}^{KLSTJ\pi} &= \sum_{\al'}A^{KLSTJ\pi}_{\al,\al'}
      \sum_{j_1,j_2,j_3,j_{12}}
      C^{j_1j_2j_3j_{12},J}_{\ell_4\ell_5L_3L_4LS_2S_3S_5S}
      \;\Xi^{KTJ\pi}_{\nu}(1,2,3,4,5,6)\\&=
      \sum_\nu \sum_{L_4S_3}A^{KLSTJ\pi}_{\al,\al'}
      C^{j_1j_2j_3j_{12},J}_{\ell_4\ell_5L_3L_4LS_2S_3S_5S}
      \;\Xi^{KTJ\pi}_{\nu}(1,2,3,4,5,6)\,,
    \end{aligned}
    \end{equation}
    where we transformed the sum over $\al'$ defined in Eq.~(\ref{eq:alpha})
    in the sum over $\nu$ defined in Eq.~(\ref{eq:nu}). Therefore
    \begin{equation}
      B^{KLSTJ\pi}_{\al,\nu}=\sum_{L_4S_3}A^{KLSTJ\pi}_{\al,\al'}
      C^{j_1j_2j_3j_{12},J}_{\ell_4\ell_5L_3L_4LS_2S_3S_5S}\,.
    \end{equation}

%% file: Lanczos.tex
\chapter{Lanczos algorithm}\label{app:lanczos}
    The eigenvalue problem presented in Eq.~(\ref{eq:gepb}) is solved
    using an iterative algorithm based on the Lanczos approach. In particular,
    we use the approach invented by Cullum and Willoughby~\cite{Cullum1981},
    which allows for a limited use of the computer memory.
    In this appendix we present
    the main points of this algorithm.

    Eq.~(\ref{eq:gepb}) can be rewritten as
    \begin{equation}\label{eq:evpij}
      (T+V)\,c=E\,N\,c\,,
    \end{equation}
    where $T$ is the kinetic energy matrix, $V$
    is the potential energy matrix, $N$
    is the norm matrix, and $E$ and $c$ are respectively the eigenvalue
    and the eigenstate. Moreover, we define $M$ as the dimension of the matrices.
    In order to use the Lanczos algorithm we need 
    to transform the generalized eigenvalue problem in a standard eigenvalue
    problem. To this aim, we use the Cholesky decomposition on the norm
    matrix $N$, namely
    \begin{equation}
      N=U\,U^\dag\,,
    \end{equation}
    so that we can rewrite Eq.~(\ref{eq:evpij}) as
    \begin{equation}\label{eq:evptilde}
      (\tilde{T}+\tilde{V})\,\tilde{c}=E\,\tilde{c}\,,
    \end{equation}
    where $\tilde{T}=U^{-1}T(U^\dag)^{-1}$, $\tilde{V}=U^{-1}V(U^\dag)^{-1}$
    and $\tilde{c}=U^\dag c$. The calculation of $\tilde{T}$ and $\tilde{V}$
    results to be  simplified, because the matrix
    $N$, and therefore the matrix $U^{-1}$, are block diagonal.

    The problem of finding the eigenvalue and the eigenvector of
    Eq.~(\ref{eq:evptilde}) can be solved iteratively. Let us select a
    possible value of the eigenvalue $E_1$. Eq.~(\ref{eq:evptilde})
    reduces to the form
    \begin{equation}\label{eq:evlanc0}
      (\tilde{T}-E_1)\,\tilde{c}=-\lambda(E_1)
      \, \tilde{V}\,\tilde{c}\,,
    \end{equation}
    where $\lambda(E_1)$ is equal to 1 only if  $E_1$ is the exact eigenvalue of
    Eq.~(\ref{eq:evptilde}).
    Rewriting Eq.~(\ref{eq:evlanc0}) as
    \begin{equation}
      (\tilde{T}-E_1)^{-1}(-\tilde{V})\,\tilde{c}
      =\frac{1}{\lambda(E_1)}\tilde{c}\,,
    \end{equation}
    the idea now is to select $E_1$ so that $\lambda(E_1)=1$.
    The problem is now reduced to find the eigenvalues of
    $(\tilde{T}-E_1)^{-1}(-\tilde{V})$.

    We can now use the Lanczos algorithm to compute the value $\lambda(E_1)$
    in the form presented in Ref.~\cite{Cullum1981}.
    Let us select a value of $E_1$ and try to solve the system
    \begin{equation}\label{eq:evlanc1}
      A(E_1)\tilde{c}=\frac{1}{\lambda(E_1)}\tilde{c}\,,
    \end{equation}
    where $A(E_1)=(\tilde{T}-E_1)^{-1}(-\tilde{V})$.
    In order to do that, we build a biorthonormal basis with dimension $\nu\ll M$.
    First of all we select two initial vectors $a_1$ and $b_1$ of dimension $M$ 
    such that $b_1^\dag a_1=1$. The rest of the not normalized vector of the
    basis (with a tilde) are defined by
    \begin{equation}\label{eq:til1}
      \begin{cases}
      \tilde{a}_{i+1}=A(E_1)a_i-\al_ia_i-\beta_ia_{i-1}\,,\\
      \tilde{b}_{i+1}=A(E_1)^\dag b_i-\al_ib_i-\gamma_ib_{i-1}\,,
      \end{cases}
    \end{equation}
    with
    \begin{equation}
      \alpha_i=b_i^\dag A(E_1)a_i\,, \qquad
      \gamma_i=\sqrt{|\tilde{b_i}^\dag\tilde{a_i}|}\,,\qquad
      \beta_i =\frac{\tilde{b_i}^\dag\tilde{a_i}}
           {\sqrt{|\tilde{b_i}^\dag\tilde{a_i}|}}\,,
    \end{equation}
    where the vectors $a_0=b_0=0$ by definition.
    The vectors $\tilde{a}_i$ and $\tilde{b}_i$ are then orthonormalized
    by imposing $b_{i+1}^\dag a_{i+1}=1$ and so we
    obtain
    \begin{equation}
      \begin{cases}
      a_{i+1}=\frac{\tilde{a}_{i+1}}{\sqrt{|\tilde{b}_{i+1}^\dag
          \tilde{a}_{i+1}|}}\\
      b_{i+1}=\frac{\tilde{b}_{i+1}\sqrt{|\tilde{b}_{i+1}^\dag
          \tilde{a}_{i+1}|}}{\tilde{b}_{i+1}^\dag\tilde{a}_{i+1}}
      \end{cases}\,.
    \end{equation}
    
    The eigenvector $\tilde{c}$ can be expanded in term of the basis
    vector $a_i$, namely
    \begin{equation}
      \tilde{c}=\sum_{i=1}^\nu \epsilon_ia_i\,,
    \end{equation}
    and so the eigenproblem of Eq.~(\ref{eq:evlanc1}) can be rewritten as
    \begin{equation}
      \sum_{i=1}^\nu\epsilon_iA(E_1)a_i=\frac{1}{\lambda(E_1)}
      \sum_{i=1}^\nu \epsilon_ia_i\,.
    \end{equation}
    If now we project this equation on $b_j^\dag$ and we use Eq.~(\ref{eq:til1})
    combined with the orthonormal properties of the basis,
    it results
    \begin{equation}
      \sum_{i=1}^\nu\epsilon_i\left(\delta_{j,i+1}\gamma_{i+1}+
      \alpha_i\delta_{j,i}+\beta_i\delta_{j,i-1}\right)=
      \frac{1}{\lambda(E_1)}\epsilon_j
    \end{equation}
    which is a new eigenvalue problem almost
    equivalent to Eq.~(\ref{eq:evlanc1}), where
    the eigenvector is $\epsilon$ and the eigenvalue is $1/\lambda(E_1)$.
    In a more familiar form, it can be written as
    \begin{equation}
      \begin{pmatrix}
        \al_1&\be_2&  0  &0&\dots &0\\
        \gm_2&\al_2&\be_3&0& \dots&0\\
        \vdots&\ddots&\ddots&\ddots&\vdots&\vdots\\
        0&\dots&0&\gm_{\nu-1}&\al_{\nu-1}&\be_\nu\\
        0&\dots&0&0&\gm_\nu & \al_\nu
      \end{pmatrix}
      \begin{pmatrix}
        \epsilon_1\\ \epsilon_2 \\ \vdots \\ \epsilon_{\nu-1}
        \\ \epsilon_\nu
      \end{pmatrix}
      =\frac{1}{\lambda(E_1)}
      \begin{pmatrix}
        \epsilon_1\\ \epsilon_2 \\ \vdots \\ \epsilon_{\nu-1}
        \\ \epsilon_\nu
      \end{pmatrix}\,.
    \end{equation}
    The eigenvalue obtained from this equation are approximately
    the eigenvalue of Eq.~(\ref{eq:evlanc1}). Increasing the
    dimension of the biorthonormal basis $\nu$, the value of the
    eigenvalue becomes closer and closer to the real value.
    Usually, a value of $\nu=20-30$ is enough to obtain a good accuracy.

    In order to obtain the eigenvalue $E$ of Eq.~(\ref{eq:evpij}) we
    need to extrapolate it. We select two energies $E_1$ and
    $E_2$ and we compute the corresponding $\lambda(E_1)$ and
    $\lambda(E_2)$ by using the algorithm explained above.
    At this point we evaluate a third values $E_3$
    using a linear extrapolation,
    \begin{equation}
      E_{i+1}=E_i+\frac{E_i-E_{i-1}}{\lambda(E_i)-\lambda(E_{i-1})}
      \left[1-\lambda(E_i)\right]\,,
    \end{equation}
    and using the Lanczos algorithm a new value $\lambda(E_3)$. The process
    is then iterated until $|\lambda(E_i)-1|<\eta$,
    where $\eta$ is usually $10^{-5}$.
    The number of iterations needed for reach such accuracy
    depends drastically on the choice of $E_1$ and
    $E_2$. The furthest they are from the real eigenvalue $E$, the
    largest will be the number of iterations needed to obtain the
    desired precision.

    On the other hand this algorithm results to be very time-saving, because
    there are no matrix operations except for
    the inversion of $(\tilde{T}-E_i)$,
    which in any case results very fast because this matrix is block diagonal.
    All the other operations in the algorithm are vector-vector or matrix-vector
    products which are very easy to parallelize.
    For example, the calculation of the eigenstate of  a matrix of
    size $M\sim70.000$ using the parallelized {\it LAPACK} libraries
    requires a computational time is $\sim4$ hours.
    The same system solved with the parallelized version of this
    algorithm requires $\sim30$ minutes. 

%% file: validation.tex
\chapter{Validation of the results}\label{app:validation}

In this appendix  we perform a benchmark with the results presented in
Ref.~\cite{Gattobigio2011} obtained with the NSHH approach.
Moreover,  we present the convergence of the results as function
of the maximum number of Laguerre polynomials ($l_{max}$) used to expand the hyperradial
part of the $\Li$ wave function. We also show the stability of the results
by changing the parameter $\gamma$ [see Eq.~(\ref{eq:fllag})].

In order to perform these tests, we use the Volkov potential
\begin{equation}
  V(r)=V_R{\rm e}^{-r^2/R_1^2}+V_A{\rm e}^{-r^2/R_2^2}\,,
\end{equation}
where $V_R=144.86$ MeV, $R_1=0.82$ fm, $V_A=-83.34$ MeV and $R_2=1.6$ fm. Moreover,
for the kinetic energy we consider $\hbar^2/m=41.47$ MeV fm$^2$.
Since, the Volkov potential is a central potential, it does not couple the
different partial wave components of the wave function. Therefore, we consider only
the $L=0$, $S=1$ and $T=0$ component. In such a way, we
are exactly in the same conditions of Ref.~\cite{Gattobigio2011}.
Since, in both the works we are using
the same expansions basis, we expect to obtain exactly the same results.

\begin{table}
  \centering
  \begin{tabular}{rccc}
    \hline
    \hline
    $K$ & $\ell_{sum}=0$ & $\ell_{sum}=2$ & Ref.~\cite{Gattobigio2011}\\
    \hline
    2  & $-61.142$ &  $-61.142$ & $-61.142$ \\  
    4  & $-62.015$ &  $-62.015$ & $-62.015$ \\ 
    6  & $-63.377$ &  $-63.377$ & $-63.377$ \\ 
    8  & $-64.415$ &  $-64.437$ & $-64.437$ \\ 
    10 & $-65.310$ &  $-65.354$ & $-65.354$ \\ 
    12 & $-65.823$ &  $-65.884$ & $-65.886$ \\ 
    \hline
    \hline
  \end{tabular}
  \caption{\label{tab:volkov}Binding energy of the bound state of $A=6$
    as function of the grandangular momentum $K$
    obtained with the Volkov potential.
    The first two columns are the results obtained in this work considering
    states up
    with $\ell_{sum}=0$ and $\ell_{sum}=2$ respectively. In the third column
    we report the results of Ref.~\cite{Gattobigio2011}.}
\end{table}

In Table~\ref{tab:volkov}
we report the binding energy of the bound state of $A=6$ as function
of the grandangular momentum $K$ and the maximum total angular momentum of
the states used in the basis expansion, i.e.
$\ell_{sum}=\ell_1+\ell_2+\ell_3+\ell_4+\ell_5$.
For this calculation we use $l_{max}=16$ and
$\gamma=4$ fm$^{-1}$.
As it can be seen from the table, if we use only the HH states with $\ell_{sum}=0$
we are not able to reproduce the results of Ref.~\cite{Gattobigio2011}.
The missing energy is due to
the fact we are not considering the states with $\ell_{sum}=2$
(states with $\ell_{sum}>2$ does not appear for $K\leq12$).
When we add them, our results coincide with those of Ref.~\cite{Gattobigio2011}
except for the case $K=12$, for which
a 2 keV difference remains.
Indeed, while in Ref.~\cite{Gattobigio2011} the entire HH
basis is used, in our calculation for $K=10$ and $12$ we need
to perform a precision truncation.
In our approach we need to eliminate the states with tiny
orthogonal component (see  Section~\ref{app:tc3} for more details),
since the numerical precision on this small components is not sufficient,
when we diagonalize the Hamiltonian. If we leave these states,
the numerical errors produced by the inversion
on the norm matrix [see Eq.~(\ref{eq:evpij})] generates
spurious bound states that we cannot control.
Therefore, we perform a truncation of the basis
in the case $K=10$ and $K=12$, which
avoids the generation of these spurious bound states. 
We want to remark that despite this truncation, only 2 keV are missing for $K=12$,
well below the precision of the convergence on $K$.
The same truncation is used also for the results presented in the main text.

\begin{table}[h]
  \centering
  \begin{tabular}{rcccc}
    \hline
    \hline
    $K$ & $\gamma=2$ fm$^{-1}$ & $\gamma=3$ fm$^{-1}$ & $\gamma=4$ fm$^{-1}$ &
    $\gamma=5$ fm$^{-1}$\\
    \hline
    2  & $-61.13930$ & $-61.14173$ & $-61.14179$ & $-61.14179$\\
    4  & $-62.01158$ & $-62.01498$ & $-62.01505$ & $-62.01505$\\
    6  & $-63.37271$ & $-63.37729$ & $-63.37738$ & $-63.37738$\\
    8  & $-64.43157$ & $-64.43714$ & $-64.43725$ & $-64.43725$\\
    10 & $-65.34743$ & $-65.35338$ & $-65.35356$ & $-65.35356$\\
    12 & $-65.87837$ & $-65.88431$ & $-65.88443$ & $-65.88444$\\
    \hline
    \hline
  \end{tabular}
  \caption{\label{tab:gamma}Binding energy of the bound state of $A=6$
    as function of the grandangular momentum $K$ obtained with the Volkov potential
    by varying the parameter $\gamma$. For all the
    cases we use $l_{max}=16$.}
\end{table}

In Table~\ref{tab:gamma}
we show the values of the binding energy as function of $K$ obtained,
using different values of the parameter $\gamma$, with $l_{max}=16$.
As it can be seen, the results are very stable changing $\gamma$,
and we obtain five digits of stability after the decimal point
for values of $\gamma=4-5$ fm$^{-1}$.
We performed a similar analysis also for the other potentials
used in this work, and the value $\gamma=4$ fm$^{-1}$ turned out
to be the natural choice in all cases.

\begin{table}[h]
  \centering
  \begin{tabular}{rcccc}
    \hline
    \hline
    $K$ & $l_{max}=4$ & $l_{max}=8$ & $l_{max}=12$ & $l_{max}=16$ \\
    \hline
    2  & $-60.9702$ & $-61.1403$ & $-61.1418$ & $-61.1418$ \\
    4  & $-61.8599$ & $-62.0129$ & $-62.0150$ & $-62.0150$ \\
    6  & $-63.2324$ & $-63.3745$ & $-63.3773$ & $-63.3774$ \\
    8  & $-64.2347$ & $-64.4336$ & $-64.4372$ & $-64.4373$ \\
    10 & $-65.1077$ & $-65.3495$ & $-65.3534$ & $-65.3536$ \\
    12 & $-65.6212$ & $-65.8804$ & $-65.8843$ & $-65.8844$ \\ 
    \hline
    \hline
  \end{tabular}
  \caption{\label{tab:lmax}Binding energy of the bound state of $A=6$
    as function of the grandangular momentum $K$
    obtained with the Volkov potential
    by varying the maximum number of Laguerre polynomials ($l_{max}$)
    used to expand the
    hyperradial part of the function. For all the
    cases we use $\gamma=4$ fm$^{-1}$.}
\end{table}

In Table~\ref{tab:lmax} we study the convergence of the binding energy
as function of $l_{max}$.
For these results we use $\gamma=4$ fm$^{-1}$.
As it can be seen by inspecting the table, we obtain the
convergence to 1 keV with $l_{max}=16$.
We studied also the convergence in the case of
the SRG and the NNLO$_{sat}$(NN) potentials, obtaining similar results.

%% file: em_calculation.tex
\chapter{Details of the calculation of $^6$Li electromagnetic static properties}
In this appendix we discuss the explicit method we have used to
calculate the mean value of the charge radius, the magnetic dipole
moment and the electric quadrupole moment of $\Li$. The definition
of the mean value of a generic operator ${\hat O}$ for $\Li$ is
\begin{equation}
  \bra  O\ket =\bra\Psi_{\Li}| {\hat O}|\Psi_{\Li}\ket\,,
\end{equation}
where ${\hat O}$ is ${\hat r_c}^2$ defined in Eq.~(\ref{eq:rcdef}),
${\hat S_z}$ and ${\hat L_z}$ defined in Eqs.~(\ref{eq:sz0}) and~(\ref{eq:lz0}),
${\hat Q}$ as defined in Eq.~(\ref{eq:qdef}). Finally, $\Psi_{\Li}$
is the $\Li$ wave function as defined in Eq.~(\ref{eq:wf}).
The computational approaches presented in this Appendix were
validated performing the same calculations with Monte Carlo integration
techniques.

\section{Charge radius}\label{app:charge_radius}
We need to compute the mean value of the proton point charge radius, which,
by using Eq.~(\ref{eq:r_prho}), translates into
\begin{equation}\label{eq:rp21}
  \bra r_p^2 \ket=\frac{1}{12}\bra\Psi_{\Li}|\rho^2|\Psi_{\Li}\ket_{\Omega,\rho}\,,
\end{equation}
where $\rho$ is the hyperradius and $\bra\cdots|\cdots|\cdots\ket_{\Omega,\rho}$
represents the integral over the hyperangular and the hyperradius variables
and the trace over the spin and isospin degrees of freedom.
By using the orthogonality properties of the HH states, Eq.~(\ref{eq:rp21}) reduces to
\begin{equation}
  \bra r_p^2\ket=
  \frac{1}{12}\sum_{l,l'}I_{l,l'}\sum_{KLST}\sum_{\alpha,\alpha'}
  c^{KLST}_{l,\alpha}N^{KLST1+}_{\al,\al'}c^{KLST}_{l',\alpha'}\,,
\end{equation}
where $N^{KLST1+}_{\al,\al'}$ is the norm of the HH states as
defined in Eq.~(\ref{eq:norm}), $c^{KLST}_{l,\alpha}$ are the variational
parameters as given in Eq.~(\ref{eq:wf}), and
\begin{equation}\label{eq:intllp}
  I_{l,l'}=\int_0^\infty d\rho \,\rho^{14} f_l(\rho)\rho^2f_{l'}(\rho)\,,
\end{equation}
where functions $f_l(\rho)$ are defined in Eq.~(\ref{eq:fllag}).

\section{Magnetic dipole moment}\label{app:mdm}

The calculation of the magnetic dipole moment can be easily performed through the
use of the auxiliary operator ${\hat S_z}$ and ${\hat L_z}$ defined in
Eqs.~(\ref{eq:sz0}) and~(\ref{eq:lz0}) respectively. In fact, the $\Li$ wave
function can be easily rewritten in term of its $L$ and $S$ components
$\Psi_{LM,SM_s}$, namely
\begin{equation}
  \Psi_{\Li}(J_z=+1)=\sum_{L,S}\sum_{M,M_s}( L M, S M_s | 1 1 )
  \Psi_{LM,SM_s}\,.
\end{equation}
By computing now the mean value of ${\hat S_z}$ and ${\hat L_z}$, we obtain
\begin{equation}
  \bra S_z\ket= \sum_{L,S}\sum_{M,M_s} M_s ( L M, S M_s | 1 1 )^2
  P_{LS}\,,
\end{equation}
and
\begin{equation}
  \bra L_z\ket= \sum_{L,S}\sum_{M,M_s} M ( L M, S M_s | 1 1 )^2
  P_{LS}\,,
\end{equation}
where $P_{LS}$ is the percentage of the ${}^{2S+1}L$ component in the 
$\Li$ wave function, which is defined as
\begin{equation}
  P_{LS}=\bra \Psi_{LM,SM_S}|\Psi_{LM,SM_S}\ket\,.
\end{equation}

\section{Electric quadrupole moment}\label{app:eqm}
The calculation of the electric quadrupole moment is performed by
using the auxiliary operator ${\hat Q}(6)$ defined in Eq.~(\ref{eq:qzz6}),
so that
\begin{equation}
  \bra Q \ket=\bra  \Psi_{\Li}| {\hat Q}(6)|\Psi_{\Li}\ket_{\Omega,\rho}\,.
\end{equation}
As already discussed in Section~\ref{sec:eqm}, the calculation of this operator is
much simplified by using the $Q$ hyperangular variables defined in Eq.~(\ref{eq:phiangq}). Therefore, we obtain 
which gives
\begin{equation}
  \bra Q \ket=2\sqrt{5\pi}
  \bra  \Psi^Q_{\Li}|\rho^2\cos \ph_5^Q\, Y_{20}(\hat{x}_5^Q)
  |\Psi_{\Li}^Q\ket_{\Omega^Q,\rho}\,,
\end{equation}
where $\Psi^Q_{\Li}$ is the $\Li$ wave function expressed in terms of the
new set of variables ``Q'' and the integration is performed over the hyperangular
variables $\Omega^Q$.
This change of variables represents an
orthonormal rotation in the coordinate space.
Therefore the independent states, with which we write the $\Li$ wave function
in the ``standard'' set of
coordinate, will remain independent without changing the expansion
of the wave function and the value of the variational parameters.
We  can rewrite the HH states of Eq.~(\ref{eq:hhst0}) as
\begin{equation}
  \Psi^{KLSTJ\pi}_\al(\Omega_5)=\sum_{\al'}A^{KLSTJ\pi,Q}_{\al,\al'}
  \Phi^{KLSTJ\pi}_\alpha(\Omega_5^Q)\,,
\end{equation}
where $A^{KLSTJ\pi,Q}_{\al,\al'}$ are new TC which permits to rewrite
the HH antisymmetrized states constructed using the ``standard'' set of hyperangular
variables in terms of the HH states build with the ``Q'' set. The calculation of these
is discussed in Appendix~\ref{app:transform}.
The functions $\Phi^{KLSTJ\pi}_\al(\Omega_5^Q)$ are the one
defined in Eq.~(\ref{eq:hhst}) written in term of the ``Q'' variables.
In such a way, Eq.~(\ref{eq:defquad}) can be rewritten as
\begin{equation}
\begin{aligned}
  \bra Q \ket&=2\sqrt{5\pi}\sum_{\xi,\xi'}
  c_{\xi}c_{\xi'}\sum_{\nu,\nu'}
  A^{KLSTJ\pi,Q}_{\al,\nu}A^{K'L'S'T'J\pi,Q}_{\al',\nu'}\\
  &\times
  \bra f_l(\rho)\Phi^{KLSTJ\pi}_\alpha|\rho^2\cos \ph_5^Q\, Y_{20}(\hat{x}_5^Q)
  |f_{l'}(\rho)\Phi^{K'L'S'T'J\pi}_{\alpha'}
  \ket_{\Omega^Q,\rho}\,,
\end{aligned}
\end{equation}
where $\xi\equiv\{KLST\al l\}$.
Thanks to the use of the ``Q''-variables this integral,
can be easily factorized in pieces, obtaining
\begin{equation}
\begin{aligned}
  \bra f_l(\rho)\Phi^{KLSTJ\pi}_\alpha|&\rho^2\cos \ph_5^Q\, Y_{20}(\hat{x}_5^Q)
  |f_{l'}(\rho)\Phi^{K'L'S'T'J\pi}_{\alpha'}
  \ket_{\Omega^Q,\rho}=\\
  &\bra f_l(\rho)|\rho^2|f_l'(\rho)\ket_\rho \bra \Phi^{KLSTJ\pi}_\alpha|
  \cos \ph_5^Q\, Y_{20}(\hat{x}_5^Q)|\Phi^{KLSTJ\pi}_\alpha\ket_{\Omega^Q}\,,
\end{aligned}
\end{equation}
where the integrals $\bra f_l(\rho)|\rho^2|f_l'(\rho)\ket_\rho=I_{l,l'}$ as
given in Eq.~(\ref{eq:intllp}), and
\begin{equation}
\begin{aligned}
\bra \Phi^{KLSTJ\pi}_\alpha|
&\cos \ph_5^Q Y_{20}(\hat{x}_5^Q)|\Phi^{K'L'S'T'J\pi}_\alpha\ket_{\Omega^Q}=\\
  &C^{LL'SJ}_{\ell_5\ell_5'L_4}\,\,{\cal I}(K_4,\ell_5n_5,\ell_5'n_5')
  \prod_{i=2}^4\delta_{\ell_i\ell_i'}\delta_{L_iL_i'}\delta_{n_in_i'}
  \prod_{i=2}^6\delta_{S_iS_i'}\delta_{T_iT_i'}\delta_{SS'}\delta_{TT'}\,,
\end{aligned}
\end{equation}
with
\begin{equation}
  C^{LL'SJ}_{\ell_5\ell_5'L_4}=\sqrt{\frac{3}{8\pi}}
  (-1)^{1+L_4+S}\hat{L}\hat{L}'\hat{\ell_5}\hat{\ell_5}'
 \begin{Bmatrix}
    2 & L & L' \\
    S & 1 & 1
  \end{Bmatrix}
  \begin{Bmatrix}
    2 & \ell_5 & \ell_5' \\
    L_4 & L' & L
  \end{Bmatrix}
  \begin{pmatrix}
    2 & \ell_5 & \ell_5' \\
    0 & 0   & 0
  \end{pmatrix}\,,
\end{equation}

and
\begin{equation}
\begin{aligned}\label{eq:intqdm}
  {\cal I}(K_4,\ell_5n_5,\ell_5'n_5')&=\frac{{\cal{N}}^{\nu_4,\ell_5}_{n_5}
    {\cal{N}}^{\nu_4,\ell_5'}_{n_5'}}{2^8\sqrt{2}}\\
  &\times\int_{-1}^1 dy\, (1+y)^{\frac{\ell_5+\ell_5'+3}{2}}
  (1-y)^{K_4+5}P_{n_5}^{\nu_4,\ell_5+1/2}(y)P_{n_5}^{\nu_4,\ell'_5+1/2}(y)\,.
\end{aligned}
\end{equation}
The integrals in Eq.~(\ref{eq:intllp})
and~(\ref{eq:intqdm}) can be calculated using the Gauss
integrations technique in an essentially exact way (only limited by the internal
numerical accuracy of the computer system). Therefore, the only approximation
of the computation of the quadrupole moment is related to the accuracy of the
wave function.

The use of the ``Q'' variables permits to reduce drastically the
number of coupling between the states. In fact,  all the quantum number
that describe a states except for $l,\ell_5,n_5,L$, are constrained by
Kronecker's $\delta$-functions. In such a way,
also the computational time required
is quite reduced. In order to test the change of variables, we computed
also the quadrupole moment by using the standard hyperspherical angle
variables for the first values of $K$,
obtaining exactly the same value calculated using the $Q$ variables.

%% file: app_cff.tex
  \chapter{Calculation of the $\alpha+d$ cluster form factor}
  \label{app:cff_cal}

  In this appendix we derive the explicit expression of the $\alpha+d$
  CFF defined in Eq.~(\ref{eq:overlap_df}) in term of the HH functions.
  We start from Eq.~(\ref{eq:overlap_df3}), which is written using
  the Jacobi coordinates of the set ``B''.
  This appendix is organized as follows. In the first section
  we rewrite the $\Li$, $\alpha$ and $d$ wave functions in terms of the
  Jacobi coordinates ``B''.
  In the second section we derive the
  explicit form of the CFF in term of the HH functions.

  \section{The wave function in term of the set ``B''}\label{app:overl1}

  To compute the integral of Eq.~(\ref{eq:overlap_df3}) we need to
  express the wave function of $\Li$ as given in Eq.~(\ref{eq:wf}) in term of the
  Jacobi coordinates of set ``B'' given in Eq.~(\ref{eq:jacvecB}).
  We define the hyperangles relative to the set ``B'' as
  \begin{equation}\label{eq:phib}
  \begin{aligned}
      \cos \ph_{2B}&=\frac{x_{4B}}{\sqrt{x_{3B}^2+x_{4B}^2}}\,,\\
      \cos \ph_{3B}&=\frac{x_{5B}}{\sqrt{x_{3B}^2+x_{4B}^2+x_{5B}^2}}\,,\\
      \cos \ph_{4B}&=\frac{x_{1B}}{\sqrt{x_{3B}^2+x_{4B}^2+x_{5B}^2+x_{1B}^2}}\,,\\
      \cos \ph_{5B}&=\frac{x_{2B}}{\sqrt{x_{3B}^2+x_{4B}^2+x_{5B}^2+x_{1B}^2+x_{2B}^2}}\,.
  \end{aligned}
  \end{equation}
  Note the definition of $\phi_{2B}$ and $\phi_{3B}$. In this way they are
  identical to the $A=4$ hyperangles used to construct the wave function
  of the $\alpha$ particle.
  The corresponding six-body HH functions are
  \begin{equation}
    \begin{aligned}\label{eq:hh6B}
    {\cal Y}^{KLM}_{\mu}(\Omega_B)&=
    \left(\left(\left(\left( Y_{\ell_1}(\hat x_{1B})
      Y_{\ell_2}(\hat x_{2B})\right)_{L_2} Y_{\ell_3}(\hat x_{3B}) \right)_{L_3}
      Y_{\ell_4}(\hat x_{4B}) \right)_{L_4} Y_{\ell_5}(\hat x_{5B}) \right)_{LM}\\
    &\times {\cal P}^{\ell_3,\ell_4,\ell_5,\ell_1,\ell_2}
         _{n_2,n_3,n_4,n_5}(\ph_{2B},\ph_{3B},\ph_{4B},\ph_{5B})\,,
    \end{aligned}
    \end{equation}
  where
  \begin{equation}
  \begin{aligned}\label{eq:hh6Sb}
    {\cal P}^{\ell_{3},\ell_{4},\ell_{5},\ell_{1},\ell_{2}}
    _{n_2,n_3,n_4,n_5}&(\ph_{2B},\ph_{3B},\ph_{4B},\ph_{5B})=\\
    &{\cal N}_{n_2}^{\ell_4,\nu_2}(\cos\ph_{2B})^{\ell_4}
    (\sin\ph_{2B})^{\ell_3}P_{n_2}^{\ell_3+1/2,\ell_4+1/2}(\cos 2\ph_{2B})\\
       &\times{\cal N}_{n_3}^{\ell_5,\nu_3}(\cos\ph_{3B})^{\ell_5}
    (\sin\ph_{3B})^{K_2}P_{n_3}^{\nu_2,\ell_5+1/2}(\cos 2\ph_{3B})\\
       &\times{\cal N}_{n_4}^{\ell_1,\nu_4}(\cos\ph_{4B})^{\ell_1}
    (\sin\ph_{4B})^{K_3}P_{n_4}^{\nu_3,\ell_1+1/2}(\cos 2\ph_{4B})\\
       &\times{\cal N}_{n_5}^{\ell_2,\nu_5}(\cos\ph_{5B})^{\ell_2}
       (\sin\ph_{5B})^{K_4}P_{n_5}^{\nu_4,\ell_2+1/2}(\cos 2\ph_{5B})\,,
  \end{aligned}
  \end{equation}
  while
  \begin{equation}\label{eq:Kalpha}
    \begin{aligned}
    K_1&=\ell_3\,,\\
    K_2&=\ell_3+\ell_4+2n_2\,,\\
    K_3&=\ell_3+\ell_4+\ell_5+2n_2+2n_3\,,\\
    K_4&=\ell_3+\ell_4+\ell_5+\ell_1+2n_2+2n_3+2n_4\,,\\
    K_5&=\ell_3+\ell_4+\ell_5+\ell_1+\ell_2+2n_2+2n_3+2n_4+2n_5\,,
    \end{aligned}
  \end{equation}
  and 
  \begin{equation}\label{eq:nualpha}
  \begin{aligned}
    \nu_1&=\ell_3+\frac{1}{2}\,,\\
    \nu_2&=\ell_3+\ell_4+2n_2+2\,,\\
    \nu_3&=\ell_3+\ell_4+\ell_5+2n_2+2n_3+\frac{7}{2}\,,\\
    \nu_4&=\ell_3+\ell_4+\ell_5+\ell_1+2n_2+2n_3+2n_4+5\,,\\
    \nu_5&=\ell_3+\ell_4+\ell_5+\ell_1+\ell_2+2n_2+2n_3+2n_4+2n_5+\frac{13}{2}\,.
  \end{aligned}
  \end{equation}
  This particular choice of the hyperangular coordinates and definition of the HH
  function
  is different compare
  to the standard  Zernike and Brinkman
  representation~\cite{Zernike1935,Fabre1983} but it is very advantageous when we
  will compute the overlap as it will be clear in the following.
  
  We need now to find a way to express the HH functions
  of Eq.~(\ref{eq:hh6}), written in term of the standard coordinates,
  in terms of those defined in
  Eq.~(\ref{eq:hh6B}), written with set ``B'' of coordinates.
  This transformation is performed by using new TC
  $A_{\beta,\beta'}^{KL_5S_6T_6,B}$
  (see Appendix~\ref{app:transform} for the details),
  with which we can rewrite the wave function of $\Li$ as
  \begin{equation}
  \begin{aligned}
      \Psi_{\Li}&
      =\sum_{KL_5S_6T_6\beta}\sum_l a_{\beta,l}^{KL_5S_6T_6}f_{l}(\rho)
      \sum_{\beta'}A_{\beta,\beta'}^{KL_5S_6T_6,B}
      {\cal P}^{\ell_3,\ell_4,\ell_5,\ell_1,\ell_2}_{n_2,n_3,n_4,n_5}
      (\ph_{2B},\ph_{3B},\ph_{4B},\ph_{5B})\\
      &\times\biggr\{\left(\left(\left( \left(Y_{\ell_1}(\hat x_{1B})
        Y_{\ell_2}(\hat x_{2B})\right)_{L_2} Y_{\ell_3}(\hat x_{3B})\ \right)_{L_3}
        Y_{\ell_4}(\hat x_{4B})\right)_{L_4}Y_{\ell_5}(\hat x_{5B})\right)_{L_5}\\
      &\times\left[\left((\sg_1\sg_2)_{S_2}\sg_3\right)_{S_3}
        \left((\sg_4\sg_5)_{S_4}\sg_6\right)_{S_5}\right]_{S_6}
      \bigg\}_{J_6J_{6z}}
      \left[\left((\tp_1\tp_2)_{T_2}\tp_3\right)_{T_3}
        \left((\tp_4\tp_5)_{T_4}\tp_6\right)_{T_5}\right]_{T_6T_6z}\,,
  \end{aligned}
  \end{equation}
  where $f_l(\rho)$ is defined in Eq.~(\ref{eq:fllag}) and $\beta$ is the complete
  set of quantum numbers for the six-body system as given in Eq.~(\ref{eq:alpha}).
  Here we use the notation $L_5$, $S_6$, $T_6$ and $J_6$ for $L$, $S$, $T$ and
  $J$ of Eq.~(\ref{eq:hhst}) in order to distinguish them from the $L$ and $S$ of
  the cluster wave function [see Eq.~(\ref{eq:clusterw})].
  
  The deuteron wave function, in the integral of Eq.~(\ref{eq:overlap_df3})
  is calculated  assuming that it is composed of
  particles $5$ and $6$. Expressing it in the set ``B'' of Jacoby coordinates,
  we have
    \begin{equation}\label{eq:psidp}
    \Psi_d(\xx_{1B})=\sum_{\ell_d=0,2}u_{\ell_d}(x_{1B})\left[Y_{\ell_d}
      (\hxx_{1B})(\sg_5\sg_6)_{S_d}\right]_{J_dJ_{dz}}(\tp_5\tp_6)_{T_dT_{dz}},
  \end{equation}
  where $J_d=1$, $S_d=1$ and $T_d=0$ and the $u_{\ell_d}(x_{1B})$ functions
  are determined solving the Schr\"odinger equation for the deuteron.
  The wave function of the $\alpha$ particle depends on the coordinates of
  particles $(1,2,3,4)$. Written using the Jacobi coordinates ``B''
  $\xx_{3B}$, $\xx_{4B}$ and $\xx_{5B}$, it reads
  \begin{equation}
  \begin{aligned}\label{eq:psialphap}
      \Psi_\alpha(\xx_{3B},&\xx_{4B},\xx_{5B})
      =\sum_{K_\al L_\al S_\al T_\al}\sum_{\beta_\al,l_\al}
      a^{K_\al L_\al S_\al T_\al}_{\beta_\al,l_\al}
      f_{l_\al}(\rho_4)\sum_{\beta'_\al}
      A^{K_\al L_\al S_\al T_\al,(4)}_{\beta_\al,\beta'_\al}\\
      &\times
      {\cal P}^{\ell_{1\al}',\ell_{2\al}',\ell_{3\al}'}_{n_{2\al}',n_{3\al}'}
    (\ph_{2B},\ph_{3B})
   \Bigg\{\left(\left( Y_{\ell_{1\al}'}(\hat x_{3B})
      Y_{\ell_{2\al}'}(\hat x_{4B})\right)_{L_{2\al}'}
      Y_{\ell_{3\al}'}(\hat x_{5B}) \right)_{L_{\al}'}\\
    &\times\left[\left((\sg_1\sg_2)_{S_{2\al}'}\sg_3\right)_{S_{3\al}'}
      \sg_4\right]_{S_{\al}'}
  \Bigg\}_{J_\al J_{\al z}}
  \left[\left((\tp_1\tp_2)_{T_{2\al}'}\tp_3\right)_{T_{3\al}'}
    \tp_4\right]_{T_\al T_{\al z}}\,,
  \end{aligned}
  \end{equation}
  where $J_\al=0$, $\rho_4=\sqrt{x_{3B}^2+x_{4B}^2+x_{5B}^2}$,
  $A^{K_\al L_\al S_\al T_\al,(4)}_{\beta_\al,\beta_\al'}$
    are the TC for the four-body system,
  \begin{equation}
    f_{l_\al}(\rho_4)=\gamma_\al^{\frac{9}{2}}\sqrt{\frac{l_\al!}{(l_\al+8)!}}\,
    L^{(8)}_{l_\al}(\gamma_\al\rho_4)e^{-\frac{\gamma_\al\rho_4}{2}}\,,
  \end{equation}
  \begin{equation}
    \beta_\al
    \equiv\{\ell_{1\al},\ell_{2\al},\ell_{3\al},L_{2\al},n_{2\al},n_{3\al},
    S_{2\al},S_{3\al},T_{2\al},T_{3\al}\}\,,
  \end{equation}
  and
    \begin{align}
      {\cal P}^{\ell_{1\al},\ell_{2\al},\ell_{3\al}}_{n_{2\al},n_{3\al}}
      (\ph_{2B},\ph_{3B})&=
    {\cal N}_{n_{2\al}}^{\ell_{2\al},\nu_{2\al}}(\cos\ph_{2B})^{\ell_{2\al}}
    (\sin\ph_{2B})^{\ell_{1\al}}P_{n_{2\al}}^{\ell_{1\al}+1/2,\ell_{2\al}+1/2}
    (\cos 2\ph_{2B})
       \nonumber\\
       &\times{\cal N}_{n_{3\al}}^{\ell_{3\al},\nu_{3\al}}(\cos\ph_{3B})^{\ell_{3\al}}
    (\sin\ph_3)^{K_{2\al}}P_{n_{3\al}}^{\nu_{2\al},\ell_{3\al}+1/2}(\cos 2\ph_{3B})\ .
  \end{align}
    We remark again that the coordinates $\xx_{3B}$, $\xx_{4B}$, $\xx_{5B}$
    and hyperangles $\ph_{2B}$, $\ph_{3B}$
    correspond exactly to the standard Jacobi coordinate and hyperangles
    we use to compute the $\alpha$ wave function.
    Therefore no transformation of the four-body HH states is needed in this case.
    
    \section{Explicit calculation of the overlap integral}\label{app:overlapint}
    By using the form of the wave functions discussed in the previous Section,
    it is now possible to compute the integral of Eq.~(\ref{eq:overlap_df3}).
    In this section we will drop the ``B'' index
    on the Jacobi and hyperangular
    variables, since we will consider only these  variables. Moreover, also the
    quantum numbers $\nu_i$ and $K_i$ which will appear in this Section are the
    ones defined in Eqs.~(\ref{eq:Kalpha}) and~(\ref{eq:nualpha}), respectively.
    
    The isospin term of the $\psi_{\alpha+d}$ wave function can be rewritten as
    \begin{equation}
    \begin{aligned}
      &\left[\left((\tp_1\tp_2)_{T_{2\al}}\tp_3\right)_{T_{3\al}}
        \tp_4\right]_{T_\al T_{\al z}}
      (\tp_5\tp_6)_{T_d T_{d z}}\\
      &\qquad\qquad=\sum_{t}C(T_{3\al},t)
      \left[\left((\tp_1\tp_2)_{T_{2\al}}\tp_3\right)_{T_{3\al}}
        \left((\tp_4\tp_5)_{t}\tp_6\right)_{1/2}\right]_{T_\al,T_{\al z}}\,,
    \end{aligned}
    \end{equation}
    where
    \begin{equation}
      C(T_{3\al},t)=(-1)^{2T_{3\al}+t}\frac{\hat t}{2}\,,
    \end{equation}
    with $\hat t=\sqrt{2t+1}$ and we use the fact that $T_d=0$.
    The trace on the isospin in Eq.~(\ref{eq:overlap_df3}) results in
    \begin{equation}
    \begin{aligned}\label{eq:isotrace}
      &\left[\left((\tp_1\tp_2)_{T_{2\al}}\tp_3\right)_{T_{3\al}}
        \tp_4\right]^\dag_{T_\al T_{\al z}}
      (\tp_5\tp_6)_{T_d T_{d z}}^\dag
      \left[\left((\tp_1\tp_2)_{T_2}\tp_3\right)_{T_3}
        \left((\tp_4\tp_5)_{T_4}\tp_6\right)_{T_5}\right]_{T_6T_{6z}}\\
      &\qquad\qquad\qquad
      =C(T_3,T_4)\delta_{T_{2\al} T_2}\delta_{T_{3\al} T_3}\delta_{T_5 1/2}
      \delta_{T_\al T_6}\delta_{T_{\al z} T_{6z}}\,.
    \end{aligned}
    \end{equation}

    For the spin part of the cluster wave function,
    we recouple the angular and spin terms such as
    \begin{align}
      &\Bigg\{\Big\{\left[\left(\left( Y_{\ell_{1\al}}(\hat x_3)
          Y_{\ell_{2\al}}(\hat x_4)\right)_{L_{2\al}} Y_{\ell_{3\al}}
          (\hat x_5) \right)_{L_\al}
        \left[\left((\sg_1\sg_2)_{S_{2\al}}\sg_3\right)_{S_{3\al}}
          \sg_4\right]_{S_\al}
        \right]_{0}\nonumber\\
      &\qquad\qquad\times\big[Y_{\ell_d}(\hxx_1)
      (\sg_5\sg_6)_{S_d}\big]_{1}\Big\}_1
      Y_L(\hxx_2)\Bigg\}_{JJ_z}\nonumber\\
      &\qquad\qquad=\sum_{\substack{L_2'L_3'L_4'L_5'\\S_4'S_5'S_6'}}
      A^{\ell_d L \ell_{1\al} \ell_{2\al} \ell_{3\al}
        L_{2\al} L_\al}_{ S_{3\al} S_\al J_\al J}
      (L_2'L_3'L_4'L_5'S_4'S_5'S_6')\nonumber\\
      &\qquad\qquad\times\Bigg\{\left[\left(\left( \left(Y_{\ell_d}(\hat x_1)
        Y_{L}(\hat x_2)\right)_{L_2'} Y_{\ell_{1\al}}(\hat x_3)\ \right)_{L_3'}
        Y_{\ell_{2\al}}(\hat x_4)\right)_{L_4'}Y_{\ell_{3\al}}
        (\hat x_5)\right]_{L_5'}\nonumber\\
      &\qquad\qquad\times\left[\left((\sg_1\sg_2)_{S_{2\al}}\sg_3\right)_{S_{3\al}}
        \left((\sg_4\sg_5)_{S_4'}\sg_6\right)_{S_5'}\right]_{S_6'}
      \Bigg\}_{JJ_{z}}\,,
    \end{align}
    where
    \begin{align}
      &A^{\ell_d L \ell_{1\al} \ell_{2\al} \ell_{3\al}
        L_{2\al} L_\al}_{ S_{3\al} S_\al J_\al J}
      (L_2'L_3'L_4'L_5'S_4'S_5'S_6')\nonumber\\
      &\qquad=
      (-)^{1+\ell_d+L+\ell_{1\al}+\ell_{2\al}+\ell_{3\al}+L_\al+L_{2\al}
        +L_4'+S_{3\al}+S_5'+S_6'}\nonumber\\
      &\qquad\times
      3\hat J\hat L_5' \hat L_4' \hat L_3' \hat L_2'\hat S_6'
      \hat S_5' \hat S_4'
      \hat L_\alpha \hat L_{2\alpha} \hat S_\alpha
      \begin{Bmatrix}
        L_\alpha & S_\alpha & 0 \\
        L_2' & 1 & J \\
        L_5' & S_6' & J
      \end{Bmatrix}
      \begin{Bmatrix}
        L & \ell_d & L_2' \\
        1 & J & 1
      \end{Bmatrix}\nonumber\\
      &\qquad\times\begin{Bmatrix}
        L_2' & L_{2\alpha} & L_4' \\
        \ell_{3\al} & L_5' & L_\alpha
      \end{Bmatrix}
      \begin{Bmatrix}
        L_2' & \ell_{1\al} & L_3' \\
        \ell_{2\al} & L_4' & L_{2\alpha}
      \end{Bmatrix}
      \begin{Bmatrix}
        S_{3\alpha} & 1/2 & S_\alpha \\
        1 & S_6' & S_5'
      \end{Bmatrix}
      \begin{Bmatrix}
        1/2 & 1/2 & S_4' \\
        1/2 & S_5' & 1
      \end{Bmatrix}
      \,.
    \end{align}
    It is now possible to compute the integrals over the angular variables
    $\hat x_{i}$ and the trace on the spins
    \begin{equation}
    \begin{aligned}\label{eq:spintrace}
      \int &\left(\prod_{i=1}^5d\hat x_i\right)
      \bigg\{\Big\{\left(\left(\left( Y_{\ell_{1\al}}(\hat x_3)
          Y_{\ell_{2\al}}(\hat x_4)\right)_{L_{2\al}} Y_{\ell_{3\al}}
          (\hat x_5) \right)_{L_\al}
        \left[\left((\sg_1\sg_2)_{S_{2\al}}\sg_3\right)_{S_{3\al}}
          \sg_4\right]_{S_\al}
        \right]_{0}\\
      &\times\left[Y_{\ell_d}(\hxx_1)(\sg_5\sg_6)_{S_d}\right]_{1}\Big\}_1
      Y_L(\hxx_2)\bigg\}_{JJ_z}^\dag\\
      &\times\biggr\{\left[\left(\left( \left(Y_{\ell_1}(\hat x_1)
        Y_{\ell_2}(\hat x_2)\right)_{L_2} Y_{\ell_3}(\hat x_3)\ \right)_{L_3}
        Y_{\ell_4}(\hat x_4)\right)_{L_4}Y_{\ell_5}(\hat x_5)\right]_{L_5}\\
      &\times\left[\left((\sg_1\sg_2)_{S_2}\sg_3\right)_{S_3}
        \left((\sg_4\sg_5)_{S_4}\sg_6\right)_{S_5}\right]_{S_6}
      \bigg\}_{J_6J_{6z}}=
      A^{\ell_1 \ell_2 \ell_3 \ell_4 \ell_5
        L_{2\al} L_\al}_{ S_{3\al} S_\al J_\al J}(L_2L_3L_4L_5S_4S_5S_6)\\
      &\times
      \delta_{\ell_1\ell_d}\delta_{\ell_2L}\delta_{\ell_3\ell_{1\al}}
      \delta_{\ell_4\ell_{2\al}}\delta_{\ell_5\ell_{2\al}}
      \delta_{S_2S_{2\al}}\delta_{S_3S_{3\al}}\delta_{J_6J}\delta_{J_{6z}J_z}\,.
    \end{aligned}
    \end{equation}

        We can now use these results to rewrite
        Eq.~(\ref{eq:overlap_df3}) as
        \begin{equation}
        \begin{aligned}
          \frac{f_L(r)}{r}&=\sqrt{15}
          \left(\frac{\sqrt{6}}{4}\right)^{\frac{3}{2}}
          \sum_{K_\al L_\al S_\al T_\al}\sum_{\beta_\al,l_\al}
          \sum_{KL_5S_6T_6}\sum_{\beta,l}
          a_{\beta_\al,l_\al}^{K_\al L_\al S_\al T_\al}
          a_{\beta,l}^{KL_5S_6T_6}
          \int \left(\prod_{i=1}^5dx_i\,x_i^2\right)\\
          &\times\delta\left(\sqrt{\frac{3}{8}}x_2-r\right)
          f_{l_\al}(\rho_4)f_{l}(\rho)
          \sum_{\ell_d}u_{\ell_d}(x_1)
          \sum_{\beta'_\al,\beta'}
          A^{K_\al L_\al S_\al T_\al,(4)}_{\beta_\al,\beta'_\al}
          A_{\beta,\beta'}^{KL_5S_6T_5,B}\\
          &\times
              {\cal P}^{\ell_{1\al}',\ell_{2\al}',\ell_{3\al}'}_{n_{2\al}',n_{3\al}'}(\ph_2,\ph_3)
              {\cal P}^{\ell_3',\ell_4',\ell_5',\ell_1',\ell_2'}_{n_2',n_3',n_4',n_5'}
              (\ph_2,\ph_3,\ph_4,\ph_5)\\
              &\times A^{L \ell_d \ell_{1\al}' \ell_{2\al}' \ell_{3\al}'
                L_{2\al}' L_\al}_{ S_{3\al}' S_\al J_\al J}(L_2'L_3'L_4'L_5S_4'S_5'S_6)
              C(T_3',T_4')\delta_{\tilde{\beta'_\al}\tilde{\beta'}}\delta_{\ell_1'\ell_d}
              \delta_{\ell_2'L}\,,
        \end{aligned}
        \end{equation}
        where
        \begin{equation}
          \tilde{\beta'}
          \equiv\{\ell'_3,\ell'_4,\ell'_5,S'_2,S'_3,T'_2,T'_3,T'_5,T'_6\}\,
        \end{equation}
        and
        \begin{equation}
          \tilde{\beta'_\alpha}
          \equiv\{\ell'_{1\alpha},\ell'_{2\alpha},\ell'_{3\alpha},
          S'_{2\alpha},S'_{3\alpha},T'_{2\alpha},T'_{3\alpha},1/2,T'_\alpha\}\,.
        \end{equation}

        We can now rewrite the integral variables as
        \begin{align}
          \int \left(\prod_{i=1}^5dx_i\,x_i^2\right)=
          \int_{0}^{\infty}dx_2\, x_2^2\int_0^\infty d\rho_5\, \rho_5^{11}
          \int_0^{\frac{\pi}{2}}
          \left(\prod_{i=2}^4d\ph_i \cos^2\ph_i\sin^{3i-4}\ph_i\right)\,.
        \end{align}
        In this way, thanks to the  definition of the $\ph_i$ angles
        in Eq.~(\ref{eq:phib}), the integrals on $d\ph_2$ and $d\ph_3$ are
        reduced by using the orthonormal properties of the Jacobi polynomials, i.e.
        \begin{equation}
        \begin{aligned}\label{eq:jpolort}
          \int_0^{\frac{\pi}{2}}&
          \left(\prod_{i=2,3}d\ph_i \cos^2\ph_i\sin^{3i-4}\ph_i \right)
              {\cal P}^{\ell_3,\ell_4,\ell_5}_{n_{2\al},n_{3\al}}(\ph_2,\ph_3)
              {\cal P}^{\ell_3,\ell_4,\ell_5,\ell_1,\ell_2}_{n_2,n_3,n_4,n_5}
              (\ph_2,\ph_3,\ph_4,\ph_5)\\
              &=\delta_{n_2n_{2\al}}\delta_{n_3n_{3\al}}
              {\cal N}_{n_4}^{\ell_1,\nu_4}(\cos\ph_4)^{\ell_1}
              (\sin\ph_4)^{K_3}P_{n_4}^{\nu_3,\ell_1+1/2}(\cos 2\ph_4)\\
              &\times{\cal N}_{n_5}^{\ell_2,\nu_5}(\cos\ph_5)^{\ell_2}
              (\sin\ph_5)^{K_4}P_{n_5}^{\nu_4,\ell_2+1/2}(\cos 2\ph_5)\,.
        \end{aligned}
        \end{equation}
        The variables $\rho_4$ and $x_1$ explicitly depend on $\ph_4$.
        In fact, by using Eq.~(\ref{eq:phib}) and the definition of $\rho_4$,
        we have $\cos \ph_4=x_1/\sqrt{\rho_4^2+x_1^2}$, and therefore
        $\rho_4=\rho_5\sin \ph_4$ and $x_1=\rho_5\cos \ph_4$.
        Therefore the integral on $\ph_4$ reads
        \begin{equation}
        \begin{aligned}
          I^{K_3,\ell_1,n_4}_{\ell_d,l_\al}(\rho_5)&={\cal N}_{n_4}^{\ell_1,\nu_4}
          \int_0^{\frac{\pi}{2}}d\ph_4\,(\cos\ph_4)^{2+\ell_1}
          (\sin\ph_4)^{8+K_3}\\
          &\times f_{l_\al}(\rho_5\sin\ph_4)\,u_{\ell_d}(\rho_5\cos\ph_4)
          P_{n_4}^{\nu_3,\ell_1+1/2}(\cos 2\ph_4)\,,
        \end{aligned}
        \end{equation}
        which can be evaluated using the Gauss-Tchebyshev quadrature.
        What remains to be calculated is the integration over $\rho_5$
        which explicitly reduce to
        \begin{equation}
        \begin{aligned}
          I^{K_3,\ell_1,n_4,\ell_2,n_5}_{\ell_d,L,l_\al,l}&(r)
          ={\cal N}_{n_5}^{\ell_2,\nu_5}
           \int_0^\infty d\rho_5\,\rho_5^{11}\,I^{K_3,\ell_1,n_4}_{\ell_d,l_\al}(\rho_5)
           f_l(\sqrt{\rho_5^2+8/3r^2})\\
           &\times\left(\frac{\sqrt{8/3}\,r}
           {\sqrt{\rho_5^2+8/3\,r^2}}\right)^{\ell_2}\left(\frac{\rho_5}
           {\sqrt{\rho_5^2+8/3\,r^2}}\right)^{K_4}
           P_{n_5}^{\nu_4,\ell_2+1/2}
           \left(\frac{8/3\,r^2-\rho_5^2}{\rho_5^2+8/3\,r^2}
           \right)\,.
        \end{aligned}
        \end{equation}
        Also in this case the integral can be computed using
        Gauss-Legendre quadrature. The final expression for
        the CFF is then obtained by eliminating the integral
        on $x_2$ through the $\delta$-function and it reads
        \begin{equation}
        \begin{aligned}
      \frac{f_L(r)}{r}&=
      \sqrt{15}\left(\frac{2\sqrt{2}}{\sqrt{3}}\right)^{\frac{3}{2}}r^2
      \sum_{K_\al L_\al S_\al T_\al}\sum_{\beta_\al,l_\al}
      \sum_{KL_5S_6T_6}\sum_{\beta,l}
      a_{\beta_\al,l_\al}^{K_\al L_\al S_\al T_\al}a_{\beta,l}^{KL_5S_6T_6}\\
      &\times\sum_{\beta'_\al,\beta',\ell_d}
      A^{K_\al L_\al S_\al T_\al,(4)}_{\beta_\al,\beta'_\al}
      A_{\beta,\beta'}^{KL_5S_6T_6,B}
      I^{K_3',\ell_1,n_4',\ell_2,n_5'}_{\ell_d,L,l_\al,l}(r)\\
      &\times A^{L \ell_d \ell_{1\al} \ell_{2\al} \ell_{3\al}
        L_{2\al}' L_\al}_{ S_{3\al} S_\al J_\al J}(L_2'L_3'L_4'L_5S_4'S_5'S_6)
      C(T_3',T_4')\\
      &\times\delta_{\tilde{\beta'_\al}\tilde{\beta}'}\delta_{\ell_1'\ell_d}
      \delta_{\ell_2'L}\delta_{n_{2\al}'n_2'}\delta_{n_{3\al}'n_3'}\,.
        \end{aligned}
        \end{equation}

        Some final comments are needed before concluding.
        The use of the variables ``B''
    increases drastically the precision of the overlap calculation.
    In fact, by passing from the ``standard'' set to the  ``B'' set,
    the dimension of the integrals to be computed diminishes
    from five to two, decreasing the numerical errors.
    The final accuracy in the calculation
    of these integrals depends on the number of Tchebyshev and
    Legendre points that we use. For the calculation of these integrals,
    we used up to 300 points obtaining a precision of the order
    of $\sim10^{-7}-10^{-8}$. We note that the integrals involve functions
    with exponential tail multiplied by Jacobi polynomials, having
    an oscillating behavior. The accuracy reached is in any case sufficient
    to have stable results for the ANCs.

%% file: app_cc.tex
    \chapter{Calculation of the coefficients $c^{\kb\,\lb\,\sbb\,\tb}_{\lbb,\bb}(\xi)$}
    \label{app:coeffcbar}
    
    The coefficients $c^{\kb\,\lb\,\sbb\,\tb}_{\lbb,\bb}(\xi)$ given in
    Eq.~(\ref{eq:cbar0}) can be evaluated by
    using the orthonormality properties of the HH basis, i.e.
    \begin{equation}\label{eq:cbarapp}
      c^{\kb\,\lb\,\sbb\,\tb}_{\lbb,\bb}(\xi)=\bra \Phi^{\kb\,\lb\,\sbb\,\tb}_{\lbb,\bb}(p=1)|
      \Psi^{LSJ}_{\alpha+d}(p=1)\ket_{\Omega_B,\rho}\,,
    \end{equation}
    where $\bra\cdots|\cdots\ket_{\Omega_B,\rho}$ represents the integration over all
    the hypercoordinates and the trace over spin and isospin
    [see Eq.~(\ref{eq:cbar})].

    As regarding the trace over the isospin, it is easy to show that we
    obtain exactly the result given in Eq.~(\ref{eq:isotrace}). Also for the
    integration over the variables $\hat x_{iB}$ and the spin trace the result
    is equal to the one given in Eq.~(\ref{eq:spintrace}). By using these
    results and the orthonormality properties of the Jacobi polynomials
    as discussed in Eq.~(\ref{eq:jpolort}), Eq.~(\ref{eq:cbar}) reduces to
    \begin{equation}\label{eq:pjint0}
      c^{\kb\,\lb\,\sbb\,\tb}_{\lbb,\bb}(\xi)
      =\int_0^\infty d\rho\, \rho^{14}f_\lbb(\rho)
      f^{\kb\,\lb\,\sbb\,\tb}_{\bb}(\rho,\xi)\,,
    \end{equation}
    where
    \begin{equation}
    \begin{aligned}
      f^{\kb\,\lb\,\sbb\,\tb}_{\bb}(\rho,\xi)&=\sum_{\ell_d}
      \sum_{\beta_\al,l_\al}a_{\beta_\al,l_\al}
      I^{\kb_3\bar \ell_1\bar n_4 \bar \ell_2 \bar n_5}_{L \ell_d l_\al}(\rho,\xi)\\
      &\times A^{L\ell_d \ell_{1\al} \ell_{2\al}\ell_{3\al}
        L_{2\al} L_\al}_{S_{3\al} S_\al J_\al J}(\bar L_2
      \bar L_3\bar L_4\bar L_5\bar S_4\bar S_5\bar S_6)
      C(\bar T_3,\bar T_4)\delta_{\bb_\al
        \tilde{\beta}_\al}\delta_{\bar \ell_1\ell_d}
      \delta_{\bar \ell_2L}\,.
    \end{aligned}
    \end{equation}
    Here we have defined
    \begin{equation}
      \tilde{\beta_\alpha}
      \equiv\{\ell_{1\alpha},\ell_{2\alpha},\ell_{3\alpha},
      S_{2\alpha},S_{3\alpha},T_{2\alpha},T_{3\alpha},1/2,T_\alpha\}\,,
    \end{equation}
    and
    \begin{equation}
      \overline{\beta}_\al
      \equiv\{\bar \ell_3,\bar \ell_4,\bar \ell_5,\bar S_2,
      \bar S_3,\bar T_2,\bar T_3,\bar T_5,\bar T\}\,.
    \end{equation}
    Moreover, $\bar \beta$ is the complete set of quantum numbers which describe
    a six-body quantum state as defined in Eq.~(\ref{eq:alpha}), and
    $a_{\beta_\al,l_\al}$ are the variational parameter of the $\alpha$-particle
    wave function computed by solving the four-body problem
    [see Eq.~(\ref{eq:psialpha})].
    The function
    $I^{\kb_3\bar \ell_1\bar n_4 \bar \ell_2 \bar n_5}_{L \ell_d l_\al}(\rho,\xi)$
    is given by
    \begin{equation}
    \begin{aligned}\label{eq:pjint1}      
      I^{\kb_3\bar \ell_1\bar n_4 \bar \ell_2 \bar n_5}_{L \ell_d l_\al}(\rho,\xi)&=
      2^{\frac{\bar \ell_2 +\bar K_4+15}{2}}
        {\cal N}_{\bar n_5}^{\bar \ell_2, \bar \nu_5}
        \int_{-1}^1 dy (1+y)^\frac{\bar \ell_2+1}{2}(1-y)^\frac{\bar K_4+10}{2}
        P_{\bar n_5}^{\bar \nu_4,\bar \ell_2+1/2}(y)\\
        &\times F\left(\sqrt{\frac{3}{8}}\rho\sqrt{\frac{1+y}{2}},\xi\right)
        I^{\bar K_3 \bar \ell_1 \bar n_4 }_{\ell_d l_\al}(y,\rho)\,,
    \end{aligned}
    \end{equation}
    where
    \begin{equation}
    \begin{aligned}\label{eq:pjint2}            
      I^{\bar K_3 \bar \ell_1 \bar n_4 }_{\ell_d l_\al}(y,\rho)&=
      2^{\frac{\bar \ell_1 +\bar K_3+12}{2}}
        {\cal N}_{\bar n_4}^{\bar \ell_1, \bar \nu_4}
        \int_{-1}^1 dx (1+x)^\frac{\bar \ell_1+1}{2}(1-x)^\frac{\bar K_3+7}{2}
        P_{\bar n_4}^{\bar \nu_3,\bar \ell_1+1/2}(x)\\
        &\times u_{\ell_d}\left(\rho\sqrt{\frac{1-y}{2}}\sqrt{\frac{1+x}{2}}\right)
        f_{l_\al}\left(\rho\sqrt{\frac{1-y}{2}}\sqrt{\frac{1-x}{2}}\right)\,,
    \end{aligned}
    \end{equation}
    and $F$ is a generic intercluster function which can depend on additional
    quantum number $\xi$.
    The integrals of Eqs.~(\ref{eq:pjint1})
    and~(\ref{eq:pjint2}) are computed using the Gauss-Tchebyshev quadrature.
    In order to obtain an accuracy of $\sim 10^{-7}-10^{-8}$ for values of
    $\rho$ between 0 and 10 fm,
    we need to use 300 points for the integration. We do not notice any
    significant improvement
    in the precision by using a larger number of integration points.
    Moreover, we notice that the accuracy decreases rapidly when $\rho>10$ fm.
    The integral in Eq.~(\ref{eq:pjint0}) is then computed by using an exact
    Lagrange-Gauss quadrature formula, with no lose of accuracy.

%% file: cs_formula.tex
\chapter{Derivation of the cross section and angular distribution expressions
for $A_1+A_2\rightarrow A_3+\gamma$}\label{app:sig_formula}
In this appendix we explicitly derive the formulas for 
the total cross section, given in
Eq.~(\ref{eq:sigtot}), and the angular distribution
of the emitted photon, given in Eq.~(\ref{eq:angulardistr}).

Let us consider the generic reaction $A_1+A_2\rightarrow A_3+\gamma$ where
$A_1$ has total angular momentum $J_1,M_1$ and so on, while with $\lambda$ we
indicate the polarization of the photon.
By using the Fermi Golden rule it is easy to derive the explicit expression
of the differential cross section by evaluating the phase space, namely
\begin{equation}\label{eq:i1}
  \frac{d\sigma_{J_3}}{d\Omega_q}
  =\frac{e^2}{8\pi v_{\rm rel}}\frac{q}{1+q/m_3}
  \frac{1}{(2J_1+1)(2J_2+1)}
  \sum_{M_1M_2}\sum_{\lambda M_3}|M_{M_1,M_2,M_3,\lambda}^{J_3}(\bmq)|^2\,,
\end{equation}
where $v_{\rm rel}$ is the relative velocity of the
two incoming particles, $q$ is the photon momentum and $m_3$ is the mass of
$A_3$ nucleus.
The matrix element $M^{J_3}_{M_1,M_2,M_3,\lambda}(\bmq)$ is given by~\cite{Walecka}
\begin{equation}
  M_{M_1,M_2,M_3,\lambda}^{J_3}(\bmq)= {}_{\hat{p}}
  \bra J_3 M_3|H_I(\lambda,\bmq)|J_1M_1,J_2M_2\ket_{\hat{p}}
\end{equation}
where $|J_1M_1,J_2M_2\ket_{\hat p}$ and  $|J_3M_3\ket_{\hat p}$
are the wave functions of
the scattering and the bound state as defined in
Eqs.~(\ref{eq:psi12}) and~(\ref{eq:bound}),  respectively (quantized along the
${\hat p}$ axis, where $\boldsymbol{p}$
is the momentum of the incident particles) and
$H_I$ is the interaction Hamiltonian given by
\begin{equation}
  H_I(\lambda,\bmq)
  =\sum_{\Lambda\geq 1} (-i)^\Lambda (2\Lambda+1)
  \left[E_{\Lambda,-\lambda}(\bmq)+\lambda M_{\Lambda,-\lambda}(\bmq)\right]\,.
\end{equation}
The above expression is valid in the frame where $\boldsymbol{q}$ is along z and
$E_{\Lambda,-\lambda}(M_{\Lambda,-\lambda})$ are the
electric (magnetic) operator multipole as defined in Ref.~\cite{Walecka}.
We have now to remember that the photon is emitted in an arbitrary direction, while
our states are quantized along the direction of the incoming momentum.
Because in the
experiment we are measuring the photon and consequently its direction,
we need to rotate our nuclear state of a $\theta$ angle in order to report
it on the photon direction.
Formally this is equivalent to rotate back the photon direction on the
$\hat{p}$-axis, which reduces the interaction
Hamiltonian in the form
\begin{equation}
  H_I(\lambda,\bmq,\theta)
  =\sum_{\Lambda\geq 1}\sum_M (-i)^\Lambda (2\Lambda+1)
  \left[E_{\Lambda,-\lambda}(\bmq)+\lambda M_{\Lambda,-\lambda}(\bmq)\right]
   {\cal D}^{\Lambda}_{M,-\lambda}(-\theta)\,,
\end{equation}
where ${\cal D}^{\Lambda}_{M,-\lambda}(-\theta)$ is the rotation matrix.
Hence, the matrix element $M^{J_3}_{M_1,M_2,M_3,\lambda}(\bmq)$ reduces to
\begin{equation}
\begin{aligned}\label{eq:M01}
  &|M_{M_1,M_2,M_3,\lambda}^{J_3}(\bmq)|^2
  =\frac{1}{{\hat J}_1{\hat J}_2{\hat J_3}^2}
  \sum_{LSJJ_z\Lambda M}\sum_{L'S'J'J'_z\Lambda' M'}
  {\hat L}{\hat L}'{\hat \Lambda}{\hat \Lambda}'(-i)^{\Lambda-\Lambda'+L'-L}\\
  &\qquad\times ( J_1 M_1, J_2 M_2|S J_z)( J_1 M_1, J_2 M_2|S' J'_z)
  ( S J_z, L 0|J J_z)( S' J'_z, L' 0|J' J'_z)\\
  &\qquad\times
  ( J J_z, \Lambda M|J_3 M_3)( J' J'_z, \Lambda' M'|J_3 M_3)
       {\cal D}^{\Lambda}_{M,-\lambda}(-\theta)
       {\cal D}^{\Lambda'*}_{M',-\lambda}(-\theta)\\
  &\qquad\times\left[E^{LSJ,J_3}_\Lambda(q)+\lambda M^{LSJ,J_3}_\Lambda(q)\right]
       \left[E^{L'S'J',J_3}_{\Lambda'}(q)+\lambda M^{L'S'J',J_3}_{\Lambda'}(q)
         \right]^*\,,
\end{aligned}
\end{equation}
where $\hat{j}=\sqrt{2j+1}$
we used the form of the wave functions given in Eqs.~(\ref{eq:psi12})
and~(\ref{eq:bound}) and  $E^{LSJ,J_3}_\Lambda (M^{LSJ,J_3}_\Lambda)$ is
the reduced electric (magnetic) matrix element, as defined in
Eq.~(\ref{eq:tlambda}).
We now use the following properties
\begin{equation}
  \sum_{M_1M_2}( J_1 M_1, J_2 M_2|S J_z)( J_1 M_1, J_2 M_2|S' J'_z)
  =\delta_{SS'}\delta_{J_zJ_z'}\,,\label{eq:m1m2}\\
\end{equation}
\begin{equation}
  \begin{aligned}
  {\cal D}^{\Lambda}_{M,-\lambda}(-\theta)
  {\cal D}^{\Lambda'*}_{M',-\lambda}(-\theta)&=
  (-)^{M'-L}\sum_k d^k_{M-M',0}(-\theta)\\
  &\times( \Lambda M, \Lambda' -M'|k M-M')
  ( \Lambda -\lambda, \Lambda' +\lambda | k 0)\,,
\end{aligned}
\end{equation}
and the fact that due to the $\delta$-function on $J_z,J'_z$
in Eq.~(\ref{eq:m1m2}), we get that
$M=M'$. Since $M=M'$ we have that
\begin{equation}
  d^k_{M-M',0}(-\theta)=P_k(\cos \theta)\,,
\end{equation}
where  $P_k(\cos\theta)$ are the Legendre polynomials.
By using these formulas the matrix element reduces to
\begin{align}\label{eq:mtok}
  \sum_{M_1M_2 M_3\lambda}&|M^{J_3}_{M_1,M_2,M_3,\lambda}(\bmq)|^2=
  \sum_k a^{J_3}_k(q) P_k(\cos \theta)\,,
\end{align}
where
\begin{equation}
\begin{aligned}\label{eq:akapp}
  a_k^{J_3}(q)&=
  \frac{1}{{\hat J}_1{\hat J}_2{\hat J_3}^2}
  \sum_{LL'SJJ'\Lambda\Lambda'}\sum_{\lambda M_3 M J_z}
  {\hat L}{\hat L}'{\hat \Lambda}{\hat \Lambda}'(-i)^{\Lambda-\Lambda'+L'-L}(-)^{M-\lambda}\\
  &\times 
  ( S J_z, L 0|J J_z)( S J_z, L' 0|J' J_z)
  ( J J_z, \Lambda M|J_3 M_3)( J' J_z, \Lambda' M|J_3 M_3)\\
    &\times ( \Lambda M, \Lambda' -M|k 0)
     ( \Lambda -\lambda, \Lambda' +\lambda | k 0)\\
     &\times\left[E^{LSJ,J_3}_\Lambda(q)+\lambda M^{LSJ,J_3}_\Lambda(q)\right]
       \left[E^{L'SJ',J_3}_{\Lambda'}(q)+\lambda M^{L'SJ',J_3}_{\Lambda'}(q)\right]^*\,.
\end{aligned}
\end{equation}
It is easy then to bring Eq.~(\ref{eq:akapp}) in the form of Eq.~(\ref{eq:ak})
by using the relation between the Clebsch-Gordon
and the 6j coefficients, i.e.
\begin{equation}
\begin{aligned}
  \sum_{m_1m_2m_6}&( j_1 m_1, j_2 m_2|j_3 m_3)
  ( j_6 m_6, j_2 m_2|j_4 m_4) ( j_1 m_1, j_5 m_5|j_6 m_6)\\
  &=(-1)^{j_2+j_3+j_5+j_6}( j_3 m_3, j_5 m_5|j_4 m_4)
    \begin{Bmatrix}
     j_1 & j_2 & j_3 \\
     j_4 & j_5 & j_6
   \end{Bmatrix}\,.
\end{aligned}
\end{equation}
Since the matrix element of Eq.~(\ref{eq:mtok})
does not depend on the angle $\phi$ we
can integrate Eq.~(\ref{eq:i1}) on $d\phi$ obtaining Eq.~(\ref{eq:sigma_theta}).  

The total cross section can then be computed by integrating
Eq.~(\ref{eq:sigma_theta}) in $d(\cos\theta)$, namely
\begin{equation}
  \sigma^{J_3}(E)=
  \int_{-1}^1d(\cos\theta)\, \sigma_0(E)\sum_k a_k^{J_3}(q)P_k(\cos\theta)
  =\sigma_0(E)a_0^{J_3}(q)\,,
\end{equation}
where $\sigma_0(E)$ is given in Eq.~(\ref{eq:sig0}) and only the coefficients
$a_0^{J_3}(q)$ survives, and is given by
\begin{equation}
  a_0^{J_3}(q)=2\sum_{LSJ}|E_\Lambda^{LSJ,J_3}(q)|^2+|M_\Lambda^{LSJ,J_3}(q)|^2\,.
\end{equation}
Therefore, we reconstruct exactly Eq.~(\ref{eq:sigtot}).